\documentclass[a4paper,fleqn,usenatbib]{mnras}

\usepackage{newtxtext,newtxmath}

\usepackage[T1]{fontenc}
\usepackage{ae,aecompl}

\usepackage{graphicx}	
\usepackage{amssymb}	
\usepackage{lscape}	
\usepackage[caption=false]{subfig}

\title[Flux density, spectral index, and polarization blazar microvariability]{Multiband optical flux density and polarization microvariability study of optically bright blazars}
 
\author[M. Pasierb et al.]
{Magdalena~Pasierb,$^{1}$\thanks{E-mail: M.Pasierb@oa.uj.edu.pl (MP)}
Arti~Goyal,$^{1}$\thanks{E-mail: arti@oa.uj.edu.pl (AG)}
Micha{\l}~Ostrowski,$^{1}$
{\L}ukasz~Stawarz,$^{1}$
\newauthor
Paul~J.~Wiita,$^{2}$
Gopal-Krishna,$^{3,4}$
Valeri~M.~Larionov,$^{5,6}$
Daria~A.~Morozova,$^{5}$
\newauthor
Ryosuke~Itoh,$^{7}$
Fahri~Alicavus,$^{8}$
Ahmet~Erdem,$^{8}$
Santosh~Joshi,$^{3}$
Staszek~Zola$^{1,9}$
\newauthor
Georgy~A.~Borman,$^{10}$
Tatiana~S.~Grishina,$^{5}$
Evgenia~N.~Kopatskaya,$^{5}$
\newauthor
Elena~G.~Larionova,$^{5}$
Sergey~S.~Savchenko,$^{5}$
Anna~A.~Nikiforova,$^{5,6}$
\newauthor
Yulia~V.~Troitskaya,$^{5}$
Ivan~S.~Troitsky,$^{5}$
Hiroshi~Akitaya,$^{11}$
Miho~Kawabata,$^{12,11}$
\newauthor
Tatsuya~Nakaoka$^{11}$
\\
$^{1}$Astronomical Observatory of the Jagiellonian University, ul. Orla 171, 30-244 Krak{\'o}w, Poland\\
$^{2}$Department of Physics, The College of New Jersey, 2000 Pennington Rd., Ewing, NJ 08628-0718, USA \\
$^{3}$Aryabhatta Research Institute of Observational Sciences (ARIES), Manora Peak, Nainital 263002, India\\
$^{4}$UM-DAE Centre for Excellence in Basic Sciences, Univ. of Mumbai, Mumbai-400098, India \\
$^{5}$Astronomical Institute of St.\ Petersburg State University, Petrodvorets 198504, Russia\\
$^{6}$196140, Pulkovskoye chaussee 65, Saint-Petersburg, Russia\\
$^{7}$Bisei Astronomical Observatory, 1723-70 Ookura, Bisei-cho, Ibara, Okayama 714-1411, Japan\\
$^{8}$Department of Physics, Faculty of Arts and Sciences, Canakkale Onsekiz Mart University, Canakkale~TR-17100, Turkey\\
$^{9}$Mt. Suhora Observatory, Pedagogical University, ul. \ Podchorazych 2,  Krak\'ow 30-084, Poland\\
$^{10}$Astrophysical Observatory, P/O Nauchny, Crimea, 298409, Russia\\
$^{11}$Hiroshima Astrophysical Science Center, Hiroshima University, Higashi-Hiroshima, Hiroshima 739-8526, Japan\\
$^{12}$Department of Astronomy, Kyoto University, Kitashirakawa-Oiwakecho, Sakyo-ku, Kyoto 606-8502, Japan\\
}

\date{Accepted XXX. Received YYY; in original form ZZZ}

\pubyear{2019}

\begin{document}
\label{firstpage}
\pagerange{\pageref{firstpage}--\pageref{lastpage}}
\maketitle

\begin{abstract}

We present the results of flux density, spectral index, and polarization intra-night monitoring studies of a sample of eight optically bright blazars, carried out by employing several small to moderate aperture (0.4\,m to 1.5\,m diameter) telescopes fitted with CCDs and polarimeters located in Europe, India, and Japan. The duty cycle of flux variability for the targets is found to be $\sim 45$ percent, similar to that reported in earlier studies. The computed two-point spectral indices are found to be between 0.65 to 1.87 for our sample, comprised of low- and intermediate frequency peaked blazars, with one exception; they are also found to be statistically variable for about half the instances where `confirmed' variability is detected in flux density. In the analysis of the spectral evolution of the targets on hourly timescale, a counter-clockwise loop (soft-lagging) is noted in the flux--spectral index plane on two occasions, and in one case a clear spectral flattening with the decreasing flux is observed. In our data set, we also observe a variety of flux--polarization degree variability patterns, including instances with a relatively straightforward anti-correlation, correlation, or counter-clockwise looping. These changes are typically reflected in the flux--polarization angle plane: the anti-correlation between the flux and polarization degree is accompanied by an anti-correlation between the polarization angle and flux, while the counter-clockwise flux--PD looping behaviour is accompanied by a clockwise looping in the flux--polarization angle representation. We discuss our findings in the framework of the internal shock scenario for blazar sources. 


\end{abstract}




\section{Introduction} 
\label{sec:intro}

Characterized by large flux density and polarization variability, blazars form a major class of active galactic nuclei (AGN), for which the total radiative output is dominated by magnetized, relativistic plasma outflows --- jets --- launched from the center of massive elliptical galaxies \citep[e.g.,][]{Urry95, Padovani17}. The blazar family includes flat-spectrum radio-loud quasars (FSRQs) and BL Lacertae objects (BL Lacs), the former of which have prominent emission lines in their optical spectra while the latter have very weak or undetectable lines. A blazar broadband spectral energy distribution (SED) is composed of two peaks: (1) a low energy segment ranging from radio to optical frequencies (sometimes extending up to X-rays in case of BL Lac objects) which is unequivocally attributed to synchrotron radiation of charged particles accelerated up to TeV energies; and (2) a high energy segment ranging from optical/X-rays up to GeV/TeV\,$\gamma-$ray frequencies which is attributed to inverse-Compton (IC) radiation of the seed photons produced locally (Synchrotron Self-Compton; SSC) or externally (External Compton; EC) to the jet plasma within the leptonic emission scenarios  \citep[e.g.,][and references therein]{Madejski16}. Alternatively, in `hadronic' scenarios for emission, the higher-frequency radiation peak is believed to originate from protons accelerated to $\simeq$\,PeV--EeV energies which could produce $\gamma$-rays via either direct synchrotron process, or meson decay and synchrotron emission by the secondaries produced in proton-photon interactions \citep[e.g.,][]{Bottcher13}. 
 
Depending on the timescales over which the flux density variations were seen, the blazar variability is conventionally divided into long-term (years to months), short-term (months to weeks) and intra-night/day variability (INV/IDV) or microvariability \citep[e.g.,][]{Wagner95, Ghosh10}. While the long-term variability easily can be reconciled within the standard paradigms of blazar emission models with modestly Doppler boosted relativistic jets ($\delta \sim $10--20; \citealt{Maraschi92, Abdo11a, Abdo11b}), sub-hour variability, especially at $\gamma-$ray energies, could be accounted for only with extremely large $\delta$'s \citep[$>$30-50;][]{Aharonian07, Begelman08} or with non-standard interpretations \citep[i.e., a synchrotron origin of the $\gamma-$ray flare in FSRQ 3C\,279;][]{Ackermann16}. Efficient energy dissipation is needed to ensure flux density variations on the smallest spatial scales and while there is no consensus on the main energy dissipation mechanism, the most favored candidates include plasma instabilities which lead to the formation of shocks and turbulence in the jet flow \citep[e.g.,][]{Spada01, Agudo11}, or alternatively, an annihilation of magnetic field lines of opposite polarity transferring energy from the field to the particles at the magnetic reconnection sites \citep{Sironi15}. Alternatively, many rapid flux density variations could be explained by geometrical effects involving small changes in the direction of motion of the jet plasma \citep{Gopal-Krishna92, Camenzind92, Meyer18}.

The evolution of colour or the spectral index, $\alpha$, (F($\nu$) $\propto$ $\nu$$^{-\alpha}$ where $\nu$ is the radiation frequency and F($\nu$) is the flux density provides an insight into the particle distribution giving rise to the observed flux density and its variability. In particular, at the synchrotron frequencies, within the simplest scenario of single-zone emission models with homogeneous magnetic field distributions, clear patterns between the spectral index and the total intensity are predicted, i.e., a ``spectral hysteresis'', depending on the relative lengths of the radiative cooling timescale and the escape timescale of the accelerated particles from the emission zone \citep[e.g.,][]{Kirk98}. A significant fraction of long-term multiband flux monitoring studies have revealed bluer--when--brighter trends for BL Lac objects but frequently redder--when--brighter trends for FSRQs \citep{Gu06, Osterman-Meyer09, Rani10, Hao10, Ikejiri11, Bonning12, Sandrinelli14, Meng18, Li18, Gupta19} while `achromatic' flux variability \citep[no colour evolution,][]{Gaur19, Bonning12, Stalin06}, and erratic patterns  \citep{Wierzcholska15} have also been reported. It has been argued that particles accelerated to higher energies are injected at the emission zone before being cooled radiatively in BL Lac sources leading to their overall SEDs being bluer-when-brighter;  however, the `redder' and more variable jet-component can overwhelm the `bluer' contribution from the accretion disc, leading to redder-when-brighter trends for FSRQ type sources \citep{Gu06}. Achromatic variability is often ascribed to changes in the Doppler boosting factor ($\delta$) as each frequency notes the same special relativistic multiplication of flux \citep{Gaur12}. However, erratic colour trends together with the opposite behaviours, i.e., redder--when--brighter changes for BL Lacs \citep{Gu11} and bluer--when--brighter trends for FSRQs \citep{Wu11}, indicate that more complex scenarios, presumably involving the dominance of the relative contributions of the Doppler boosted jet emission component and the accretion disc component, respectively, are particularly relevant for blazars with peak synchrotron frequencies in the range of 10$^{13-15}$\,Hz \citep[low-frequency peaked blazars;][]{Isler17, GK19}.

Polarization variability, i.e., changes in the polarization degree (PD) and/or the electric vector polarization angle ($\chi$) is yet another diagnostic to probe  emission scenarios. The PD is a measure of the structure of the magnetic field and the $\chi$ traces the direction of the projected magnetic field (being perpendicular to it) on the sky. For the simple case of a single emission zone, the maximum PD is $\approx$70\% for a power-law distribution of elections with energy index $\approx$2\ and uniform pitch angle distribution, immersed in a uniform magnetic field \citep{Rybicki86}. Blazars often show $\sim$1--30\% PDs at optical frequencies \citep{Mead90, Ikejiri11, Jermak16, Angelakis16}, indicating highly ordered magnetic fields at the emission sites in some cases. Moreover, the abrupt rotation of $\chi$ observed during outbursts has been taken as a signature of shocks in the jet \citep{Marscher85, Jorstad07,  Marscher08, Hughes11, Saito15}. Indeed, statistically significant correlations between swings of $\chi$  and $\gamma-$rays flares have been noted \citep{Blinov16}. However, the distributions of Q and U Stokes intensities in the (Q, U) plane often indicate a random-walk type of behaviour, suggesting that many emission regions with different magnetic field orientations contribute to the aggregate emission over longer monitoring periods \citep{Moore82, Villforth10, Gupta19}. Therefore, strictly simultaneous polarization monitoring, coupled with flux monitoring at a few frequencies, is an important step towards understanding the physical processes in blazar jets.

Only a handful of studies have probed the colour and polarization evolutions of blazars on microvariability timescales. In particular, \citet{Stalin06} showed that BL Lacertae exhibited `bluer--when--brighter' trends while S5\,0716+714 showed achromatic trends during intra-night monitoring sessions carried out in 1996 and 2000--2001. For BL Lac alone, bluer--when--brighter trends on intra-night timescales have often been observed during different monitoring campaigns \citep[1999-2001;][]{Papadakis03}, \citep[2012-2016;][]{Meng17} and \citep[2014-2016;][]{Gaur17}. For the blazar  S5\,0716+714, \citet{Dai13} noted bluer--when--brighter trends during the intra-night monitoring carried out in 2004--2011 while \citet{Zhang18} showed that the blazar exhibited both achromatic and bluer--when--brighter trends during the monitoring session carried out in 2013--2016. As for polarization variations, the blazar population in general shows significantly variable polarization on intra-night timescales if the PD is found to be more than 5\% \citep{Villforth09}. A highly polarized (PD$\sim$50\%) microflare was observed for the blazar S5\,0716$+$714 during the Whole Earth Blazar Telescope campaign carried out in 2014 \citep[][]{Bhatta15}. An orphan flare in polarized flux density was observed for the blazar CTA\,102 \citep[][]{Itoh13a}. Intense variability in total flux density, polarized flux density, and $\chi$ was noted for BL Lacertae while the blazar PKS\,1424+240 remained steady during the study by \citet{Covino15}. In this context, the present study aims to expand efforts to probe physical conditions on microvariability timescales through systematic multiband flux and polarization monitoring for a well-defined sample of blazars with well-known microvariability properties.

Here we present the results of our monitoring campaign to characterize the flux (B, V, R, and I band), colour (or spectral index), and polarization microvariability for a sample of eight bright blazars, each observed for a continuous monitoring duration of $\geq$3 hours using several telescopes fitted with charge-coupled devices (CCDs) and polarimeters. Our aim was to obtain strictly simultaneous observations in flux and polarization by coordinated monitoring between two observatories, one serving as a photometer for flux monitoring while the other was equipped with a polarimeter to obtain PD and $\chi$. However, our observations were severely limited by weather conditions and despite several attempts at coordinated monitoring, we could obtain strictly simultaneous data only on a few occasions. The data presented in this study were obtained in the years 2014 to 2017. In Sections~\ref{sec:sample} and ~\ref{sec:observations}, we describe the sample selection and data gathering, reduction and  generation of differential light curves (DLCs). Section~\ref{sec:analysis} describes the methodology for estimation of variability parameters. Results are given in Sections~\ref{sec:results} with Section~\ref{sec:conclusions} listing the conclusions of the study.

\begin{table*}
\small
\caption{Basic parameters of the blazars studied in the present work}
\label{sample}
\begin{center}
\begin{tabular}{cccccccc }
\hline \hline
  Source name		& 	R.A. (J2000)	& 	Dec. (J2000)		& 	 B  		& 	 V 		&   R 		&	 z  		&	Reference	\\
				&	(h m s) 	 &	($^\circ$ $^\prime$ $^{\prime\prime}$ ) & (mag) & (mag) & (mag)	&			&		\\
	(1)			&	(2)			&	(3)				&	(4)		&	(5)		&	(6)  		&         (7) 	&	(8)		\\
\hline
0109+224	      	&	01 12 05.82	&    +22 44 38.7 		&	16.00	 &	15.66 	&	15.47	& 	0.265	& \citet{Healey08}			\\  

3C 66A		      	&	02 22 39.61	&    +43 02 07.7		&	15.71	 &	15.21	&	14.5		&	0.444	&\citet{Dominguez11}			\\ 

S5 0716+714		&	07 21 53.44	&    	+71 20 36.3		&	15.5		 &	14.17	&	14.27	& 	0.310	& \citet{Nilsson08}	 		\\ 

OJ 287			&	08 54 48.87 	&    	+20 06 30.6		&	15.91 	 &	15.43 	&	15.56 	&	0.306	&\citet{Nilsson10}			 \\ 

3C 279		      	&	12 56 11.16	&    	$-$05 47 21.5 		&	18.01	 &	17.75	&	15.87	&	0.536 	&\citet{Marziani96}				\\ 

PG 1553+113		&	15 55 43.04 	&	+11 11 24.3		&	14.72	&	14.57  	&	13.99	&	0.360	& \citet{Richards11}	\\ 
CTA 102		      	&	22 32 36.40	&    +11 43 50.9		&	17.75	 &	17.33	&	--		& 	1.037 	& \citet{Schmidt65}				 \\  

3C 454.3		      	&	22 53 57.74	&   +16 08 53.5 		&	16.57	 &	16.10 	&	15.22	&	0.859 	& \citet{Lynds67}			 \\  
\hline
\end{tabular}
\end{center}

Columns: (1) Most common name; (2) right ascension; (3) declination; (4) apparent magnitude in B-filter; (5) apparent magnitude in V-filter; (6) apparent magnitude in R-filter; (7) spectroscopic redshift; (8) reference for the redshift.
\end{table*}

\section{Sample selection}
\label{sec:sample}
The blazars monitored in the study are chosen from the sample of \citet{Goyal13b} aside from the addition of CTA\,102 \citep{Itoh13a}, based on their established microvariability properties. We briefly recall here the selection criteria for our monitoring.  The source must: (i) be persistently bright ($<$18 V-mag), to be able to obtain measurements with few percent accuracies in a few minutes of integration time with 0.4--1.5\,m class telescopes; (ii) have declination greater than $+$20$^\circ$, to ensure the target's visibility for a continuous monitoring duration of more than 3 hours from the telescope sites used in this study; (iii) have good (non-variable) comparison stars of comparable magnitude available within a few arcminute radius of the target blazar to ensure calibration of the instrumental magnitudes to the standard Landolt photometric system. Table~\ref{sample} gives the basic parameters of the blazars observed in the present study.  

\begin{table*}
\small
\caption{Positions and magnitudes of the comparison stars used to generate the DLCs of the seven blazars for which the multiband flux density monitoring was carried out.}\label{compstar}
\begin{center}
\begin{tabular}{lccccccc }
\hline \hline
Source Name		& 	R.A. (J2000)	& 			Dec. (J2000)					& 	 B  	& 	 V 	&   R  	& 	I	& Ref. code	\\
Comp. Star no.		&	(h m s) 		&	($^\circ$ $^\prime$ $^{\prime\prime}$ ) 	& (mag)	& (mag)	& (mag)	& (mag) 	&		\\
		(1)		&	(2)			&				(3)						&	(4)	&	(5)	&	(6)	& (7)	& (8)	\\
\hline
0109+224	      	&&&&&&&\\   
S1 [C1]			&	01 12 00.3 	&  +22 45 22.3 		& 16.30(0.10)	& 15.28(0.07)	& 14.72(0.06) & 14.22(0.08) 	& a	\\  
S2 [D]			&	01 11 53.4 	&  +22 43 17.9		& 15.19(0.06)	& 14.45(0.05)	& 14.09(0.05 )&	-- 		& a	\\  
\\
3C 66A		      	&&&&&&&\\
S1  [23]		&	02 22 44.00	&    +43 05 29.1	& 14.119(0.005)  & 13.630(0.002)	& 13.333(0.007) &	13.048(0.003)	& b	\\  
S2  [21]		&	02 22 45.13	&    +43 04 19.6	& 15.786(0.006)  & 14.780(0.004)	& 14.233(0.006) &	13.717(0.005)	& b	\\  
\\
S5 0716+714		&&&&&&&\\  
S1 [11]		&	07 21 54.36	&    +71 19 20.92		& 14.152(0.001)  & 13.552(0.004)	& 13.189(0.010) & 12.855(0.002)	& b	\\ 
S2 [18]		&	07 22 12.63	&    +71 21 14.80		& 14.246(0.002)  & 13.641(0.003)	& 13.300(0.003) &	12.972(0.001) 	& b	\\  
S3 [23]		&	07 22 18.05	&    +71 23 34.53		& 13.684(0.001)  & 13.221(0.001)	& 12.941(0.001) &	12.656(0.001) 	& b	\\  
\\
OJ 287			&&&&&&&\\   
S1 [13]		&	08 54 54.1	&    +20 06 15			& 15.141(0.003)  & 14.627(0.003)	& 14.315(0.003)	& 13.999(0.004)	& b	\\  
S2 [9]			&	08 54 52.6	&    +20 04 46		& 15.051(0.003)  & 14.192(0.003)	& 13.707(0.002)	& 13.262(0.004)	& b	\\  
S3 [12]		&	08 54 54.8	&    +20 05 46			& 15.519(0.004)  & 14.974(0.003)	& 14.632(0.003)	& 14.304(0.004)	& b	\\ 
\\

PG 1553+113		&&&&&&&\\  
S1 [2]				&	15 55 46.07	&    +11 11 19.55 & 14.543 (0.050)	& 13.923 (0.022)	&	13.582 (0.029) & -- 		& c	\\  
S2 [1]				&	15 55 52.17	&    +11 13 18.52	& 14.503 (0.047)	& 13.832 (0.027)	&	13.465 (0.032)	& -- 		& c	\\  
\\
CTA 102		      	&&&&&&&\\  
S1 [2]			&	22 32 22.27	&    	+11 42 22.3	&	16.17(0.04)	 &	14.88(0.03)	&	14.07(0.07) 	& --		& c	\\  
S2 [1]			&	22 32 24.07	&    	+11 44 21.8	&	14.77(0.04)	 &	13.98(0.03)	&	13.56(0.04)	& --		& c	\\  
S3 			& 	22 32 27.70	&	+11 42 38.0	&	--			&	--			&		--		& --		& 	\\ 
\\
3C 454.3		      	&&&&&&&\\  
S1 [14]		&	22 53 44.63	&    +16 09 08.1	& 15.551 (0.079)	&  14.610 (0.045)	& 14.126 (0.011) &	13.623(0.012) 	& b	\\  
S2  [9]			&	22 54 04.87	&    +16 07 47.1	& 16.565(0.140) 	&  15.762(0.039)	& 15.194(0.017) &	14.752(0.031) 	& b	\\  
\hline
\end{tabular}
\end{center}
\textbf{Columns:} (1) Source name and the comparison stars used (in parentheses, we note the designation assigned to the star used in the reference given in column 8);  (2) right ascension; (3) declination; (4),(5),(6),(7) - apparent magnitudes in B, V, R and I--bands, respectively; 
(8) reference for the comparison star magnitudes: (a) \citet{Ciprini03}, (b) \citet{Gonzalez-Perez01}, (c) \citet{Raiteri98}.
\end{table*}

\section{Data acquisition, reduction and analysis}
\label{sec:observations}

\subsection{Multiband flux monitoring}
We employed four telescopes to carry out the intra-night flux monitoring in at least two frequencies on a given observing session. Below we give the details on the observatories and the instruments used, with the differential photometry for seven blazars discussed in this subsection and  the polarimetry for seven blazars (six of them also in the photometry sample) in the following one.

\begin{enumerate}

\item{{\bf 50\,cm Cassegrain (50 Cass), Poland:} A significant fraction of the intra-night observations presented in this study were obtained with the 50\,cm aperture diameter Cassegrain telescope of the Astronomical Observatory of the Jagiellonian University (AOJU), located in Krak\'ow, Poland. This telescope is of the Ritchey-Chr\'etien (RC) design with f$/$6.7 beam at the Cassegrain focus. The detector was a thermoelectric cooled with variable gain settings which are selected by the observer during the observations. We used readout noise set to 2.9\,e$^{-}$/pixel and gain to 4\,e$^{-}$$/$ADU during the observations. The CCD chip is 1024$\times$1024 pixels and a corresponding image scale of 0.70\,arcsec$/$pixel which covers a total of $12^{\prime} \times 12{^\prime}$ on the sky \citep{Zola12}.}

\item{{\bf 104\,cm Sampurnanand Telescope (ST) and 130\,cm Devesthal Fast Optical Telescope (DFOT), India:} We used 104\,cm Sampurnanand telescope (ST) located at Naini Tal and 130\,cm Devesthal Optical Telescope, located at Deveshtal site of the Kumaun region. Both facilities are run by the Aryabhatta Research Institute of observational sciencES (ARIES), India. The ST has an RC optics with a f$/$13 beam \citep{Sagar99}. The detector was a cryogenically cooled $2048 \times 2048$ chip mounted at the Cassegrain focus. This chip has a readout noise of 5.3\,e$^{-}$/pixel and a gain of 10\,e$^{-}$$/$Analog to Digital Unit (ADU) in slow readout mode. Each pixel has a dimension of 24\,$\mu$m$^{2}$ which corresponds to 0.37\,arcsec$^{2}$ on the sky, covering a total field of $13^{\prime} \times 13^{\prime}$. The 130\,cm DFOT has a modified RC design with an f$/$4 beam at the focus \citep{Sagar10}. The gain and read out noise are 1.39\,e$^{-}$$/$ADU and 6.14\,e$^{-}$/pixel. Each pixel in the CCD has a dimension of 13.5\,$\mu$m which corresponds to 0.29\,arcsec$/$pixel. The 2500$\times$2500 chip covers a total of $12^{\prime} \times 12^{\prime}$.}

\item{{\bf ASTELCO 60\,cm Cassegrain telescope (IST-60), Turkey:} Lastly, we used the 60\,cm IST-60 telescope of the Ulup{\i}nar Observatory, located at \c{C}annakle, Turkey. The IST-60 has  RC optics with a f$/$8 beam. The detector was a cryogenically cooled $1024 \times 1024$ chip mounted at the Cassegrain focus. This chip has a readout noise of 10\,e$^{-}$/pixel and a gain of 2\,e$^{-}$$/$ADU. Each pixel has a dimension of 24\,$\mu$m$^{2}$ which corresponds to 0.58\,arcsec$^{2}$ on the sky, covering a total field of $13^{\prime} \times 13^{\prime}$. }

\end{enumerate}

The observing strategy consisted of obtaining consecutive images of the blazar field in at least two  optical bands (B, V, R, and I), with the integration time roughly ranging from 2--5 minutes, depending on the brightness of the blazar and the transparency of the sky from the given telescope site. In such a manner, data were gathered for a continuous monitoring period of longer than 3 hours. This criterion was chosen to homogeneously characterize flux and colour variability on microvariability timescales; however, we note that about two-thirds of these observations were conducted for over 5 hours, with the longest around 6.5 hours. The calibration images (bias and flat frames) were obtained either just before or soon after the monitoring period of each night.

The details of the reduction procedure are given in \citep[][and references therein]{Goyal12} and here we briefly recall the main features. The data reduction of raw CCD images was conducted using the Image Reduction and Analysis Facility ({\sl IRAF})\footnote{\texttt{http://iraf.noao.edu/}} software package. The procedure begins with bias subtraction, followed by flat-fielding and cosmic-ray removal of the target images. The instrumental magnitudes of the target blazar and the stars (all point-like) in the image frames were determined by aperture photometry using {\sl APPHOT}. The magnitude of the target blazars was measured relative to a few apparently steady comparison stars present on the same CCD frame (termed as S1, S2, S3). Among the few DLCs of comparison stars, we chose the star-star pair which showed the least variance during the monitoring session. In this manner, the Differential Light Curves (DLCs) for the blazar (against the selected two comparison stars) and the comparison stars (between each other) were derived. The  corresponding target-star and star-star DLCs are denoted as `BL-S1', `BL-S2', `S1-S2', respectively. Basic information about the comparison stars is given in Table~\ref{compstar}.

It has been noted that blazar DLCs could indicate spurious INV detection if the blazar differs much in brightness from the comparison star(s) used to produce the DLCs\citep[e.g.,][]{Cellone07}. Therefore, in our analysis, we used comparison stars which were within one magnitude of the target blazar. Also, spurious variability on account of different second-order extinction coefficients for the blazar and the comparison stars can also be problematic if the target blazar and the comparison stars have very different optical colours. However, we note that even for colour differences of up to 1.5 mag, the differential extinction of photons travelling through varying air masses do not influence significantly the derived INV parameters given their typical flux measurement uncertainties \citep[$\simeq$ 0.005-0.03\ mag, including the observations analysed here;][]{Stalin04, Goyal13b}. For each night, an optimum aperture radius for the photometry was chosen by identifying the minimum dispersion in the star-star DLC, starting from the median seeing (i.e., full width at half maxima) value on that monitoring session to four times that value. Typically, the selected aperture radius was $\sim 4-8^{\prime\prime}$ and the effective seeing was $\sim 3^{\prime\prime}$. 

Since we performed intra-night monitoring in multiple optical frequencies, we also derived a `{\it colour} (apparent magnitude difference in two frequency bands)' DLC (CDLC) which is defined as the difference of blazar--star light curve in one frequency band to the second frequency band. We chose the two most distant frequency bands available during the monitoring session to produce CDLCs. We note that a CDLC also contains the contribution of the steady star colour (Table~\ref{compstar}) and hence the blazar colour could be obtained by subtracting the colour of the steady star concerned (Table~\ref{compstar}). Finally, we computed the two-point spectral index, $\alpha_{\nu_1}^{\nu_2}$, which is related to {\it colour($\nu_1$, $\nu_2)$} by 

\begin{equation}
\alpha_{\nu_1}^{\nu_2} = \frac{0.4 \times colour(\nu_1,\nu_2) - \mathrm{log}_{10}(\mathrm{F}({0,\nu_1})/\mathrm{F}({0,\nu_2}))}{\mathrm{log}_{10}(\nu_1/\nu_2)} 
\end{equation}
where $\nu_1$ and ${\nu_2}$ refer to central frequencies of the passbands while $\mathrm{F}({0,\nu_1})$ and $\mathrm{F}({0,\nu_2})$ refer to zero point magnitude fluxes of the frequency bands. For the  photometric system system used in our observations, these are 4063\,Jy, 3636\,Jy, 3064\,Jy, and 2635\,Jy for the B, V, R, and I-bands, respectively \citep{Glass99}. The statistical error on $\alpha_{\nu_1}^{\nu_2}$ was derived using the standard error propagation formula \citep{Bevington03}.  

\begin{figure*}
\centering
\hbox{
\includegraphics[width=0.33\textwidth]{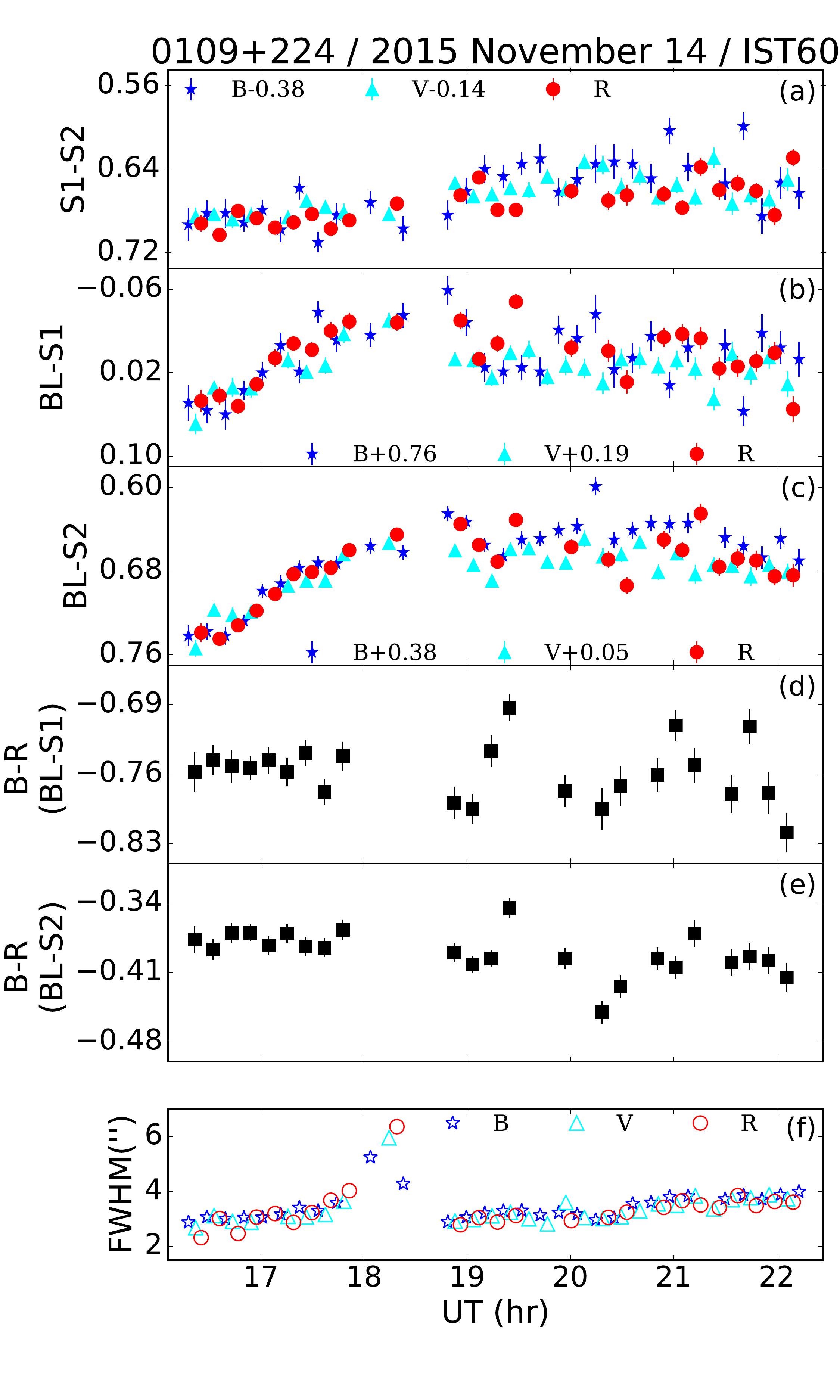}
\includegraphics[width=0.33\textwidth]{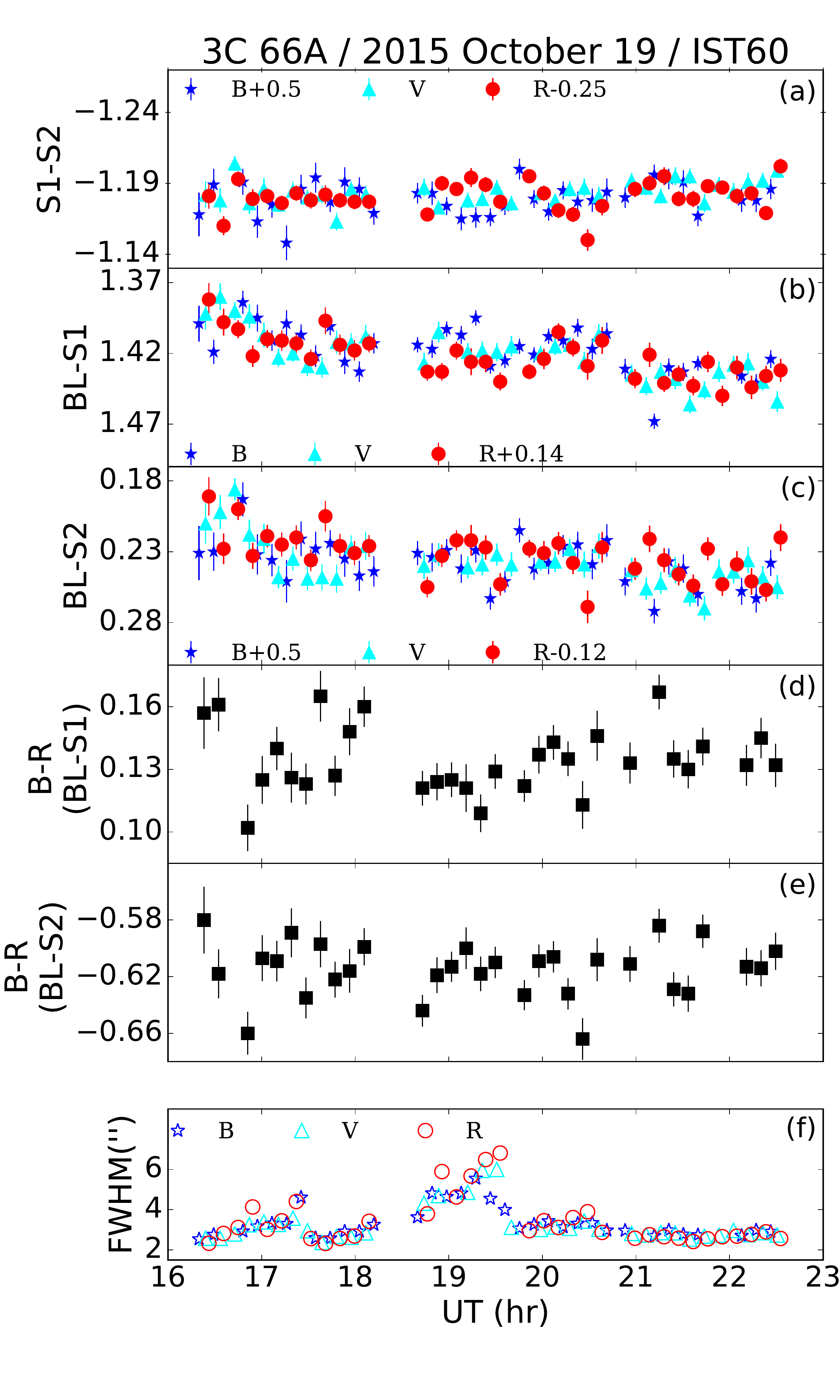}
\includegraphics[width=0.33\textwidth]{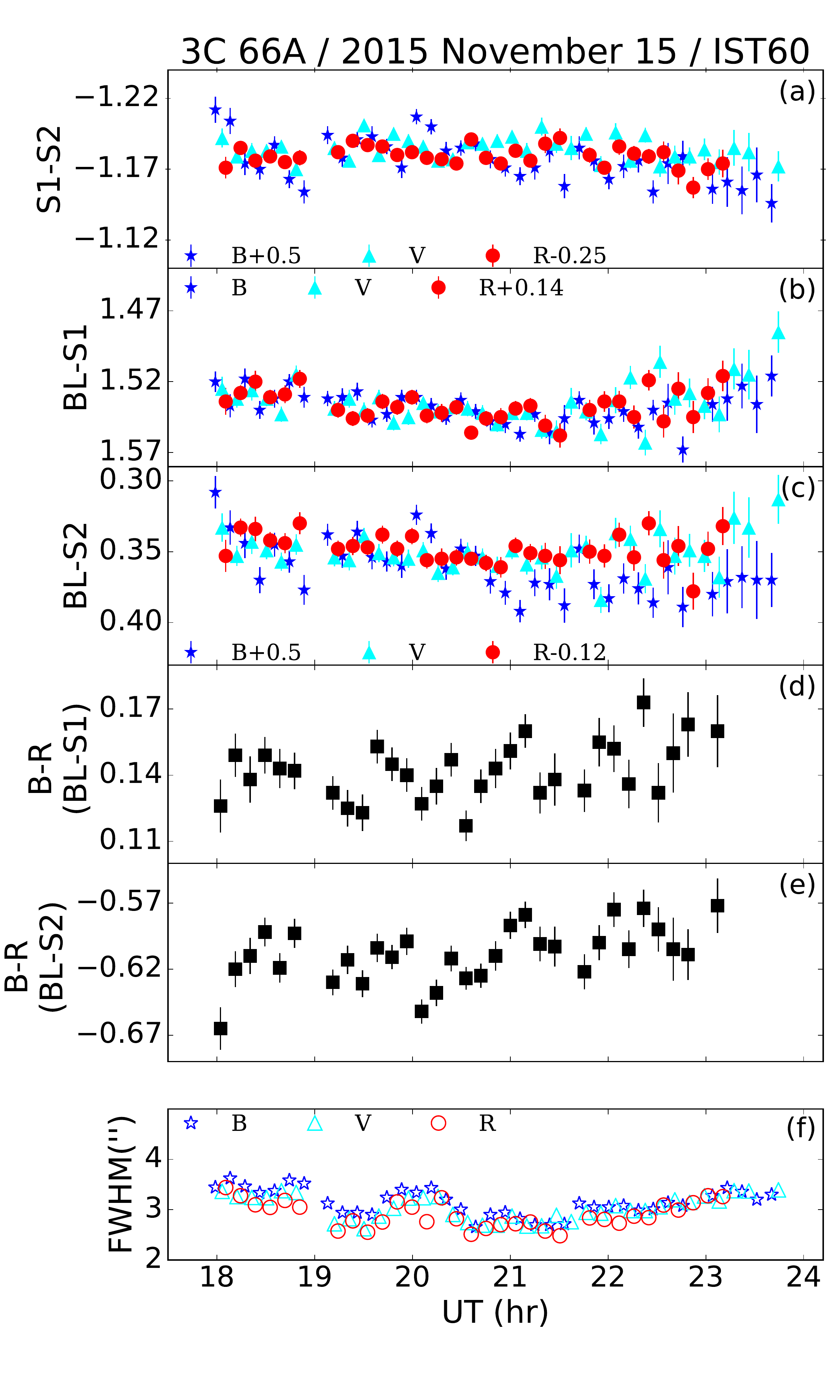}
}
\hbox{
\includegraphics[width=0.33\textwidth]{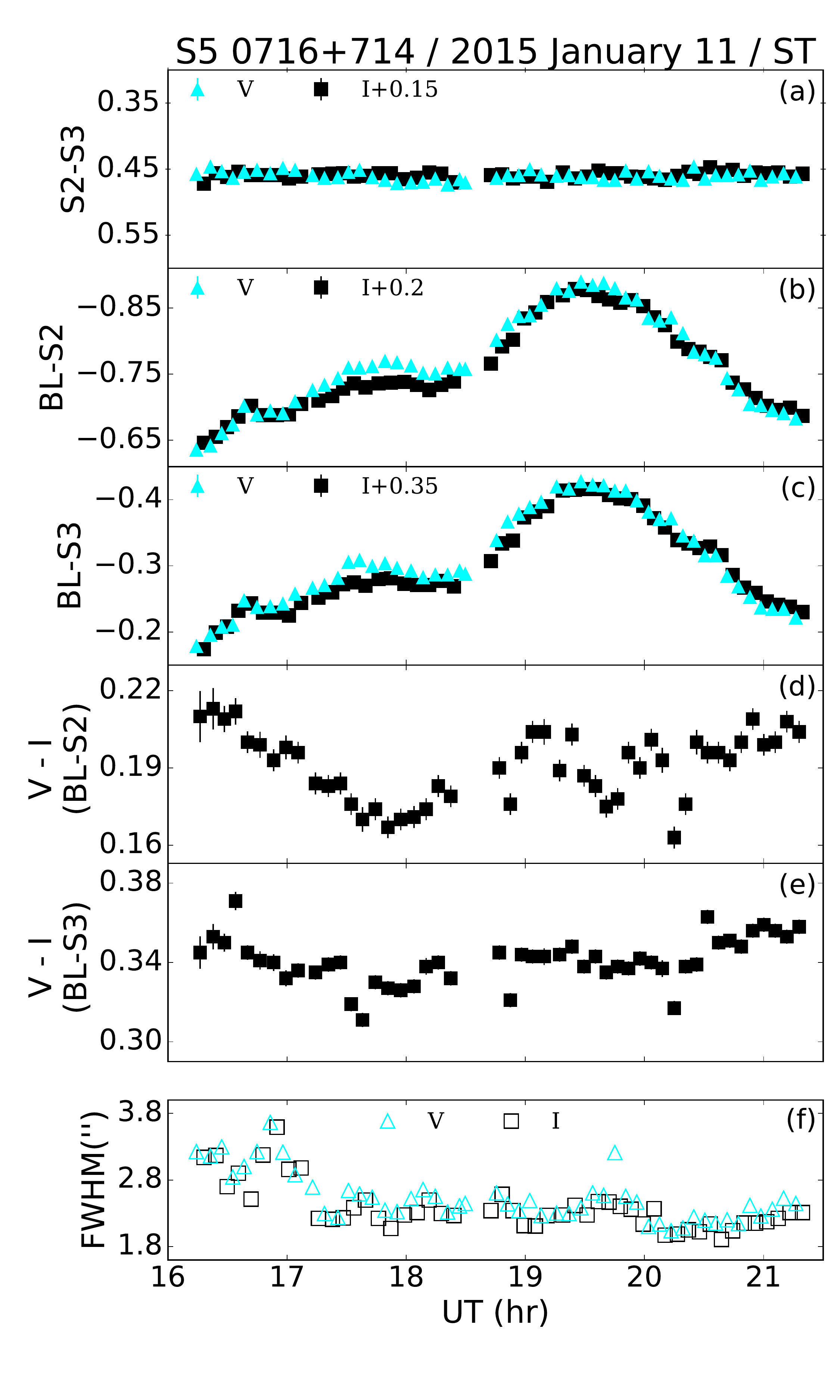}
\includegraphics[width=0.33\textwidth]{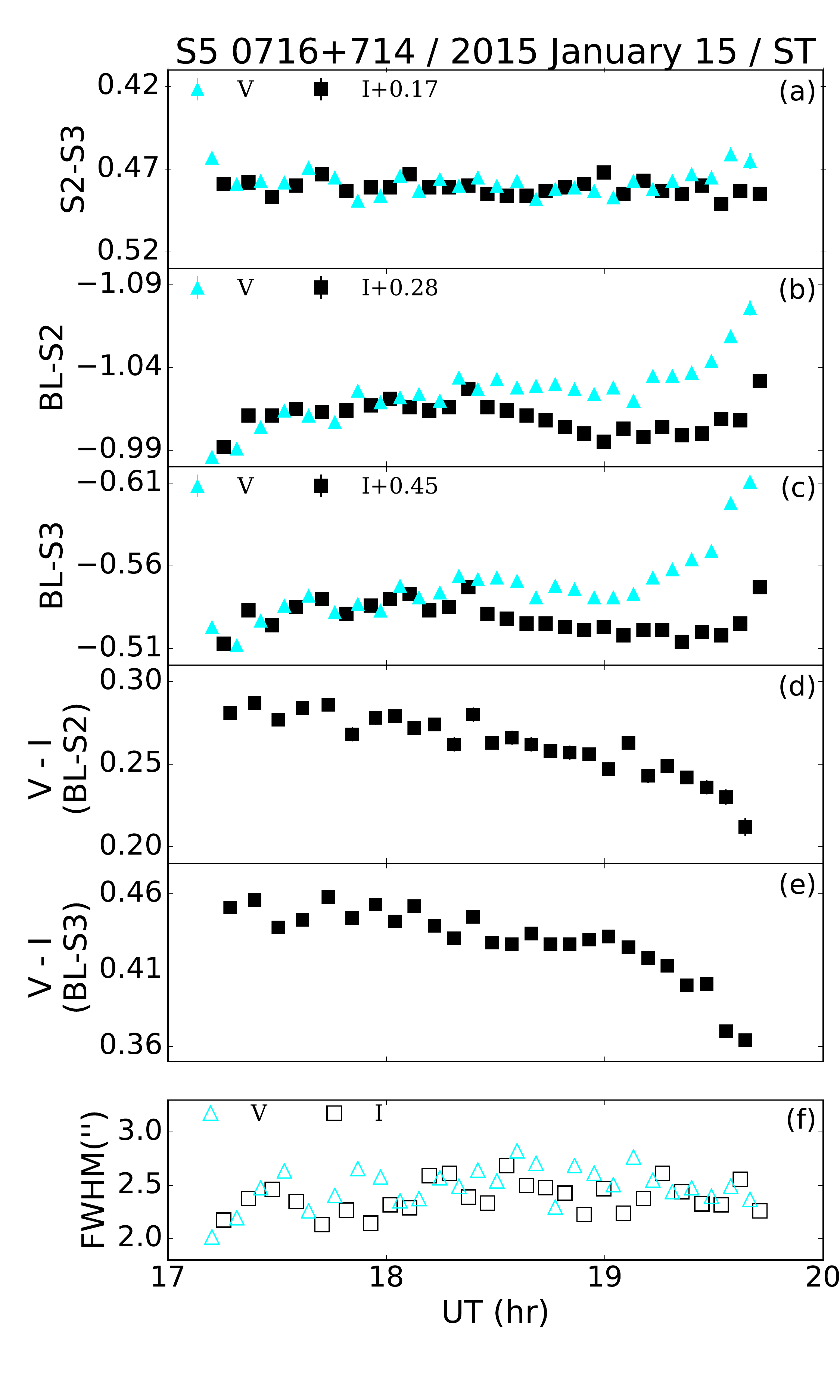}
\includegraphics[width=0.33\textwidth]{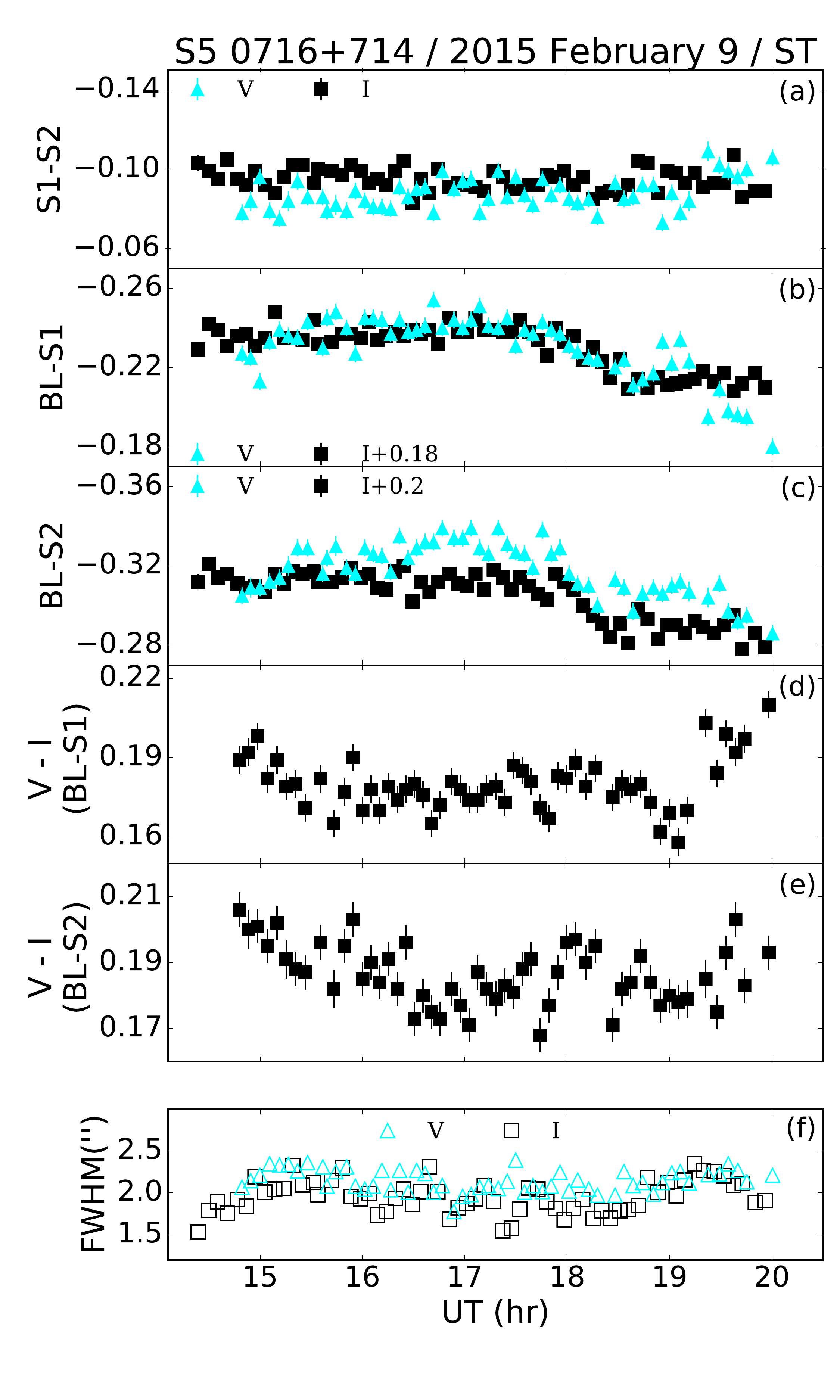}
}

\begin{minipage}{\textwidth}
\caption{Flux density and colour DLCs for the blazars in the present study: panel (a) gives the `steady' star--star DLC (all DLCs in magnitudes);  (b) and (c) give the blazar DLCs relative to steady star 1 and  steady star 2, respectively;  (d) and (e) give the CDLCs relative to steady star 1 and  steady star 2, for the two most  separated frequency bands available during the monitoring, respectively; (f) displays variation of the seeing disk during the monitoring period. For each figure, the source name, the date of observation, and the telescope used are given at the top. Relative magnitudes for the blazar and the comparison stars are shown by filled symbols in the online version: blue star (B--band), cyan triangle (V--band), red circle (R--band), and black square (I--band). In the printed version, open symbols show the relative magnitudes: cross (B--band), triangle (V--band), circle (R--band), and square (I--band). The heights of panels (a), (b) and (c) are the same for individual nights and the filter magnitudes are shifted by the amounts noted in each panel for better visibility.}
\label{fig_finv}%
\end{minipage}
\end{figure*}

Our entire intra-night data are summarized in Table~\ref{results_finv}, and the newly acquired DLCs are shown in Figure~\ref{fig_finv}.  Additionally, we also searched for the presence of colour microvariability in CDLCs using the F$-$test and the significance criterion adopted for flux microvariability detection (see Section~\ref{sec:microvar}). Table~\ref{results_cinv} provides the summary of colour microvariability for the CDLCs in our monitoring, along with the `mean' differential colour and two-point spectral index of the blazar on a given session.

\begin{table*}
\small
\caption {Summary of observational results}\label{results_finv}
\begin{center}
\begin{tabular}{ccccccccccc}
\hline \hline
  Source 	 & 	Date of obs.  	& 	Tel.		 	& Filter	& Duration  	& 	N$_p$	&$\sigma$ 	& $\psi$		& F$_{S1}$, F$_{S2}$, 	& 	 F$_{S1-S2}$ 	& 	Final 	\\
			 &				&				&  		&	(h)	    	&		  	&	(\%)		&	( \%)			&	(Status$^\ast)$		& 		(Status)		& 	status	 \\
	(1)		 &	(2)			&	(3)			&	(4)	&	(5)		&	(6)		&	(7)		&	(8)			&		(9)				&		(10)			&	(11)		 \\
\hline
0109+224	&	2015 Nov 14	&    	IST60		&	B	&	5.92		&	30		&	2.8	&	11.5		&	2.01, 7.74 (PV, V)		& 	 1.82 (N)			&		PV		\\	
			&				&				&	V	&	5.74		&	28		&	1.7	&	9.7		&	1.81, 5.21 (N, V)		& 	 1.83 (N)			&		PV		\\	
			&				&				&	R	&	5.74		&	25		&	1.9	&	10.1		&	3.79, 7.33 (V, V)		& 	 2.67 (V)			&		N		\\	
\\
3C 66A		&	2015 Oct 19$^\ast$    &	IST60		&	B	&	6.11		&	34		&	1.1	&	8.3		&	2.30, 0.91 (V,N)		& 	 0.72 (N)			&		PV		\\ 
			&				&				&	V	&	6.11		&	34		&	0.8	&	7.4		&	2.49, 1.46 (V,N)		& 	 0.74 (N)			&		PV		\\
			&				&				&	R	&	6.11		&	35		&	1.0	&	6.5		&	1.68, 1.45 (N,N)		& 	 1.38 (N)			&		N		\\
			&	2015 Nov 15   &	IST60		&	B	&	5.69		&	37		&	1.6	&	4.5		&	0.72, 1.06 (N,N)		& 	 1.25 (N)			&		N		\\ 
			&				&				&	V	&	5.69		&	36		&	0.8	&	7.3 		&	1.42, 0.74 (N,N)		& 	 0.67 (N)			&		N		\\
			&				&				&	R	&	5.08		&	32		&	0.7	&	3.6		&	0.88, 0.61 (N,N)		& 	 0.73 (N)			&		N		\\
\\
S5 0716+714	&	2015 Jan 11	&   	ST			&	V	&	5.03		&	50		&	0.7	&	25.4		&	147.38, 212.65 (V,V) 	& 	 1.20 (N)			&	 	V		\\	
			&				&				&	I	&	5.02		&	49		&	0.5	&	23.3		&	271.87, 326.90 (V,V) 	& 	 0.97 (N)			&	 	V		\\	
			&	2015 Jan 15$^\ast$ 	&	ST			&	V	&	2.35		&	27		&	0.7	&	9.0		&	10.42, 20.40 (V,V)		& 	 1.52 (N)			&		V		\\	
			&				&				&	I	&	2.46		&	27		&	0.4	&	4.0		&	6.88, 7.72 (V,V)		& 	 0.98 (N)			&		V		\\	
			&	2015 Feb 9	&   	ST 			&	V	&	5.18		&	55		&	0.5	&	7.3		&	5.91, 3.90 (V,V)		& 	 1.46 (N)			& 		V		\\ 	
			&				&				&	I	&	5.54		&	62		&	0.8	&	3.9		&	5.97, 6.82 (V,V)		& 	 1.37 (N)			&		V		\\	
\\
OJ 287		&	2014 Feb 20$^\ast$	&	DFOT			&	V	&	6.26		&	17		&	0.3	&	7.3		&	89.66, 84.64 (V,V)		& 	 1.54 (N)			&		V		\\
			&				&				&	I	&	6.48		&	17		&	0.2	&	6.6		&	89.90, 72.33 (V,V)		& 	 0.93 (N)			&		V		\\
			&	2015 Feb 12   	&    	ST			&	V	&	4.52		&	27		&	0.8	&	6.7		&	3.53, 3.61 (V,V)		& 	 1.17 (N)			&		V		\\  
			&				&				&	I	&	5.86		&	35		&	0.6	&	1.9		&	0.96, 1.43 (N,N)		& 	 1.05 (N)			&		N		\\
			&	2016 Jan 13 	&    	IST60		&	B	&	3.01		&	16		&	2.0	&	8.0		&	2.71, 1.94 (PV,N)		& 	 1.68 (N)			&		N		\\ 
			&				&				&	V	&	3.01		&	22		&	0.7	&	5.2		&	1.20, 1.59 (N,N)		&  	 0.39 (N)			&		N		\\
			&				&				&	R	&	2.90		&	22		&	1.1	&	4.6		&	0.90, 1.97 (N,N)		& 	 0.72 (N)			&		N		\\
			&	2016 Feb 6    	&    	50 Cass		&	V 	&	6.21		&	25		&	1.9	&	20.5		&	3.53, 3.01 (V,V)		& 	 1.25 (N)			&		V		\\  
			&				&				&	R	&	6.21		&	26		&	1.9	&	15.0		&	2.63, 3.02 (V,V)		& 	 1.20 (N)			&		V		\\
			&	2016 Mar 7     	&    	IST60		&	B	&	5.97		&	33		&	1.2	&	8.7		&	5.37, 7.33 (V,V)		& 	 0.85 (N)			&		V		\\ 
			&				&				&	V	&	5.80		&	27		&	0.7	&	6.6		&	5.77, 10.99 (V,V)		& 	 0.72 (N)			&		V		\\
			&				&				&	R	&	5.80		&	29		&	1.3	&	9.7		&	4.09, 10.47 (V,V)		& 	 1.06 (N)			&		V		\\
			&	2016 Apr 2 	&    	50 Cass		&	B	&	5.69		&	32		&	1.9	&	11.9		&	3.75, 6.03 (V,V)		& 	 1.34 (N)			&		V		\\  
			&				&				&	V	&	5.54		&	31		&	1.1	&	10.6		&	5.00, 7.21 (V,V)		& 	 0.86 (N)			& 		V		\\
			&				&				&	R	&	5.56		&	30		&	1.5	&	9.1		&	3.74, 6.66 (V,V)		& 	 1.40 (N)			&		V		\\
			&	2016 Apr 3$^\ast$     	&    	50 Cass		&	V	&	3.67		&	13		&	2.3	&	10.1		&	2.20, 0.99 (N,N)		& 	 1.40 (N)			&		N		\\ 
			&				&				&	R	&	3.37		&	12		&	1.7	&	6.2		&	1.46, 1.36 (N,N)		& 	 1.37 (N)			&		N		\\
			&	2016 Apr 4$^\ast$     	&    	IST60		&	V	&	5.29		&	30		&	0.9	&	7.5		&	3.00, 4.62 (V,V)		& 	 0.66 (N)			&		V		\\ 
			&				&				&	R	&	5.29		&	30		&	1.4	&	7.5		&	1.51, 5.09 (N,V)		& 	 0.83 (N)			&		PV		\\ 
\\	
PG 1553+113	&	2016 May 10	& 	50 Cass		&	B	&	3.73		&	31		&	0.9	&	3.9		&	0.82, 1.03 (N,N)		& 	 0.72 (N)			&		 N		\\
			&				&				&	V	&	3.59		&	34		&	1.0	&	2.4		&	0.58, 1.29 (N,N)		& 	 1.39 (N)			&		N		\\
			&				&				&	R	&	3.68		&	35		&	0.8	&	1.9		&	0.44, 0.70 (N,N)		& 	 0.76 (N)			&		N		\\
\\	
CTA 102		&	2015 Oct 13$^\ast$   	&	IST60		&	B	&	3.46		&	13		&	1.5	&	4.3		&	0.51, 0.58 (N,N)		& 	 0.93 (N)			&		N		\\ 
			&				&				&	V	&	3.46		&	15		&	0.6	&	8.9		&	0.97, 1.03 (N,N)		& 	 0.86 (N)			&		N		\\
			&				&				&	R	&	3.46		&	15		&	0.5	&	11.5		&	1.79, 2.03 (N,N)		& 	 0.99 (N)			&		N		\\
			&	2015 Oct 15$^\ast$  	&	IST60 		&	V	&	5.18		&	20		&	2.6	&	24.4		&	1.59, 1.25 (N,N)		& 	 0.99 (N)			&		N		\\
			&				&				&	R	&	5.19		&	24		&	1.9	&	19.0		&	1.18, 0.73 (N,N)		& 	 1.22 (N)			&		N		\\
\\
 3C 454.3	  &	2015 Oct 18			&	IST60			&	V	&	5.16		&	22		&	2.4	&	15.6		&	2.75, 2.30 (PV,PV)		& 	 1.10 (N)			&		PV		\\
			&				&				&	R	&	5.16		&	24		&	1.9	&	12.9		&	5.47, 2.96 (V,V)		& 	 1.11 (N)			&		V		\\
\hline
\end{tabular}
\end{center}
\label{tab2}
\textbf{Columns:} (1) Most common source name; (2) date of observation ($\ast$ indicates the simultaneous flux density and polarization monitoring between two observatories); (3) telescope used: ST = 104\,cm Sampurnanand Telescope (India), DFOT = 130\,cm Devesthal Fast Optical Telescope (India), 50 Cass = 50\,cm Cassegrain telescope (Poland), IST60 = 60\,cm Cassegrain telescope (Turkey); 
(4)  filter used; (5) duration of monitoring; (6) number of data points; (7) rms of the star-star DLC; (8) INV peak-to-peak amplitude ($\psi$); (9) F-values computed for the BL-S1 and BL-S2 DLCs (variability status for the corresponding DLC); (10) F-value for the (S1-S2) DLC (variability status for the DLC); (11) final variability status for the blazar (V = Variable; N = Non-variable; PV = Probable Variable).
\end{table*}

\renewcommand{\thefigure}{\arabic{figure} (Cont.)}
\addtocounter{figure}{-1}

\begin{figure*}
\centering
\hbox{
\includegraphics[width=0.33\textwidth]{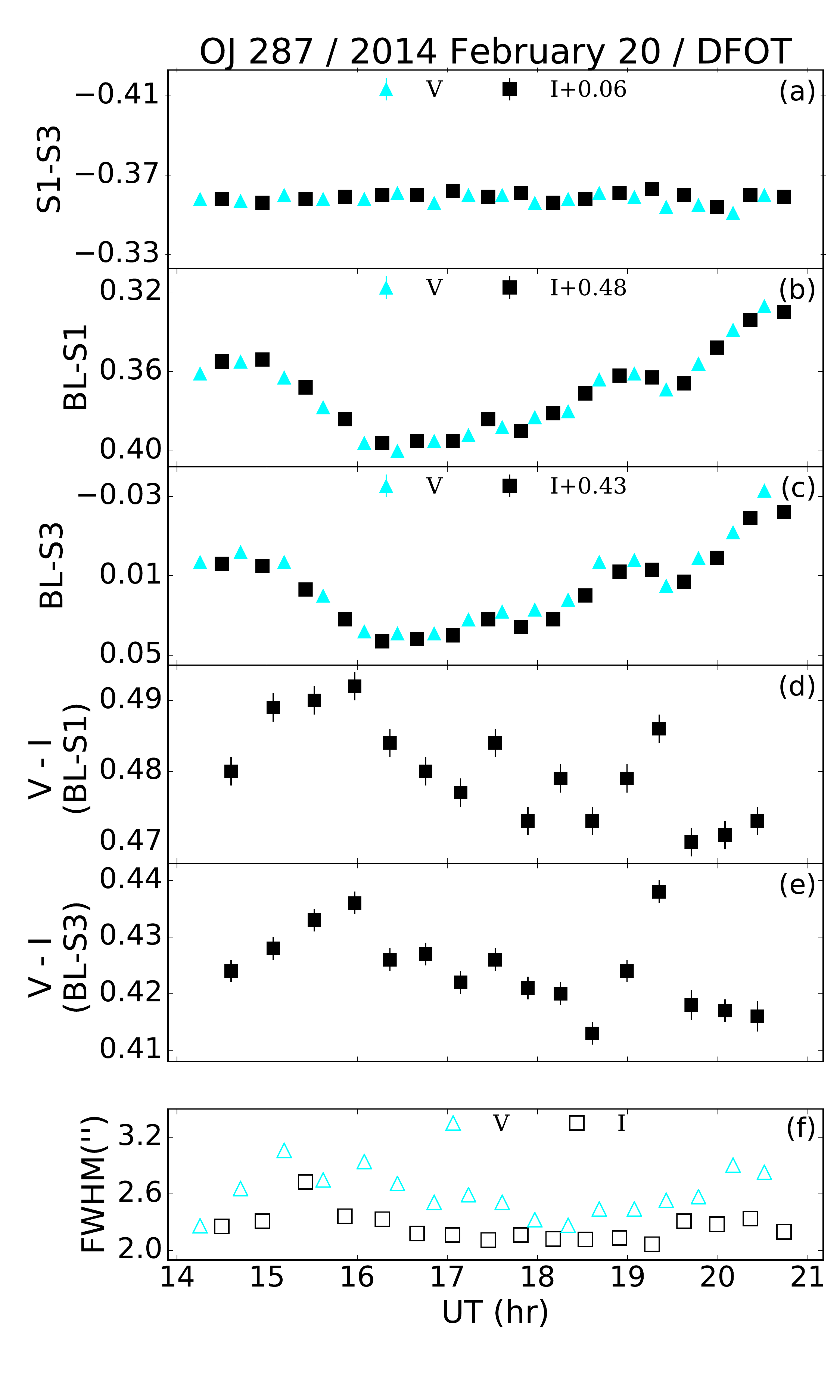}
\includegraphics[width=0.33\textwidth]{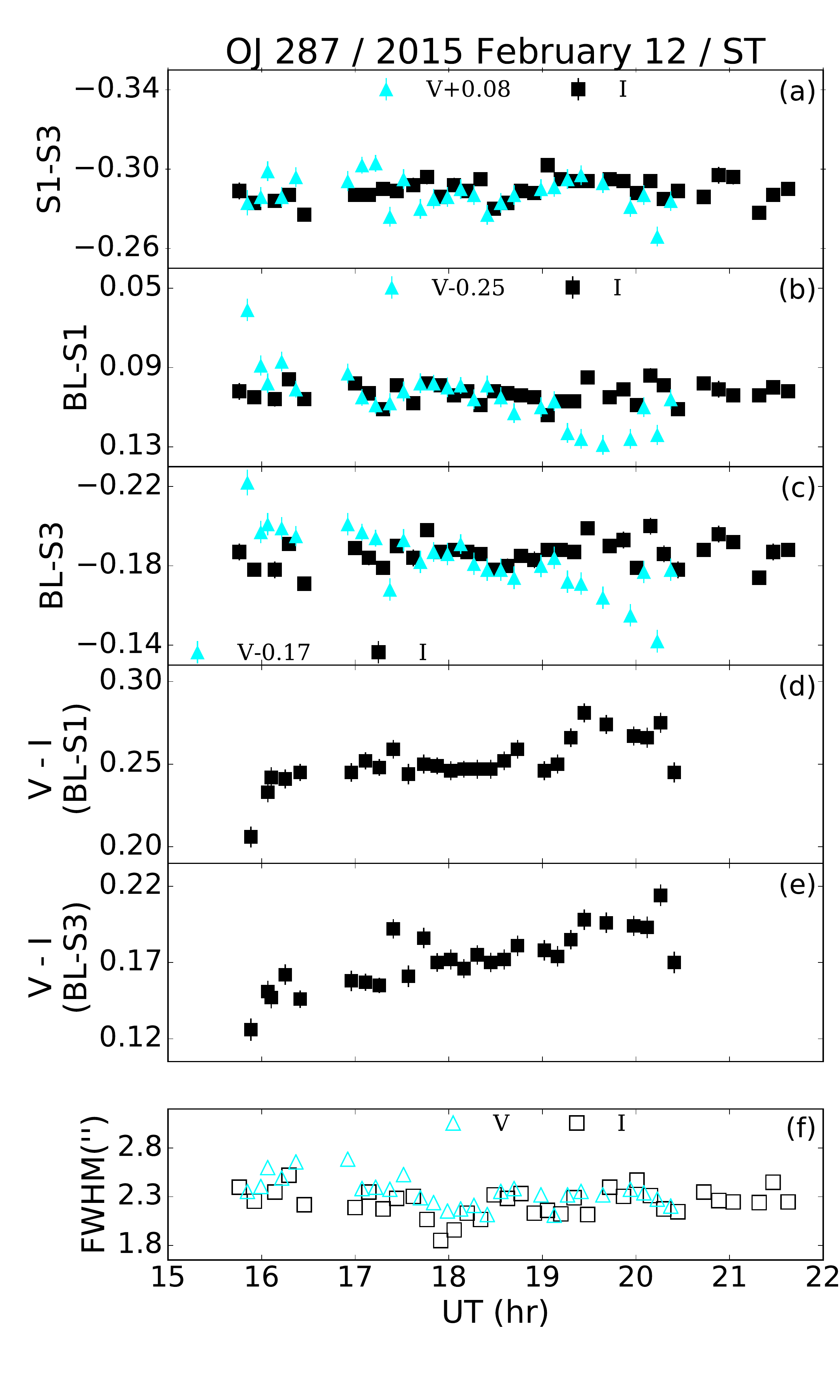}
\includegraphics[width=0.33\textwidth]{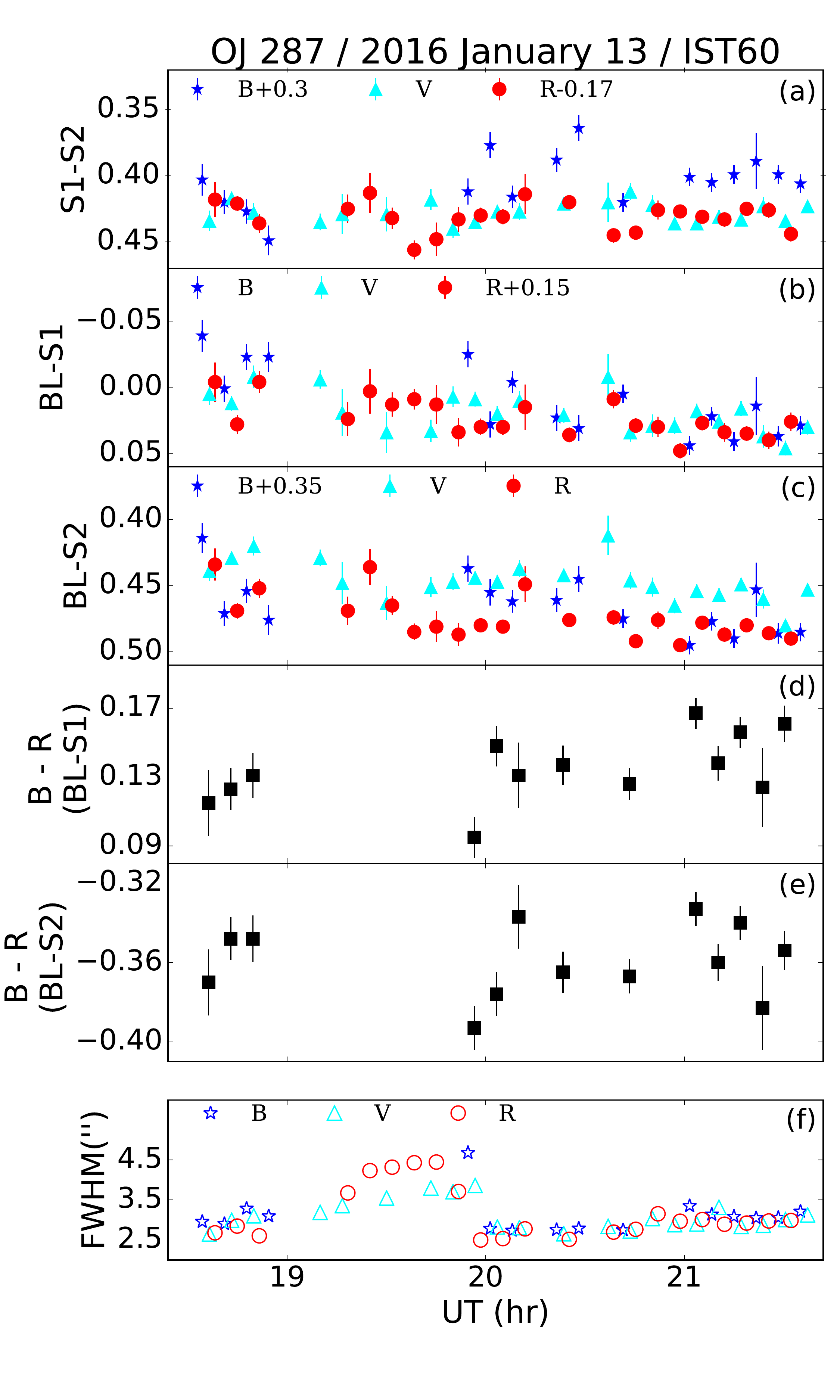}
}
\hbox{
\includegraphics[width=0.33\textwidth]{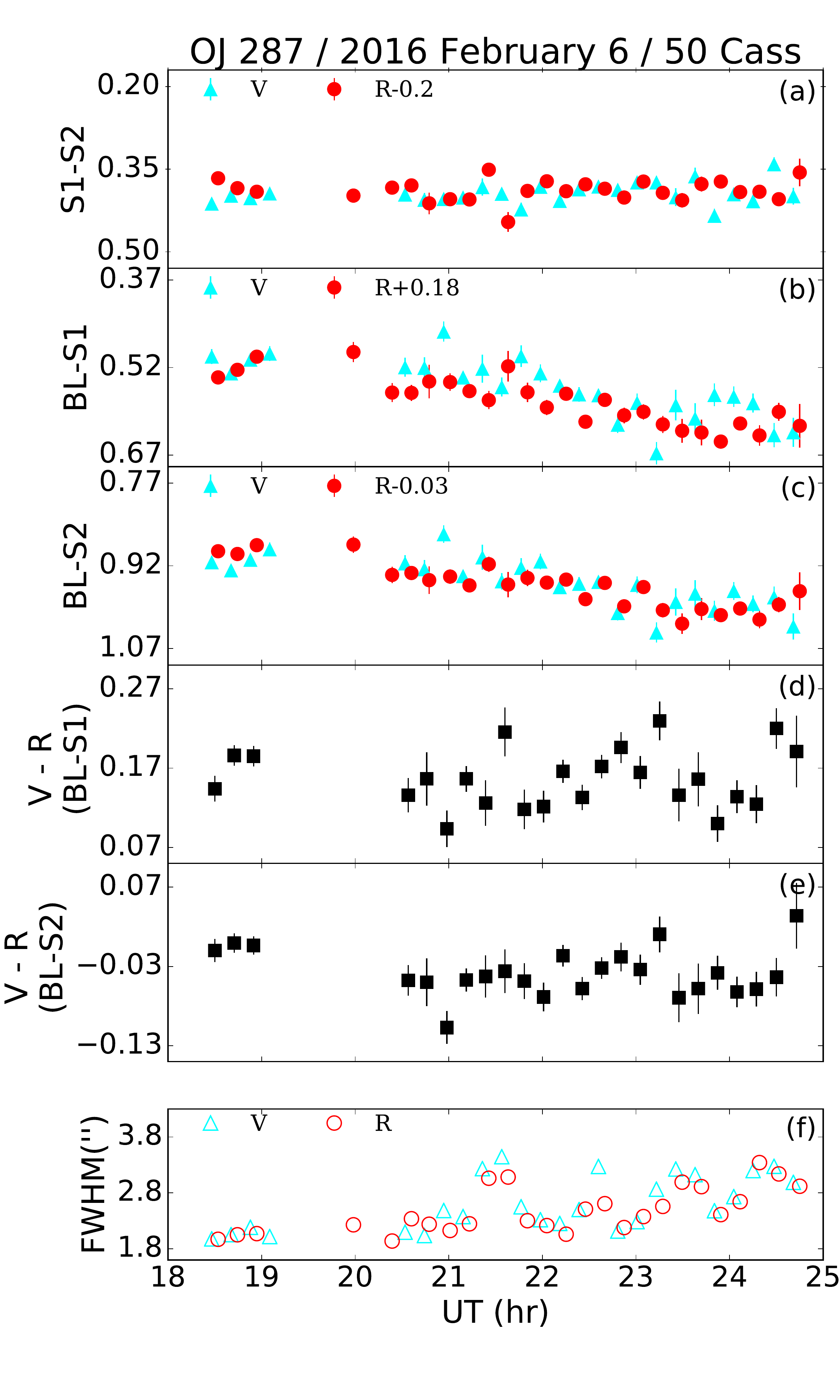}
\includegraphics[width=0.33\textwidth]{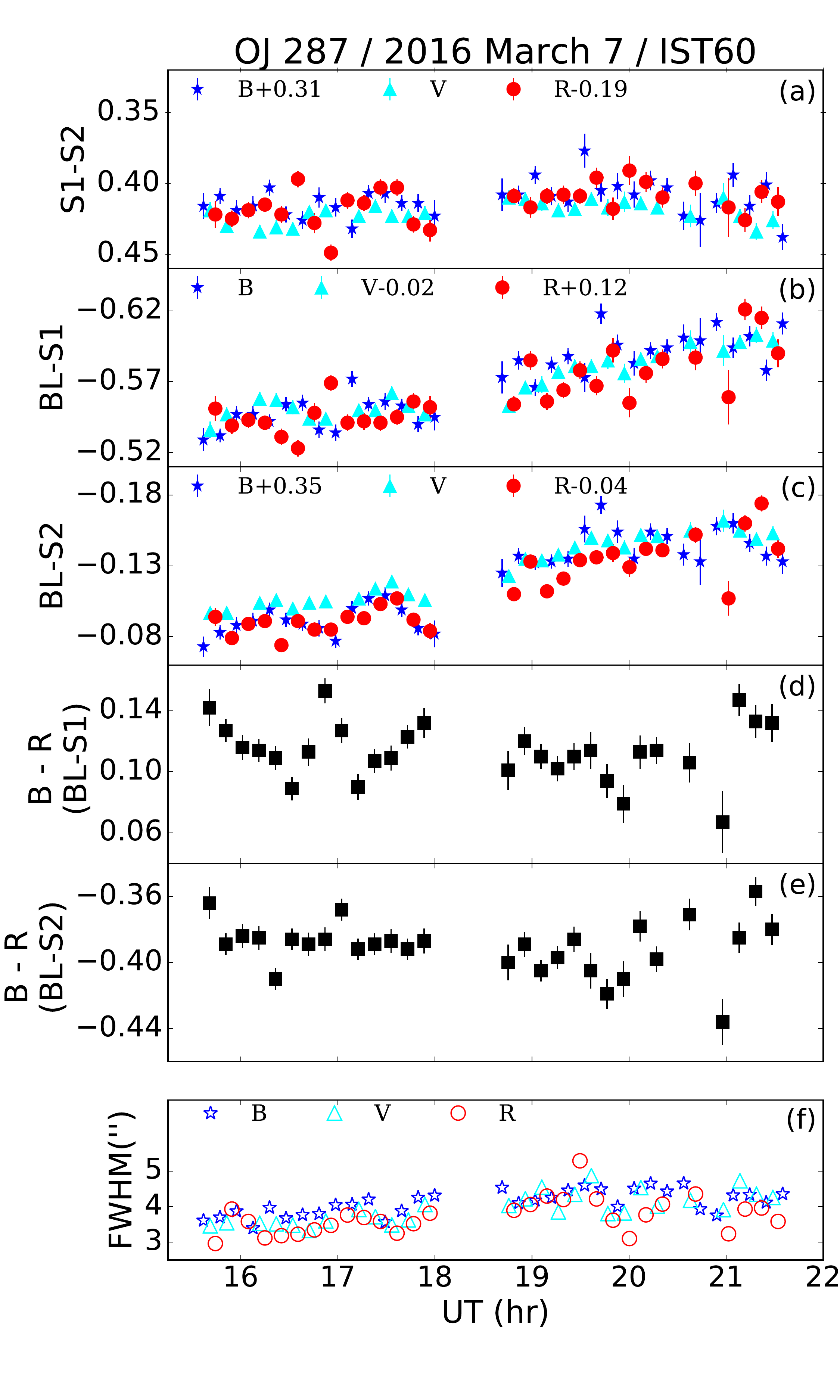}
\includegraphics[width=0.33\textwidth]{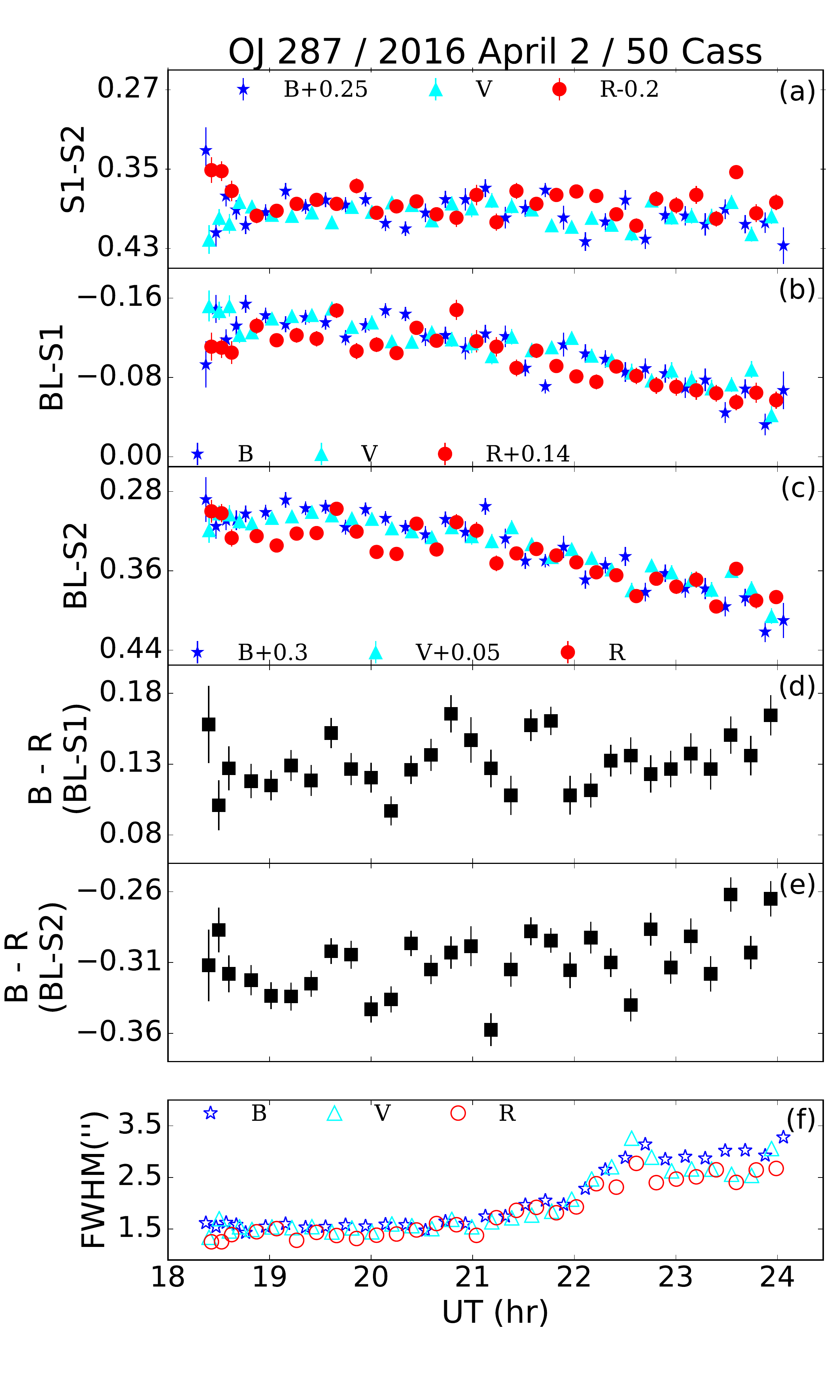}
}

\begin{minipage}{\textwidth}
\caption{}
\end{minipage}
\end{figure*}

\renewcommand{\thefigure}{\arabic{figure} (Cont.)}
\addtocounter{figure}{-1}

\begin{figure*}
\centering
\hbox{
\includegraphics[width=0.33\textwidth]{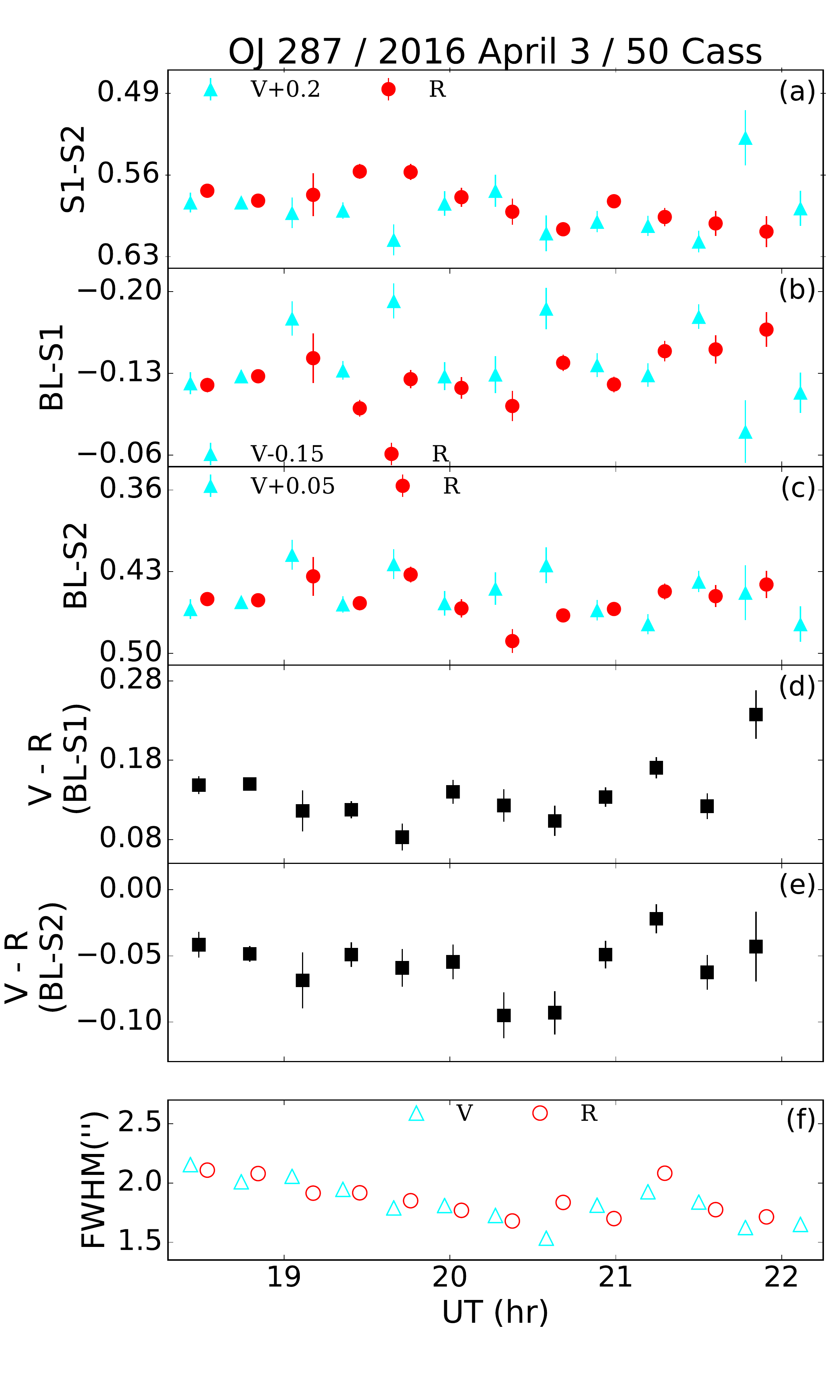}
\includegraphics[width=0.33\textwidth]{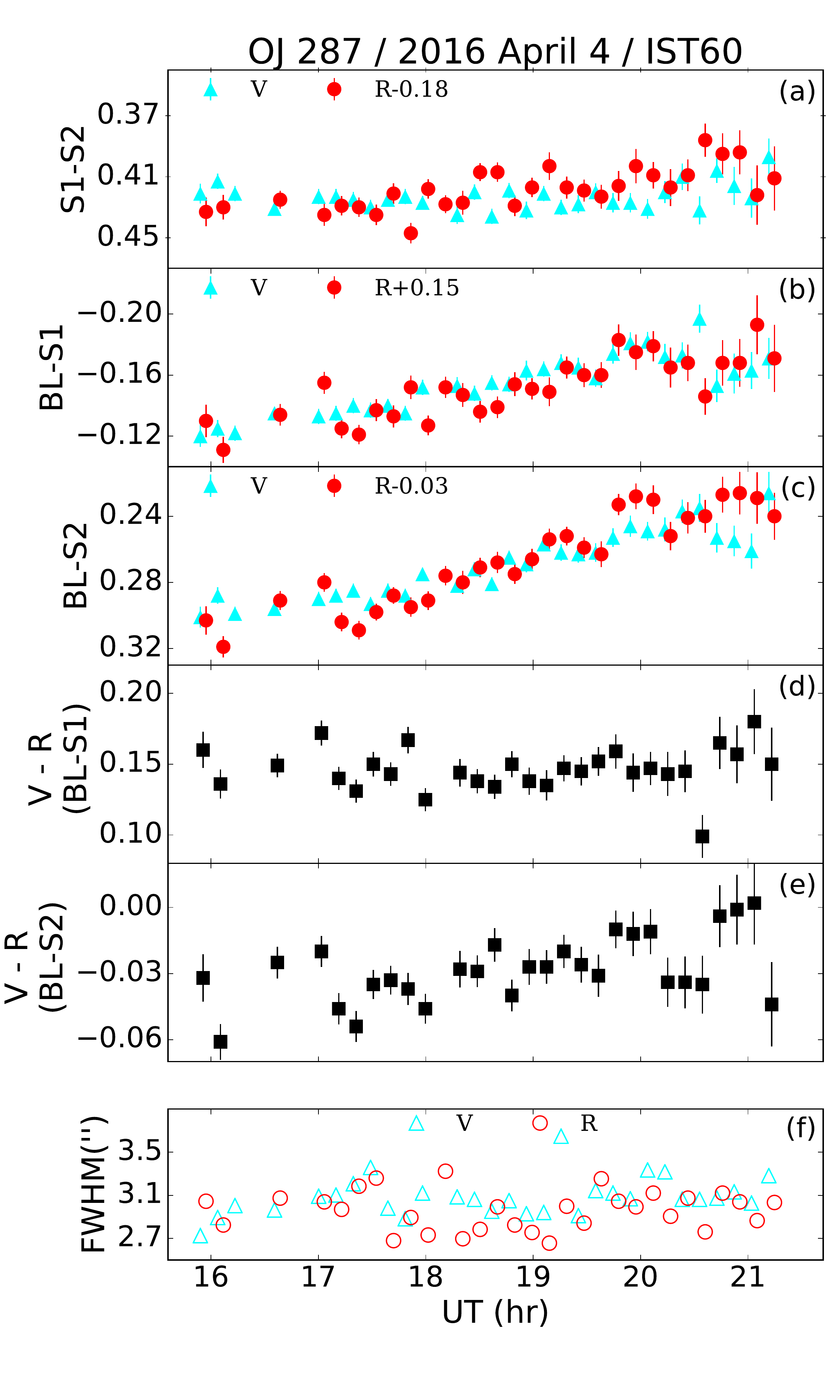}
\includegraphics[width=0.33\textwidth]{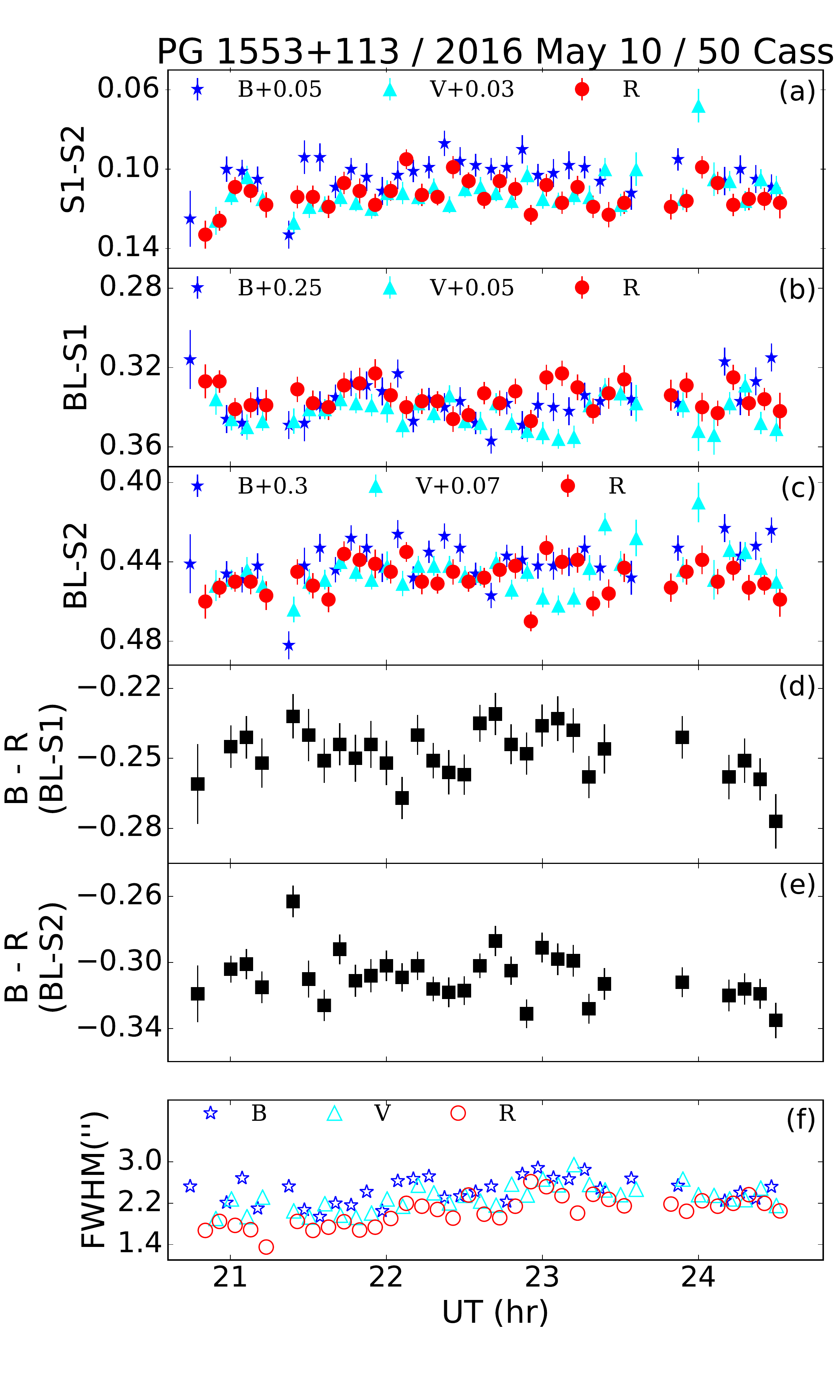}
}
\hbox{
\includegraphics[width=0.33\textwidth]{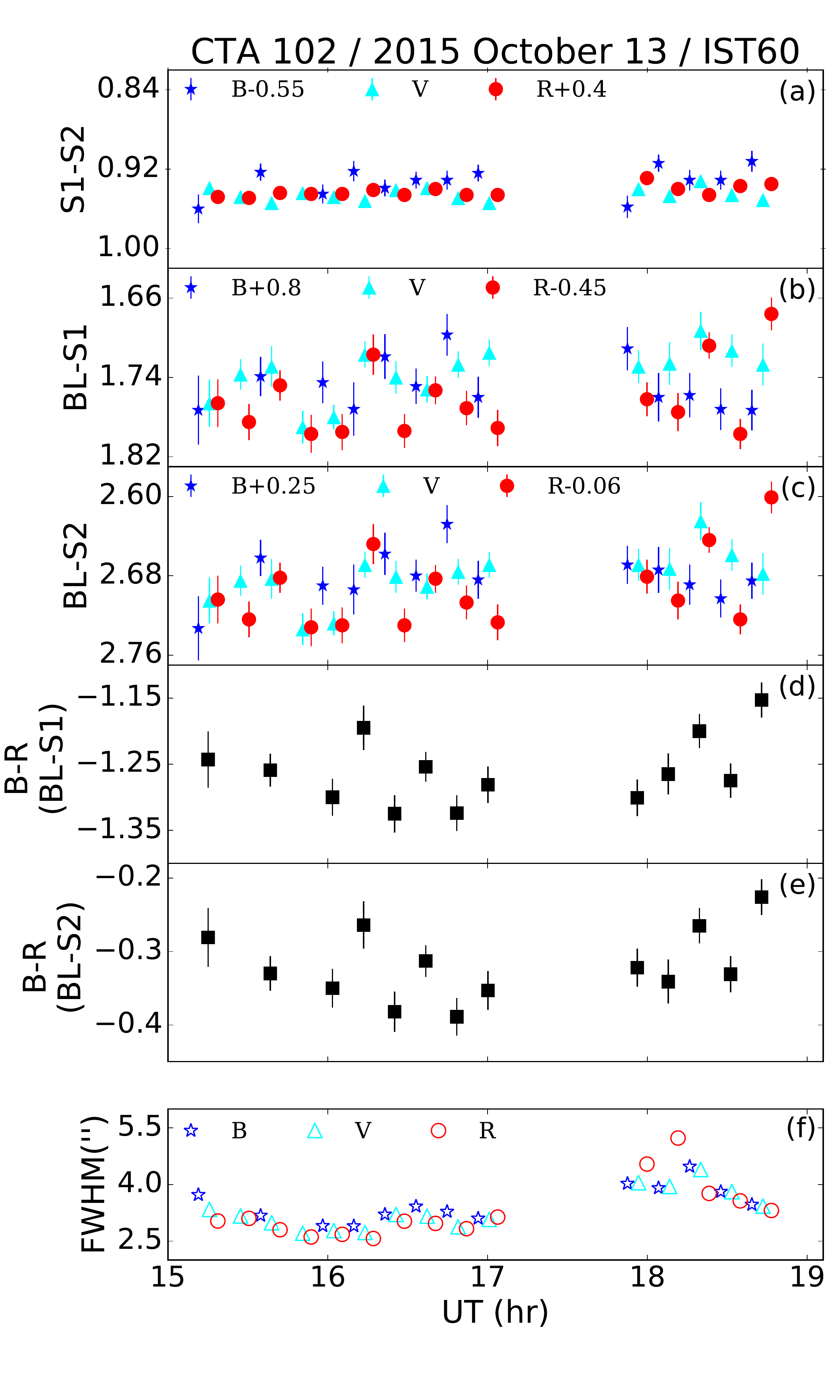}
\includegraphics[width=0.33\textwidth]{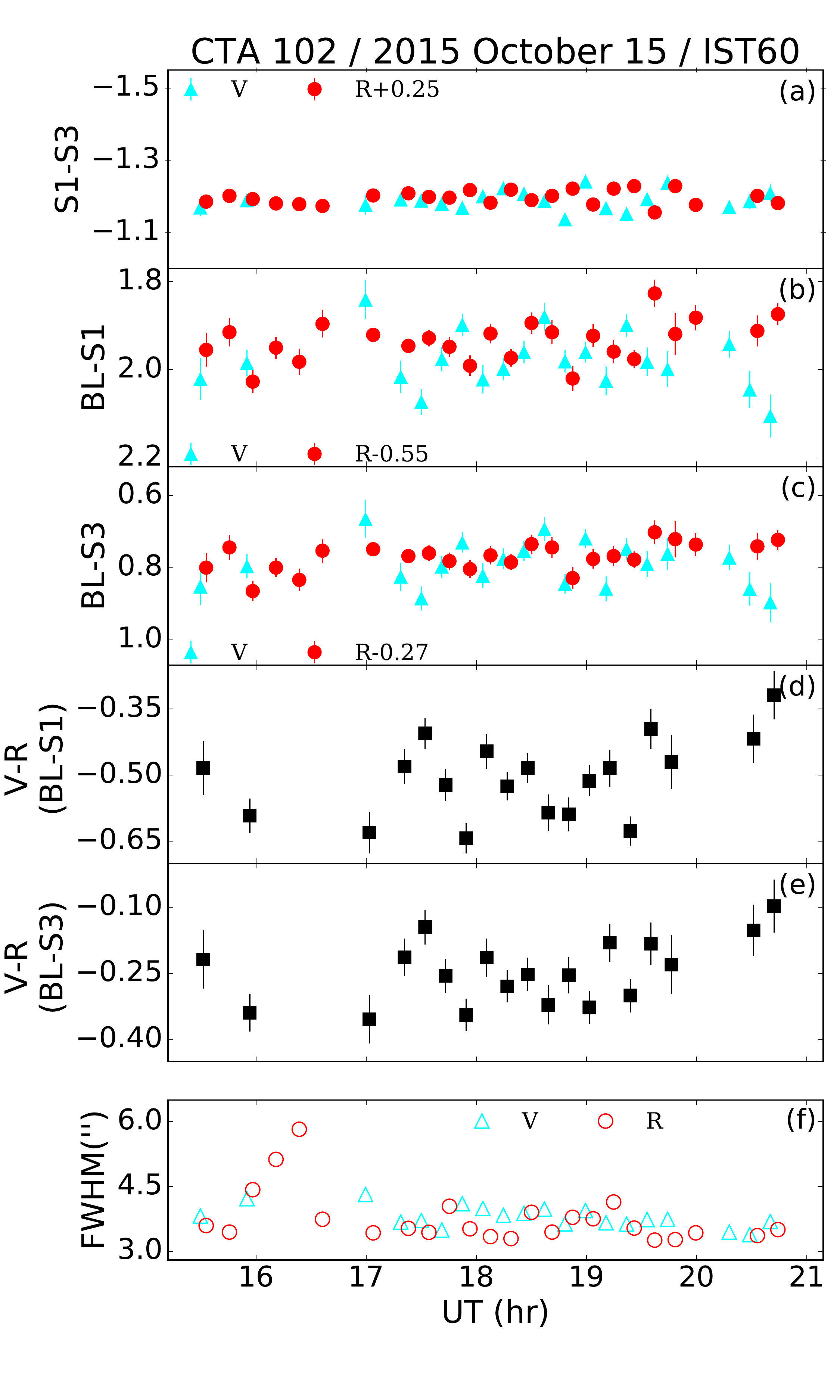}
\includegraphics[width=0.33\textwidth]{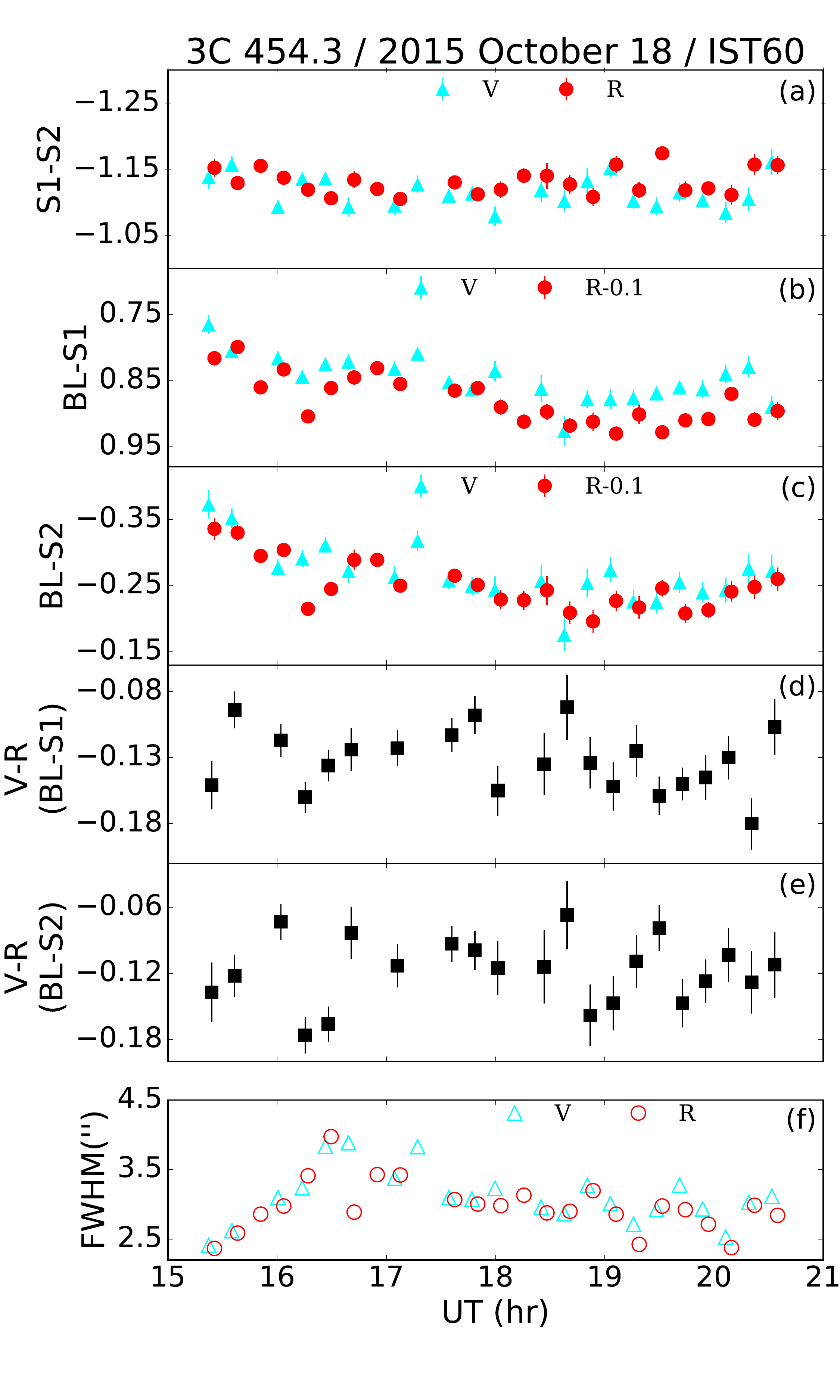}
}

\begin{minipage}{\textwidth}
\caption{}
\end{minipage}
\end{figure*}

\renewcommand{\thefigure}{\arabic{figure}}

\begin{table*}
\small
\caption{Colour microvariability results}\label{results_cinv}
\begin{center}
\begin{tabular}{cccccccc}
\hline \hline
  Source 	 & 	Date of obs.  			& Tel. 	& N$_p$ 	& Filters	&			colour				&  $\alpha_{\nu_1}^{\nu_2}$	&  F$_{cc}$ (status$^\ast$) \\
	(1)		 &	(2)					&	(3)		&     (4)	&	(5)	&			(6)					& (7) 			&	(8)					\\
\hline
0109+224	&	2015 Nov 14			&    	IST60	&	23	&B--R	&	$-$0.76$\pm$0.03, $-$0.39$\pm$0.02	& 1.11$\pm$0.06	&	1.34, 1.58 (N, N)		\\
3C 66A		&	2015 Oct 19    			&	IST60	&	31	&B--R	&	0.14$\pm$0.02, $-$0.62$\pm$0.02	& 1.51$\pm$0.04	&	1.02, 0.85 (N, N)		\\ 
			&	2015 Nov 15			&	IST60	&	31	&B--R	&	0.14$\pm$0.01, $-$0.61$\pm$0.02	& 1.53$\pm$0.04	&	0.68, 1.15 (N, N)		\\
S5 0716+714	&	2015 Jan 11$^\dag$	&   	ST 		&	48	&V--I	&	0.19$\pm$0.01, 0.34$\pm$0.01	& 1.33$\pm$0.03	&	3.47, 3.77 (V, V)		\\
			&	2015 Jan 15$^\dag$	&   	ST 		&	26	&V--I	&	0.26$\pm$0.02, 0.43$\pm$0.02	& 1.53$\pm$0.05	&	8.03, 17.31 (V, V)		\\
			&	2015 Feb 9$^\dag$		&   	ST 		&	54	&V--I	&	0.18$\pm$0.01, 0.19$\pm$0.01	& 1.28$\pm$0.02	&	1.62, 1.30 (PV, N)		\\
OJ 287		&	2014 Feb 20$^\dag$	&    	DFOT		&	16	&V--I	&	0.48$\pm$0.01,  0.42$\pm$0.01	& 1.87$\pm$0.02	&	5.14, 8.83 (V, V)		\\ 
			&	2015 Feb 12$^\dag$   	&    	ST		&	27	&V--I	&	0.25$\pm$0.02, 0.17$\pm$0.02	& 1.27$\pm$0.04	&	2.77, 3.61 (V, V)		\\
			&	2016 Jan 13    			&    	IST60	&	13	&B--R	&	0.14$\pm$0.02, $-$0.36$\pm$0.02	& 1.62$\pm$ 0.04	&	0.88, 0.92 (N, N)		\\
			&	2016 Feb 6$^\dag$    	&    	50 Cass	&	24	&V--R 	&	0.16$\pm$0.04, $-$0.04$\pm$0.03	& 1.55$\pm$0.18	&	0.96, 0.76 (N, N)		\\
			&	2016 Mar 7$^\dag$   	&    	IST60	&	29	&B--R	&	0.11$\pm$0.02, $-$0.39$\pm$0.02	&1.55$\pm$0.04	&	1.48, 161	(N, N)		\\
			&	2016 Apr 2$^\dag$     	&    	50 Cass	&	30	&B--R	&	0.13$\pm$0.02, $-$0.31$\pm$0.02	&1.67$\pm$0.04 	&	0.81, 1.41 (N, N)		\\  
			&	2016 Apr 3    			&    	50 Cass	&	12	&V--R 	&	0.14$\pm$0.04, $-$0.06$\pm$0.02	& 1.43$\pm$0.14	&	2.04, 0.81 (N, N)		\\  
			&	2016 Apr 4$^\dag$     	&    	IST60	&	29	&V--R 	&	0.15$\pm$0.02, $-$0.03$\pm$0.02	& 1.55$\pm$0.08	&	0.61, 0.93	(N, N)		\\  
PG 1553+113	&	2016 May 10			& 	50 Cass	&	30	&B--R	&	$-$0.25$\pm$0.01, $-$0.31$\pm$0.02	& 1.01$\pm$0.03	&	0.51, 0.97 (N, N)		 \\
CTA 102		&	2015 Oct 13   			&	IST60 	&	13	&B--R	&	$-$1.26$\pm$0.05, $-$0.32$\pm$0.05	& 1.37$\pm$0.12	&	1.33, 1.29 (N, N)		\\ 
			&	2015 Oct 15   			&	IST60	&	19	&V--R 	&$-$0.51$\pm$0.09 $-$0.25$\pm$0.07 & 0.65$\pm$0.51 & 1.78, 1.04 (N, N)		\\
3C 454.3		&	2015 Oct 18$^\dag$	&    	IST60	&	21	&V--R	&	$-$0.13$\pm$0.02,  $-$0.12$\pm$0.03	& 1.22$\pm$0.13	&	0.80, 0.69  (N,N)		\\ 
\hline
\end{tabular}
\end{center}
\textbf{Columns:} (1) Most common source name; (2) date of the observation/monitoring; (3) telescope used; (4) number of data points; (5) filters used for colour estimation; (6) computed mean value (colour index) for both  CDLCs; (7) spectral index; (8) F-values computed for the BL-S1 and BL-S2 CDLCs (variability status for the corresponding CDLC).\\
$^\ast$ V = Variable; N = Non-variable; PV = Probable Variable. $^\dag$ blazar showed confirmed INV in at least one frequency band.
\end{table*}

\subsection{Summary of polarization monitoring}
We employed three other telescopes to carry out the intra-night polarization monitoring for the sample. Below we give key information on these observatories and instruments.

\begin{enumerate}

\item{{\bf 150\,cm Kanata, Japan:} A portion of our polarization microvariability light curves come from  monitoring using the 150\,cm Kanata telescope, run by Higashi--Hiroshima Observatory, Japan. We used Hiroshima One--shot Wide--field Polarimeter \citep[HOWPol;][]{Kawabata08} which is installed at the Nasmyth focus of the telescope. The polarimetric  observations were performed using the Cousins R$_c$-filter.  }

\item{{\bf 70\,cm AZT-8+ST7, Russia:} We used the 70\,cm AZT-8+ST7 telescope of Crimean Astrophysical Observatory, located at Nauchnij, Russia for a large number of polarimetric measurements. The observations were carried out in R-band \citep{Larionov13, Larionov16}.}

\item{{\bf 40\,cm LX-200, Russia:} Lastly, we used the 40\,cm LX-200 telescope of St. Petersburg State University, located at  St. Petersburg, Russia {for a comparable number of observations.} The observations were carried out in white-band \citep{Larionov13, Larionov16}.}

\end{enumerate}

Our observing strategy consisted of obtaining successive images at four position angles of the half-wave plate, of 0$^\circ$,  45$^\circ$,  22.5$^\circ$, and  67.5$^\circ$, for a continuous monitoring duration lasting for more than 3 hours. The calibration images, bias, and flat-frames were obtained either just before or soon after the target observations. In addition, un-polarized and highly polarized standard stars were observed to set the instrumental polarization and the position angle of the instrument. The observations were conducted using R-band at Kanata and AZT-8+ST7 and white light at LX-200. The details of data reduction from AZT-8+ST7 and LX-200 telescopes are given in \citet{Larionov08, Larionov13, Larionov16}. 

\begin{figure*}
\centering
\hbox{
\includegraphics[width=0.33\textwidth]{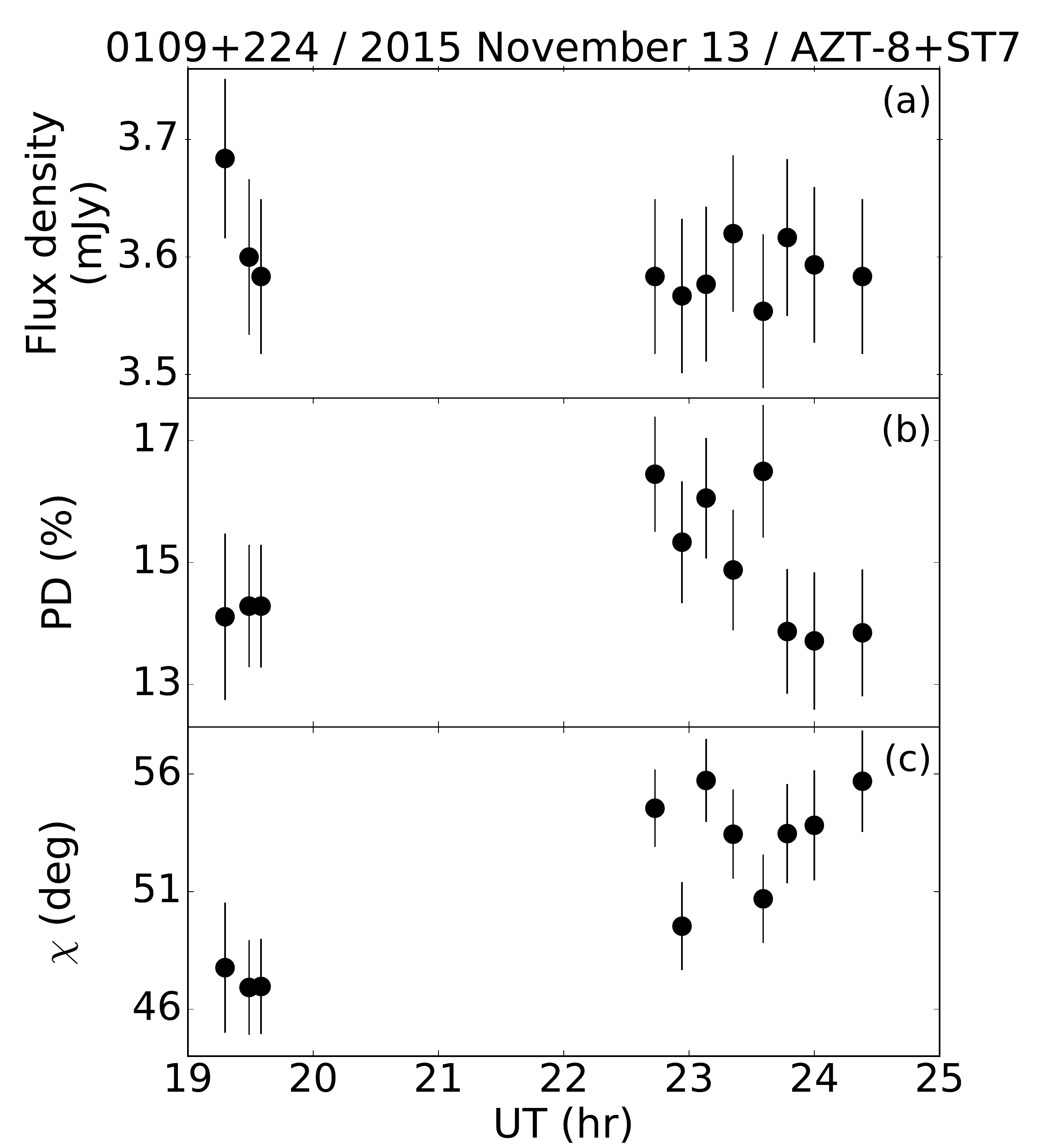}
\includegraphics[width=0.33\textwidth]{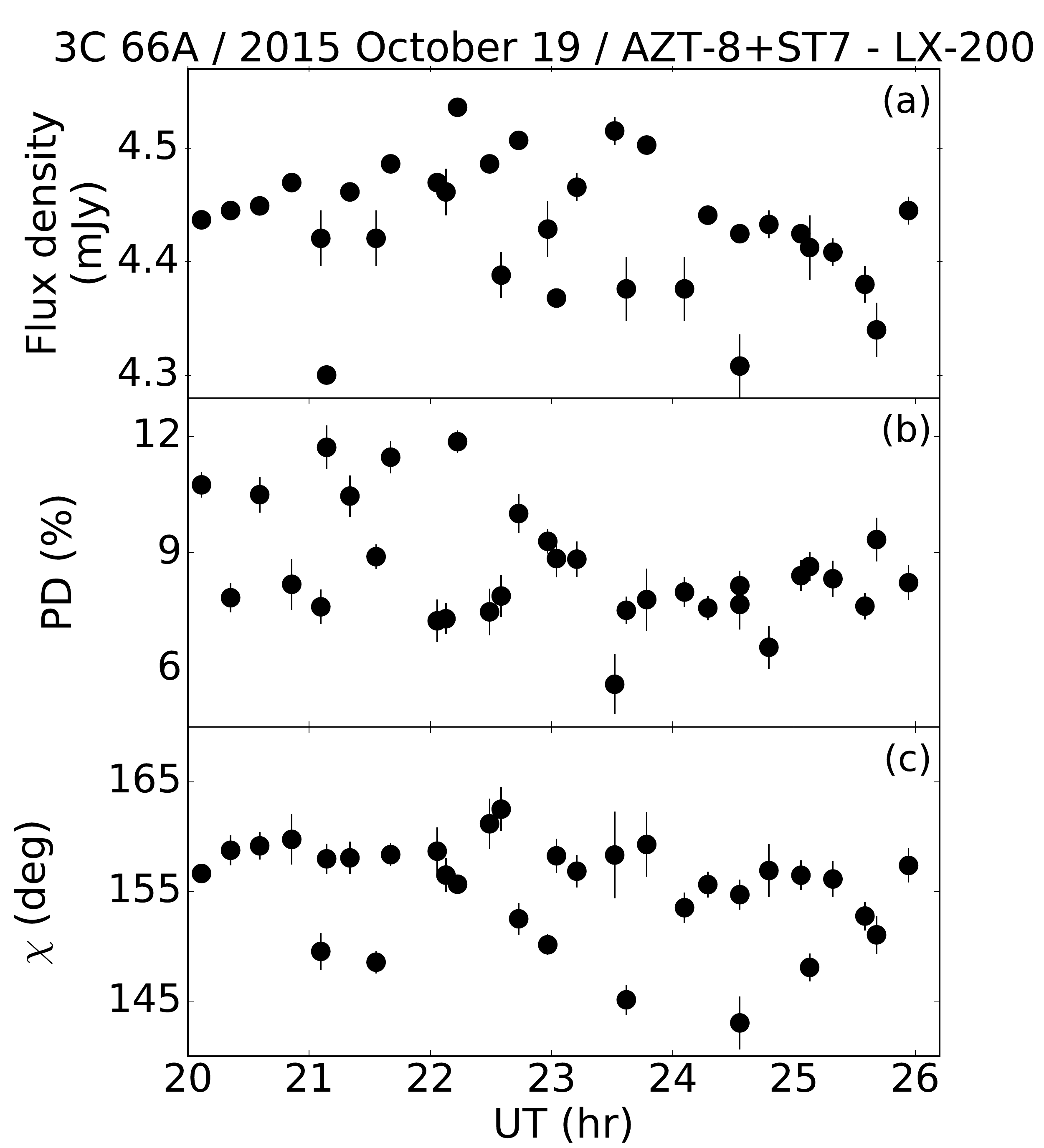}
\includegraphics[width=0.33\textwidth]{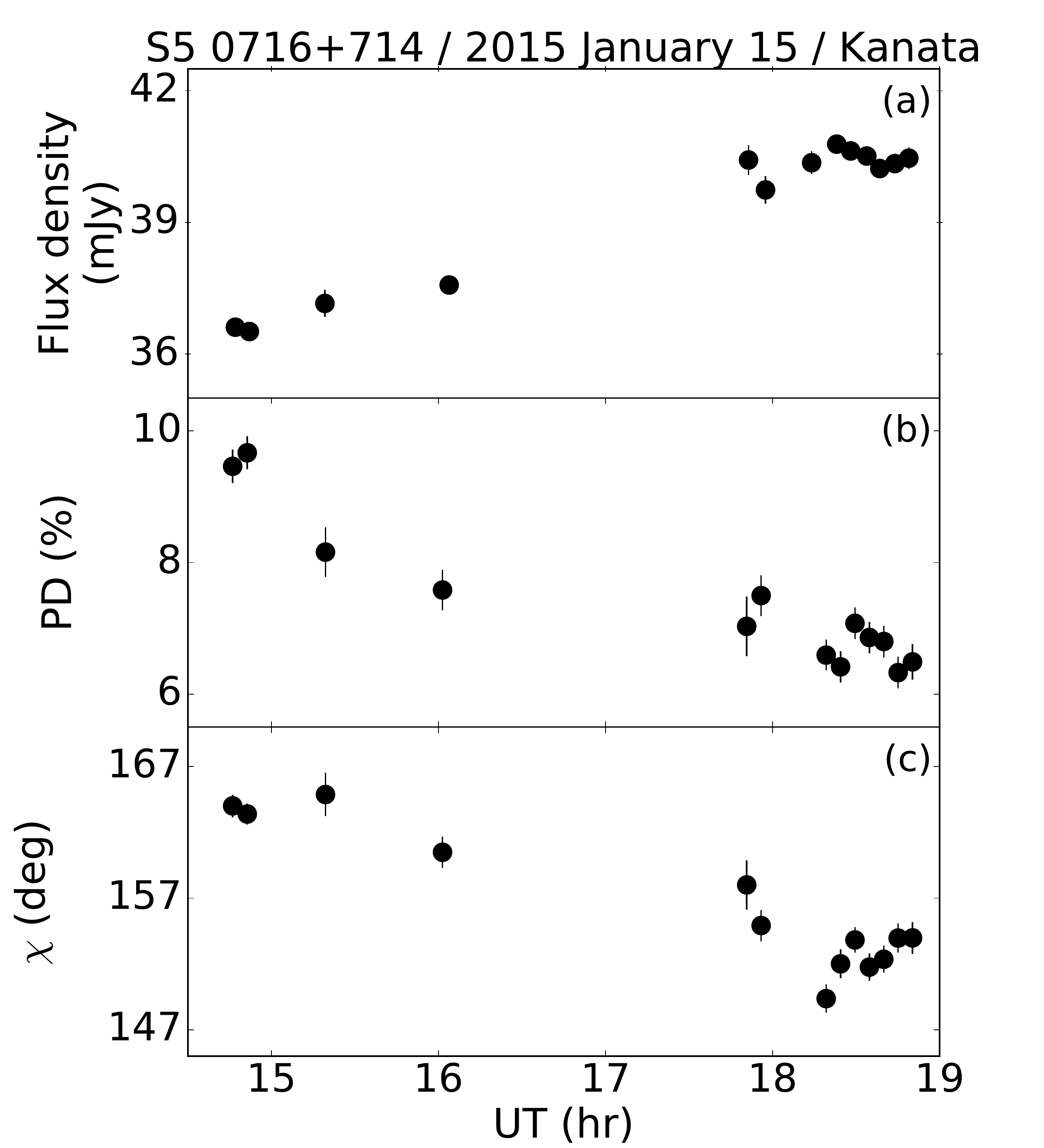}
}
\hbox{
\includegraphics[width=0.33\textwidth]{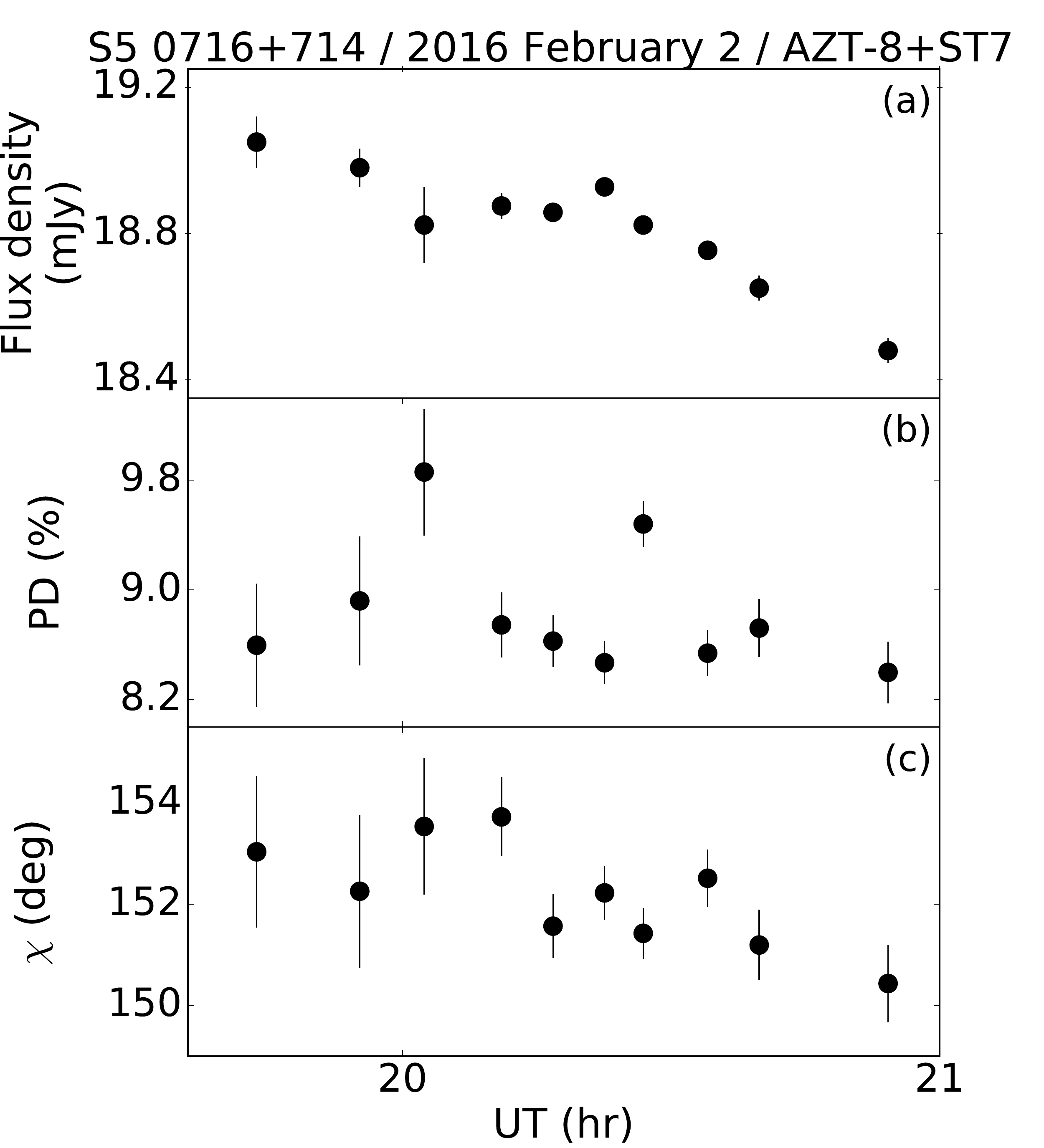}
\includegraphics[width=0.33\textwidth]{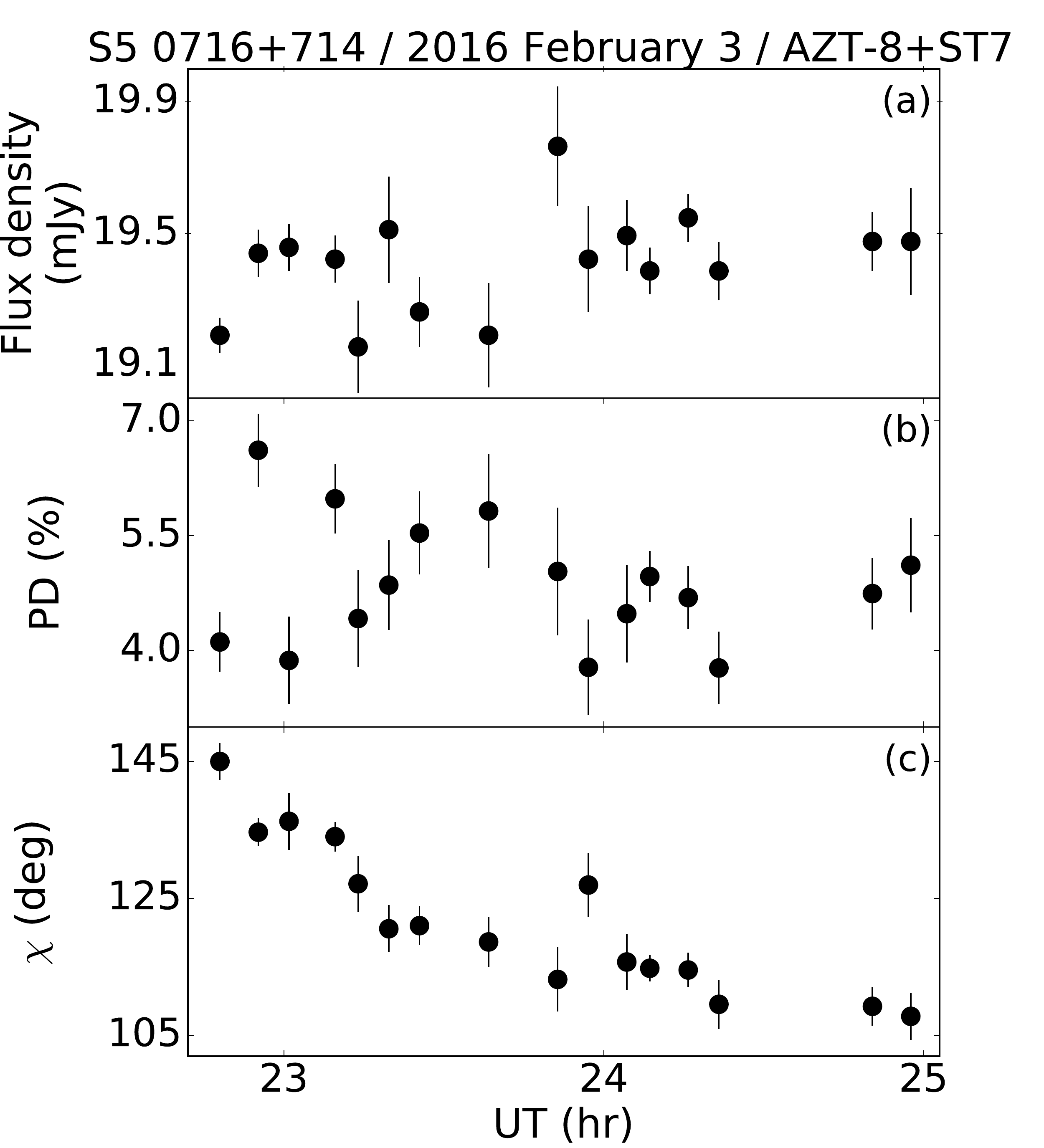}
\includegraphics[width=0.33\textwidth]{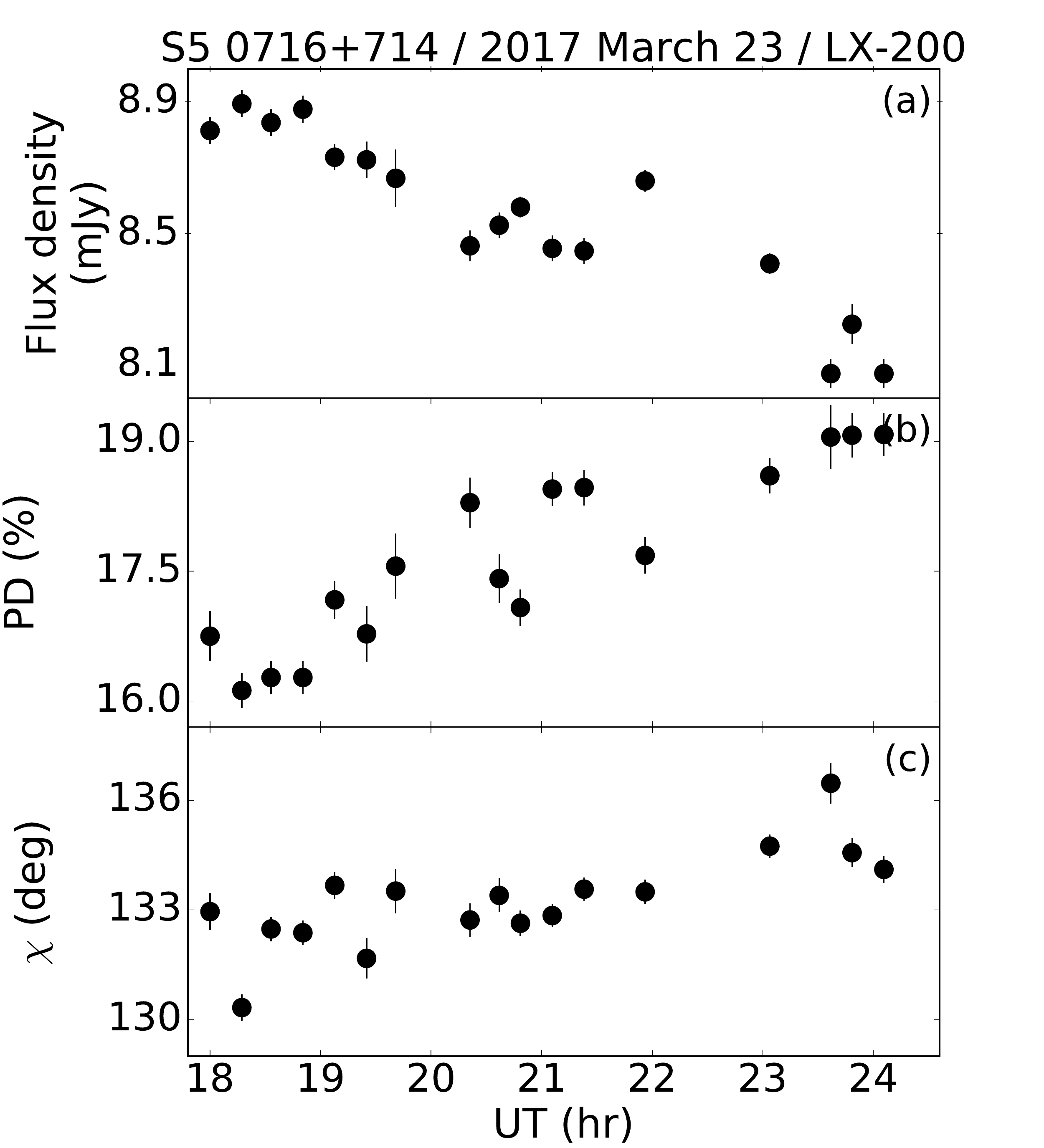}
}

\hbox{
\includegraphics[width=0.33\textwidth]{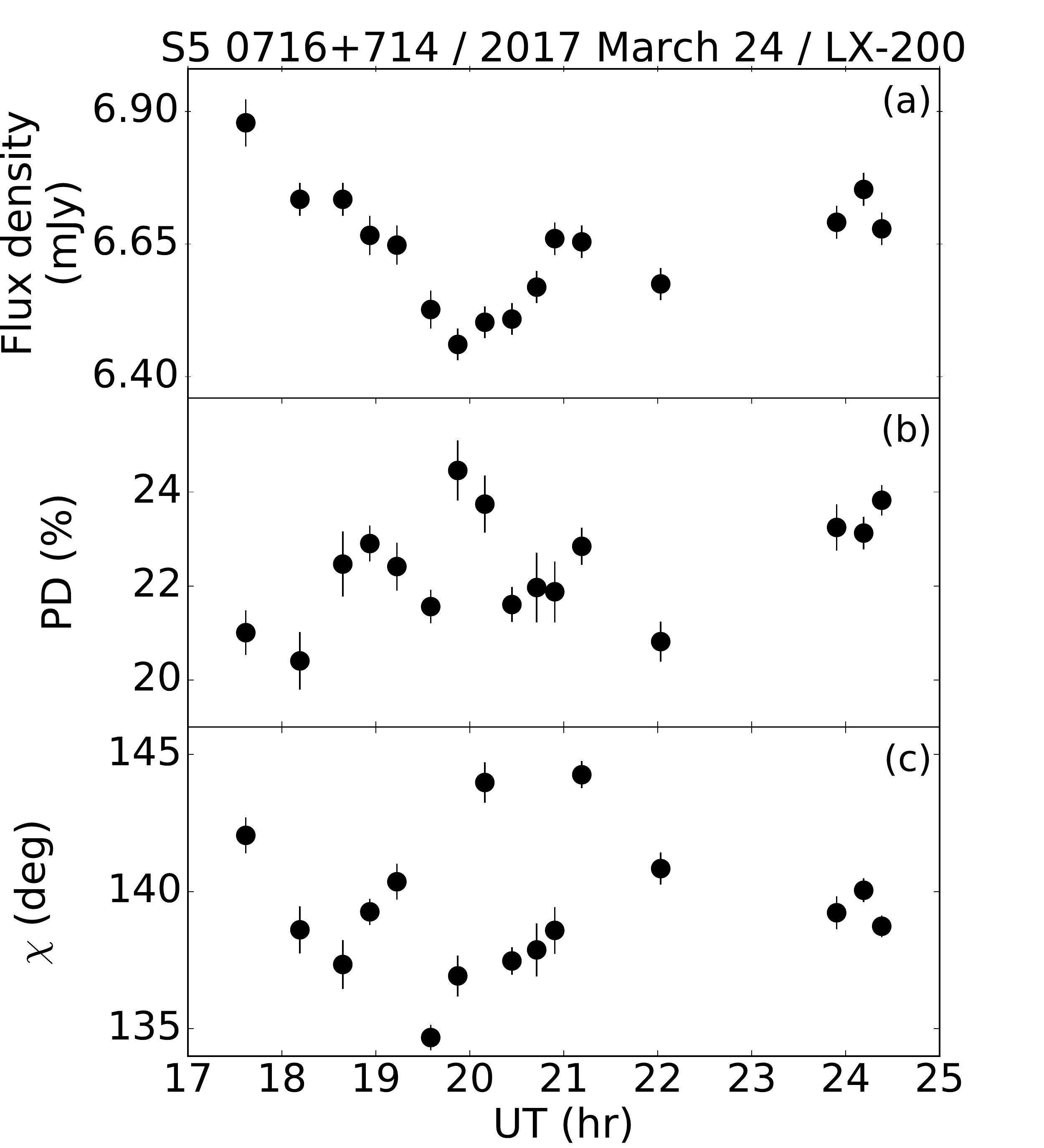}
\includegraphics[width=0.33\textwidth]{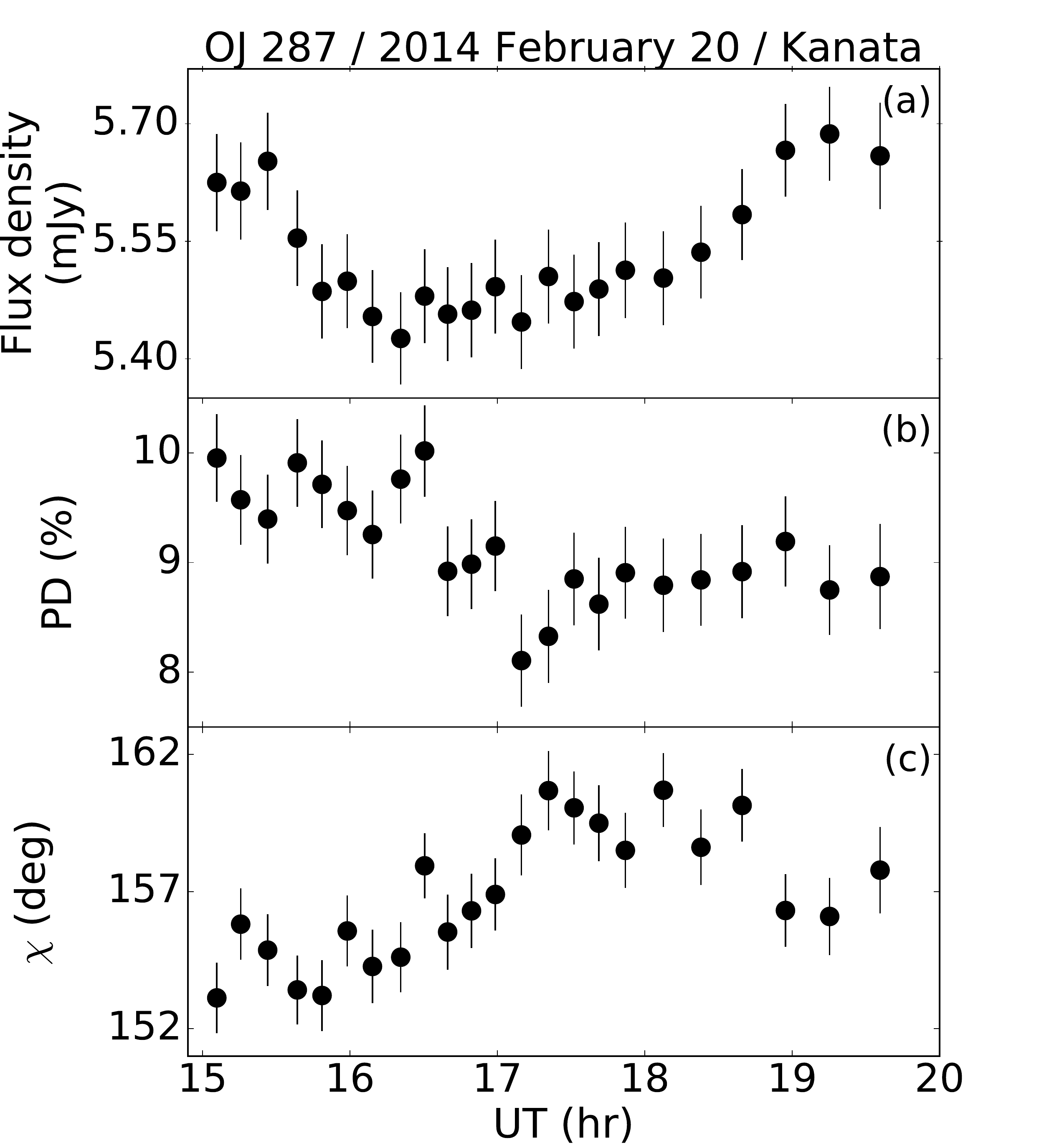}
\includegraphics[width=0.33\textwidth]{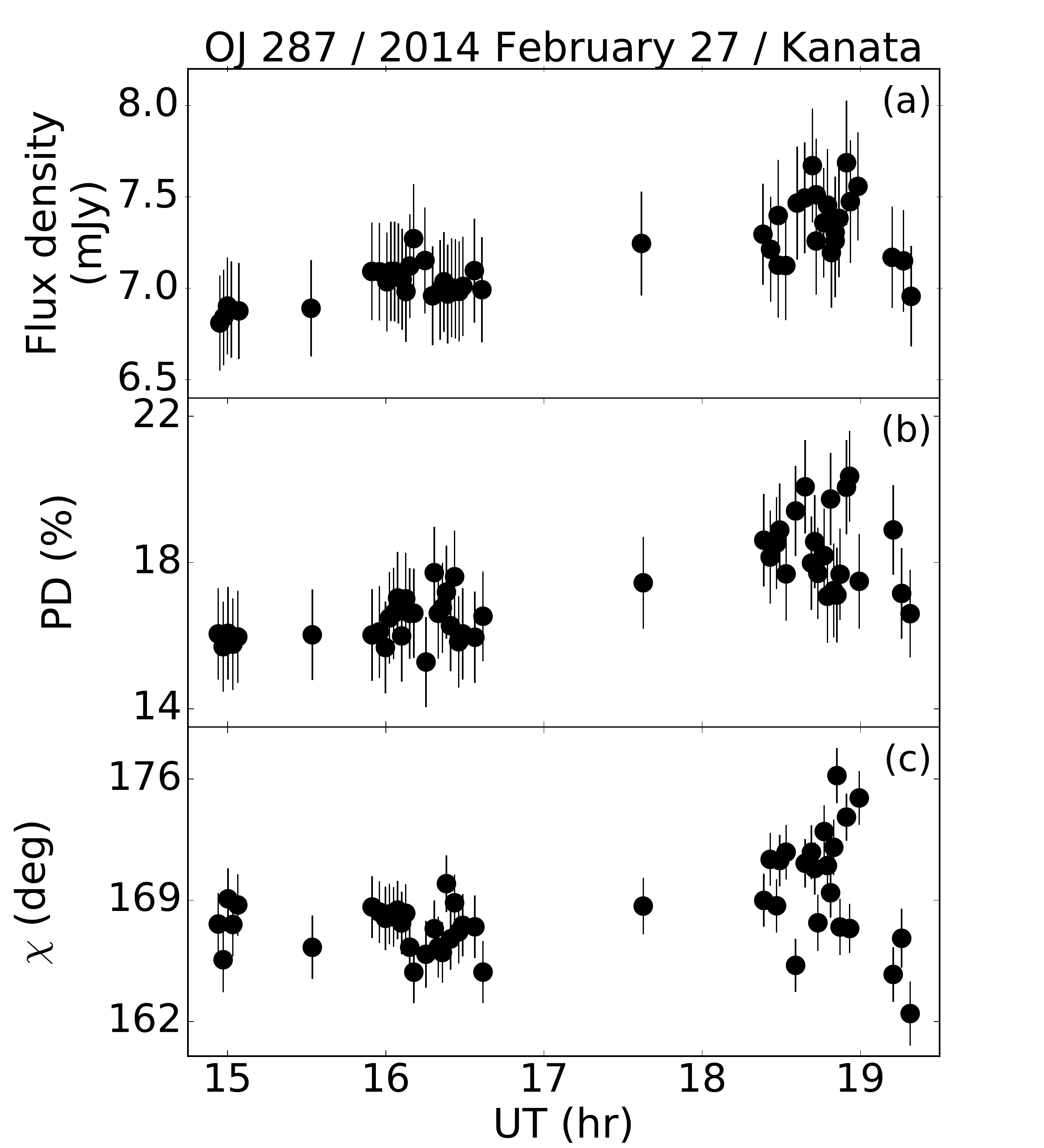}
}

\begin{minipage}{\textwidth}
\caption{Variations of flux density in R-band (for Kanata and AZT-8+ST7 telescopes) or white-band (for LX-200) (panel a), PD (panel b), and $\chi$ (panel c) in the present study. For each figure, the source name, the date of observation, and the telescope used are given at the top.}
\label{fig_polinv}%
\end{minipage}
\end{figure*}

For the Kanata data, the pre-processing (bias subtraction and flat-fielding) of the images were carried out using standard procedures in {\sl IRAF}. The flux of the target in each half-wave plate combination was gathered using the aperture photometry in the {\sl APPHOT} package. This enables the measurement of PD and $\chi$ for the target while the total intensity was computed using differential photometry using the comparison star on the same CCD frame. The polarizer splits the radiation into two parts, with orthogonal electric field components which are denoted as `ordinary',  I$_\mathrm{o}$) and `extra-ordinary'  I$_\mathrm{e}$, images. The fluxes of I$_\mathrm{o}$ and I$_\mathrm{e}$ images are thus obtained for the blazar and the comparison star using a circular aperture that is roughly 2--3 times larger than the median seeing disc on the given night. The fractional linear polarization (PD) is computed following the methodology given by \citet{Wang15}:

\begin{equation}
PD = \sqrt{Q^{2}+U^{2}}
\label{equ:pf}
\end{equation}
where, $Q$ and $U$ are the Stokes parameters which are determined from:

\begin{equation}
Q = \frac{1-\sqrt{ \frac{ \mathrm{(I_e/I_o)_{0.0 deg}}}{\mathrm{(I_e/I_o)_{45.0 deg}}} } } {1+\sqrt{ \frac{ \mathrm{(I_e/I_o)_{0.0 deg}}}{\mathrm{(I_e/I_o)_{45.0 deg}}}}  },  \quad U = \frac{1-\sqrt{ \frac{ \mathrm{(I_e/I_o)_{22.5 deg}}}{\mathrm{(I_e/I_o)_{67.5 deg}}} } } {1+\sqrt{ \frac{\mathrm{ (I_e/I_o)_{22.5 deg}}}{\mathrm{(I_e/I_o)_{67.5 deg}} }}  },
\label{equ:iratio}
\end{equation}
where $\mathrm{I_{e(0.0 deg)}}$, $\mathrm{I_{e(45.0 deg)}}$, $\mathrm{I_{e(22.5 deg)}}$, $\mathrm{I_{e(67.5 deg)}}$,  $\mathrm{I_{o(0.0 deg)}}$, $\mathrm{I_{o(45.0 deg)}}$, $\mathrm{I_{o(22.5 deg)}}$, and $\mathrm{I_{o(67.5 deg)}}$ are the fluxes of the extra-ordinary and ordinary image components for the HWP combinations at 0.0, 45.0, 22.5, and 67.5 deg., respectively. The electric vector polarization angle is

\begin{equation}
\chi = \frac{1}{2} \arctan \Big (\frac{U}{Q}\Big ).
\label{equ:chi}
\end{equation}
Statistical errors on $Q$ and $U$ are obtained following standard error propagation \citep{Bevington03} and assuming Poisson statistics, meaning $\delta \mathrm{I_e}$ $=$ $\sqrt\mathrm{I_e}$ and $\delta \mathrm{I_o}$ $=$ $\sqrt\mathrm{I_o}$, etc., for each HWP combination. Table~\ref{results_polinv} presents the  polarization data on all 30 monitoring sessions while Figure~\ref{fig_polinv} shows the light curves for which more than 10 data points, without any large gaps, were available for a monitoring session.

\begin{table*}
\scriptsize
\caption{Summary of polarization observations}
\label{results_polinv}
\begin{center}
\begin{tabular}{cccccccccc}
\hline \hline
  Source 	 & 	Date of Obs.  	& Tel.	& Dur. &	N$_p$	 &  $\psi$     	& 	Flux density      				&  	PD     					& 	$\chi$         &	F$_c$	\\
                 & 	 	& 	&  (h)&	  	 &     (\%)	& 	             (mJy)				&  	   (\%)					& 	       (deg)	 	&	      \\
	(1)		 &	(2)			& (3)	&	(4)	&	(5)	&	(6)	&  	(7)						&	(8)						&	(9)				&  (10)		 \\
\hline
0109+224	&	2015 Nov 13	&	a 	&	5.09	&	11	& 1.6 	&  3.60$\pm$0.02				&	14.85$\pm$0.32			&	51.69$\pm$0.62	&	0.28 (N)	\\
			 &	2015 Dec 16	&	b	&	0.67	&	3	& 0.5 	&  3.89$\pm$0.02				&	12.03$\pm$0.48			&	31.97$\pm$1.16	 &	0.63 (N)	\\
\\
3C 66A		&	2015 Oct 19$^\ast$	& a, b 	&	5.83	&	32	& 1.2 	&  4.30$\pm$0.01, 4.54$\pm$0.01	& 5.61$\pm$0.78, 11.87$\pm$0.30& 	143.03$\pm$2.41, 162.52$\pm$1.98 &	11.98 (V)	 \\ 
			&	2015 Nov 16	&	a	&	0.26	&	3	& 0.2 	&  4.11$\pm$0.01				&	6.23$\pm$0.20			&	169.06$\pm$0.91 &	 0.26 (N)	\\
\\
S5 0716+714	 &	2015 Jan 15$^\ast$	&	c	&	4.03	& 	13 	& 4.2	&  36.51$\pm$0.22, 40.78$\pm$0.22&	6.33$\pm$0.24, 9.67$\pm$0.25 & 	149.37$\pm$1.08, 164.87$\pm$1.65 &	45.38 (V)	 \\
			 &	2016 Feb 2	&	a	&	1.17	&   	10	&	0.8	&  18.48$\pm$0.03, 19.05$\pm$0.07&	8.40$\pm$0.23, 9.86$\pm$0.46 & 	150.43$\pm$0.77, 153.73$\pm$0.78 &	11.63 (V)	 \\
			 &	2016 Feb 3	&	a	&	2.16	&	16	&	0.5	&  19.41$\pm$0.03				&	4.86$\pm$0.14			&	121.83$\pm$0.85 	&	1.71 (N)	\\
			 &	2017 Mar 18	&	b	&	--	&	1	&	--	&  22.03$\pm$0.10				&	21.15$\pm$0.65			&	144.40$\pm$0.87	&	--	\\  
			 &	2017 Mar 19	&	a	&	--	&	1	&	 --	&  18.96$\pm$0.35				&	12.02$\pm$0.21			&	142.11$\pm$0.51	&	--	\\ 
			 &	2017 Mar 20	&	b	&	1.78	&	4	&	2.4	&  12.78$\pm$0.09				&	14.10$\pm$0.69			&	120.85$\pm$1.36	&	3.70 (N)	\\
			 &	2017 Mar 23	&	b	&	6.10	&	17	&	3.0	&  8.07$\pm$0.04, 8.89$\pm$0.04	& 16.12$\pm$0.20, 19.08$\pm$0.25 & 	130.33$\pm$0.36, 136.46$\pm$0.56 &	31.21 (V)	\\
			 &	2017 Mar 24	&	b	&	6.77	&	16	&	1.6	&  6.46$\pm$0.03, 6.88$\pm$0.04	& 20.41$\pm$0.61, 24.46$\pm$0.64 & 	134.68$\pm$0.47, 144.26$\pm$0.50 &	11.39 (V)	\\
\\ 
OJ 287		&	2014 Feb 20$^\ast$	&	c	&	4.50	&	23	& 0.9	&  5.53$\pm$0.01				&	9.14$\pm$0.09			&	 156.91$\pm$0.28 &	1.75	(N)\\
			&	2014 Feb 27	&	c	&	4.37	&	50	& 2.6	&  7.16$\pm$0.04 				&	17.24$\pm$0.18			&	 168.53$\pm$0.24 &	0.58	(N)\\
			&	2016 Jan 15    	&	b	&	2.62	& 	6 	& 1.7	&  4.99$\pm$0.02				&	2.27$\pm$0.32			&	115.01$\pm$5.53 &	3.36	(N)\\
			&	2016 Jan 15	&	a	&	--	& 	1	& --		&  5.15$\pm$0.01				&	3.32$\pm$0.29			&	49.48$\pm$2.53	&	--	\\
			&	2016 Jan 17	&	b	&	5.62	& 	13	& 3.0 	&  6.02$\pm$0.11, 6.85$\pm$0.13 & 8.56$\pm$0.88, 14.08$\pm$1.33 &	107.65$\pm$3.01, 133.49$\pm$2.18 &	4.43	(V)\\
			&	2016 Feb 7	&	a	&	5.16	& 	28	&  1.7	&  5.62$\pm$0.01, 6.05$\pm$0.01 & 11.45$\pm$0.30, 19.77$\pm$0.28&	141.25$\pm$0.49, 156.06$\pm$0.44 &	85.82 (V)	\\
			&	2016 Apr 3$^\ast$ 	&	b	&	3.36	&	11	& 0.5	&  6.51$\pm$0.02				&	11.31$\pm$0.11			&	120.19$\pm$0.27 &	1.42	(N)\\
			&	2016 Apr 4$^\ast$ 	&	b	&	3.28	&	9	& 1.5	&  7.22$\pm$0.04, 7.60$\pm$0.04	& 11.01$\pm$0.38, 13.40$\pm$0.39&	118.32$\pm$0.83, 122.62$\pm$1.24 &	8.23 (V)	\\
\\
3C 279		&	2015 Mar 30	&	c	&	3.32	&	17	&  1.4	&	 1.68$\pm$0.01				& 	18.95$\pm$0.61 			&	157.64$\pm$0.95 &	 1.33 (N)	\\
			&	2016 April 30	&	a	&	3.50	&	10	&  2.0	&	 2.28$\pm$0.01, 2.43$\pm$0.07& 4.86$\pm$0.46, 8.34$\pm$0.34 	& 	31.16$\pm$2.71, 51.23$\pm$1.89 &	22.85 (V)	\\
			&	2016 May 1	& a, b 	&	3.41	& 	14	&  0.8	&	 2.02$\pm$0.09, 2.09$\pm$0.10& 7.89$\pm$0.39, 9.33$\pm$0.40 	& 	34.13$\pm$1.44, 40.23$\pm$1.23 &	4.88	(V) \\
			&	2016 May 2	&	a	&	2.88	&	12	&  0.4	&	 1.94$\pm$0.01				&	8.61$\pm$0.51			&	35.03$\pm$1.87 &	0.97 (N)	\\
			&	2016 May 3	&	a	&	0.67	&	2	&  3.3	&	 1.92$\pm$0.04				&	8.56$\pm$0.88			&	32.83$\pm$3.02 &	2.39 (N)	\\
\\
CTA 102		&	2015 Oct 13$^\ast$	&	b	&	4.51	&	10	&  2.2	&	 0.92$\pm$0.01				&	5.80$\pm$0.72			&	177.77$\pm$3.48 &	0.694 (N)	\\
			&	2015 Oct 14	& a, b 	&	4.13	&	9	&  4.3	&	 0.76$\pm$0.01				&	4.54$\pm$0.71			&	156.98$\pm$8.87 &	1.778 (N)	\\
			&	2015 Oct 15$^\ast$	& a, b 	&	6.10	&	23	&  1.4	&	 0.71$\pm$0.02, 0.75$\pm$0.01	& 0.12$\pm$1.47, 7.08$\pm$1.55	& 	119.84$\pm$22.88, 198.97$\pm$16.04 &	0.50 (N)	\\
\\
3C 454.3		&	2015 Oct 16	& a, b 	&	6.38	&	41	&  4.0	&	 2.74$\pm$0.03, 3.24$\pm$0.01 & 3.68$\pm$0.52, 8.78$\pm$0.31 	& 	205.36$\pm$1.27, 222.15$\pm$2.71&	43.23 (V)	\\
			&	2015 Oct 17	& a, b 	&	6.77	&	22	&  4.1	&	 2.65$\pm$0.01, 2.99$\pm$0.03 & 1.33$\pm$0.42, 7.78$\pm$1.59 	& 	221.40$\pm$2.96, 245.53$\pm$1.96 &	23.70 (V)	\\
\hline
\end{tabular}
\end{center}
\textbf{Columns:} (1) most common object name; (2) date of observation ($\ast$ indicates simultaneous flux density and polarization monitoring between two observatories); (3) telescope(s) used: (a) AZT-8+ST7 (Russia), (b) LX-200 (Russia), (c) Kanata (Japan); (4) duration of the monitoring; (5) number of data points; (6) amplitude of variability is the square root of normalized excess variance, calculated using Eq.\ 2 of \citet{Abdo10c}; (7) mean of flux density; (8) and (9) mean values (in case of no change) or minimum and maximum when PD or $\chi$ variability was noted, respectively; (10) $F-$value and variability status for the total flux density measurements from the polarization monitoring (Eq. ~\ref{eq:ftest}; Section~\ref{sec:microvar}). For these measurements, however, the $\eta$ is set at 1 as the blazar flux densities are obtained using the comparison star measurements in the standard photometric system. We note that the total flux density intra-night light curves resulting from polarization monitoring are not included in the derivation of flux density microvariability DC because these measurements do not provide the comparison star light curves which are needed to obtain the expected number of false-positive, necessary for the sanity check on the analysis procedure (Section~\ref{sec:microvar}).  \\
\end{table*}

\renewcommand{\thefigure}{\arabic{figure} (Cont.)}
\addtocounter{figure}{-1}

\begin{figure*}
\centering
\hbox{
\includegraphics[width=0.33\textwidth]{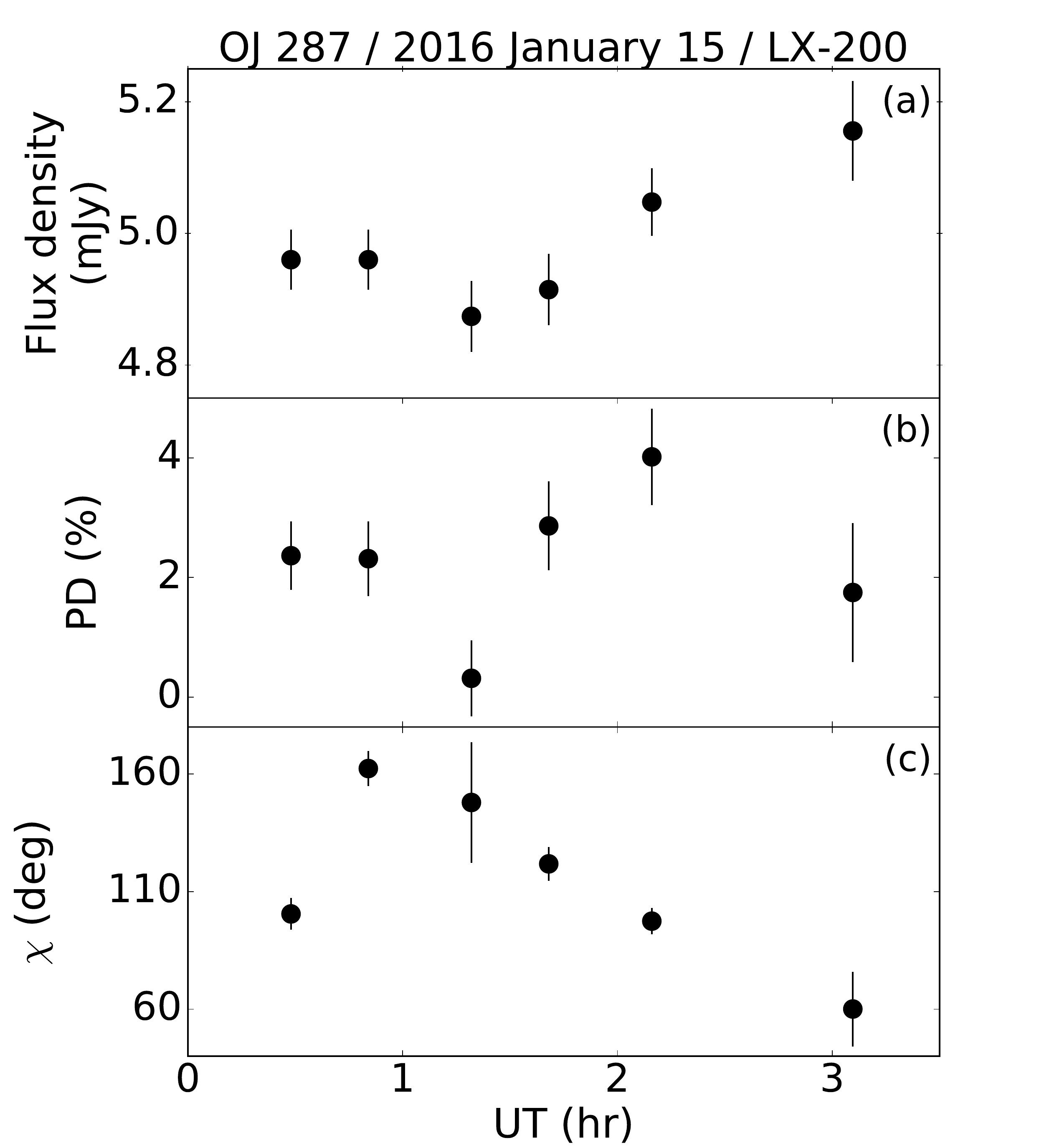}
\includegraphics[width=0.33\textwidth]{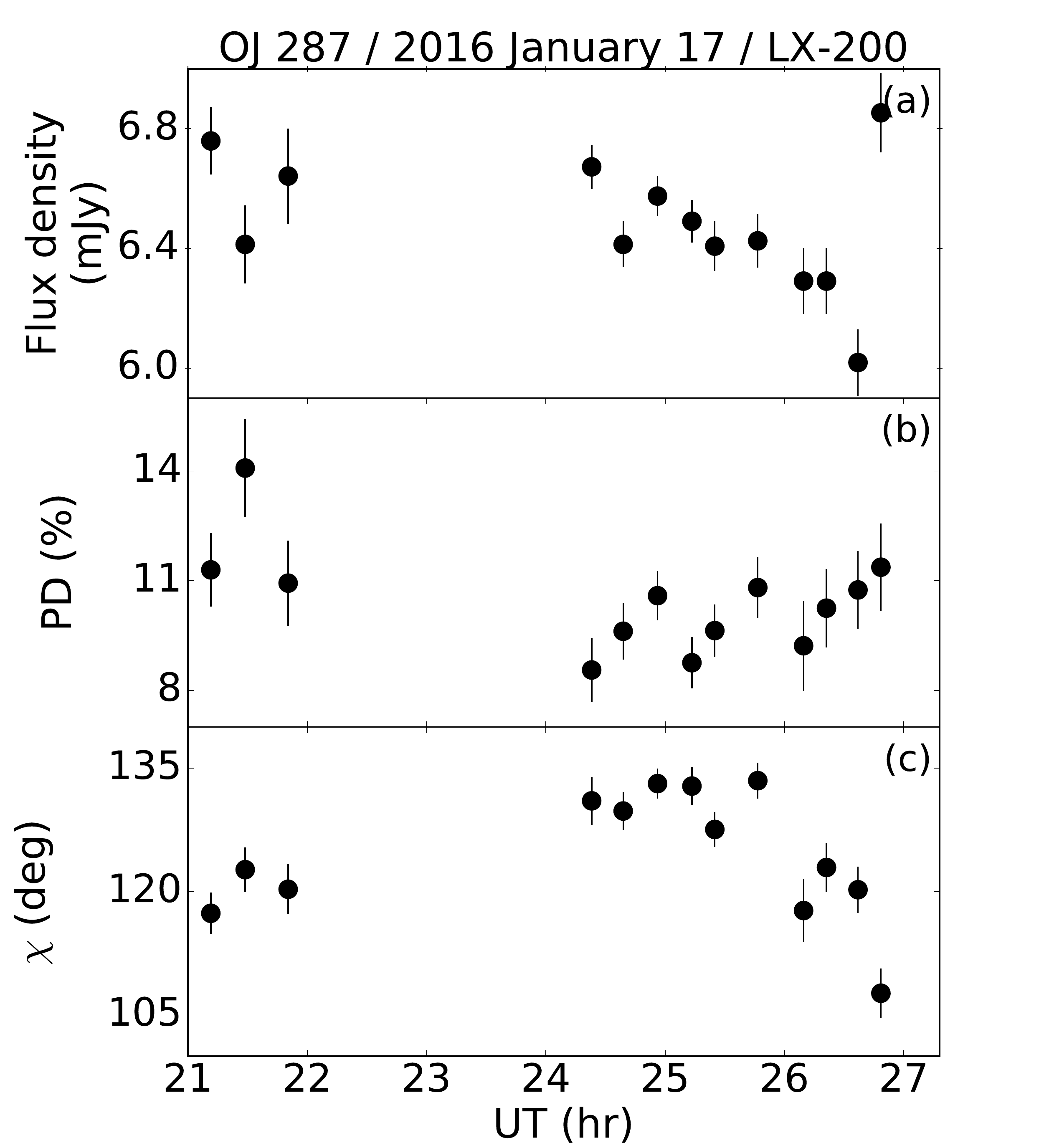}
\includegraphics[width=0.33\textwidth]{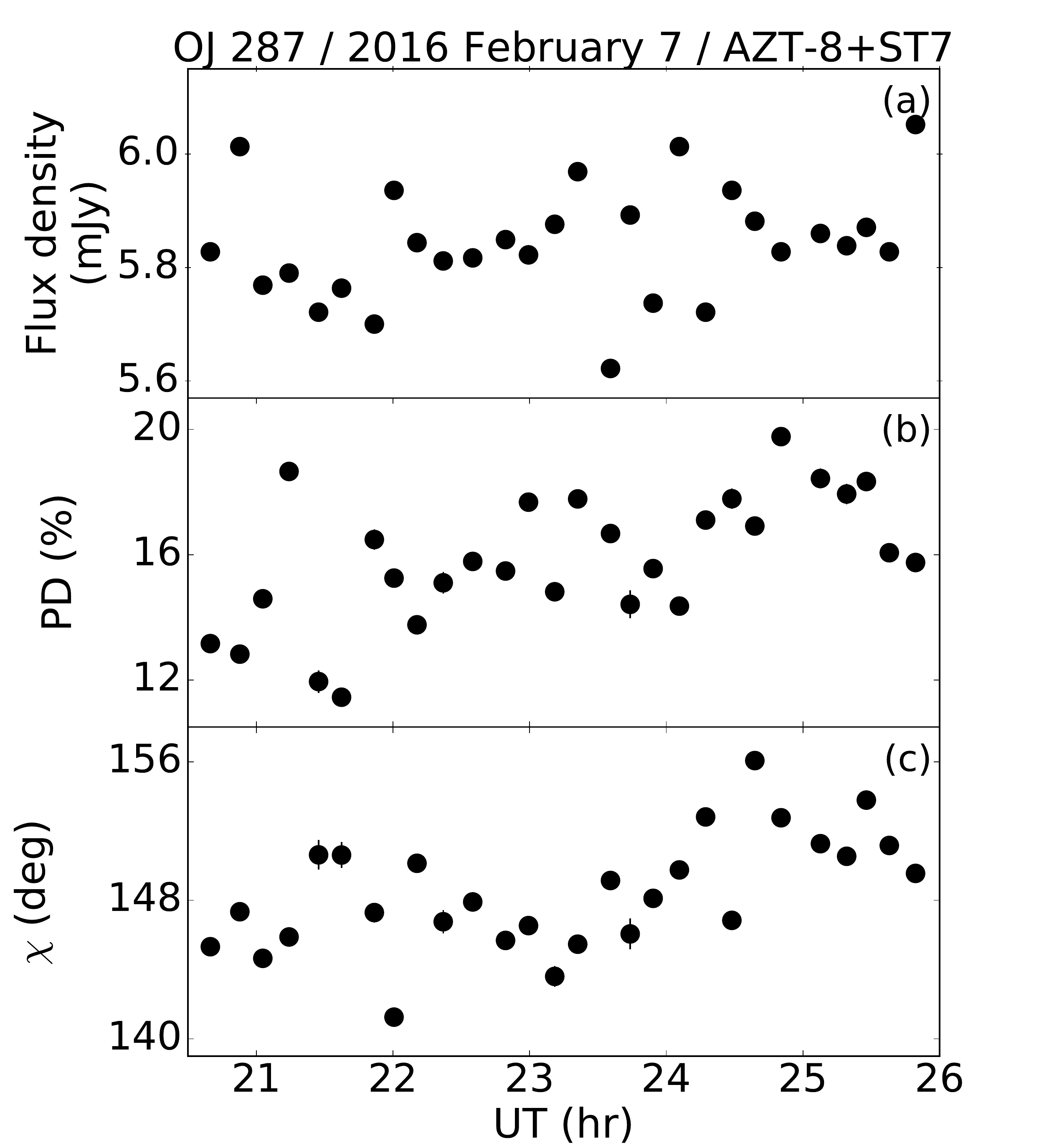}
}

\hbox{
\includegraphics[width=0.33\textwidth]{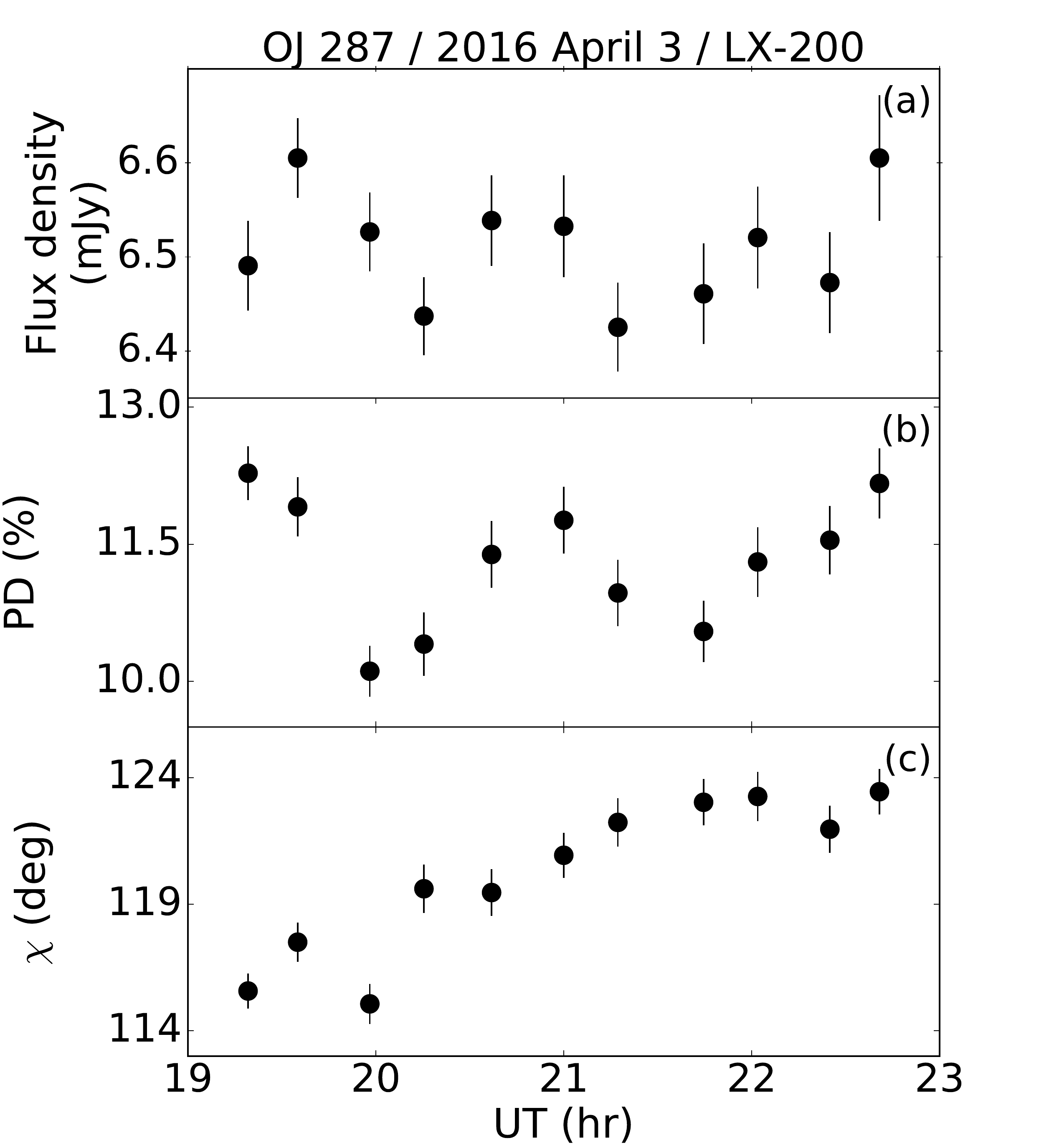}
\includegraphics[width=0.33\textwidth]{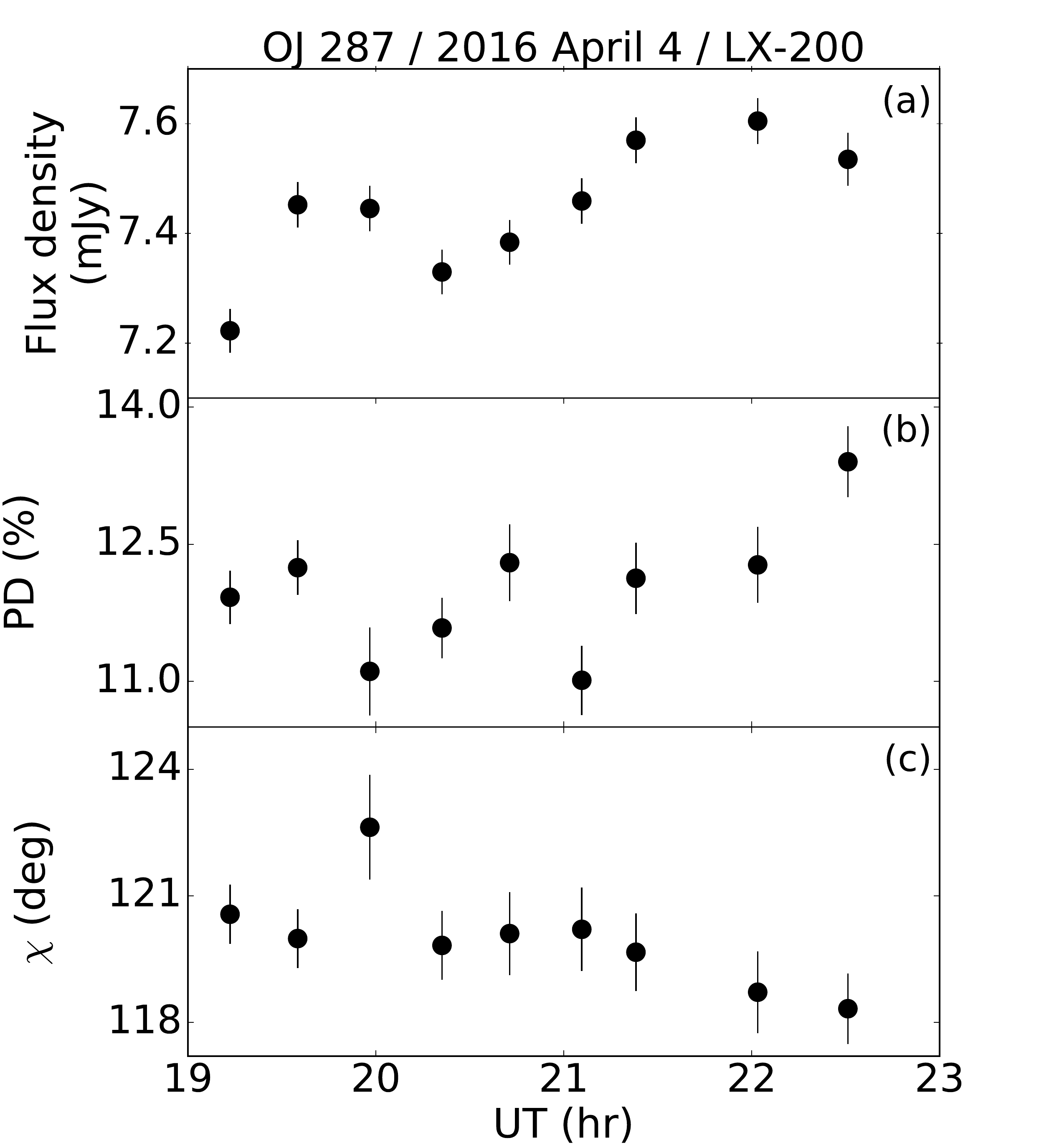}
\includegraphics[width=0.33\textwidth]{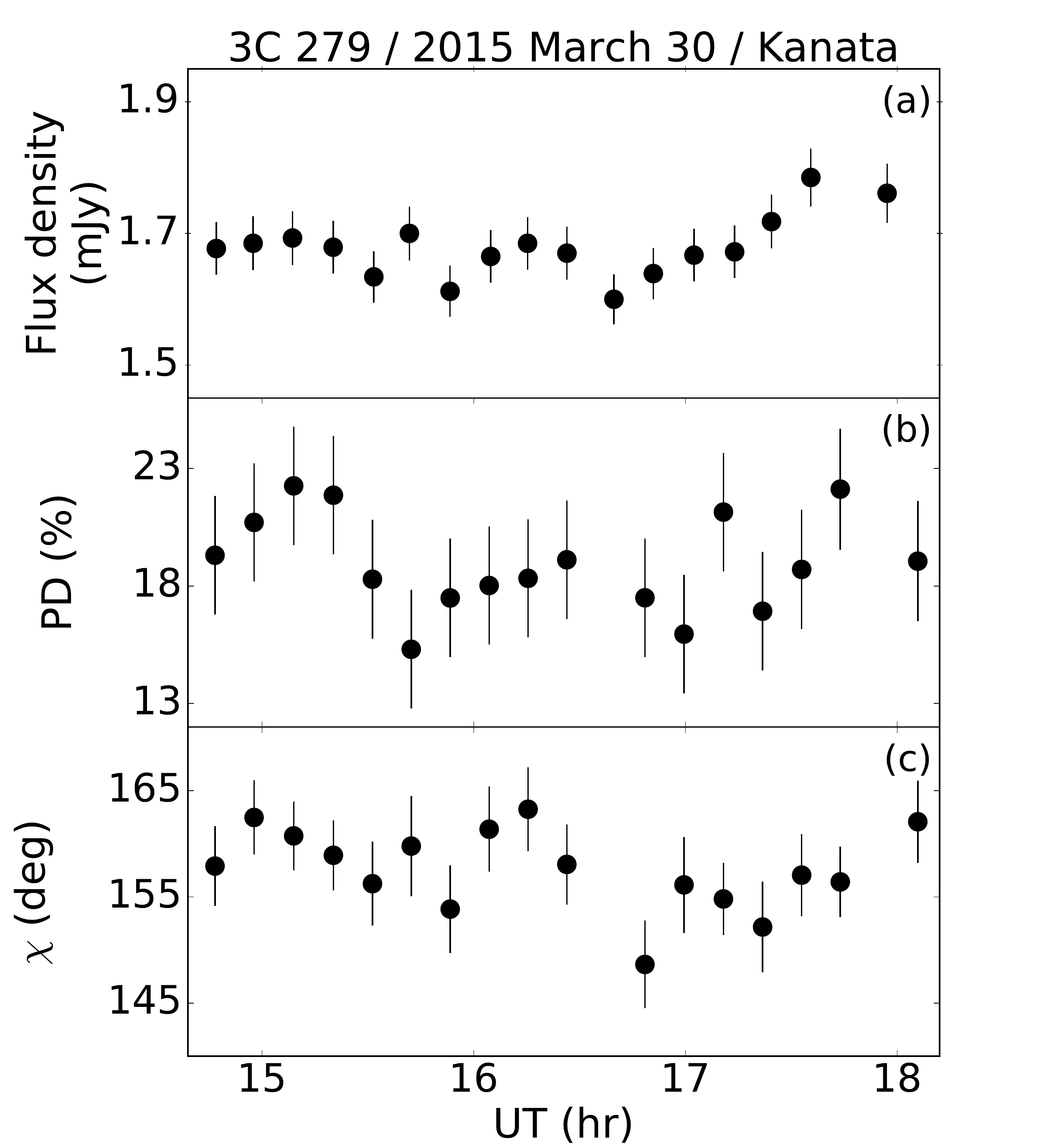}
}

\hbox{
\includegraphics[width=0.33\textwidth]{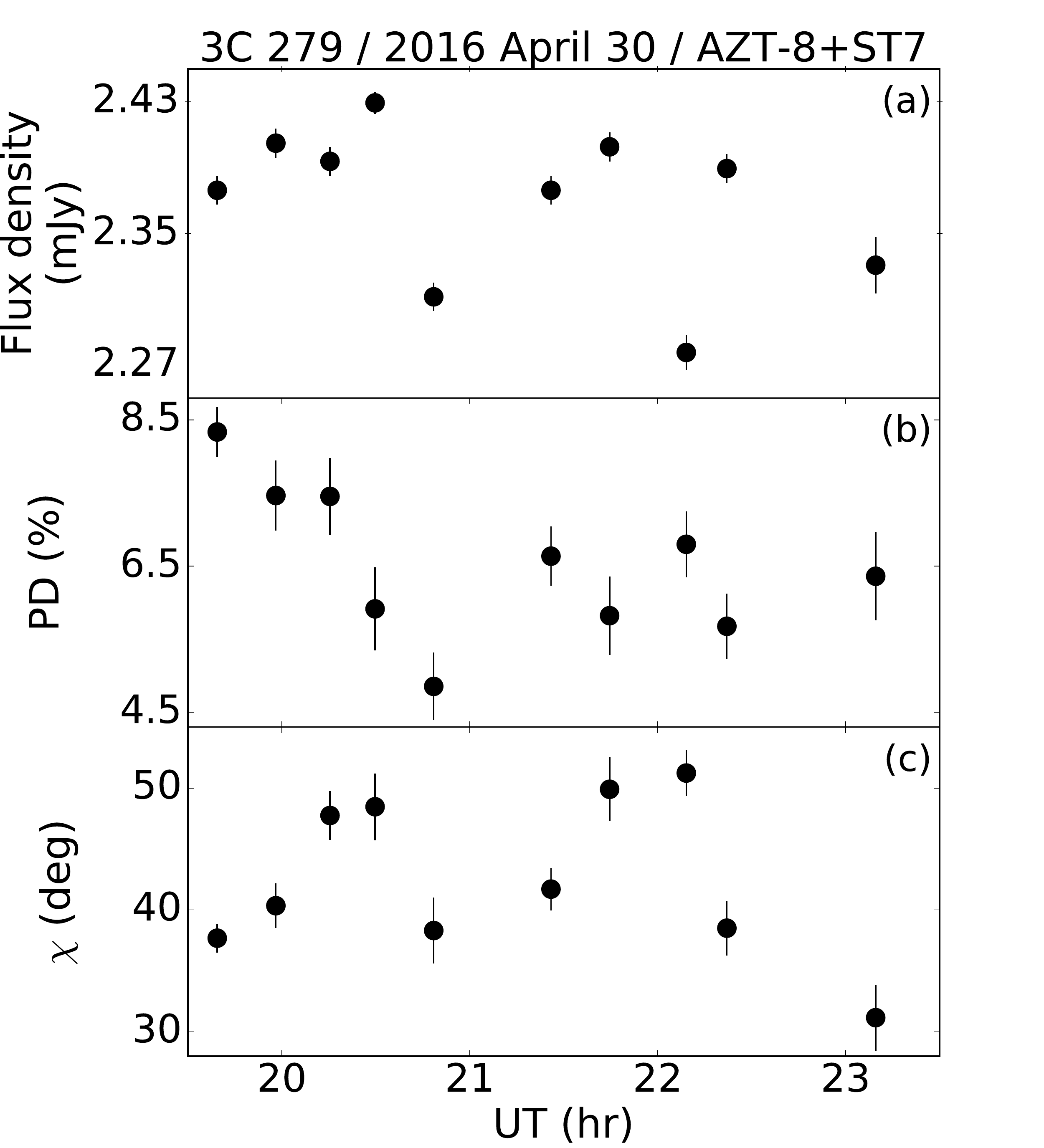}
\includegraphics[width=0.33\textwidth]{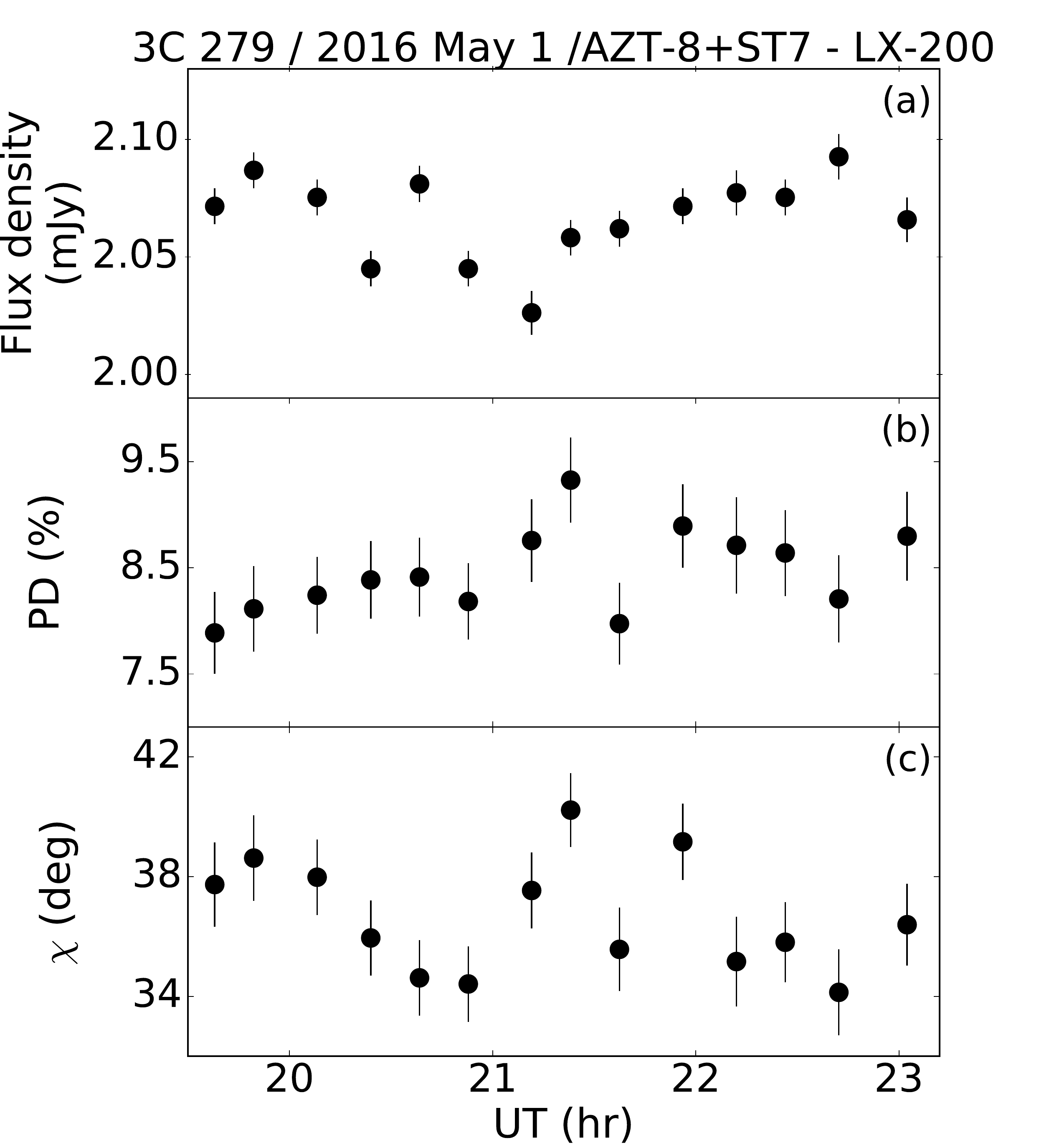}
\includegraphics[width=0.33\textwidth]{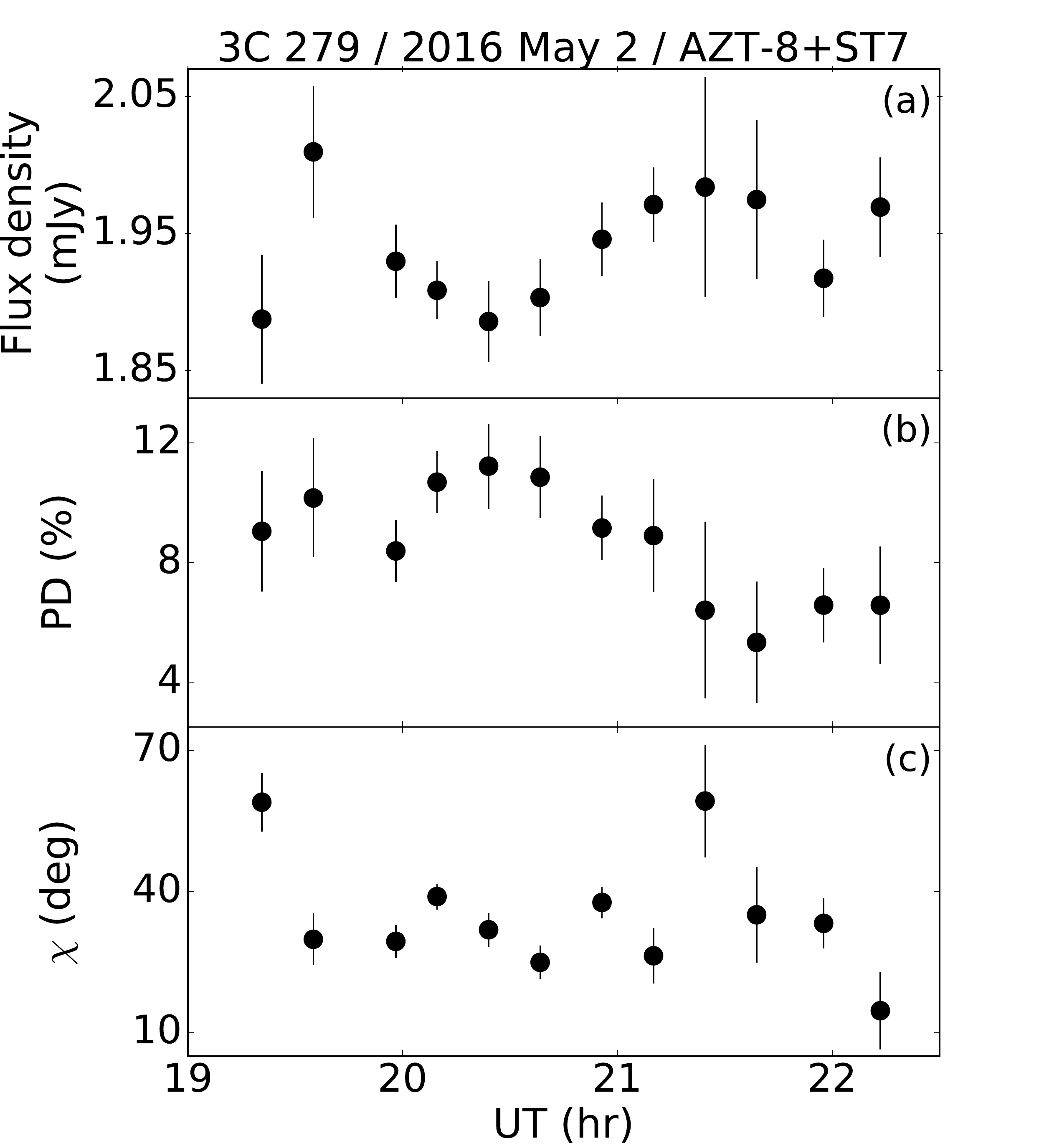}
}

\begin{minipage}{\textwidth}
\caption{
}
\end{minipage}
\end{figure*}

\renewcommand{\thefigure}{\arabic{figure} (Cont.)}
\addtocounter{figure}{-1}

\begin{figure*}
\centering
\hbox{
\includegraphics[width=0.33\textwidth]{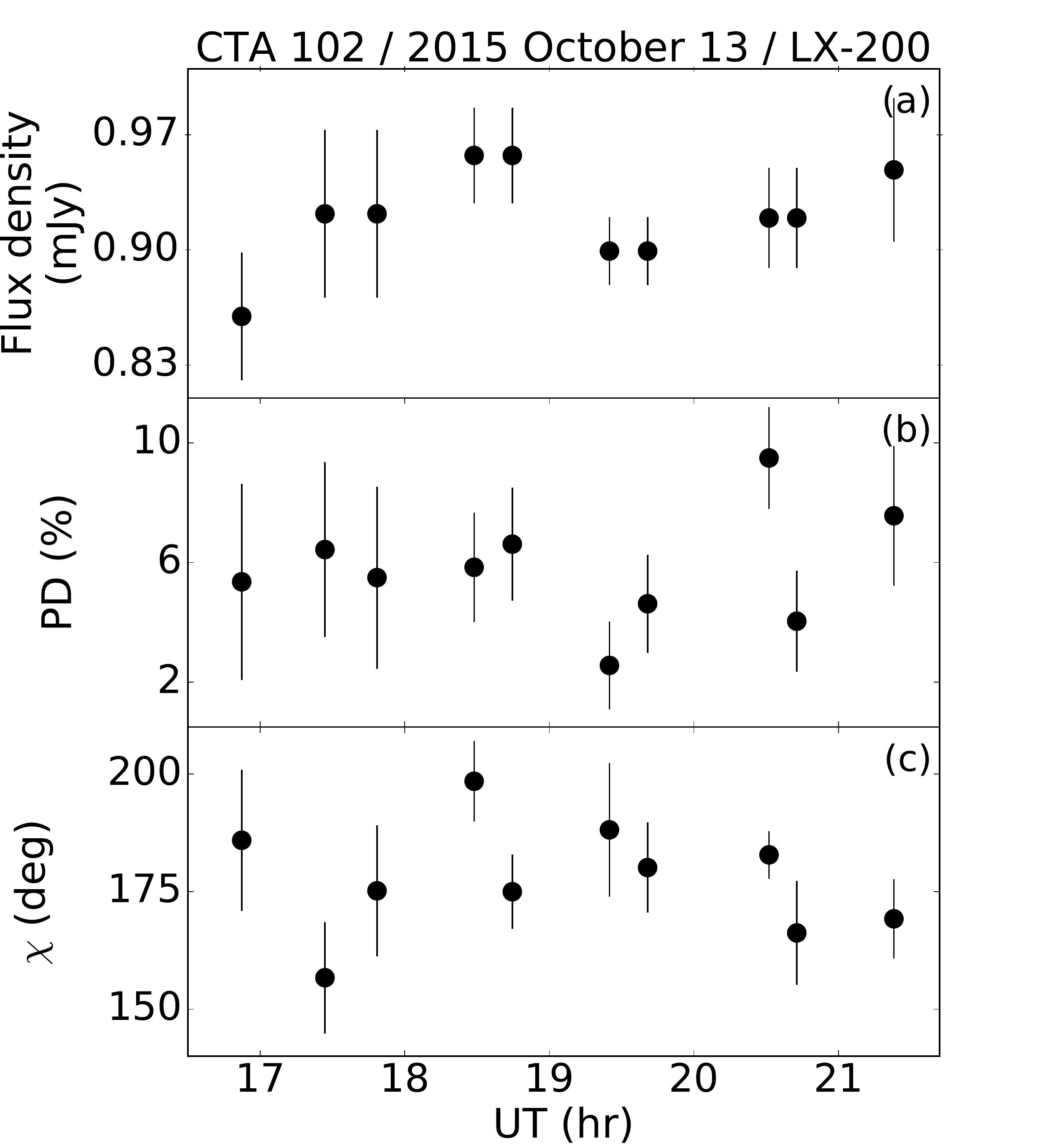}
\includegraphics[width=0.33\textwidth]{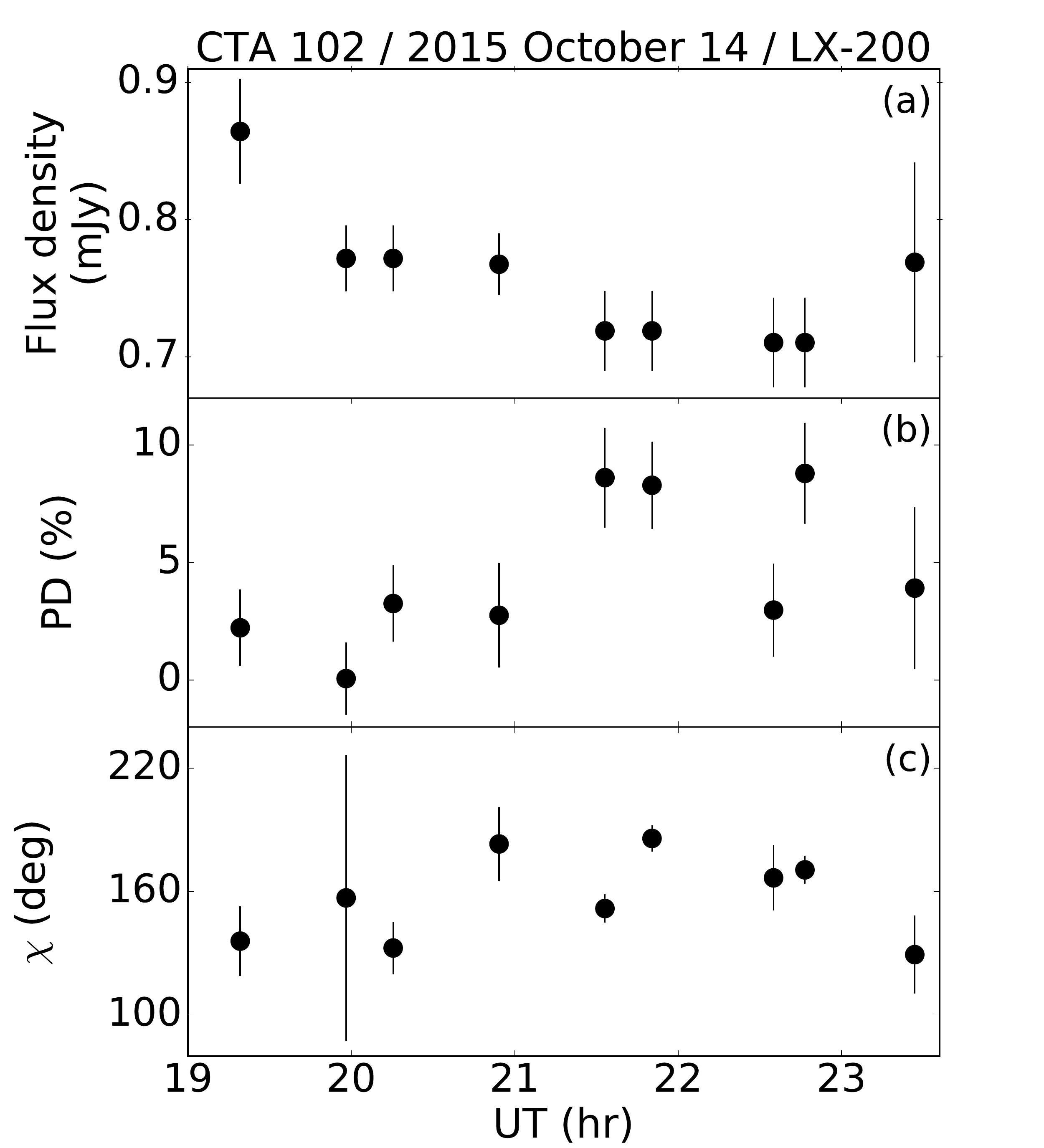}
\includegraphics[width=0.33\textwidth]{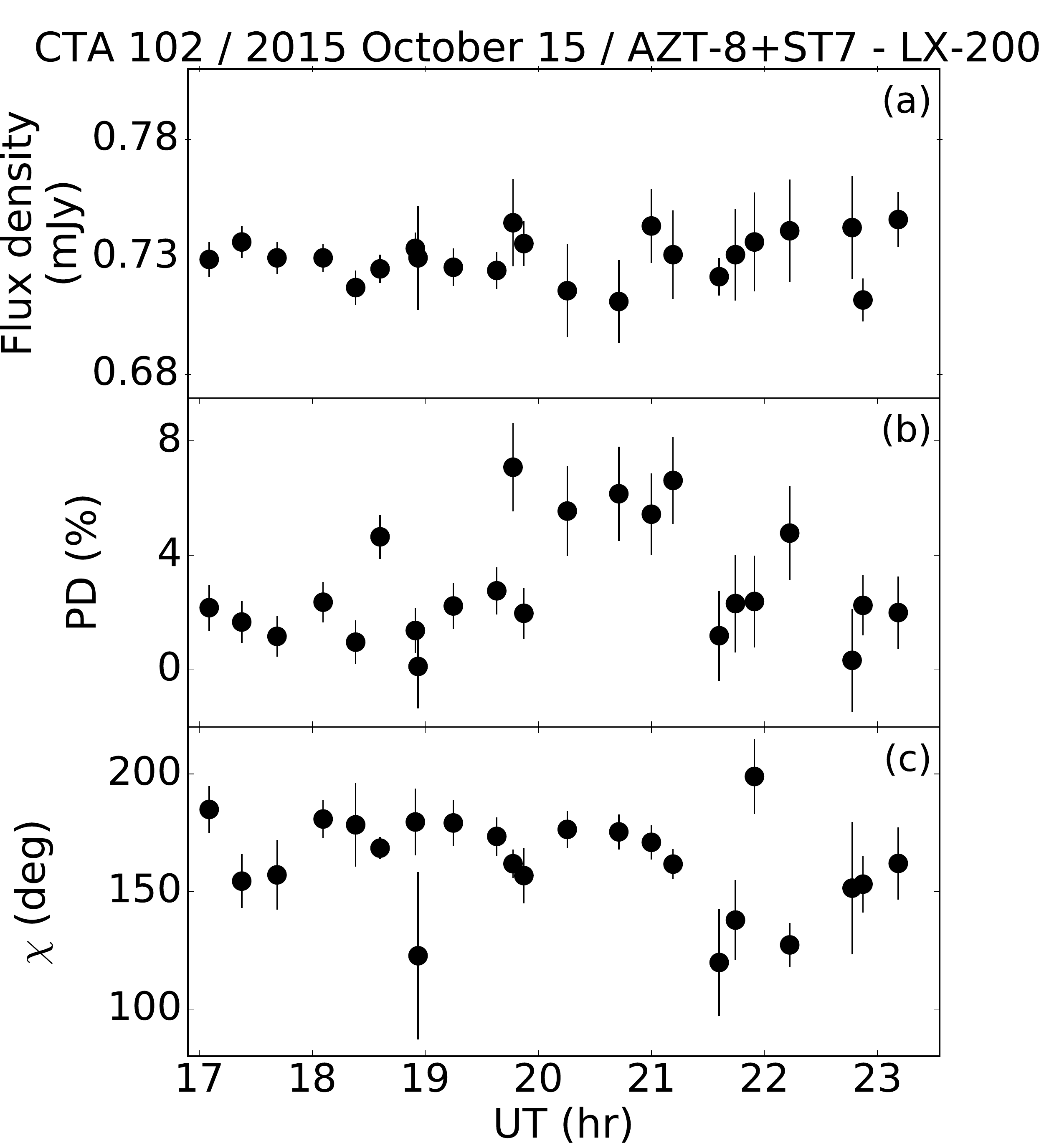}
}
\hbox{
\includegraphics[width=0.33\textwidth]{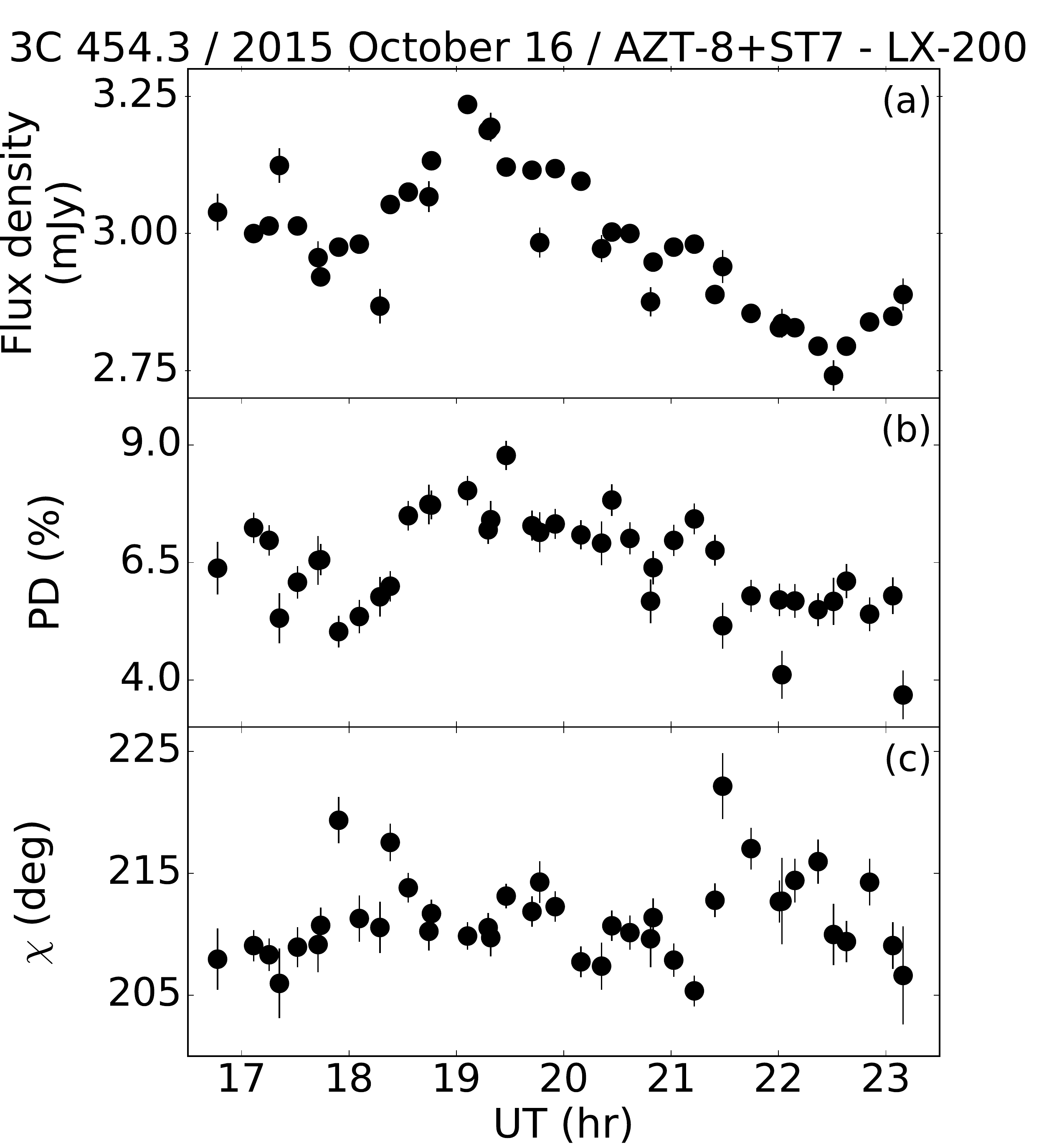}
\includegraphics[width=0.33\textwidth]{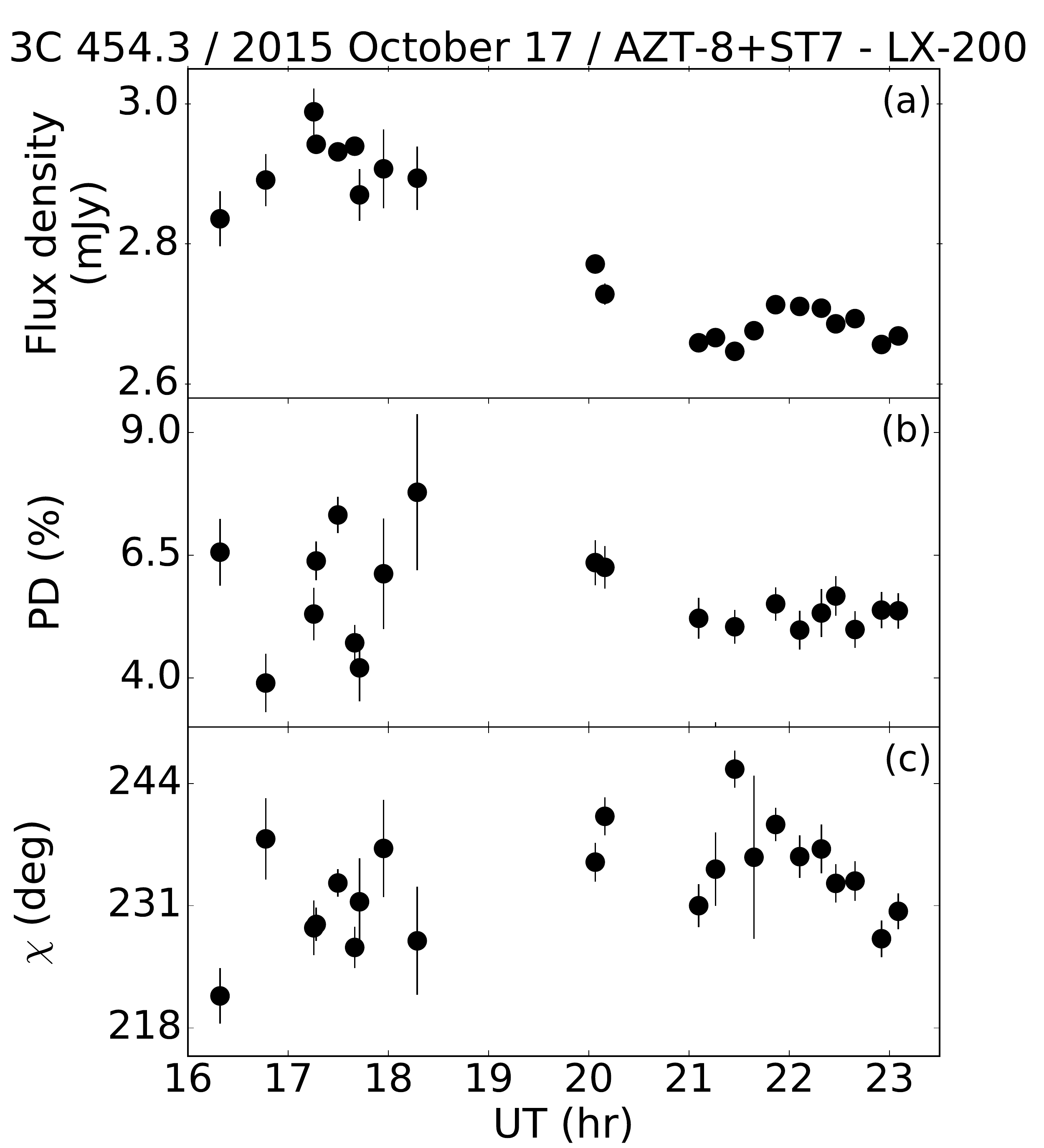}
}
\begin{minipage}{\textwidth}
\caption{}

\end{minipage}
\end{figure*}

\renewcommand{\thefigure}{\arabic{figure}}

\section{Analysis of intra-night light curves}
\label{sec:analysis}
\subsection{Determination of microvariability parameters in the DLCs}
\label{sec:microvar}

Following \citet{Goyal13b}, we used the $F-$test for assigning the INV detection significance. The $F-$statistic compares the observed variance $V_{\rm obs}$ to the expected variance $V_{\rm exp}$. The null hypothesis of no variability is rejected when the ratio
\begin{equation}
F_\nu^\alpha =\frac{V_{\rm obs}}{V_{\rm exp}} = \frac{V_{t-s}}{\langle \eta^2 \, \sigma_{t-s}^2 \rangle} ,
\label{eq:ftest}
\end{equation}
exceeds a critical value for a chosen significance level $\alpha$, for a given number of degrees of freedom (DOF) $\nu$; here $V_{t-s}$ is the variance of the `target-star' DLC, $\langle \sigma_{t-s}^2 \rangle$ is the mean of the squares of the (formal) rms errors of the individual data points in the `target-star' DLC. We note that one-way Analysis of Variance \citep[ANOVA;][see also, \citealt{deDiego14} for the enhanced $F-$test which requires usage of multiple star-star DLCs]{deDiego10} could also be used to test for microvariability and has some advantages; however,  a robust implementation of this approach requires a sufficiently large number of data points ($>$30).  Since our aim was to study colour evolution along with polarization on intra-night timescales, the number of data points in a single optical band rarely satisfied this condition. Therefore, we use the $F-$test to assign the INV detection in the DLCs. Since this method requires flux density or magnitude estimates along with their error estimates, it is important to determine the photometric errors accurately. As emphasized in several independent studies, the photometric errors returned by {\sl APPHOT} are significantly underestimated (by factors of $\eta =$ 1.3--1.7; \citealt{GK95, Garcia99, Stalin04, Bachev05, Zhang18}). \citet{Goyal13a} obtained $\eta$ =1.54$\pm$0.05, using 262 steady star-star DLCs and involving intra-night observations with three different telescopes located in India. Thus, $\eta=$1.54 has been used in the present analysis to scale up the photometric magnitude errors returned by {\sl IRAF}. 

The significance level set for a given test determines the {\it expected} number of {\it false positives}, which is an indicator of the robustness of the test. We have chosen two significance levels, $\alpha =  0.01$ and 0.05, corresponding to $p-$values of $\gtrsim 0.99$ and $\gtrsim 0.95$, respectively. Since the smaller the value of $\alpha$ is, the less likely it is for the variability to occur by chance, thus, a genuine INV detection is claimed, i.e., a `variable' designation (V) is assigned, if the computed statistic value is above the critical value corresponding to $p > 0.99$ (i.e., $\alpha=$ 0.01) for a given degree of freedom ($\nu = N_p - 1$, where $N_p$ stands for the number of data points in a given DLC). We assign a `probable variable' designation (PV) when the computed test statistic value is found to be between the critical values at $\alpha = $ 0.01 and 0.05; otherwise, a `non-variable' (N) designation is assigned to a DLC, {\it though of course variability could still be present but at a lower level of significance.} All the three DLCs, i.e., BL$-$S1, BL$-$S2, and S1$-$S2, are subjected to the $F-$test analysis. In a few cases, the  microvariability status was different for the two blazar-star DLCs, and this indicated a small amplitude variation of one or the other comparison stars was likely to be present. Since such small amplitude variations in star-star DLCs are difficult to ascertain, we conservatively only ascribed a ``V'' status if both blazar-star DLCs gave a ``V'' status; otherwise, we quote a``PV'' or ``N'' status if the star-star DLC itself turned to be variable. Table~\ref{results_finv} summarizes the analysis results for the seven blazars (except 3C\,279) for which the multiband flux density monitoring was carried out. 

Following \citet{Romero99} the peak-to-peak microvariability amplitude was calculated as 
\begin{equation} 
\psi= \sqrt{({D_{\rm max}}-{D_{\rm min}})^2-2\sigma^2} \, ,
\label{eq:psi}
\end{equation}
with $D_{\rm min/max}$ denoting the minimum/maximum in the DLC of the source, and $\sigma^2= \eta^2 \, \langle\sigma^2_{i}\rangle$ where $\eta =1.54$ \citep{Goyal13a} and $\sigma_i$ is the nominal error associated with each data point. 

The microvariability duty cycle (DC) was computed according to
\begin{equation} 
DC = 100\% \,\,\, \frac{\sum_{j=1}^n N_j \, (1/\Delta t_j)}{\sum_{j=1}^n (1/\Delta t_j)} \, ,
\label{eq:DC}
\end{equation}
where $\Delta t_j = \Delta t_{j,\, {\rm obs}} \, (1+z)^{-1}$ is the duration of the monitoring session of a source on the $j^{th}$ night, corrected for the cosmological redshift $z$, and $N_j$ is set equal to 1 if microvariability was detected, and otherwise to 0 \citep{Stalin04}. This estimate is essentially the ratio of the number of nights a source is found to be variable to the total number of nights it was monitored; $\Delta t_j$ is used to weight to the monitoring duration for the evaluation of DC. We compute the  DC using intra-night light curves in V-filter because it was the passband used for monitoring all the nights in this study. The computed microvariability DC for the light curves in V-band, consisting of 18 nights on 7 blazar sources, is 45\% (55\% if  PV cases are also included).

We have performed a sanity check by computing the number of `Type 1 errors', or the false positives, for our data set. A false positive arises due the rejection of a true null hypothesis by a test, when applied to a non-varying DLC and is solely dependent on the $\alpha$ value set for the test and the number of DLCs. We note that for our data set consisting of 18 steady star-star DLCs in V-filter, the means of the {\it expected} numbers of false positives are $\simeq 0.2$ and $\simeq 1$ for $\alpha=$ 0.01 and 0.05, respectively. Since the distribution of false positives is expectedly binomial, for $\alpha =$ 0.01 the number of false positives should in fact be scattered between 0 and 2, and for most of the cases around $\simeq 0.3\pm 0.5$. Similarly, with $\alpha = 0.05$, the number of false positives should lie between 0 and 5, and should largely cluster at $\simeq 1 \pm 1$. Meanwhile, the {\it observed} numbers of false positives reported by the application of the $F-$test (see column~10 of Table~\ref{results_finv}) is 0 for $\alpha =$ 0.01 and 0 for $\alpha =$ 0.05 (V-filter). The good agreement between the {\it expected} and the {\it observed} numbers of false positives provides validation for our analysis procedure.

\subsection{Analysis of $\alpha_{\nu_1}^{\nu_2}$ --flux density plane on intra-night timescales}
\label{sec:cmd}

\begin{figure*}
\centering
\hbox{
\includegraphics[width=0.33\textwidth]{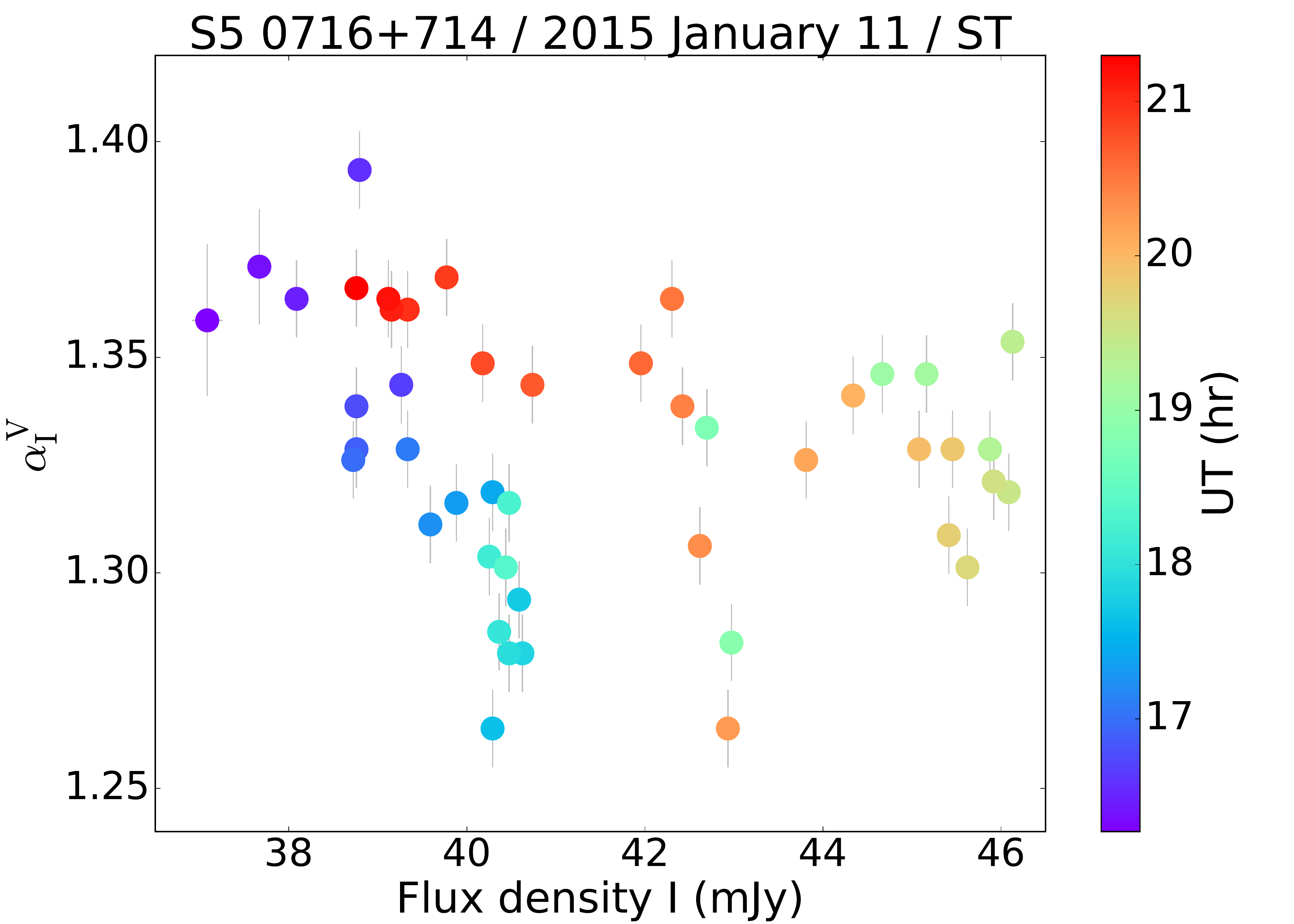}
\includegraphics[width=0.33\textwidth]{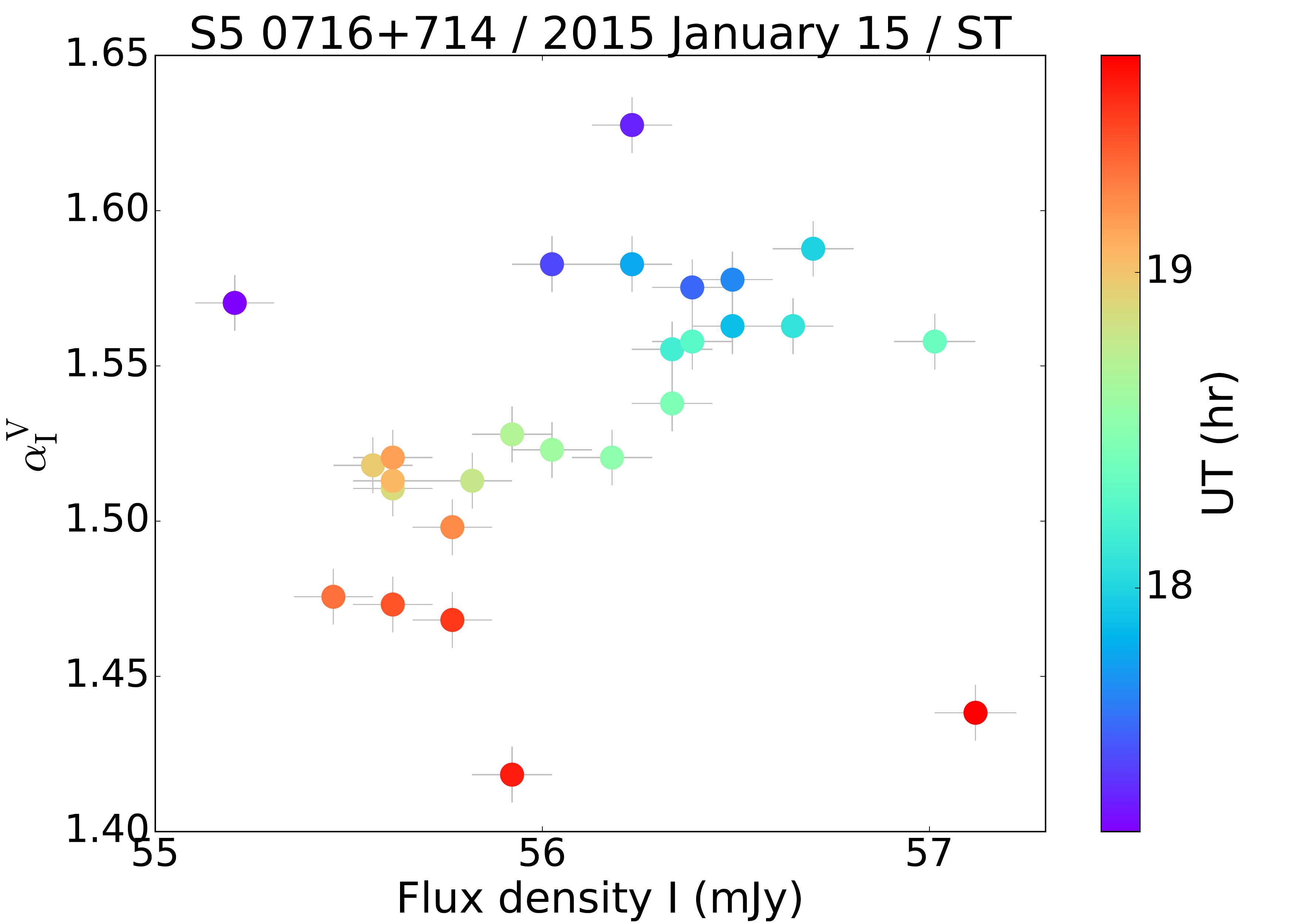}
\includegraphics[width=0.33\textwidth]{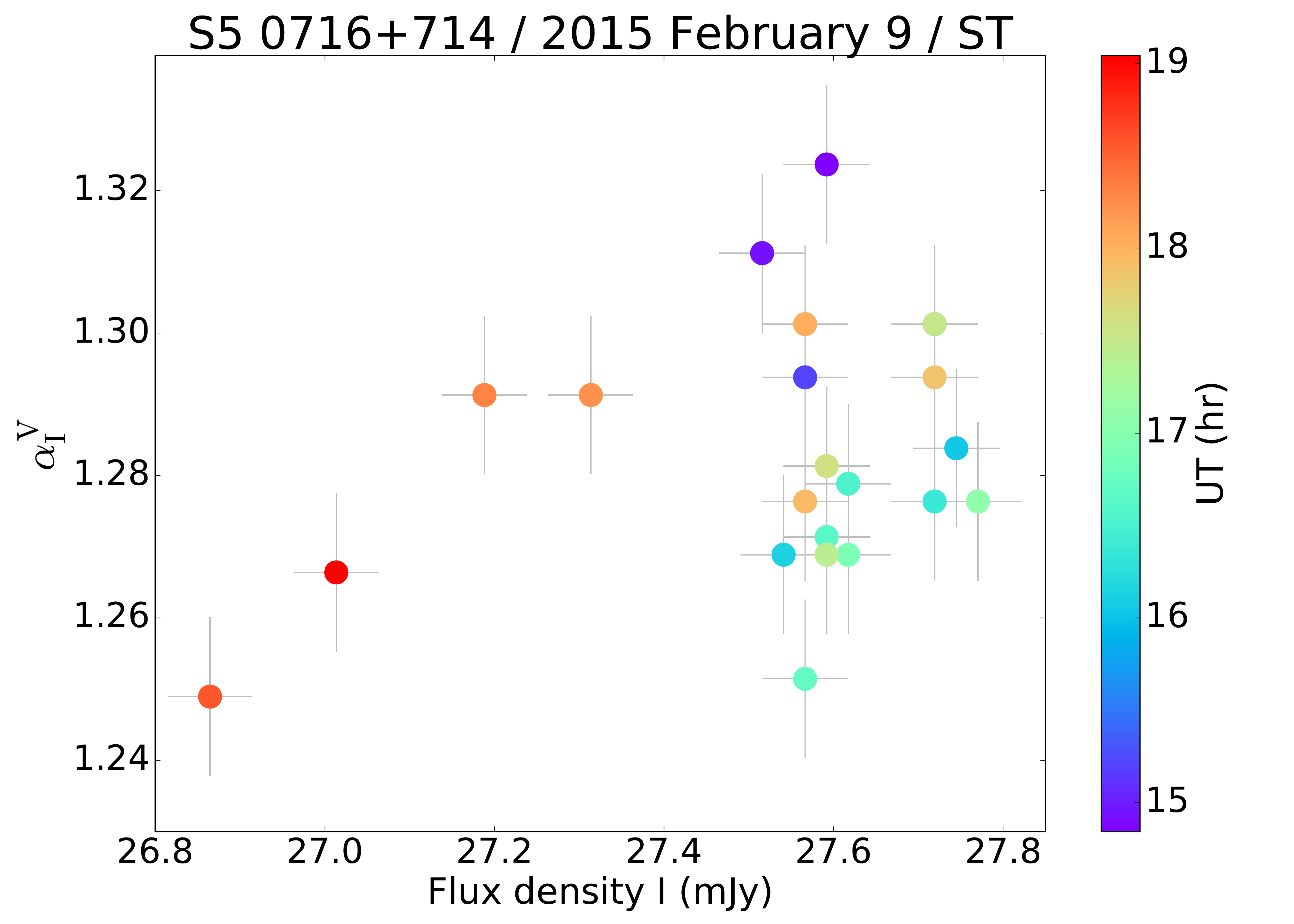}
}
\hbox{
\includegraphics[width=0.33\textwidth]{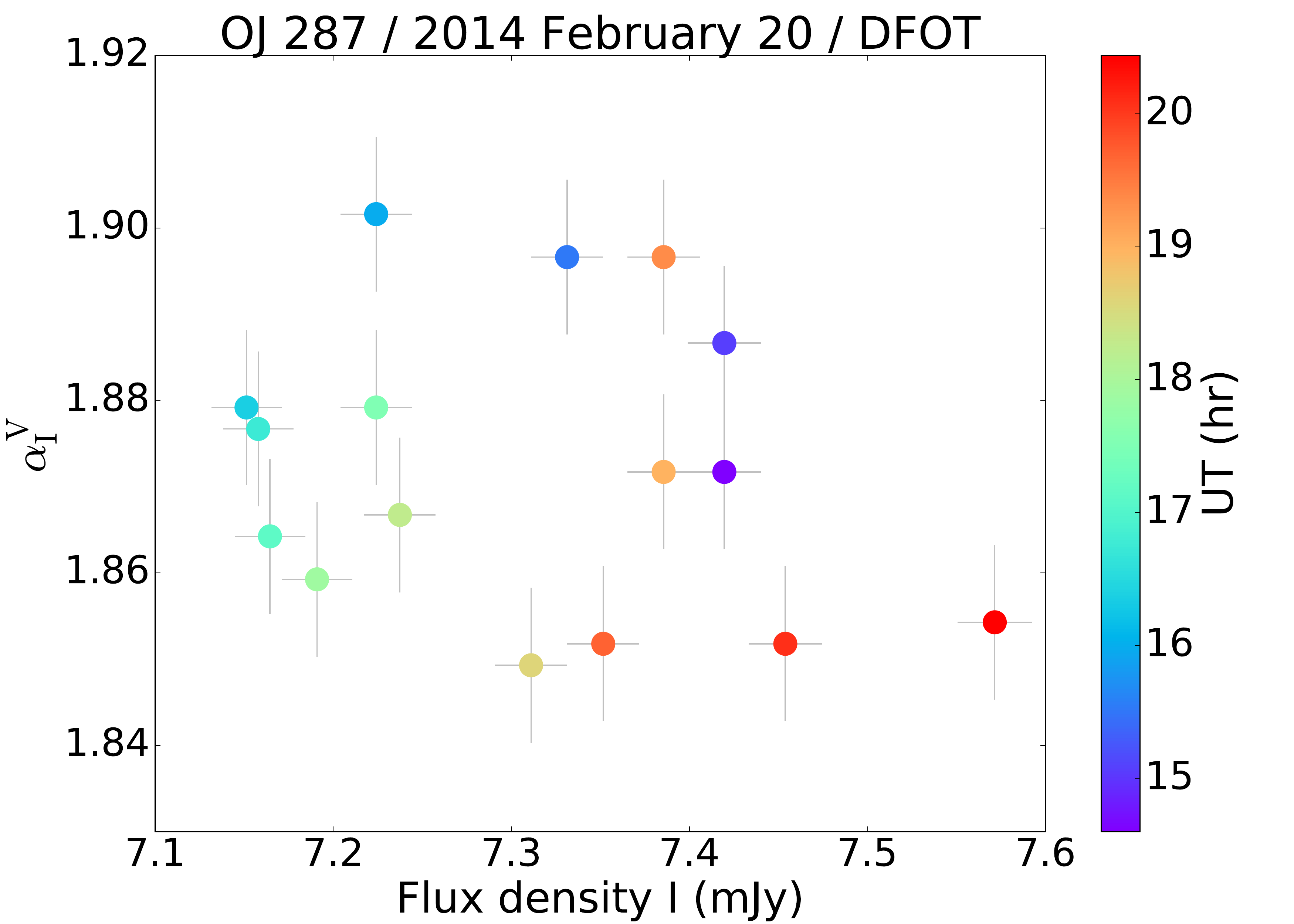}
\includegraphics[width=0.33\textwidth]{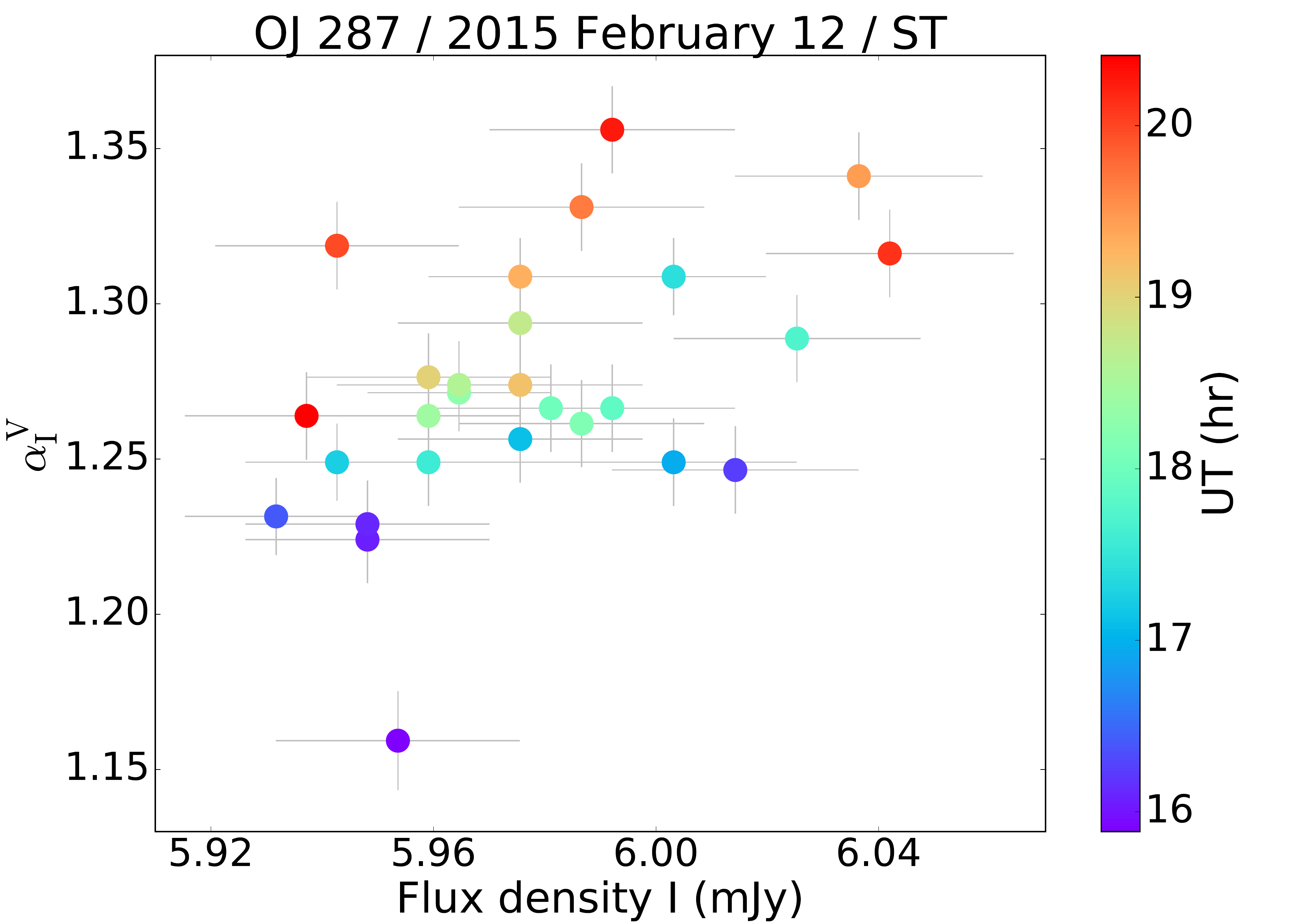}
}

\begin{minipage}{\textwidth}
\caption{$\alpha$ vs. I-band flux density evolution for sessions when colour microvariability was detected (Table~\ref{results_cinv}).}

\label{fig_cmd}%
\end{minipage}
\end{figure*}

Figure~\ref{fig_cmd} shows the evolution of spectral index as function of flux density on the monitoring sessions when significant flux variability was detected (see, Table~\ref{results_cinv}).

\subsection{Analysis of intra-night polarization light curves: flux density--PD, $\chi-$flux density, and PD--$\chi$ planes}
\label{sec:polarization}

\begin{figure*}
\hbox{
\includegraphics[width=0.33\textwidth]{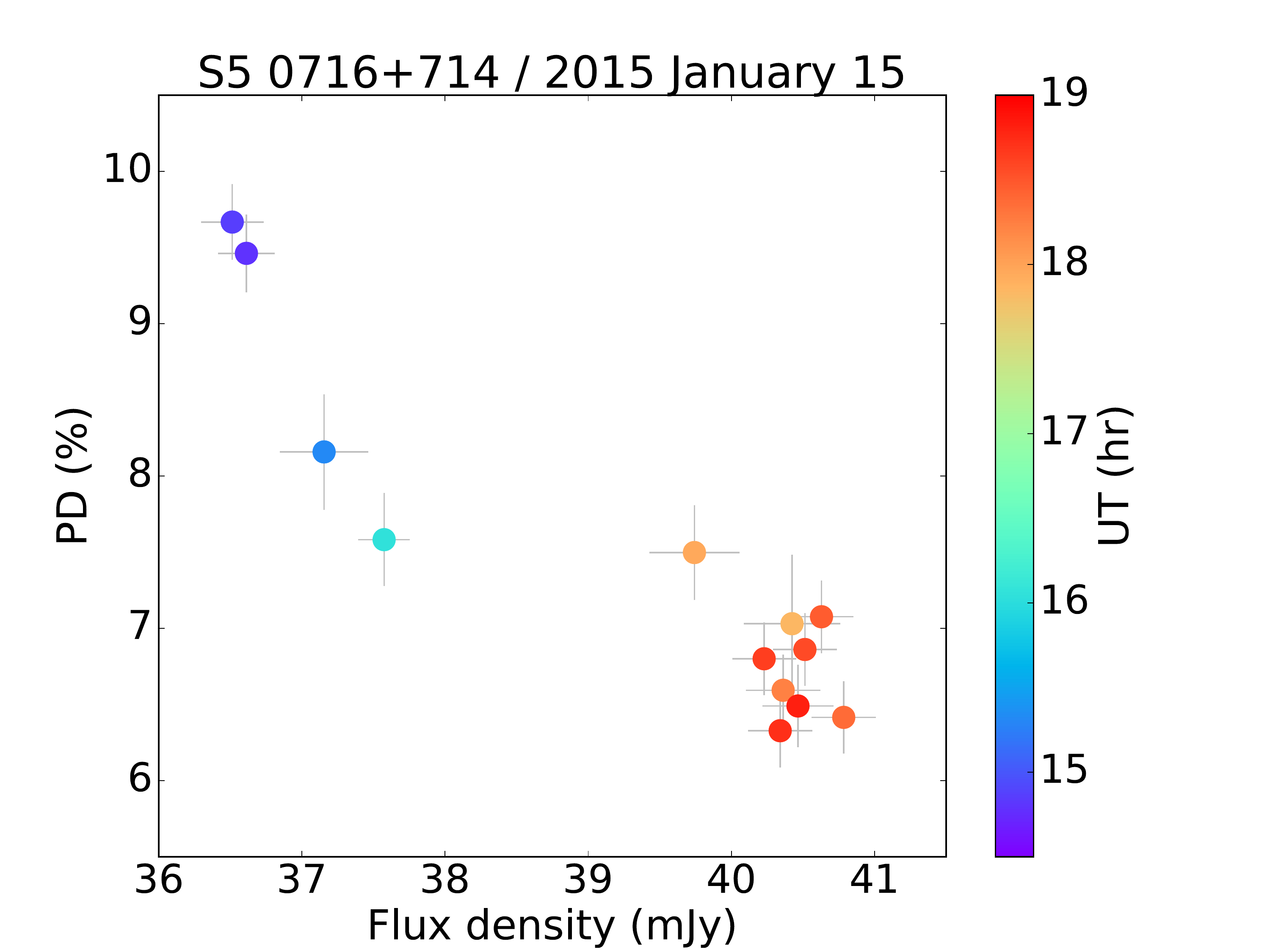}
\includegraphics[width=0.33\textwidth]{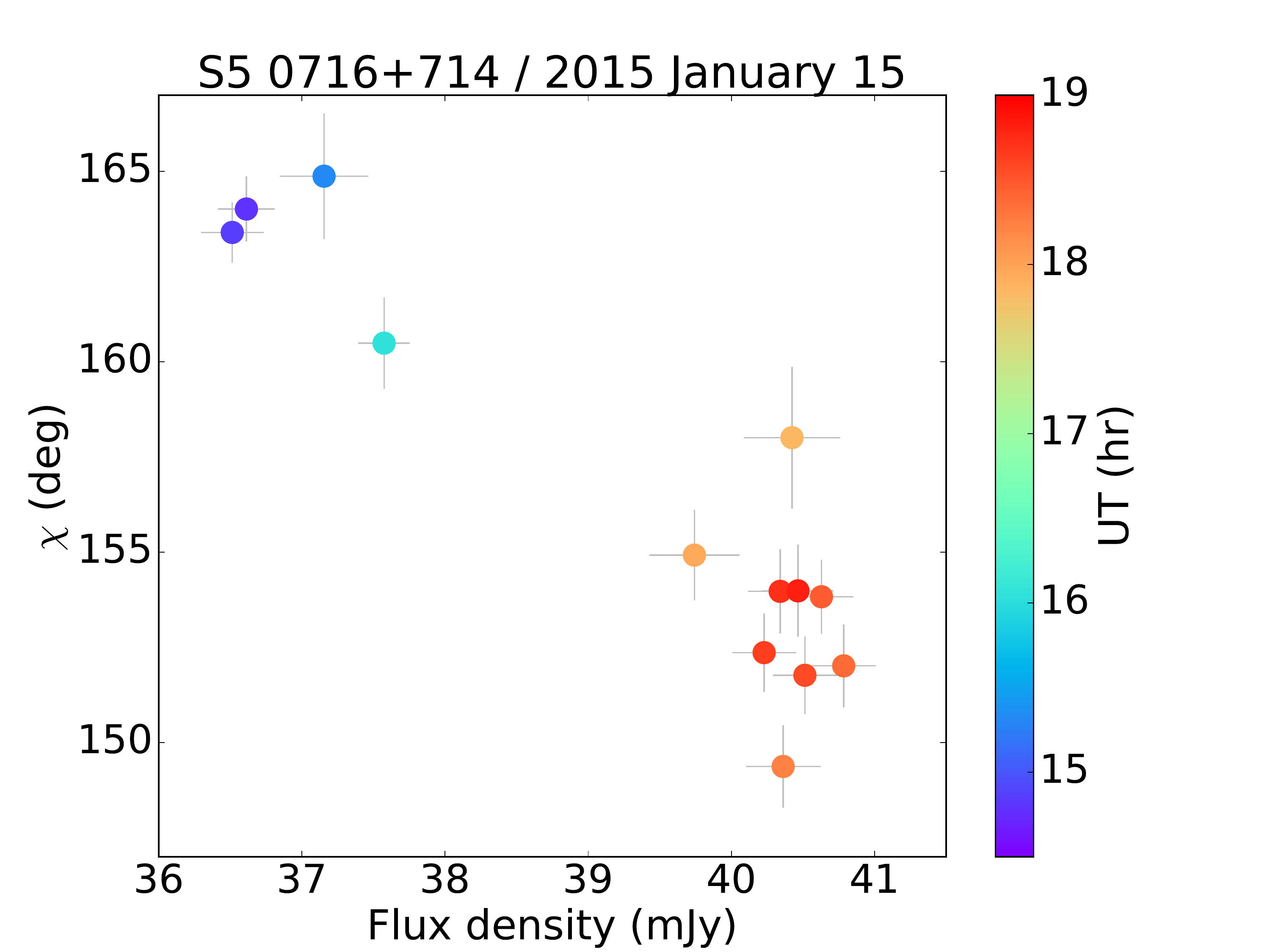}
\includegraphics[width=0.33\textwidth]{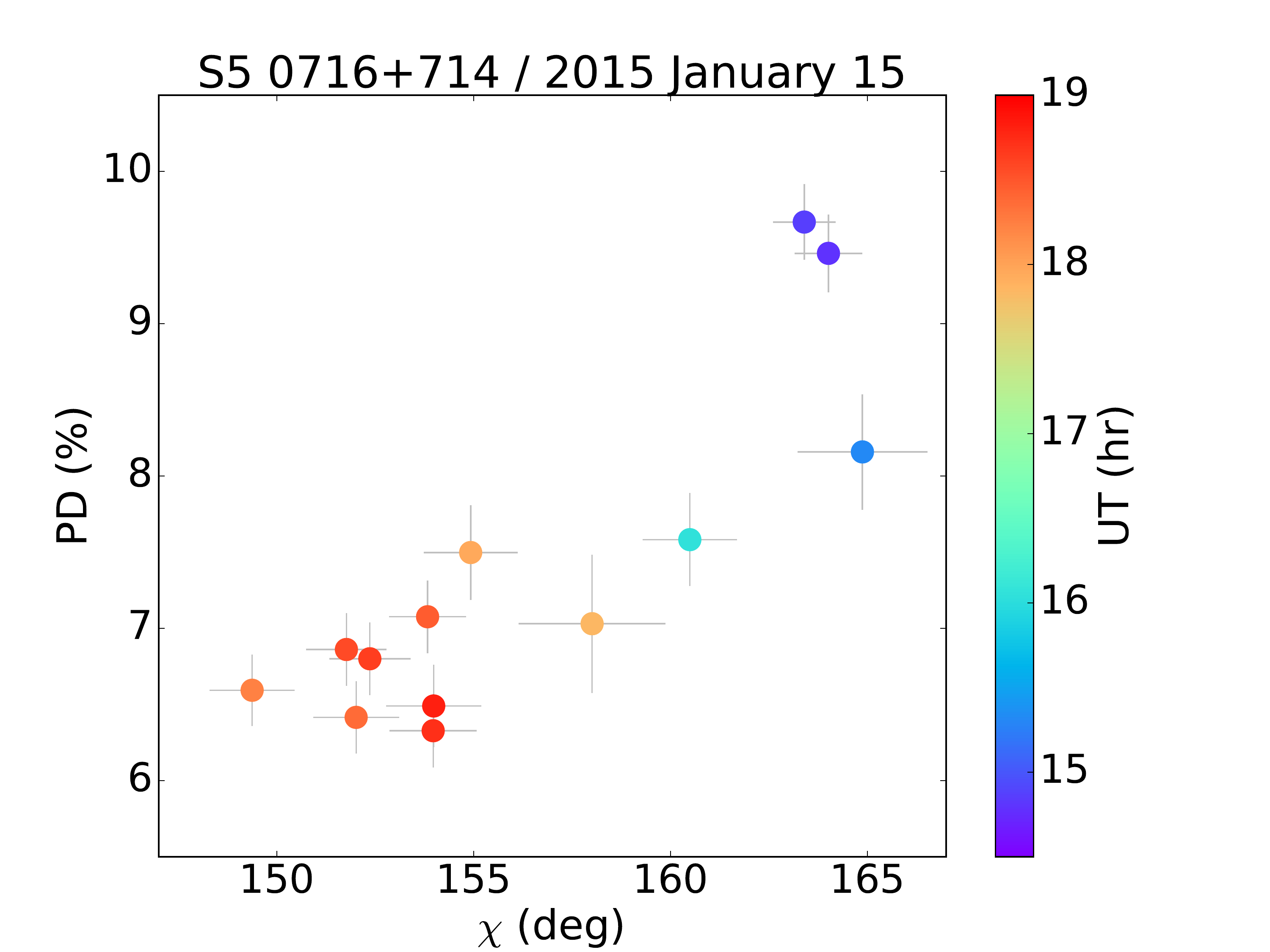}
}
\hbox{
\includegraphics[width=0.33\textwidth]{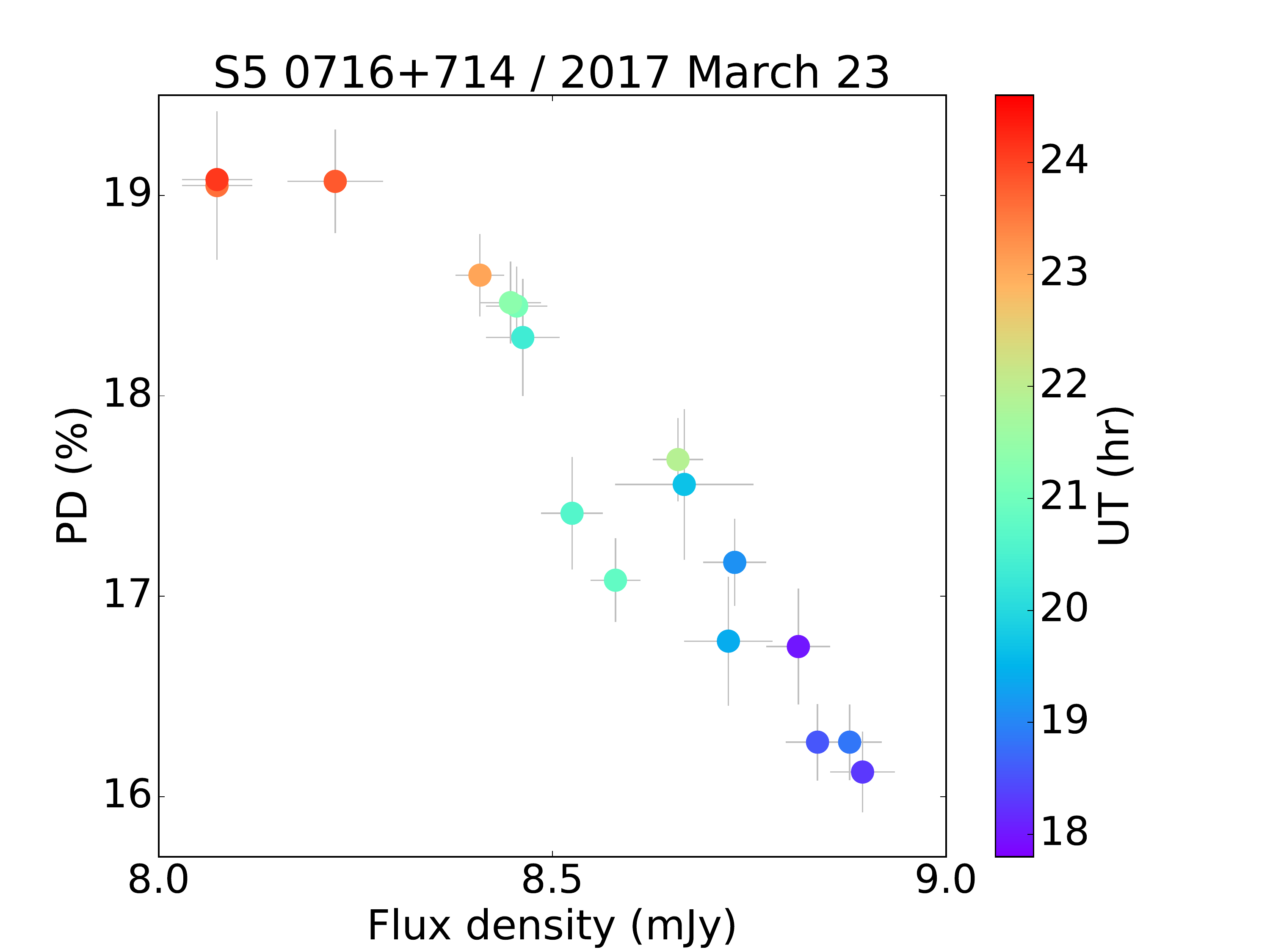}
\includegraphics[width=0.33\textwidth]{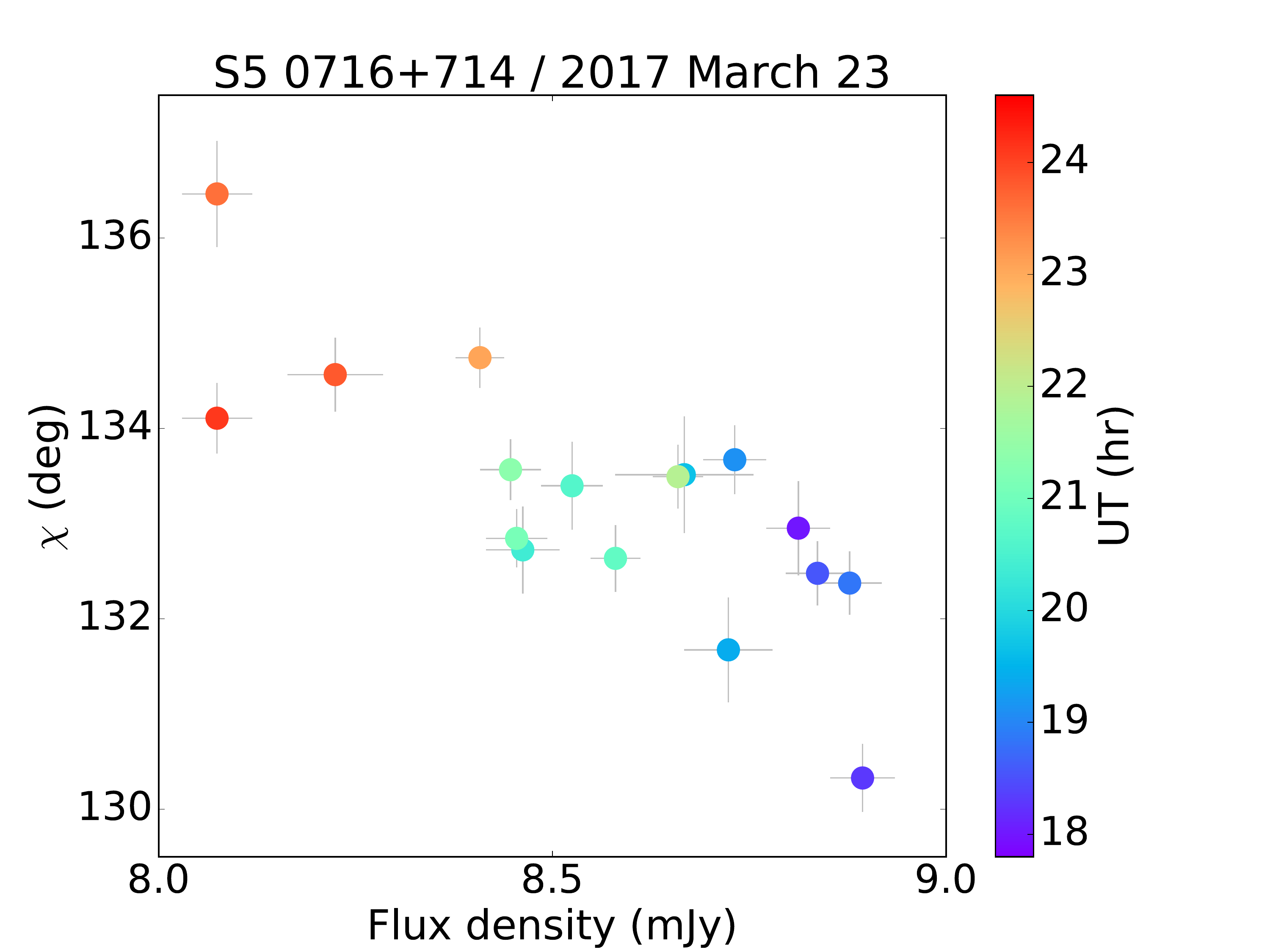}
\includegraphics[width=0.33\textwidth]{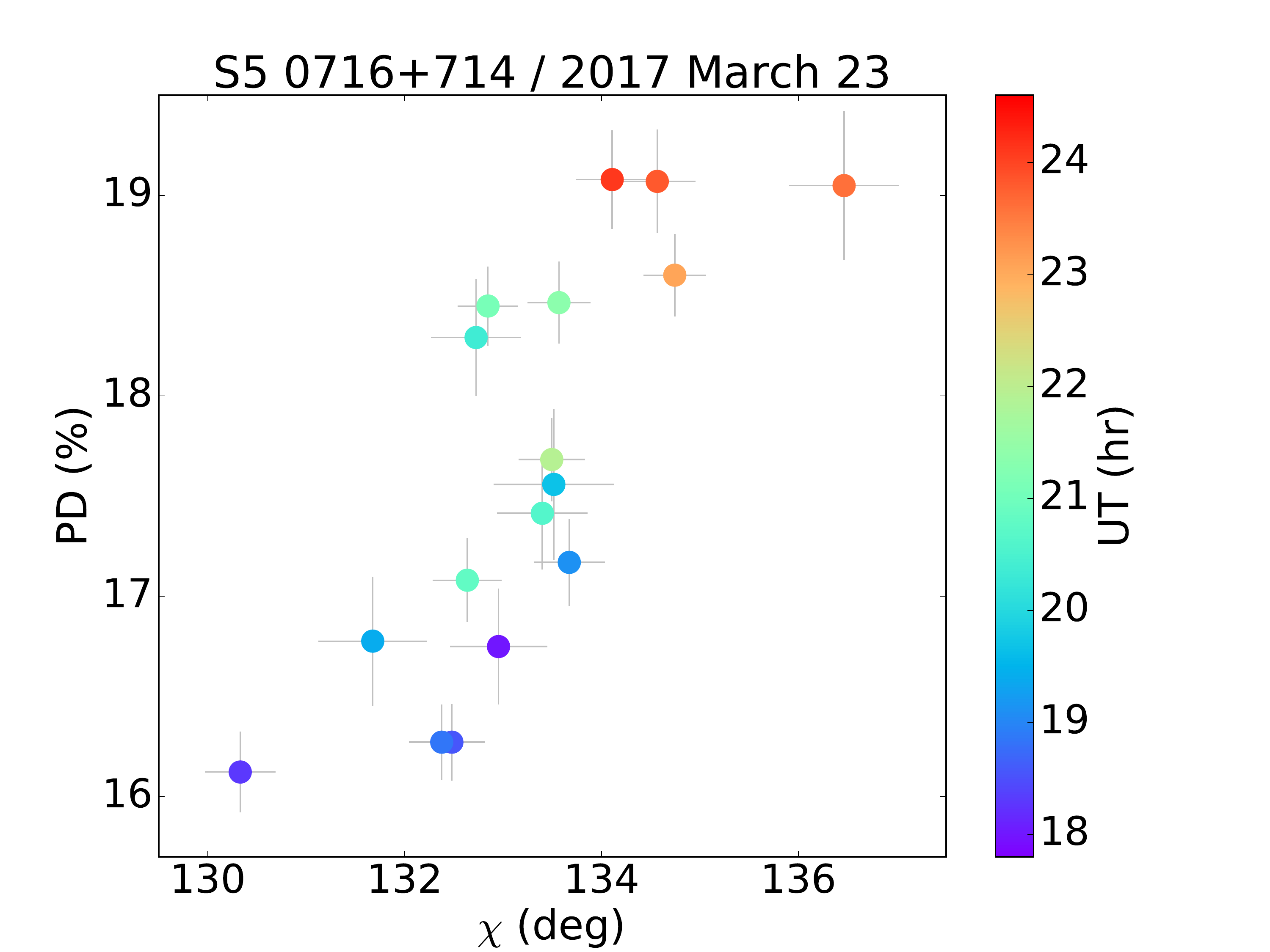}
}
\hbox{
\includegraphics[width=0.33\textwidth]{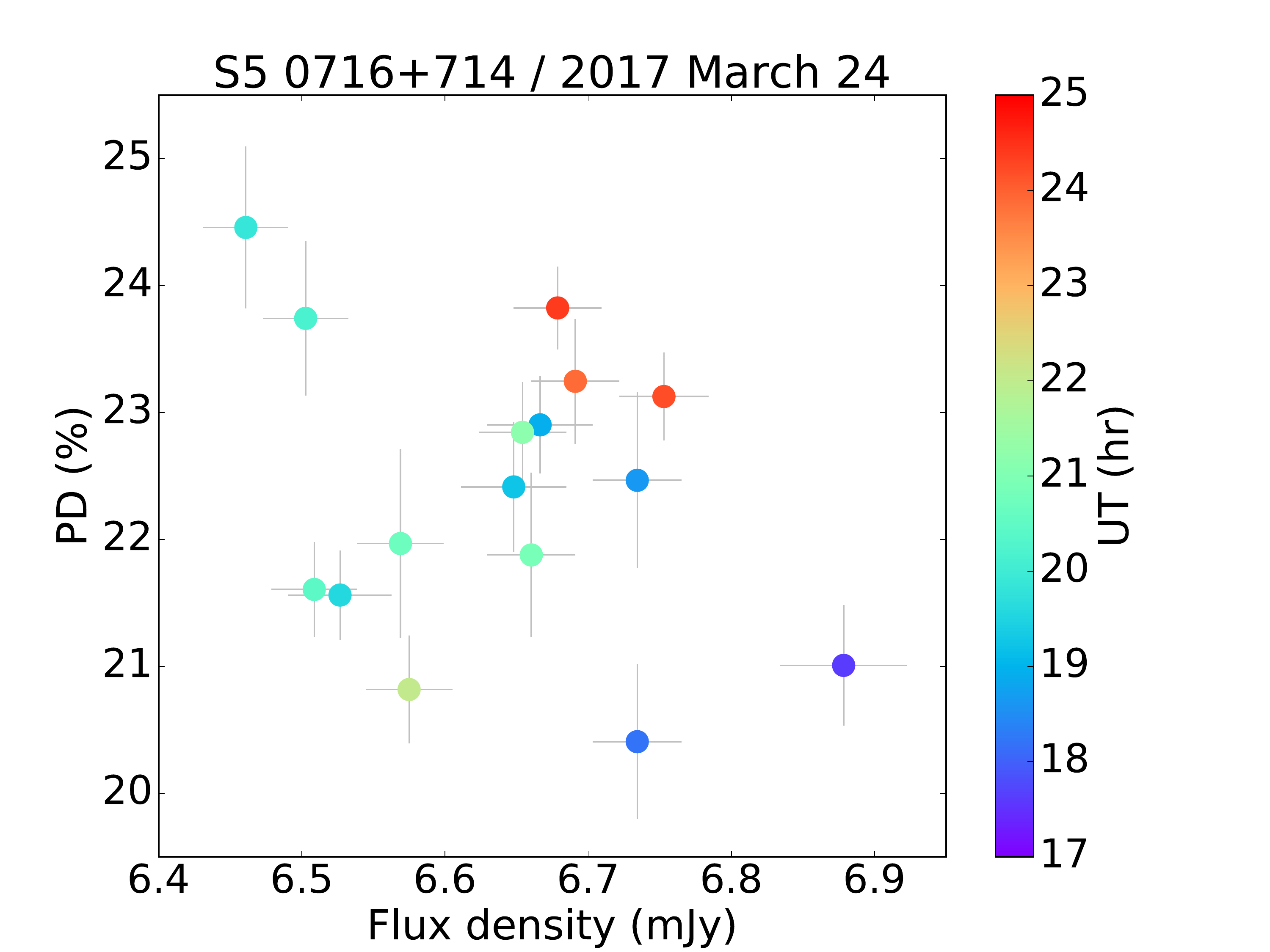}
\includegraphics[width=0.33\textwidth]{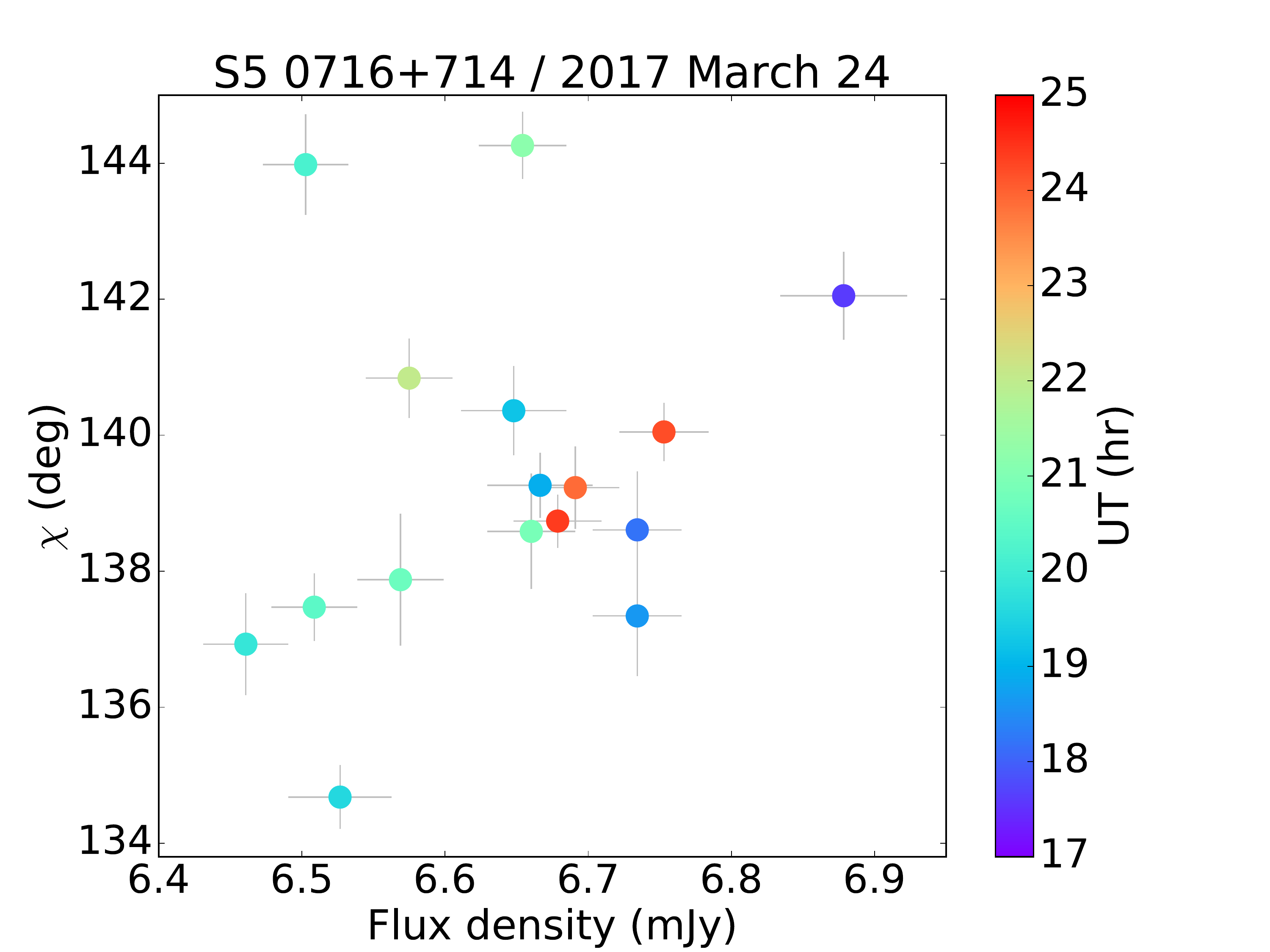}
\includegraphics[width=0.33\textwidth]{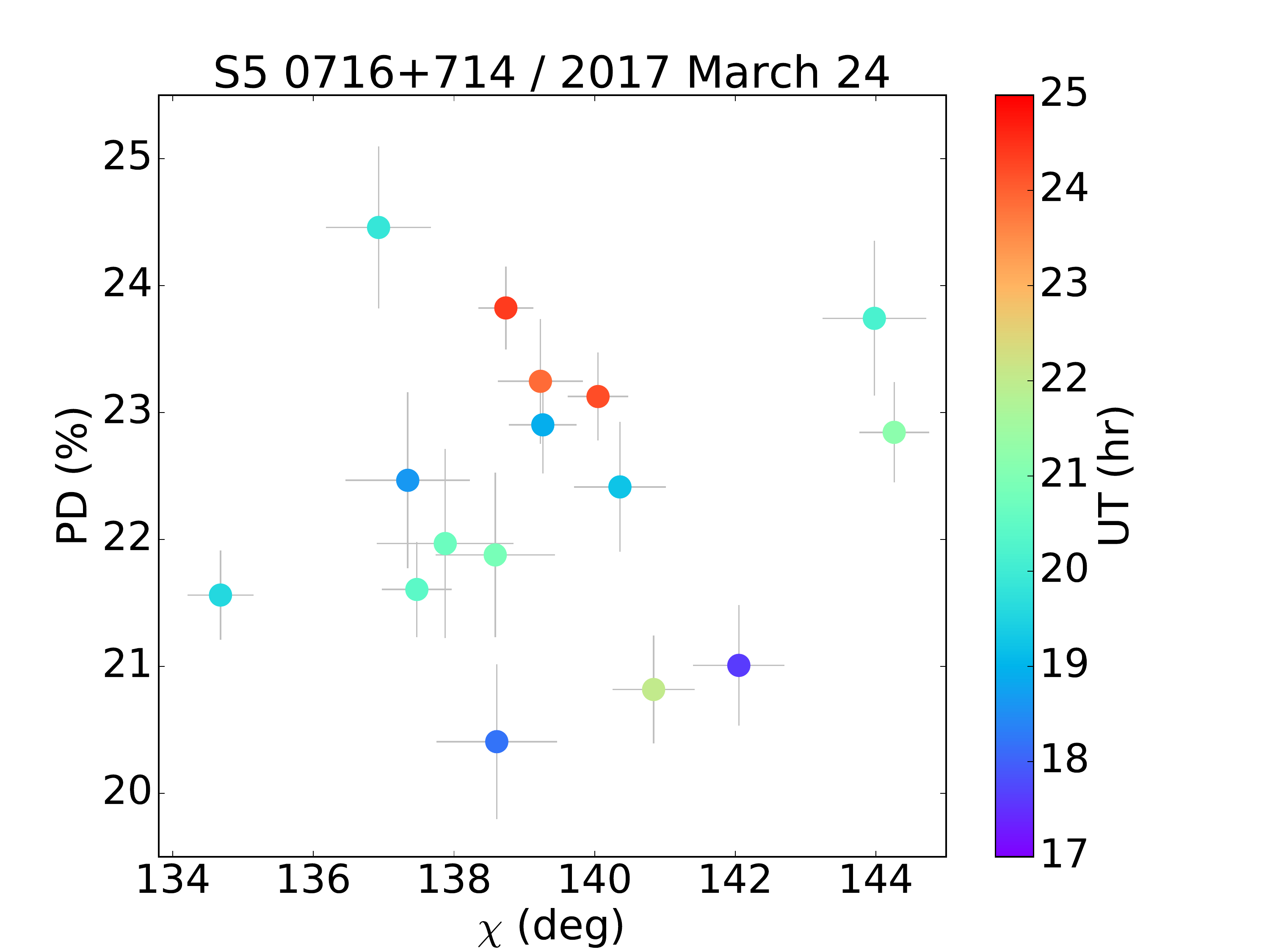}
}
\hbox{
\includegraphics[width=0.33\textwidth]{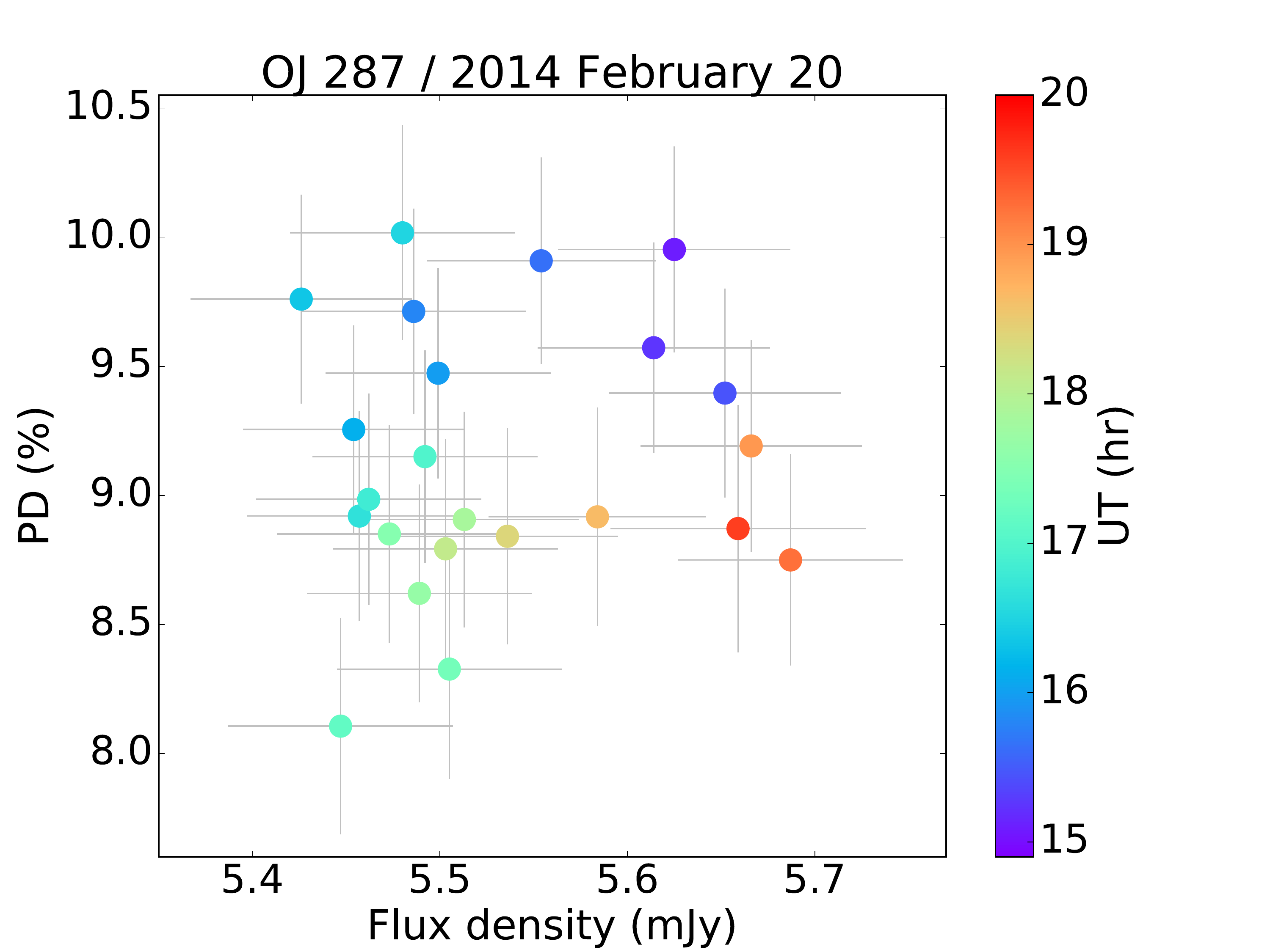}
\includegraphics[width=0.33\textwidth]{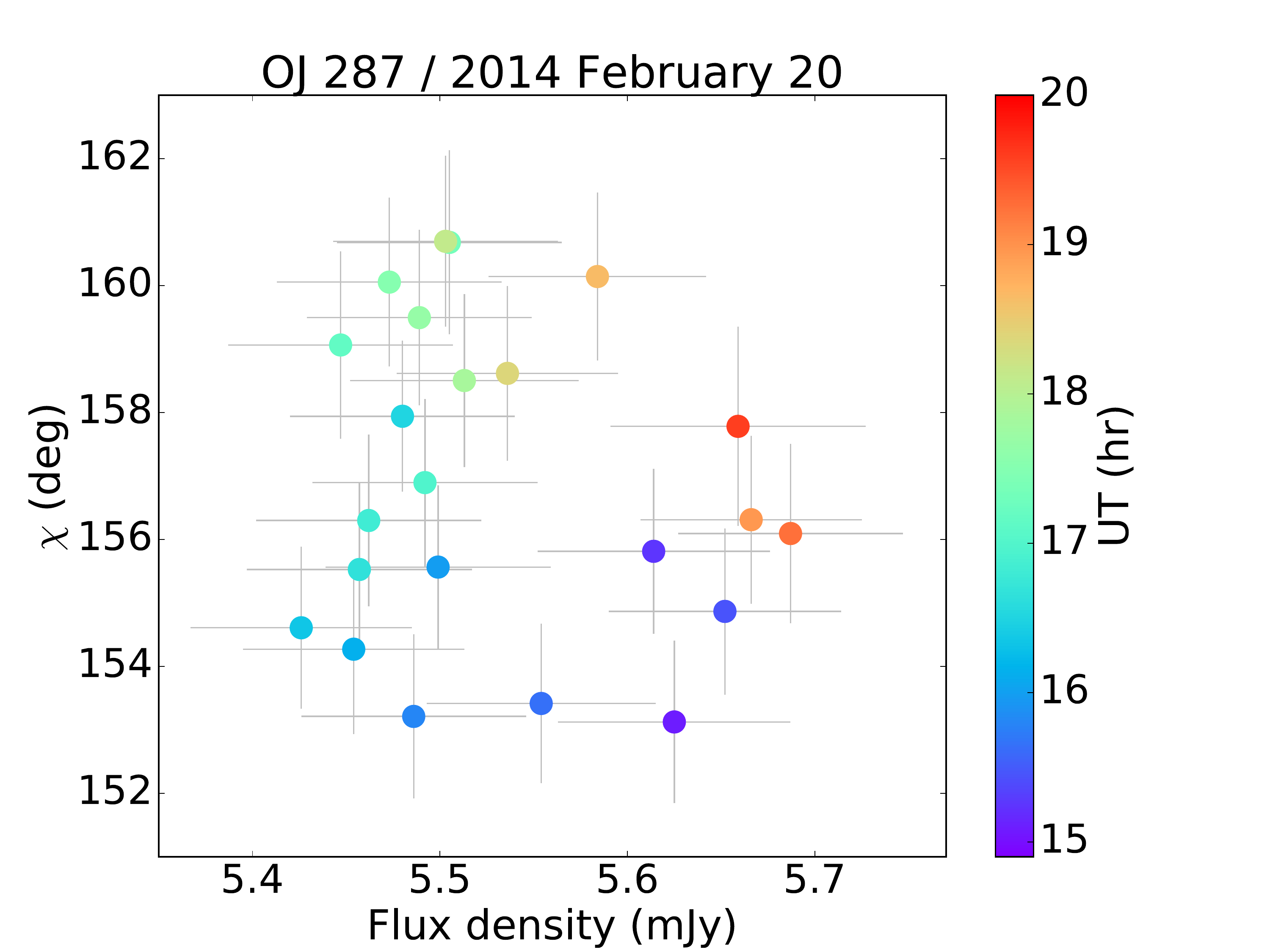}
\includegraphics[width=0.33\textwidth]{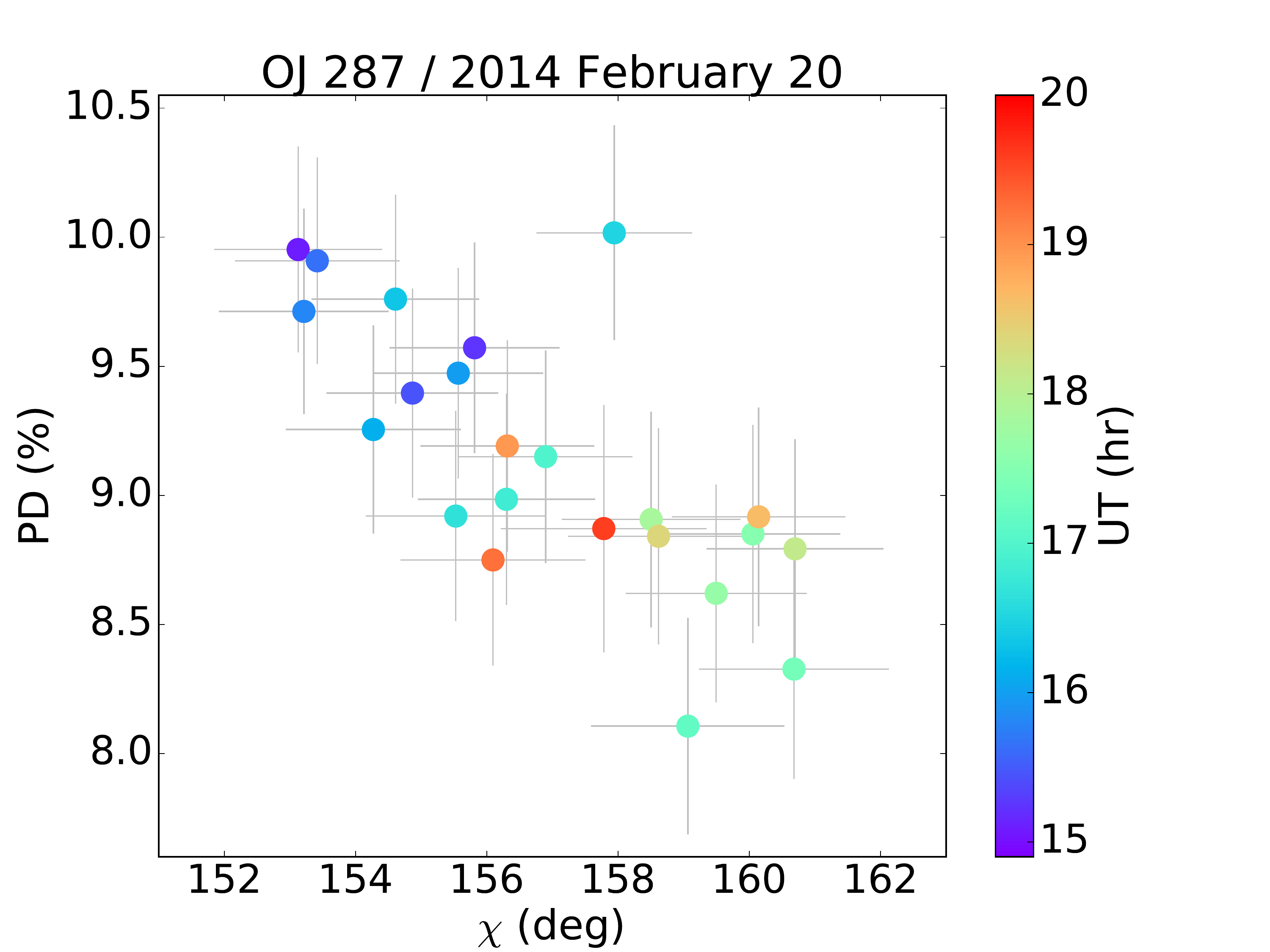}
}
\hbox{
\includegraphics[width=0.33\textwidth]{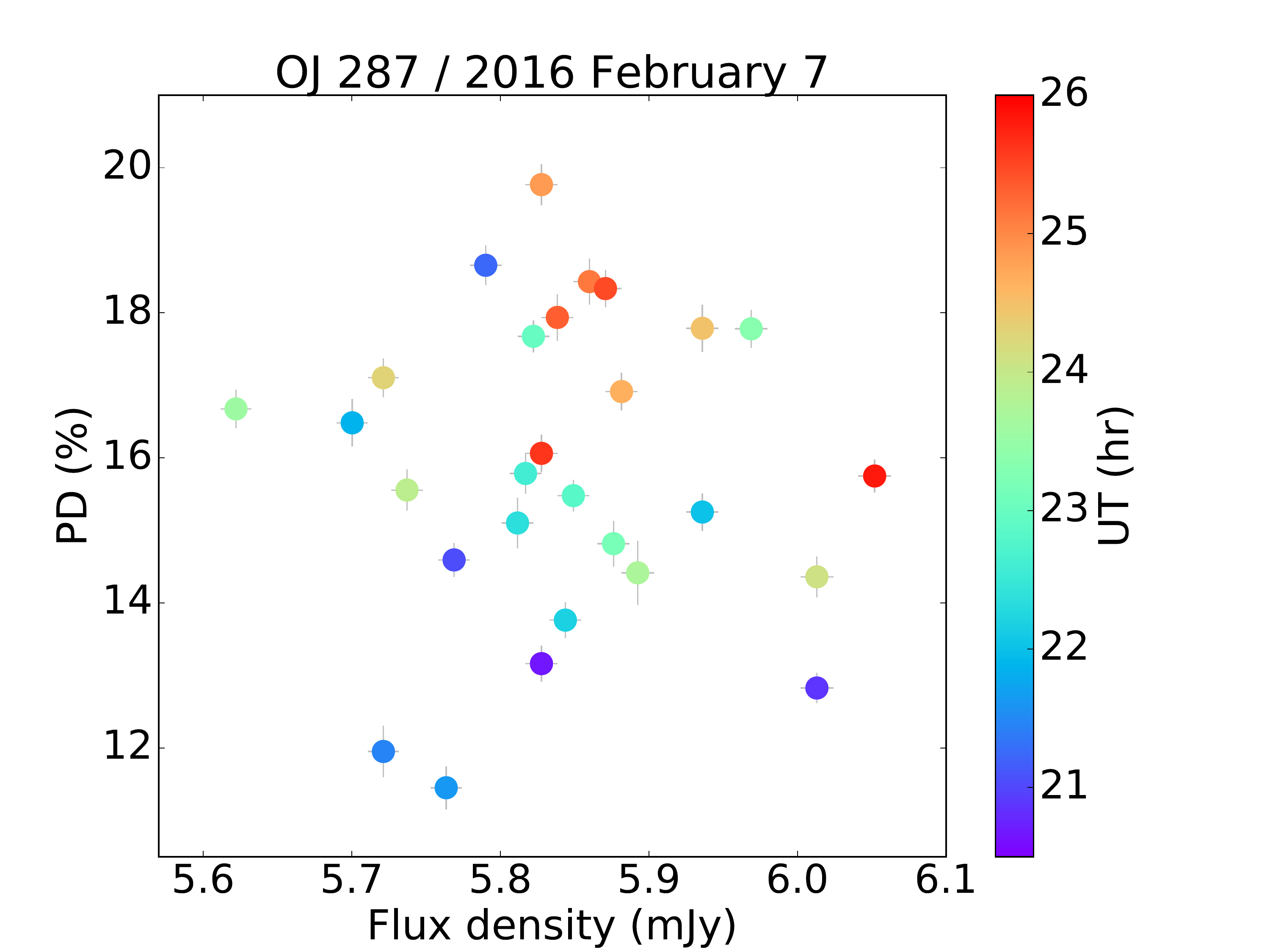}
\includegraphics[width=0.33\textwidth]{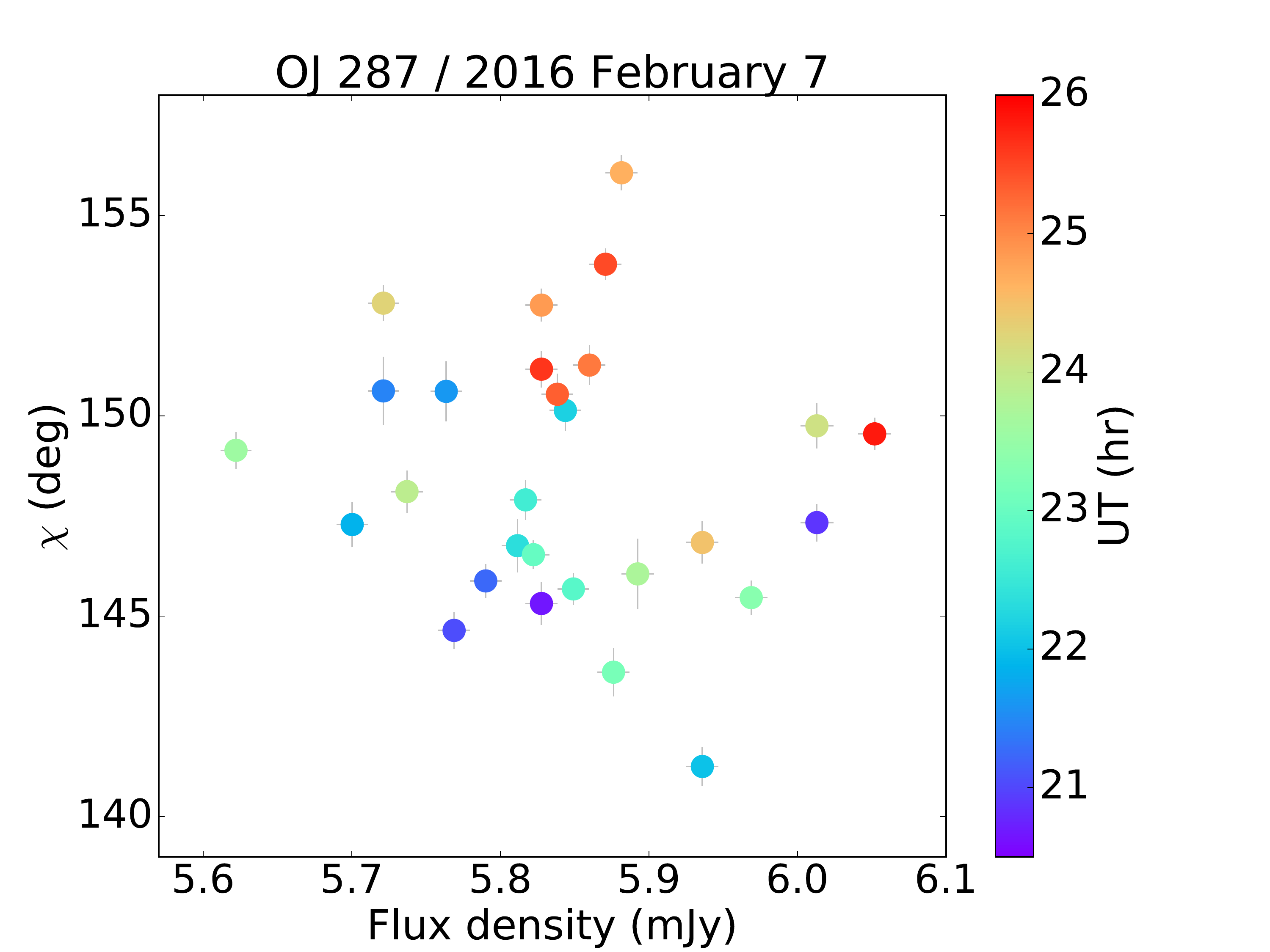}
\includegraphics[width=0.33\textwidth]{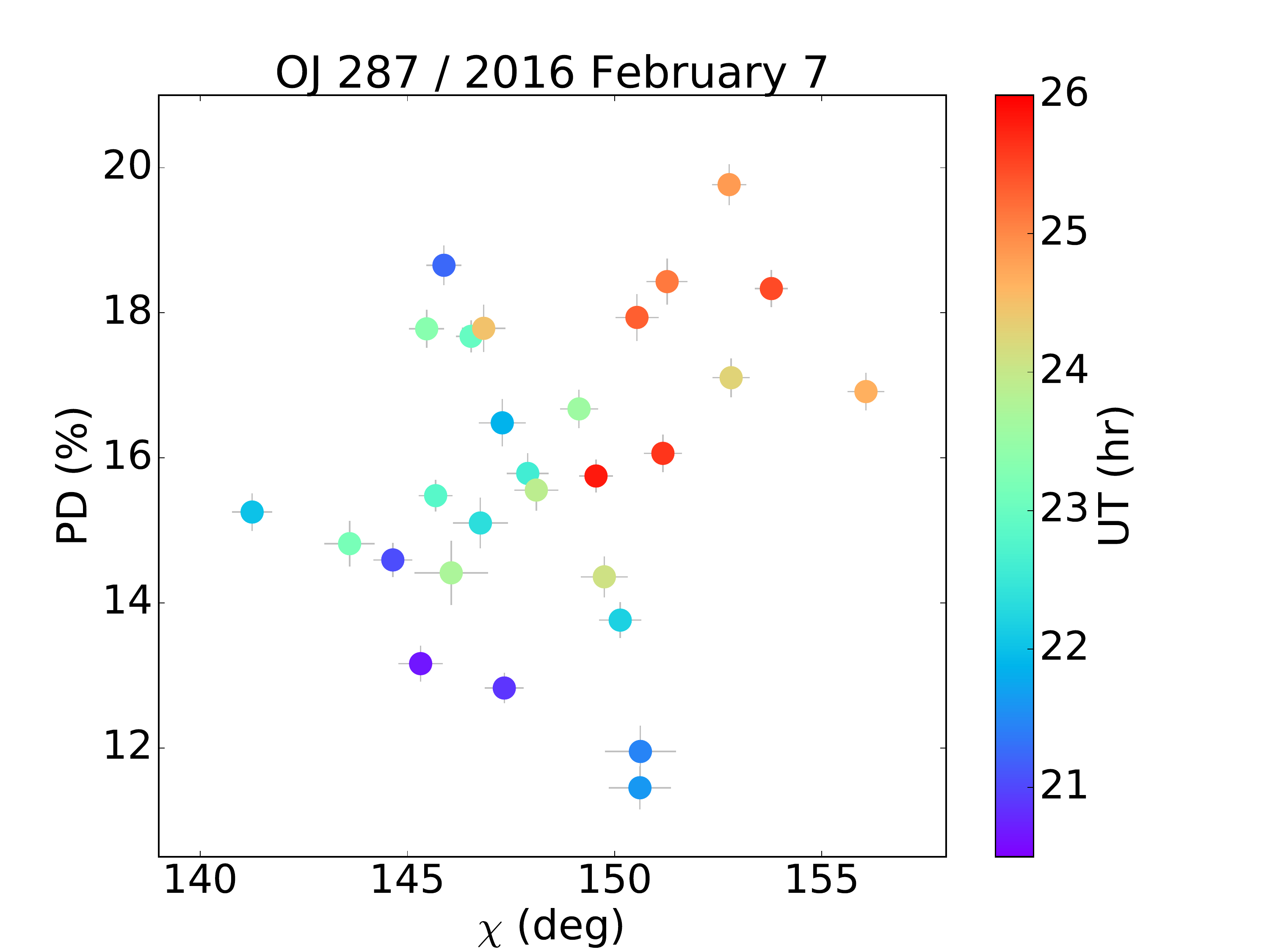}
}

\begin{minipage}{\textwidth}
\caption{Variations of PD vs. R--band flux density (left),  $\chi$ vs. R--band flux density (middle) and PD vs.  $\chi$ for blazars we monitored. Name of the blazar and the date are given at the top of each panel.}
\label{fig_poltrends}
\end{minipage}

\end{figure*}

We note that on the majority of the monitoring sessions, the polarization light curves gathered by us consist of 20 or fewer data points (Table~\ref{results_polinv}), sometimes with big gaps in the light curve and large measurement uncertainties (Table~\ref{results_polinv}; Figure~\ref{fig_polinv}) due to bad weather or the flux density states of the blazars. Therefore, the polarization behaviour is examined further only for the few sessions for which the good quality intra-night polarization monitoring data were available. Figure~\ref{fig_poltrends} presents the evolution of flux density, both as a function of PD and $\chi$ and the evolution of PD as a function $\chi$ of the sources for those selected sessions.

\renewcommand{\thefigure}{\arabic{figure} (Cont.)}
\addtocounter{figure}{-1}

\begin{figure*}
\hbox{
\includegraphics[width=0.33\textwidth]{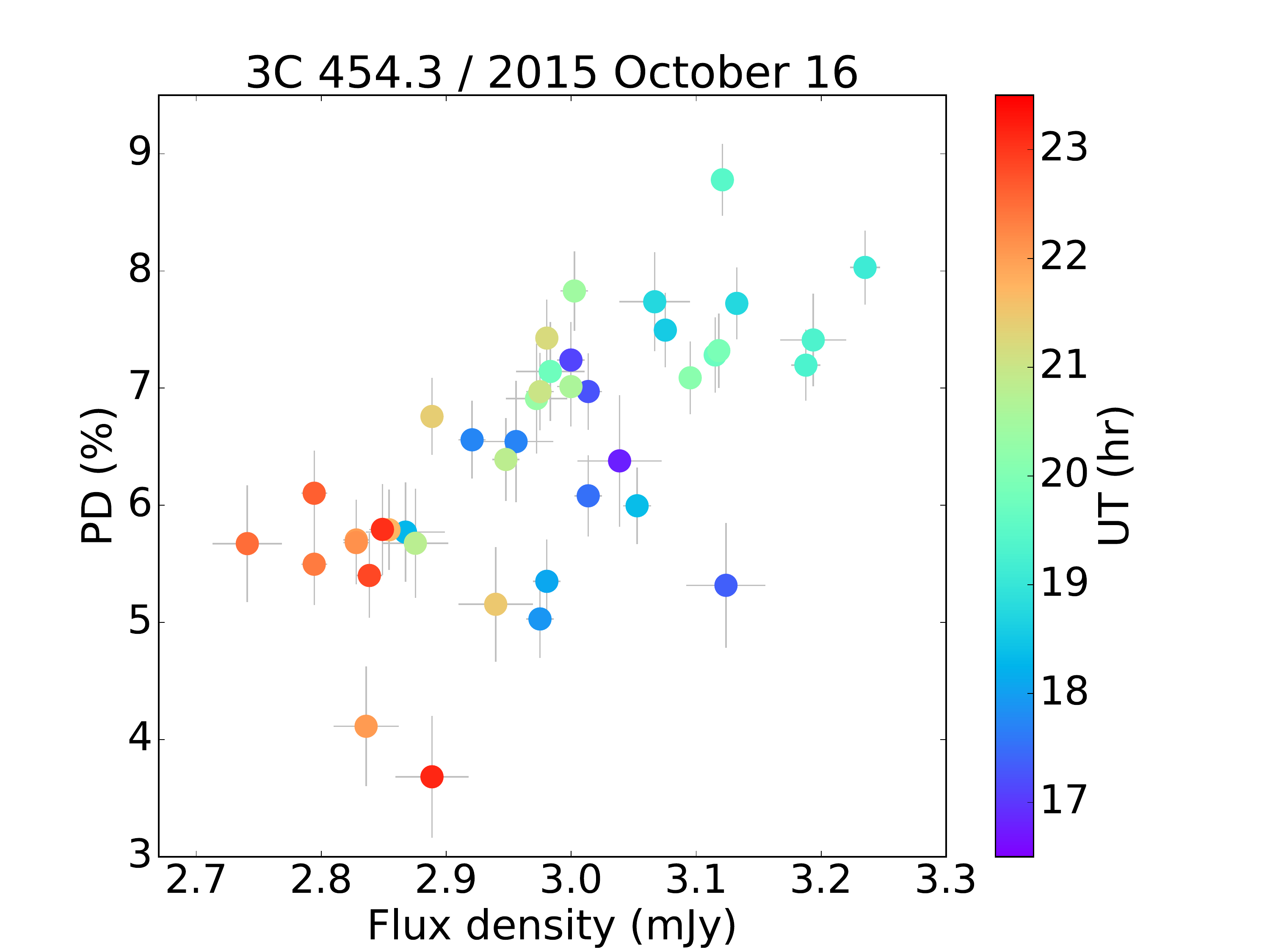} 
\includegraphics[width=0.33\textwidth]{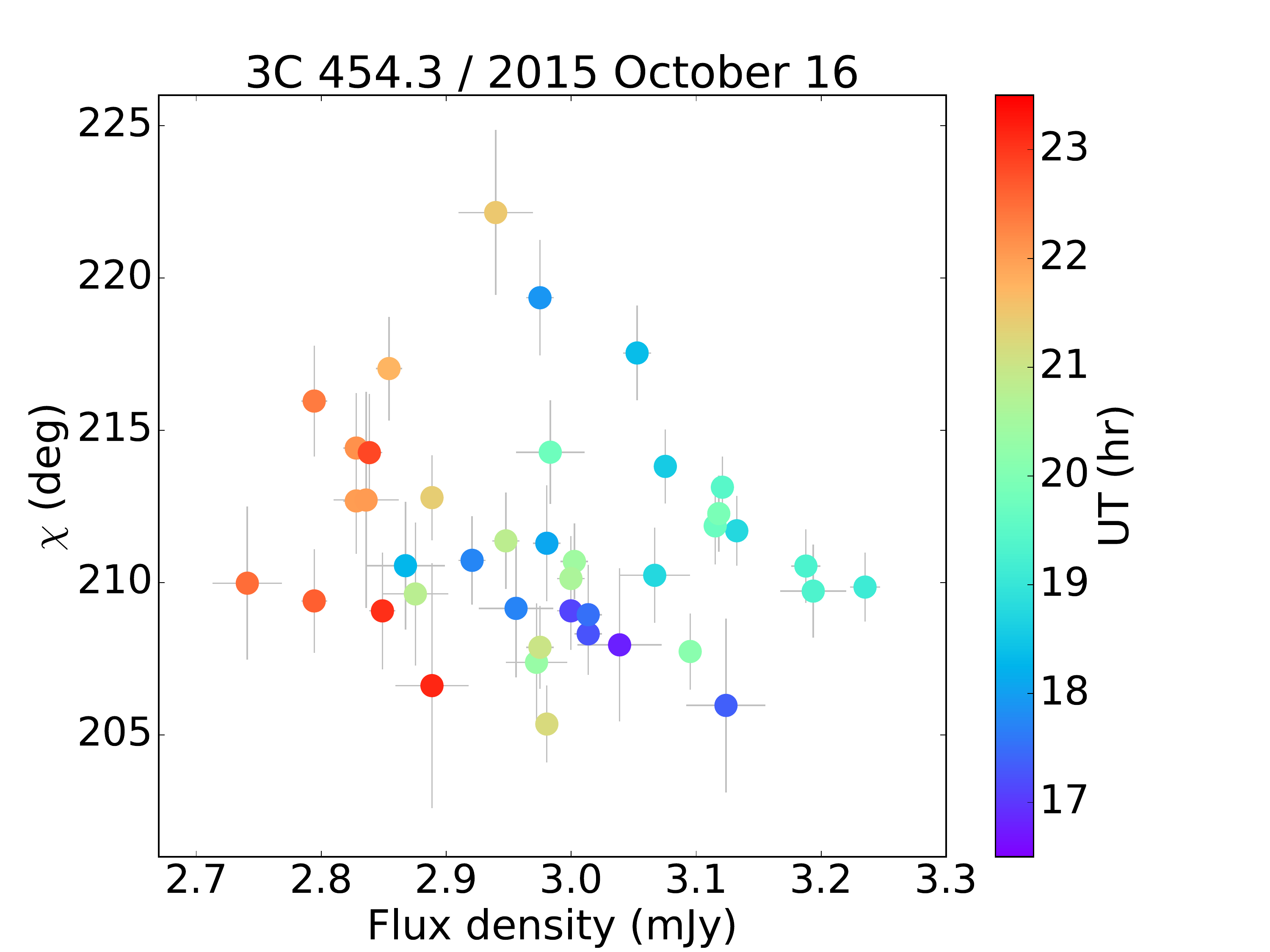} 
\includegraphics[width=0.33\textwidth]{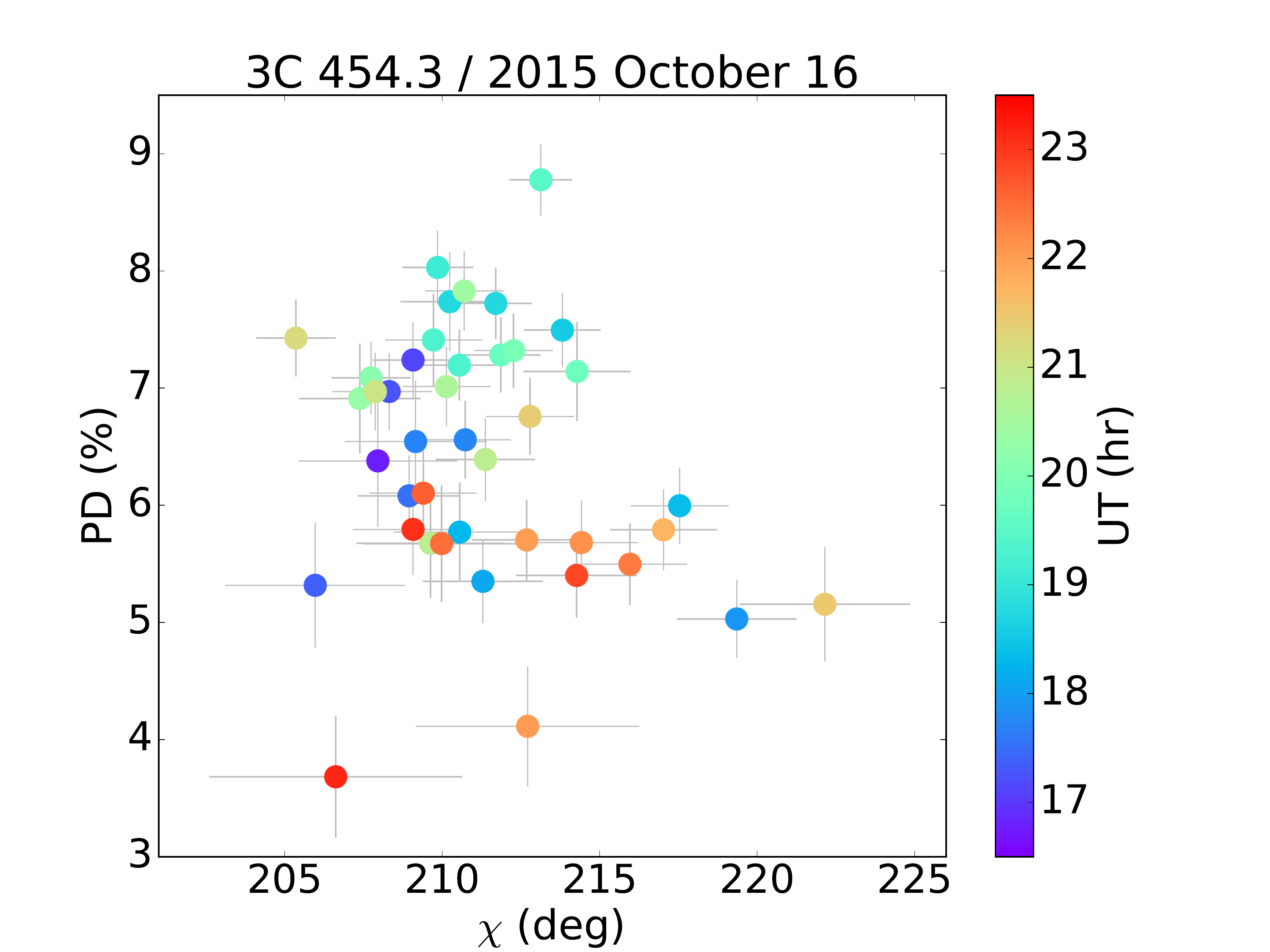} 
}
\hbox{
\includegraphics[width=0.33\textwidth]{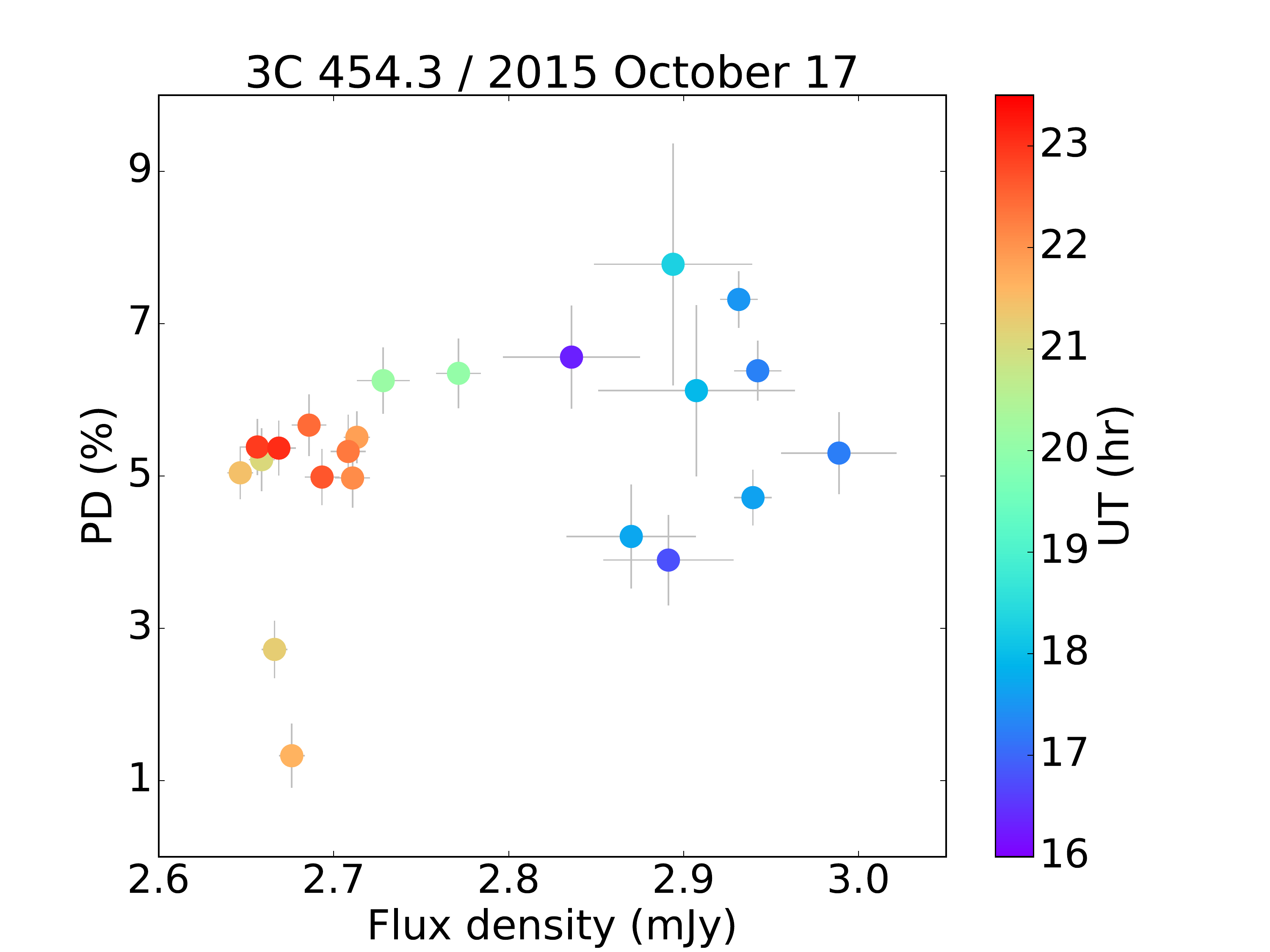} 
\includegraphics[width=0.33\textwidth]{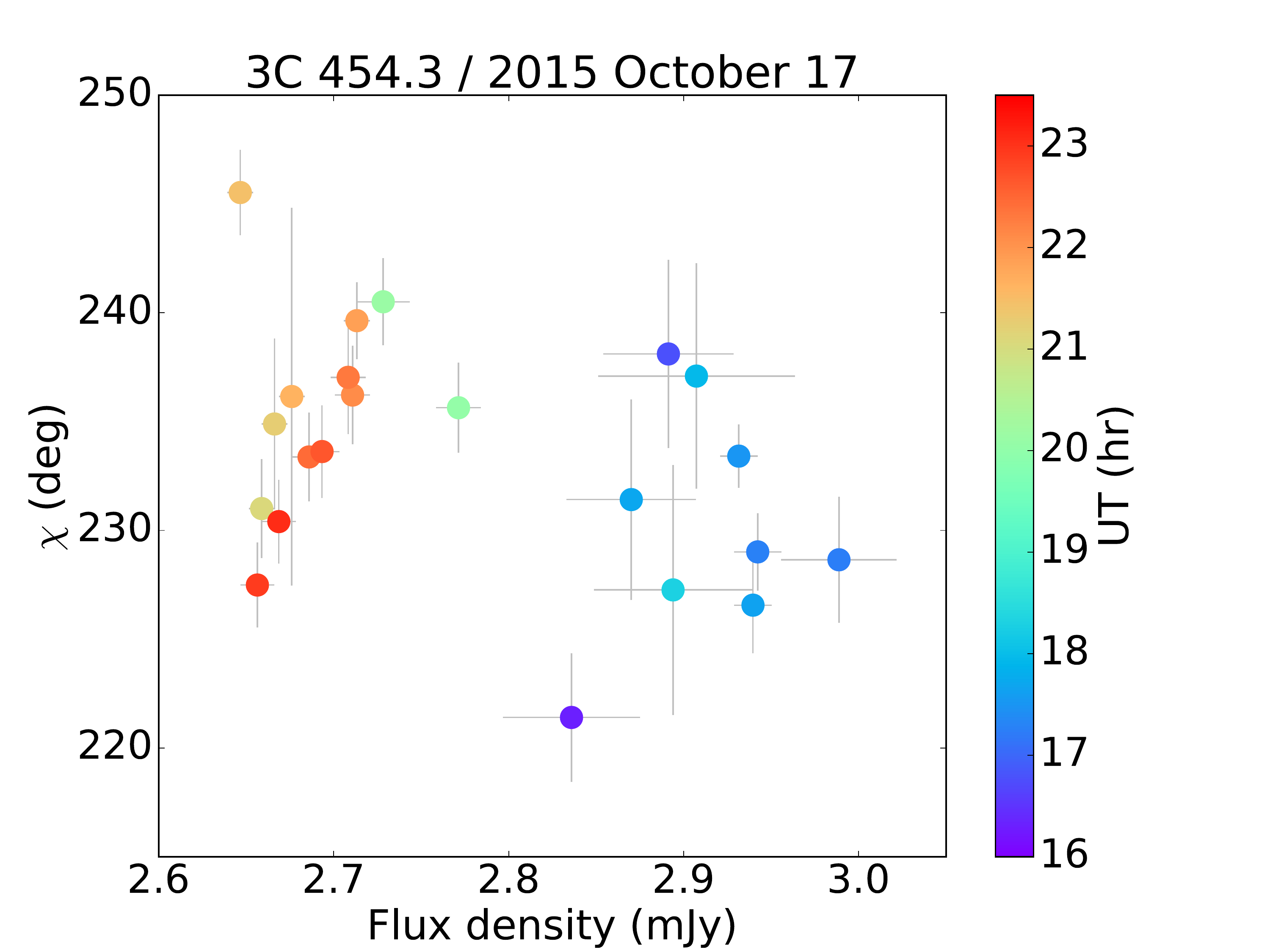} 
\includegraphics[width=0.33\textwidth]{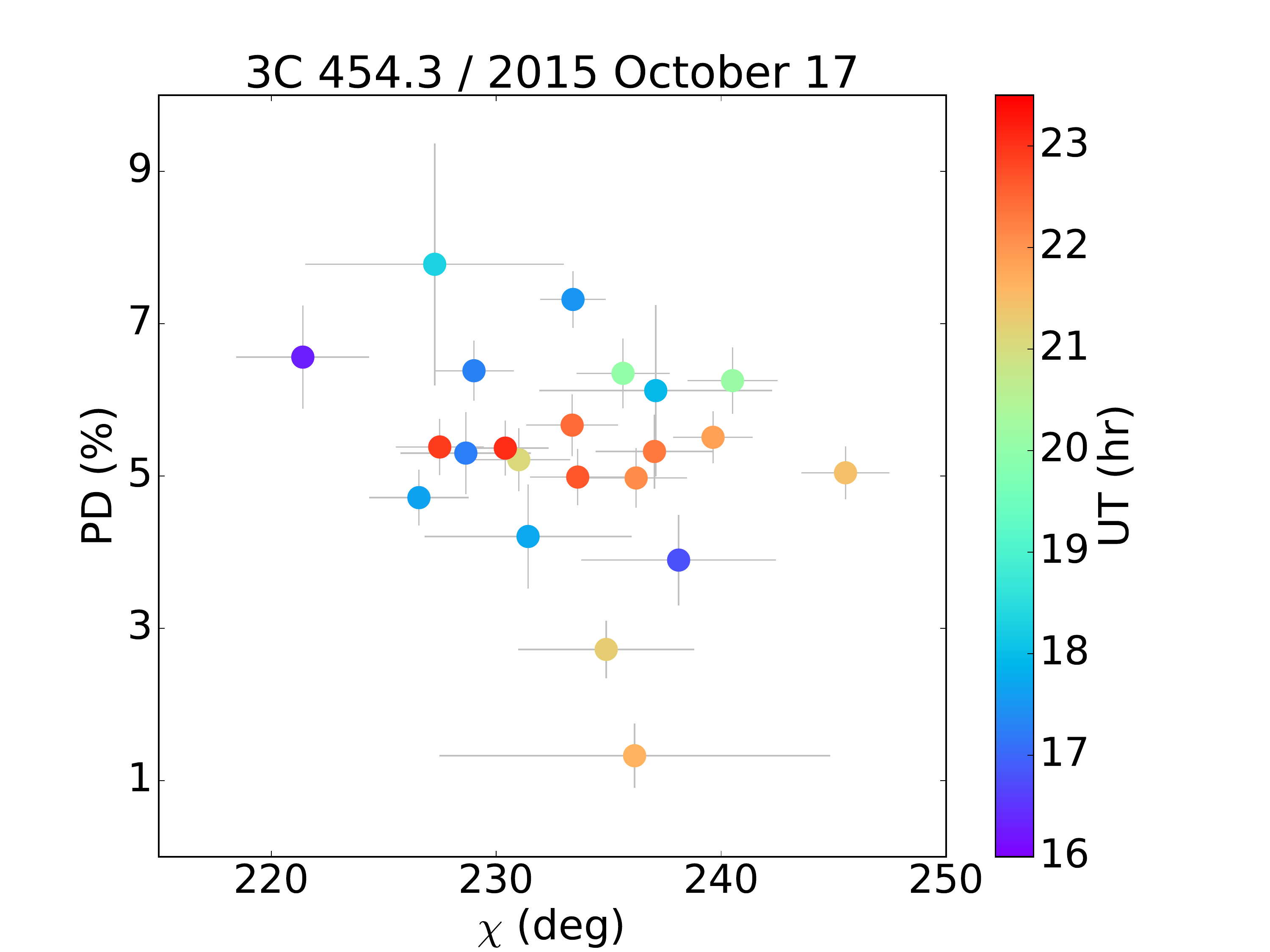} 
}

\caption{
}
\end{figure*}

\renewcommand{\thefigure}{\arabic{figure}}

\subsection{Analysis of simultaneous flux density, colour, and polarization microvariability monitoring}
\label{sec:fluxpol}

\begin{figure*}
\centering
\hbox{
\includegraphics[width=0.33\textwidth]{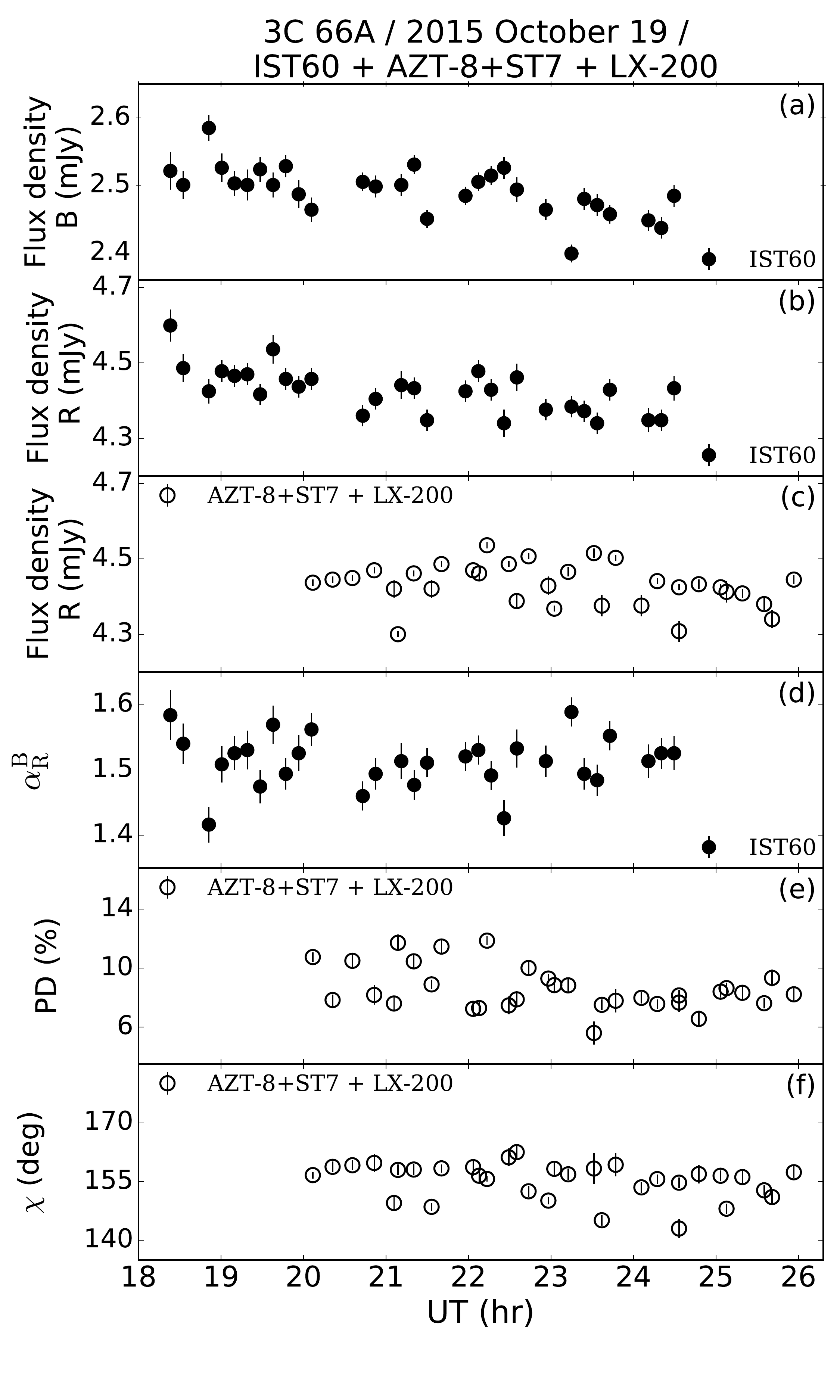} 
\includegraphics[width=0.33\textwidth]{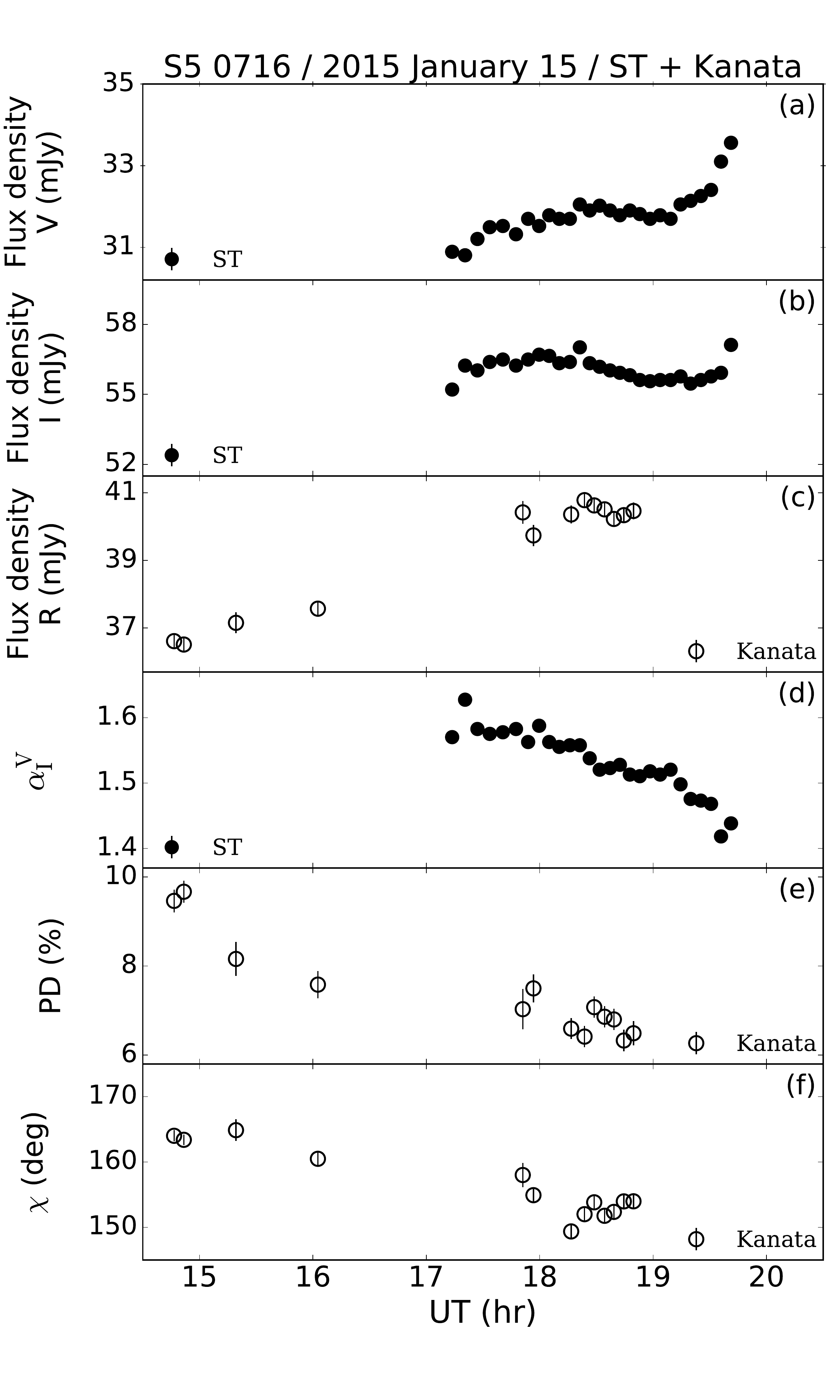} 
\includegraphics[width=0.33\textwidth]{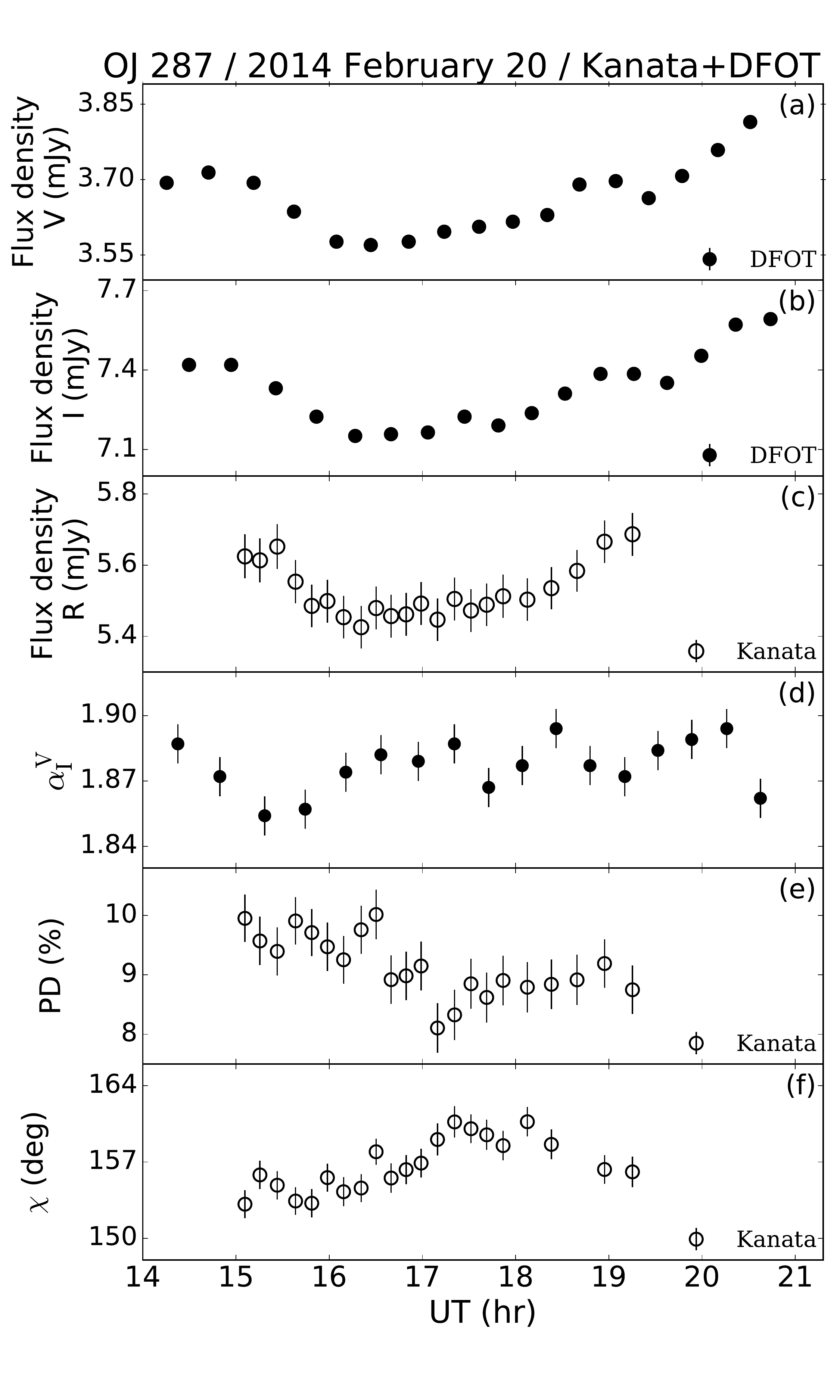}
}
\hbox{
\includegraphics[width=0.33\textwidth]{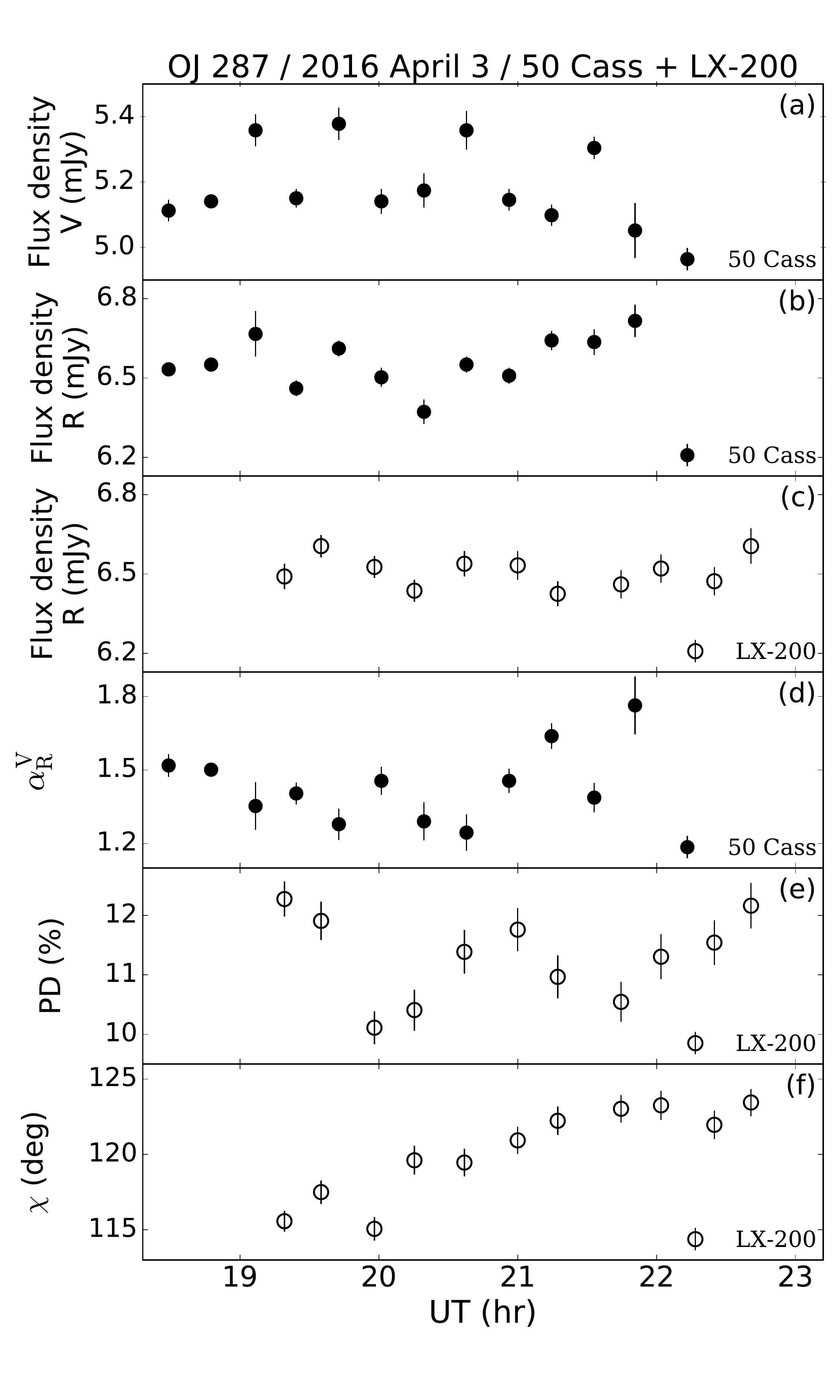}
\includegraphics[width=0.33\textwidth]{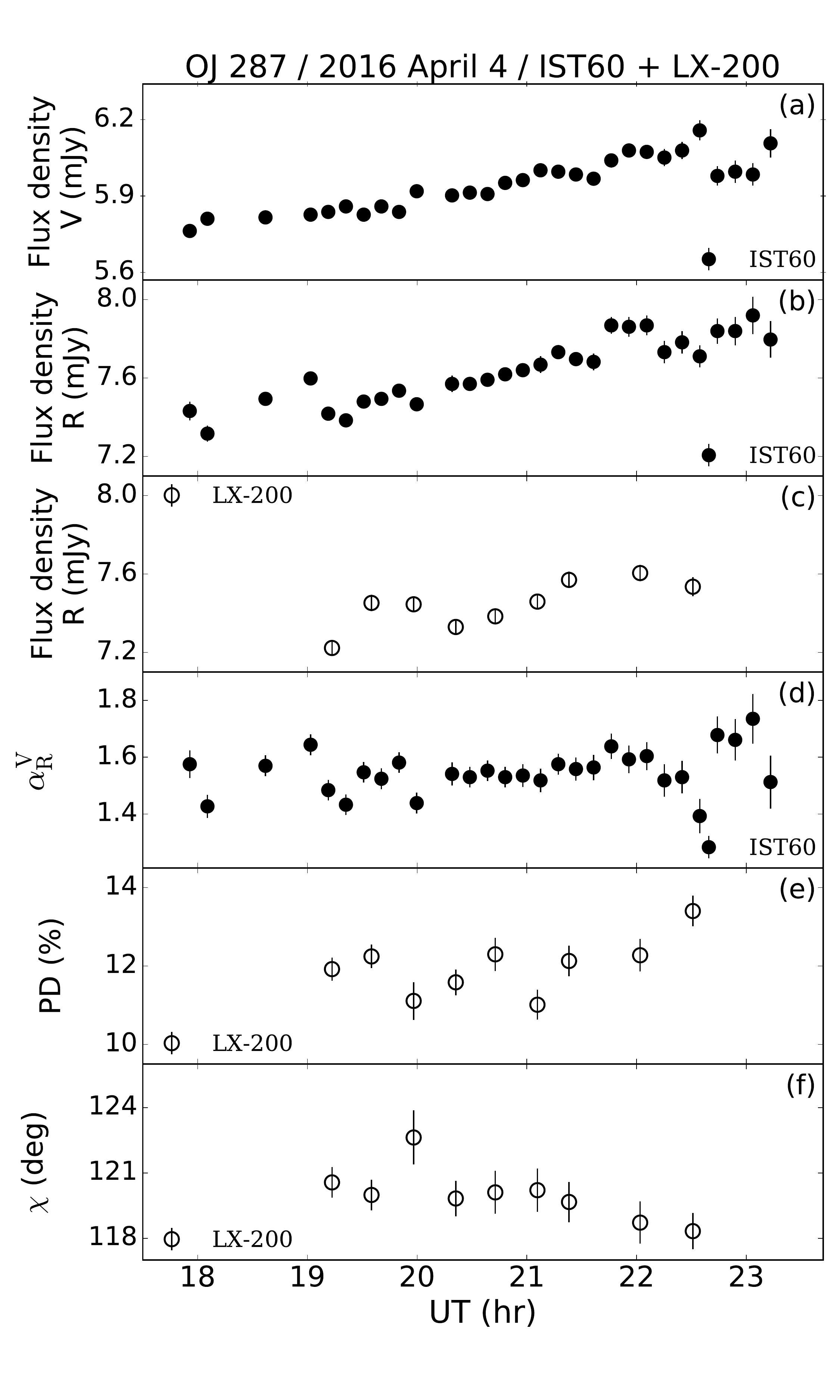}
\includegraphics[width=0.33\textwidth]{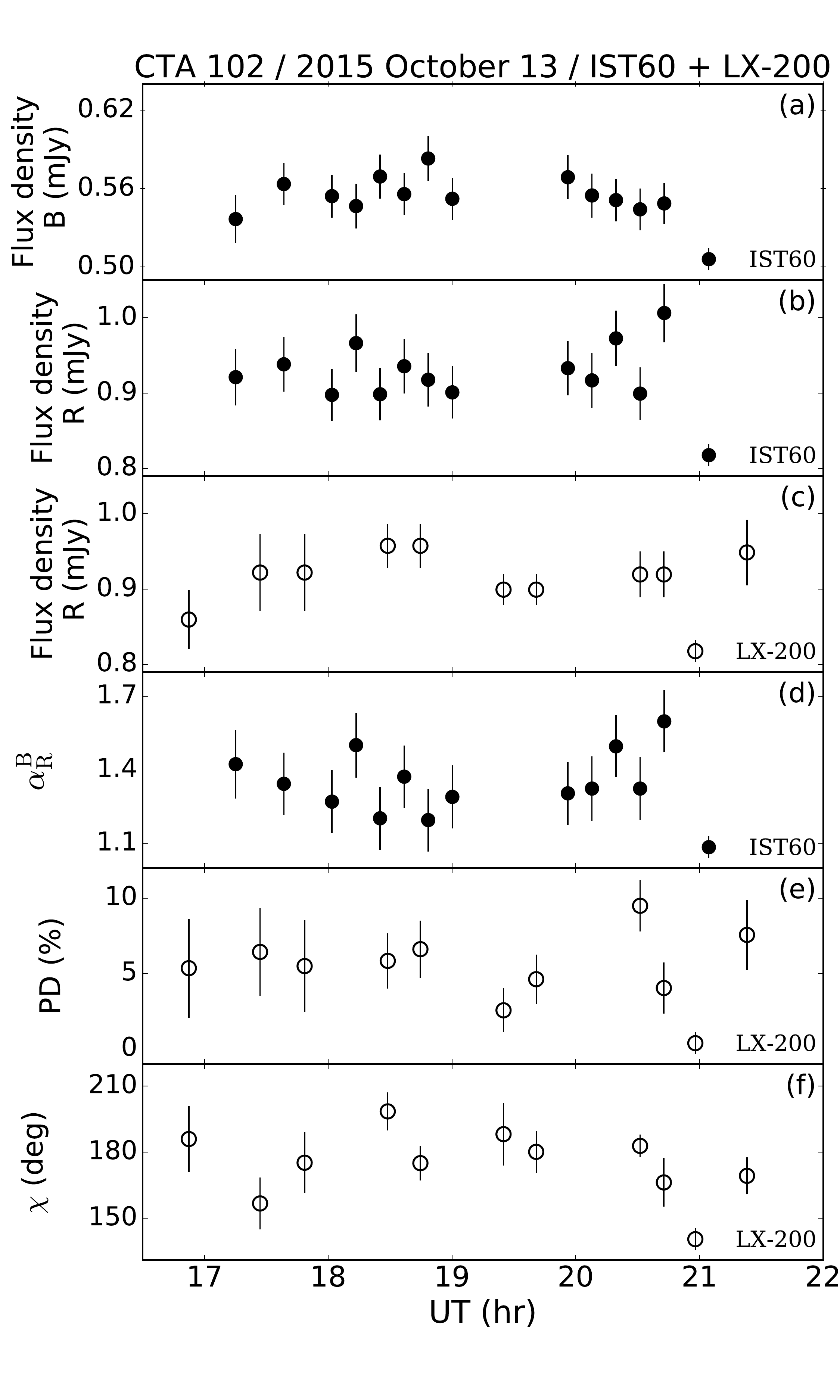}
}

\caption{Total flux density measurements at the two most distant frequency pairs available from flux monitoring (panels a and b) and total flux density measurements from polarization monitoring data usually in R-band (panel c), $\alpha$ (derived between the flux density measurements shown in the panels a and b; panel d), and polarization parameters (PD and $\chi$ in panels e and f, respectively) for strictly simultaneous observations from multiple observatories.  At the top of each panel, the blazar name, date of monitoring, and, telescopes used, are given.}

\label{fig_simultaneous}%
\end{figure*}

Figure~\ref{fig_simultaneous} shows the joint plots for the blazar sources for which multiband total flux density and polarization intra-night light curves were strictly simultaneous from different sites. These agree within the errors.  

\renewcommand{\thefigure}{\arabic{figure} (Cont.)}
\addtocounter{figure}{-1}

\begin{figure*}
\includegraphics[width=0.33\textwidth]{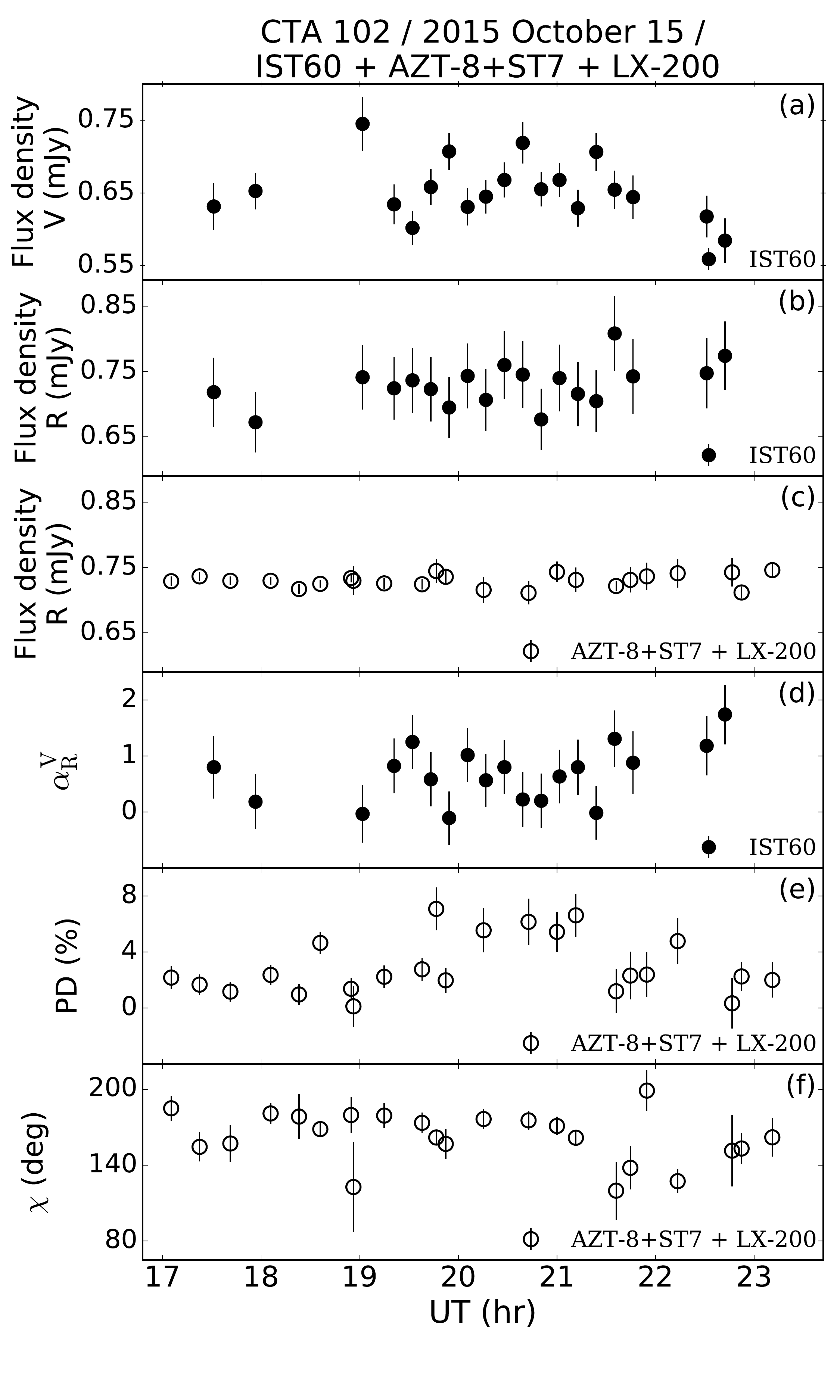}
\centering
\caption{}.
\label{}%
\end{figure*}

\renewcommand{\thefigure}{\arabic{figure}}

\subsection{Analysis of $\alpha_{\nu_1}^{\nu_2}$--flux density plane on long-term timescales }

\begin{figure*}
\hbox{
\includegraphics[angle=0, width=0.5\textwidth]{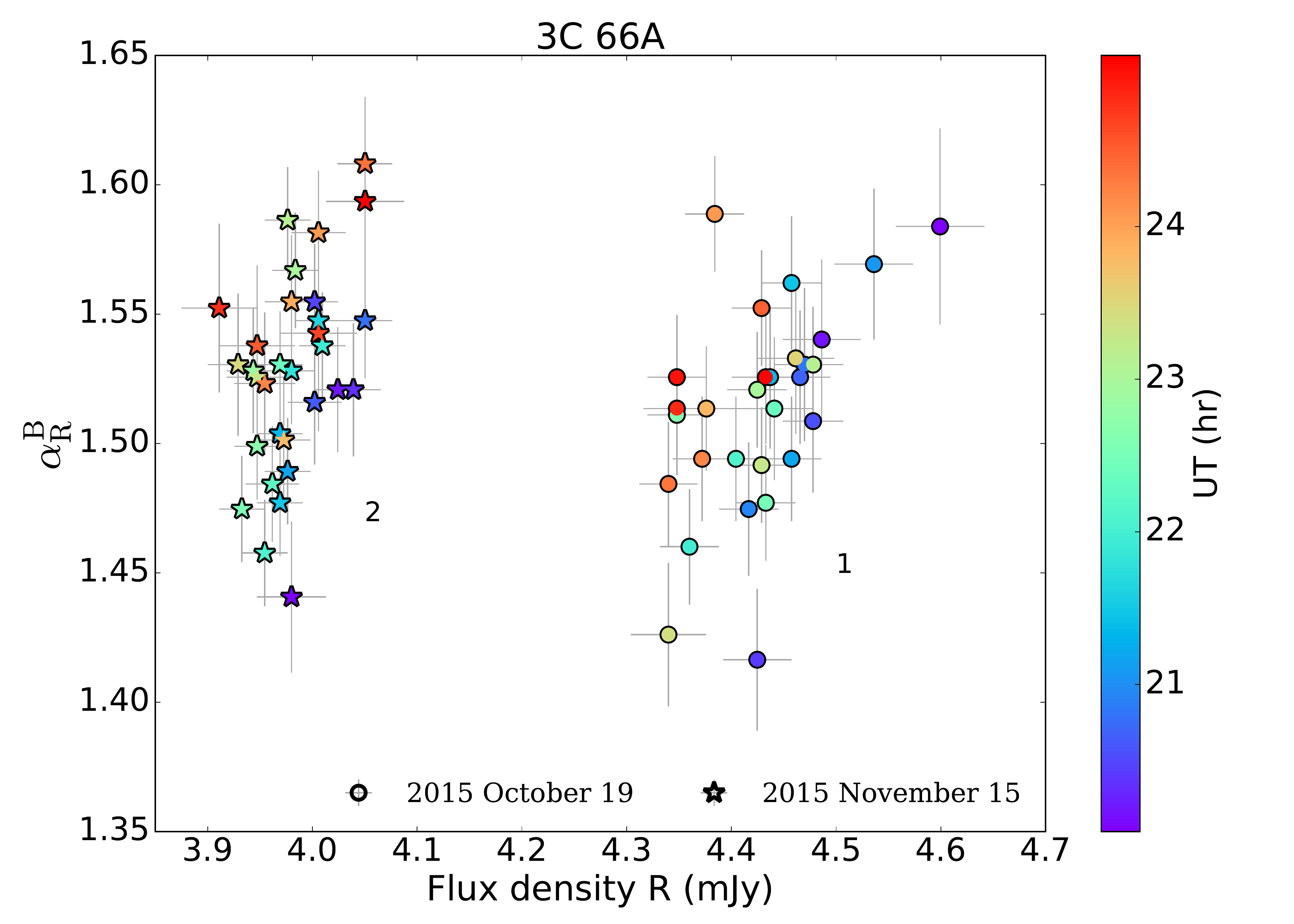}
\includegraphics[angle=0, width=0.5\textwidth]{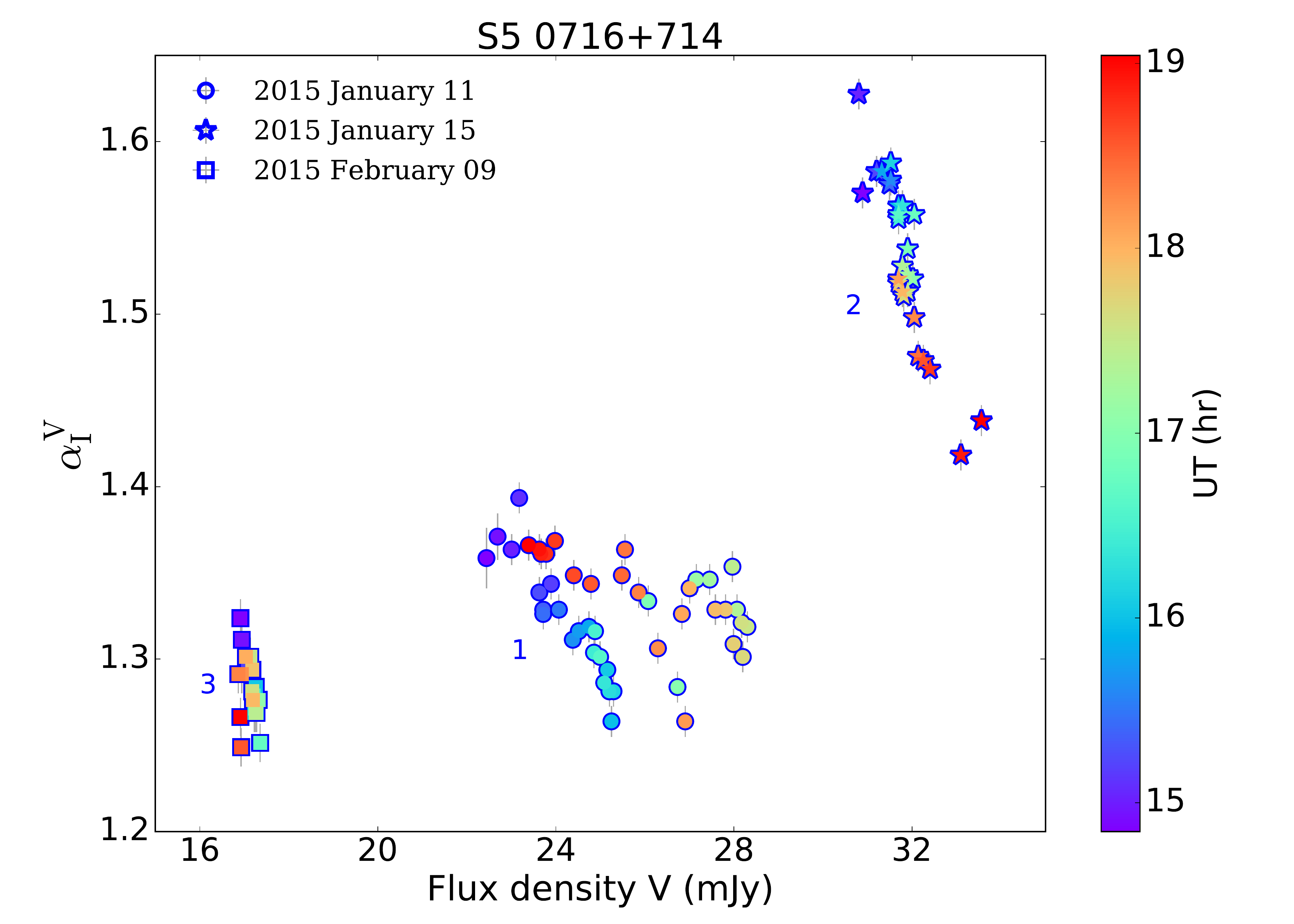}
}
\hbox{
\includegraphics[angle=0, width=0.5\textwidth]{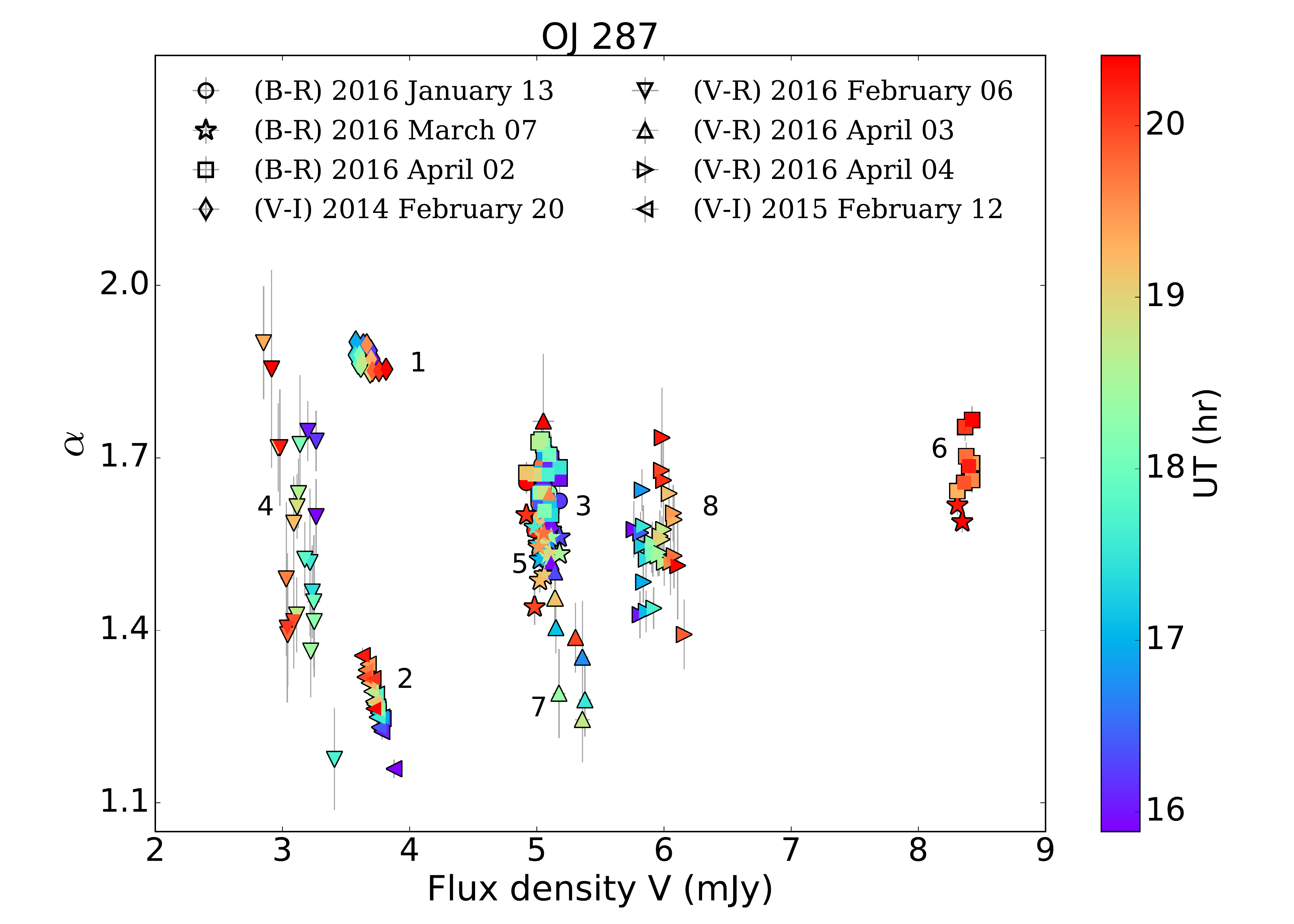}
\includegraphics[angle=0, width=0.5\textwidth]{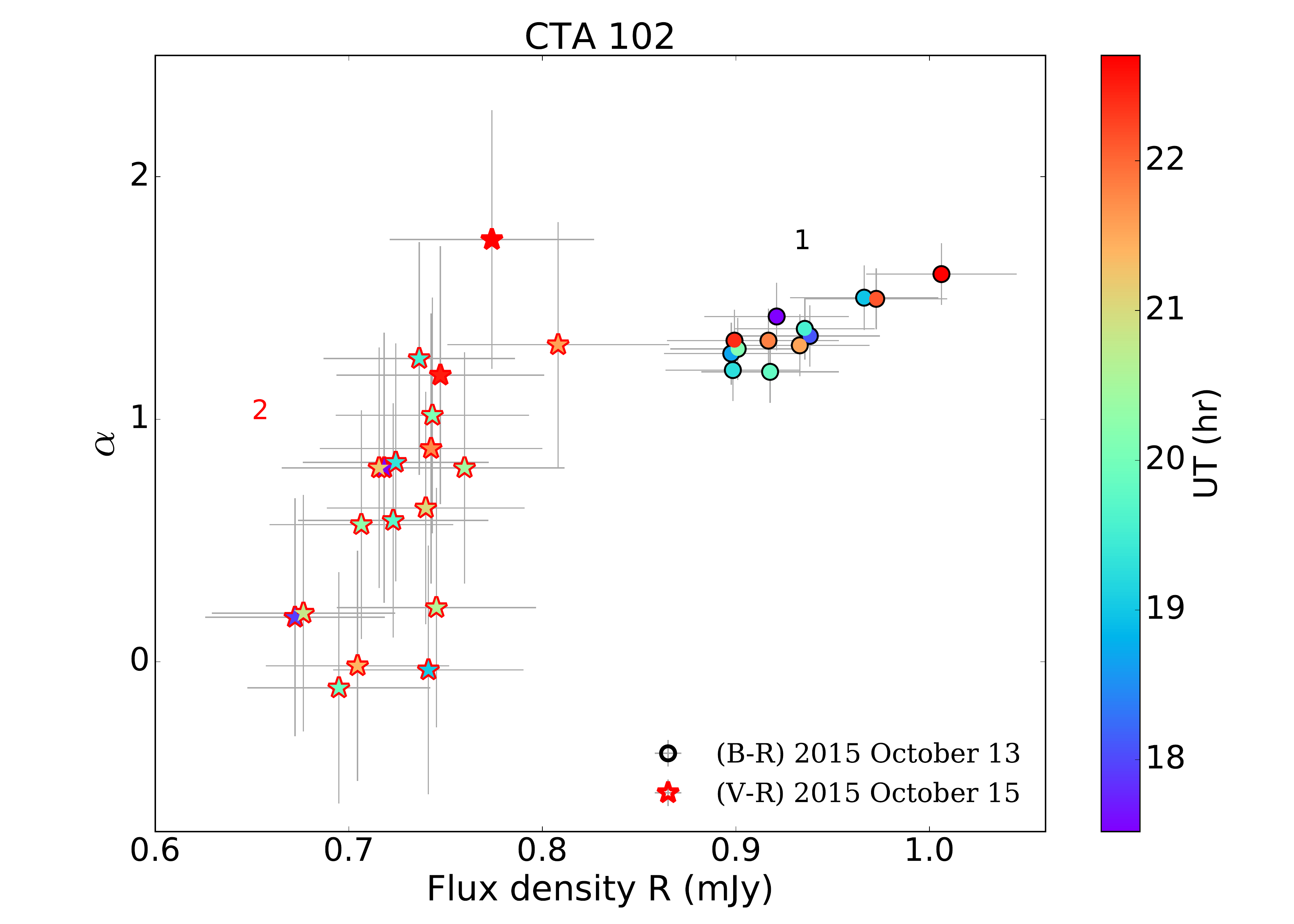}
}
\begin{minipage}{\textwidth}
 \caption{Variations of spectral index as function of flux density for the blazar sources for which the intra-night monitoring was carried out on more than one occasion. Different symbols represents different monitoring sessions while colours show the UT (hr) time during the monitoring session; the numbers give the sequence of observing nights for the blazar in chronological order (Table~\ref{results_finv}). }
\label{longterm_spflux}
\end{minipage}
\end{figure*}

Figure~\ref{longterm_spflux} shows the long-term evolution of spectral index as a function of flux density for the blazar sources which were monitored on more than one occasions during the study.    

\subsection{Analysis of long-term polarization light curves}

\begin{figure*}
\centering
\hbox{
\includegraphics[width=0.5\textwidth]{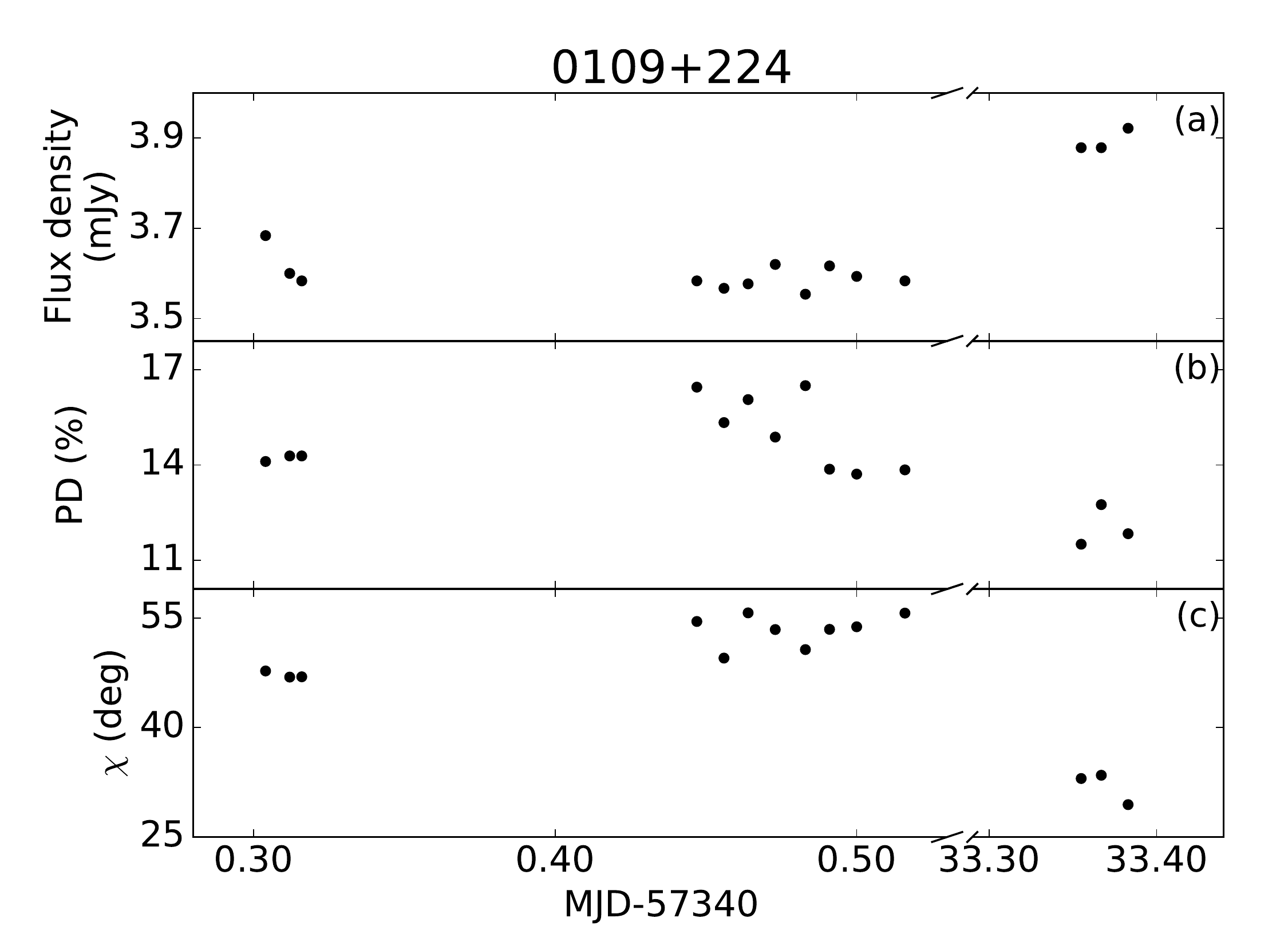}
\includegraphics[width=0.5\textwidth]{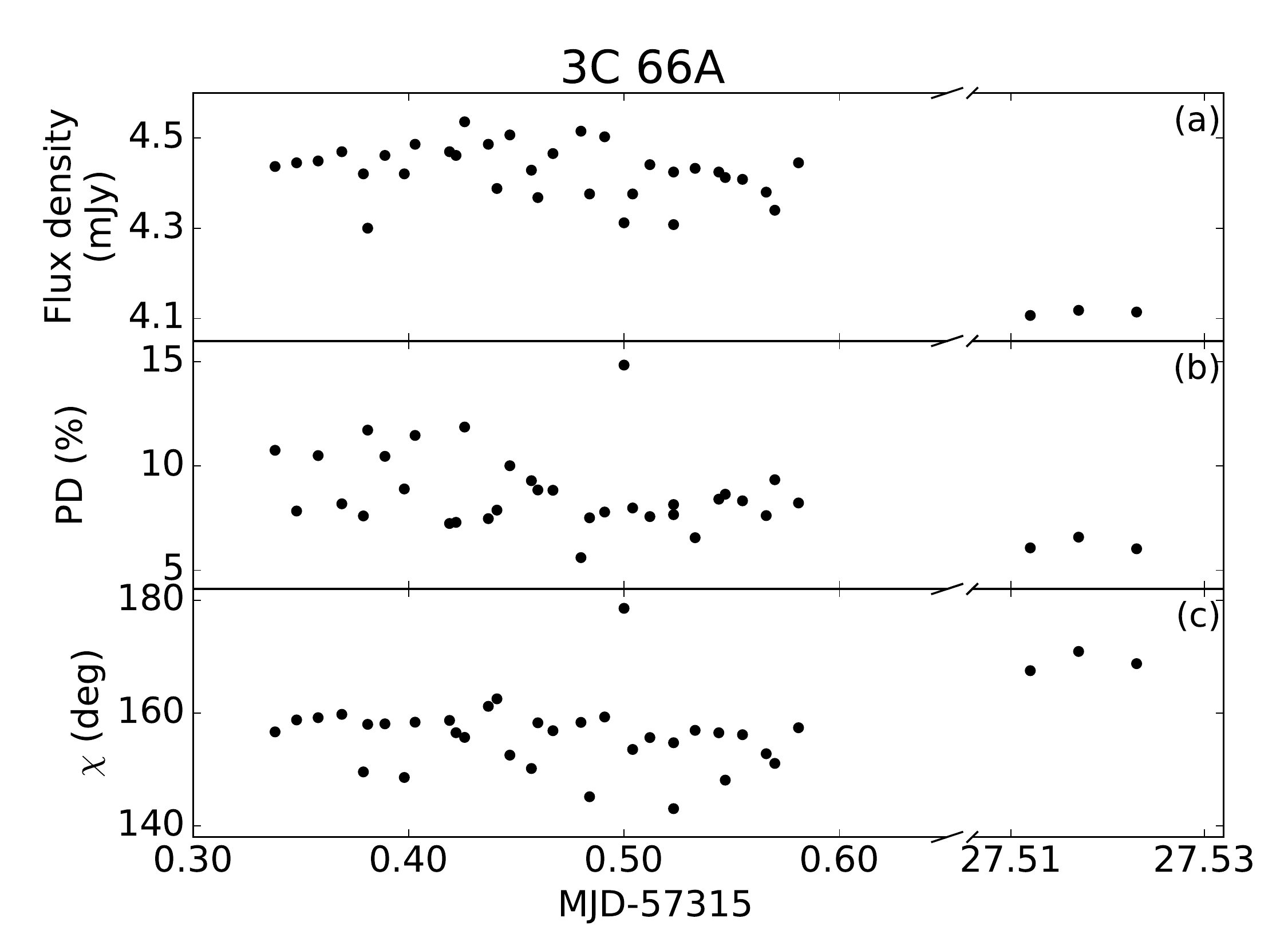}
}
\begin{minipage}{\textwidth}
 \caption{Long-term variability of total flux density, PD and $\chi$ of the seven blazars for which the intra-night monitoring was carried out on more than one occasions.}
\label{longterm_polflux}
\end{minipage}

\end{figure*}

Figure~\ref{longterm_polflux} presents the long-term polarization variability for the blazars monitored on more than one occasion. Unlike for the intra-night polarization light curves, quite substantial changes in total intensity, PD and $\chi$ are noted whenever our observations extended over timescales of months to years.

\renewcommand{\thefigure}{\arabic{figure} (Cont.)}
\addtocounter{figure}{-1}

\begin{figure*}
\centering
\hbox{
\includegraphics[width=0.5\textwidth]{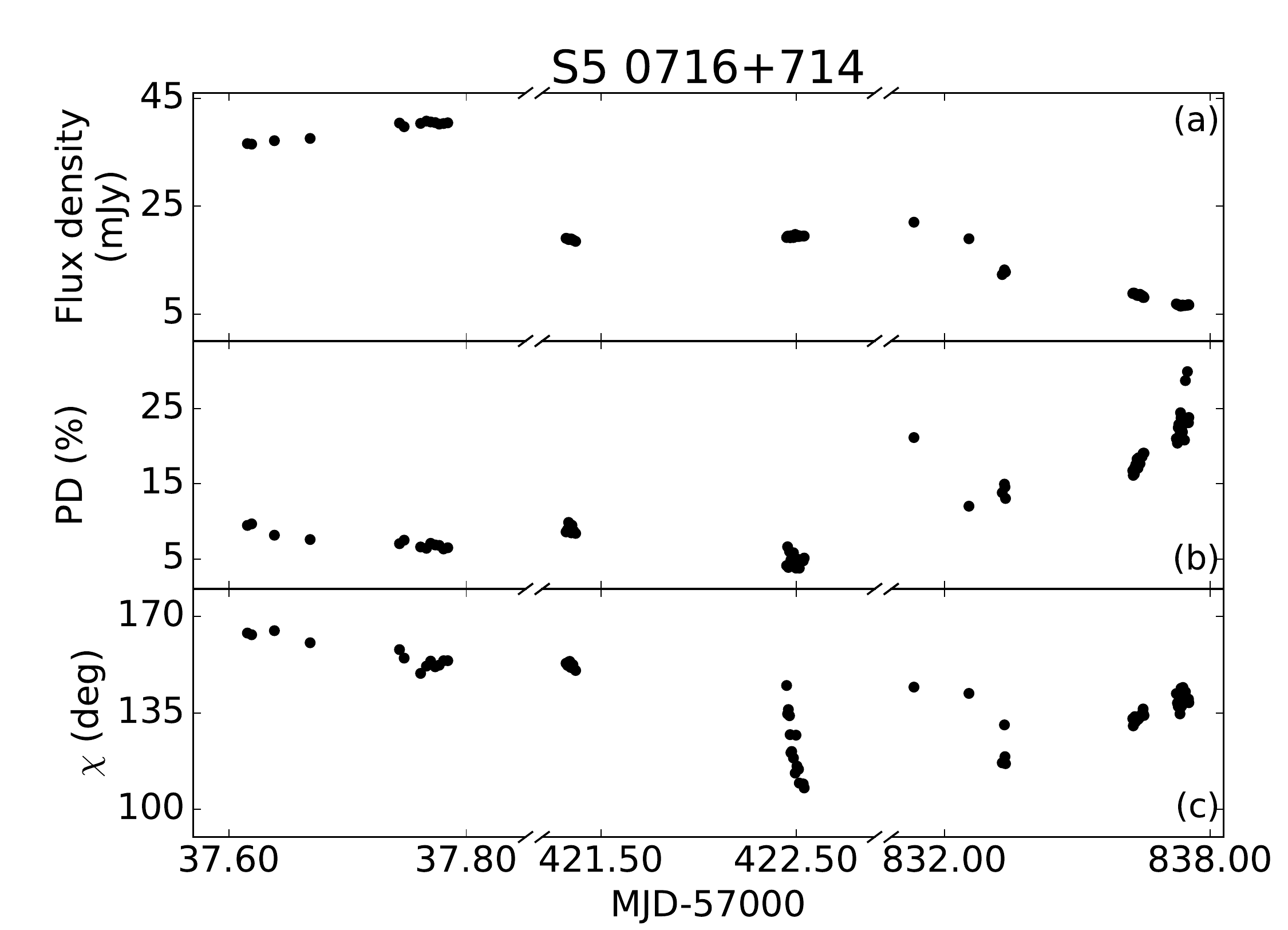}
\includegraphics[width=0.5\textwidth]{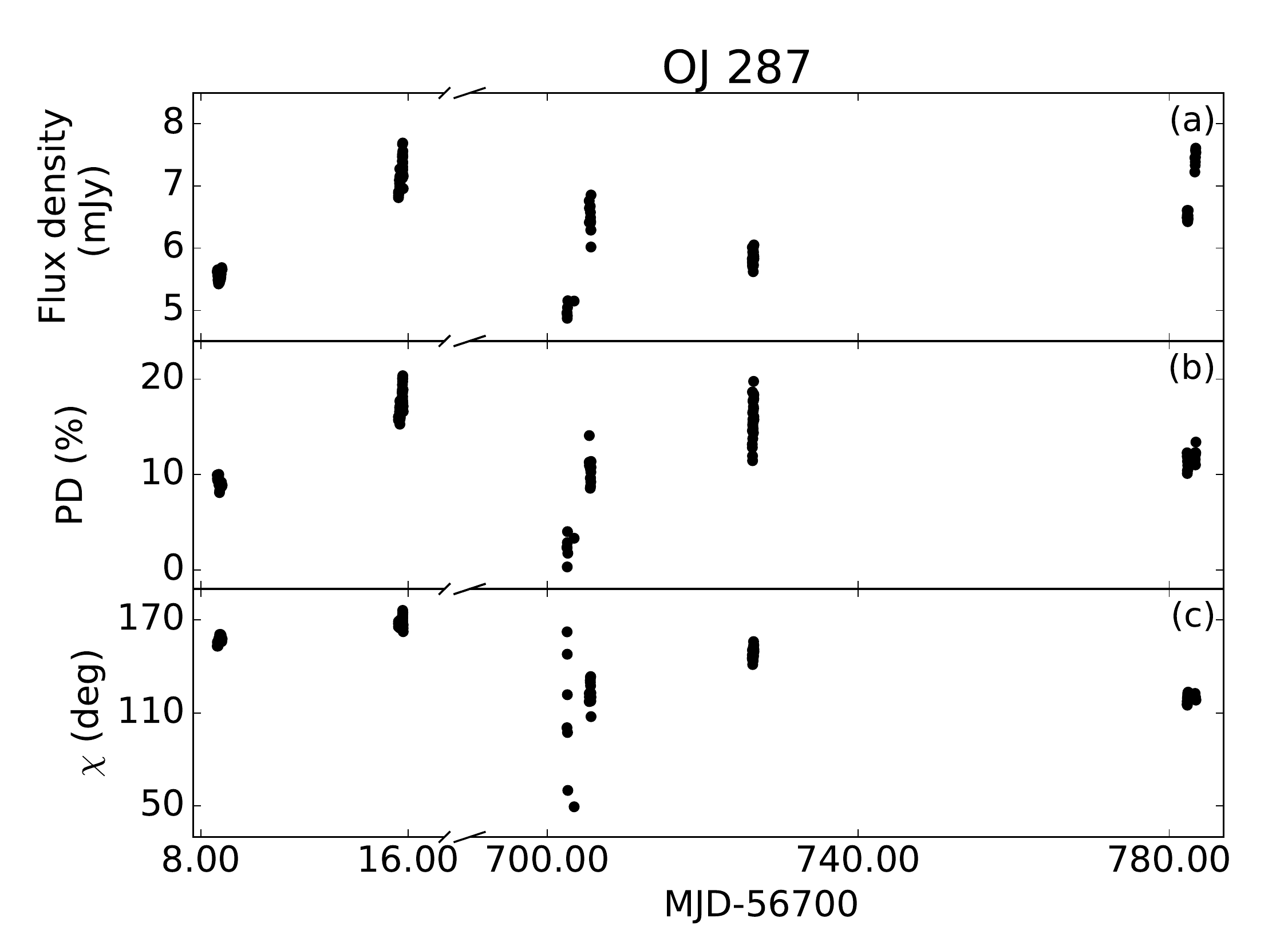}
}
\begin{minipage}{\textwidth}
 \caption{}
\end{minipage}

\end{figure*}

\renewcommand{\thefigure}{\arabic{figure} (Cont.)}
\addtocounter{figure}{-1}

\begin{figure*}
\centering
\hbox{
\includegraphics[width=0.5\textwidth]{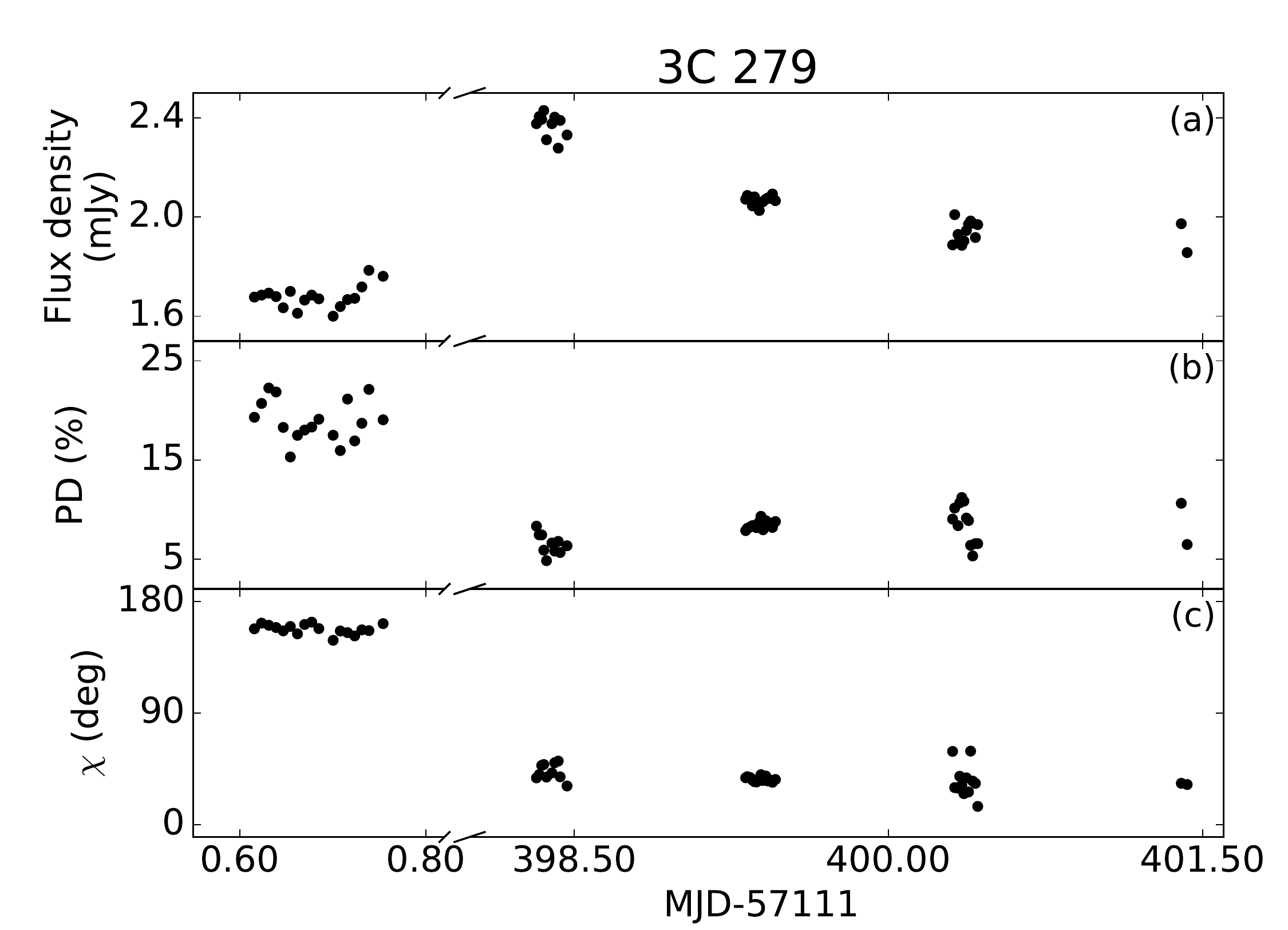}
\includegraphics[width=0.5\textwidth]{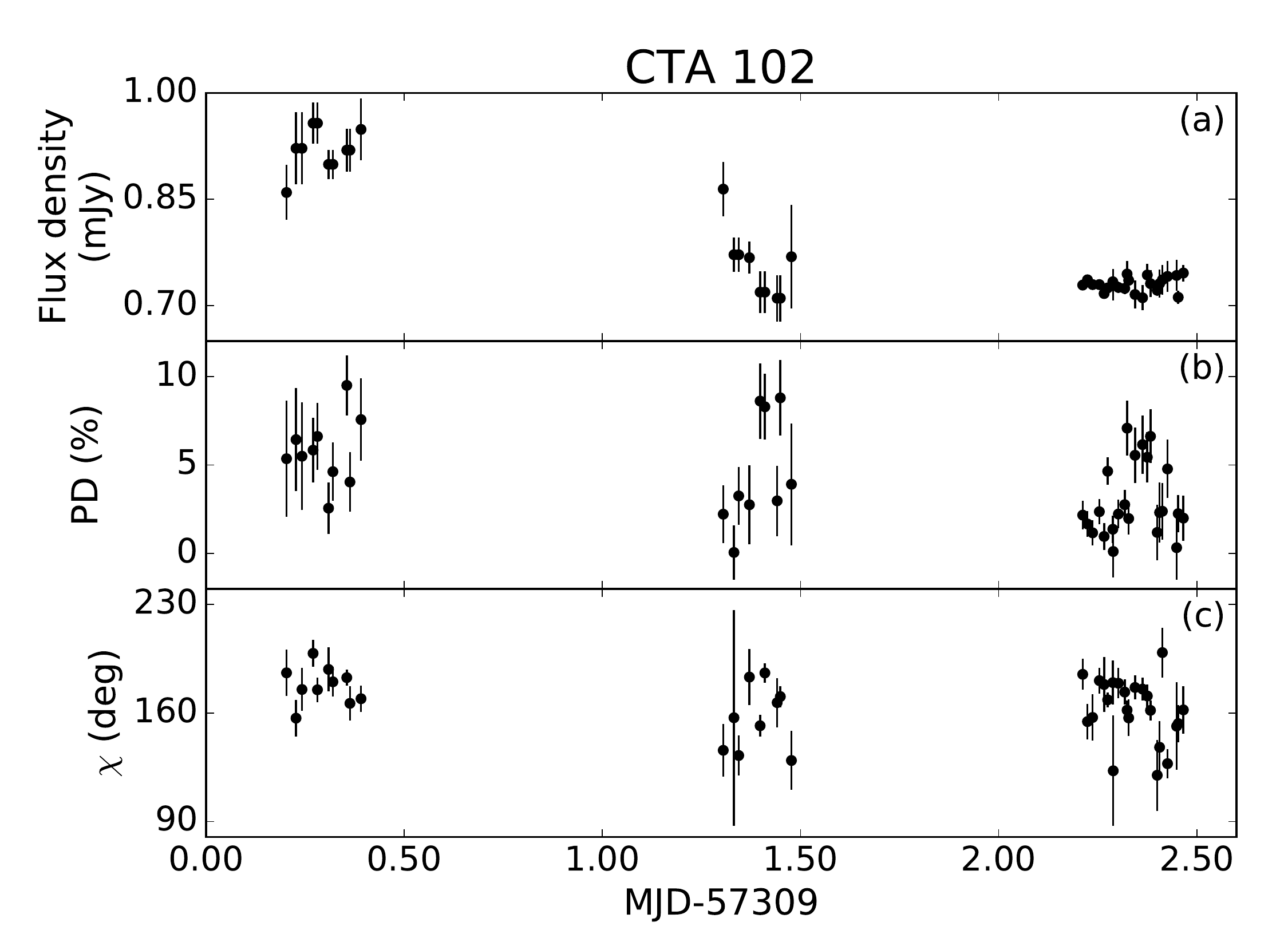}
}
\begin{minipage}{\textwidth}
 \caption{}
\end{minipage}

\end{figure*}

\renewcommand{\thefigure}{\arabic{figure} (Cont.)}
\addtocounter{figure}{-1}

\begin{figure*}
\centering
\hbox{
\includegraphics[width=0.5\textwidth]{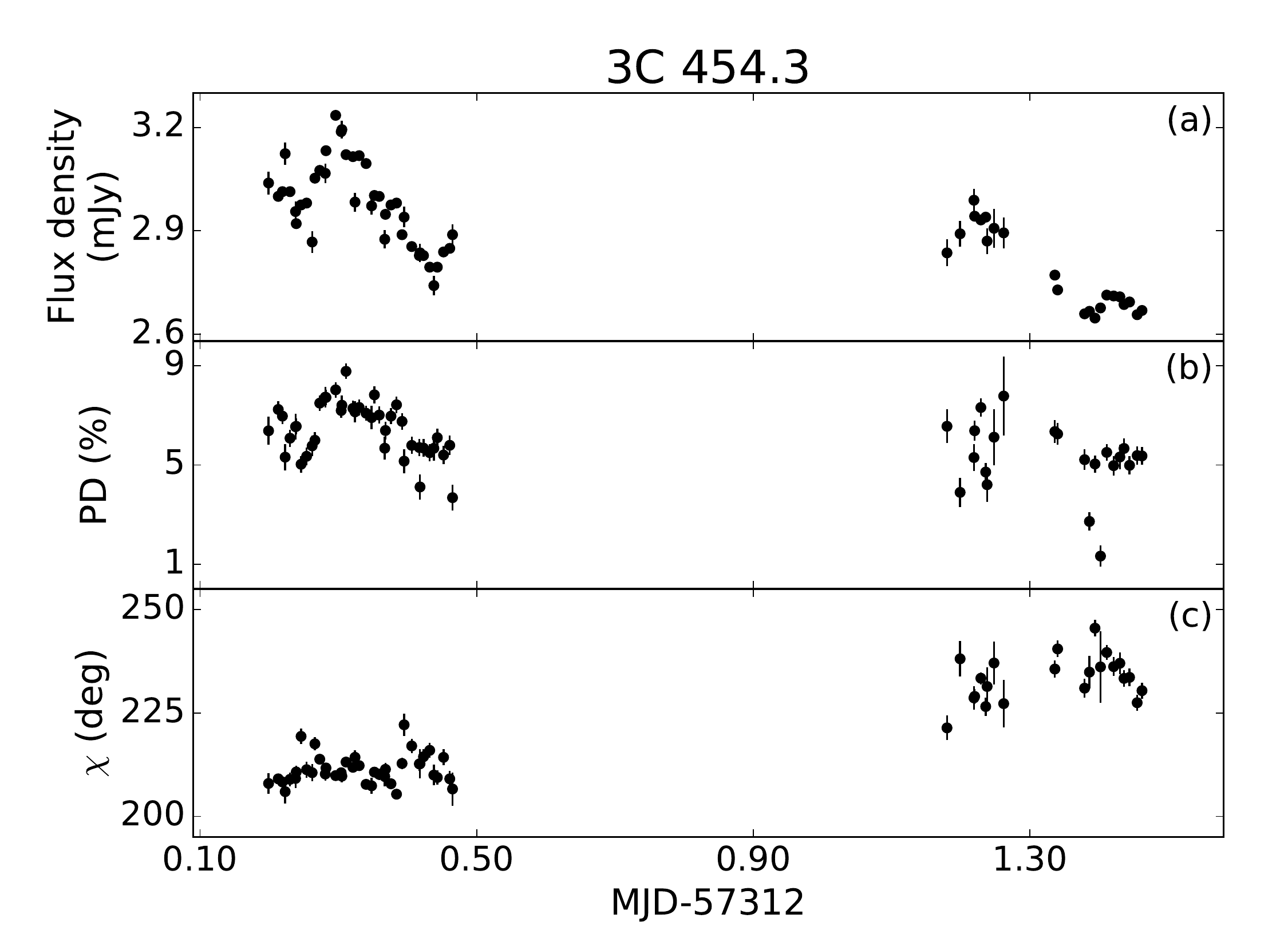}
}
\begin{minipage}{\textwidth}
 \caption{}
\end{minipage}

\end{figure*}

\section{Results} 
\label{sec:results}
Considering the myriad features shown by blazars on both intra-night and long-term timescales, we provide specific notes on variability shown by sources that emerge from this study.

\begin{enumerate}

\item{ {\bf 0109$+$224:} This is a intermediate-frequency peaked BL Lac object \citep{Abdo10b}. We could gather the total intensity data on one occasion (Table~\ref{results_finv}, Figure~\ref{fig_finv}) and polarization data on two occasions (Table~\ref{results_polinv}, Figure~\ref{fig_polinv}). The blazar appeared to show  an apparent $\approx$10 percent amplitude variability but was assigned a `probable variable' status in the B- and V-bands and a `non-variable' status in R-band because the star--star DLC itself turned out to be variable (Table~\ref{results_finv}). The polarization intra-night curve is shown only for the monitoring conducted on 2015 November 13 (Figure~\ref{fig_polinv}); as we could gather only three data points on 2015 December 16. The PD remained stable at $\approx$13 percent with about 20 deg change in polarization angle between the two occasions (Table~\ref{results_polinv}; Figure~\ref{longterm_polflux}). }

\item{ {\bf 3C\,66A:} This blazar is classified as an intermediate-frequency peaked BL Lac object \citep{Abdo10b}. It was monitored on two nights in total intensity (Table~\ref{results_finv}, Figure~\ref{fig_finv}) and two nights in polarized light (Table~\ref{results_polinv}, Figure~\ref{fig_polinv}) where only mild variations in total intensity are detected with no changes in PD or $\chi$. Simultaneous monitoring in flux density and polarization was successfully conducted on 2015 October 19 (Figure~\ref{fig_simultaneous}). On longer timescales, the blazar shows no change in the spectral index ($\sim$1.5) when the flux density decreased by $\approx$ 10 percent in about one month (Table~\ref{results_cinv}; Figure~\ref{longterm_spflux}). 3C\,66A's PD ranges from $\approx$5 percent to over 10 percent and $\chi$ fluctuates by $\sim$15 deg on the night of 2015 October 19 (Table~\ref{results_polinv}). Its flux density dropped by $\approx$7 percent between the two sessions, which were separated by 25 days, while the PD and $\chi$ remain essentially steady during the latter observation, though we note it was limited to 3 measurements}  (Figure~\ref{longterm_polflux}).

\item{ {\bf S5\,0716$+$714:} This blazar is classified as an intermediate-frequency peaked BL Lac object \citep{Abdo10b}. The total intensity monitoring was carried out in V- and I-bands on three occasions where it exhibited persistently `confirmed' variability in both bands (Table~\ref{results_finv}; Figure~\ref{fig_finv}). The $\psi$ ranged between 3 to 25 percent on a given night. Confirmed colour variability is noted on two nights while a `probable' status is assigned for  2015 February 9 (Table~\ref{results_cinv}; top panels of Figure~\ref{fig_cmd}). Polarization monitoring was carried out on eight nights with light curves presented for five occasions (Table~\ref{results_polinv}; Figure~\ref{fig_polinv}). The flux density in the R-band shows `confirmed' variations on four occasions, with changes in PD of a few percent and in $\chi$ by a few degrees being noted (Table~\ref{results_polinv}). 

The two-point spectral index exhibits a clear counter-clockwise loop on 2015 January 11 (Figure~\ref{fig_cmd}). On 2015 January 15, however, the spectral index seems to flatten as the flux density decreases (Figure~\ref{fig_cmd}). No clear pattern is noticed on 2015 February 9. In polarization data obtained on 2015 January 15, a monotonic decrease in PD (10 percent to 3 percent) with an increase in flux density (10 percent) and a rotation of $\chi$ by 8 deg is noted (Figure~\ref{fig_poltrends}). Exactly the opposite trend is noted on 2017 March 24 where the PD increased (16 percent to 19 percent) as the flux density decreased by 10 percent and $\chi$ rotated by 8 deg (Figure~\ref{fig_poltrends}). However, no trend is discernible for PD--flux density, $\chi$--flux density or PD--$\chi$ plane for 2015 March 25 (Figure~\ref{fig_poltrends}). Incidentally, 2015 January 15 is also the night when simultaneous flux density and polarization monitoring was available for the blazar for 1.5 hours (17.5 to 19 UT; Figure~\ref{fig_simultaneous}). During this time, a nearly monotonic increase of flux density (5 percent) with a flattening of spectral index (1.6 to 1.45), decrease in PD (2 percent) and a stable $\chi$ (152 deg) is observed.  On longer-term timescales, a steepening of the spectral index (1.28 to 1.53) with a 25 percent increase in flux density over four days is recorded. The spectral index flattens back to the initial value (1.3) with the flux density decreasing by 60 percent over 25 days when it was last monitored by us (Figure~\ref{longterm_spflux}). The long-term polarization variability for the blazar shows a variable PD (6--20 percent) but a stable $\chi$ $\approx$155 deg (Figure~\ref{longterm_polflux}). }

\item{ {\bf OJ\,287:} This blazar is classified as low-frequency peaked BL Lac object \citep{Abdo10a} and  was most frequently observed during our monitoring programme (Table~\ref{results_finv}; Figure~\ref{fig_finv}). It showed `confirmed' microvariability on four occasions in all the passbands (2014 February 20, 2016 February 6, 2016 March 7, 2016 April 2) and on two occasions the results of microvariability detection in passbands were different (2015 February 12 and 2016 April 4). Significant colour variability was detected on 2014 February 20 and 2015 February 12 (Table~\ref{results_cinv}).  We saw a mild hint of a counter-clockwise loop in the spectral index--flux density plane on 2014 February 20 (Figure~\ref{fig_cmd}), while no discernible trend was seen on 2015 February 12 (Figure~\ref{fig_cmd}). 

Polarization monitoring was carried out on seven occasions (Table~\ref{results_polinv}; Figure~\ref{fig_polinv}). On 2014 February 20, along with the previously mentioned  hint of counter--clockwise loop in the PD--flux density plane a clockwise loop in the $\chi$--flux density plane was seen accompanied by a small monotonic decrease in PD (2 percent) and with a rotation of $\chi$ by 8 deg (Figure~\ref{fig_poltrends}). The total intensity varied in all bands on 2014 February 20 and 2016 April 4 while the spectral index varied only on 2014 February 20 (Tables~\ref{results_finv} and ~\ref{results_cinv}). From the analysis of long-term variability, the blazar showed a variety of features in the spectral index--flux density plane (Figure~\ref{longterm_spflux}). At first, the different flux density states of OJ 287 could be characterized by a `mean' spectral index ($\approx$1.5). Later, the same flux density states appear to have significantly different spectral indices (1.9 and 1.3 on nights marked as 1 and 2). Steepening of the spectral index with the increase of flux density (between observing epochs 2 and 3)  was observed as well as no change in spectral index (between epochs 3 and 4, 4 and 5, 5 and 6; see also Table~\ref{results_cinv}). The flux density changed by $\approx$100 percent  over a timescale of  two months (between epochs 4 and 6). Similarly, the PD ranged between 2--17 percent and $\chi$ rotated by 90 deg over the two year timespan covered by us (Figure~\ref{longterm_polflux}).}

\item{ {\bf 3C\,279:} This blazar is classified as a low-frequency peaked FSRQ \citep{Abdo10a}. We could  monitor it only in polarized light, also resulting in R--band total intensity (Table~\ref{results_polinv}, Figure~\ref{fig_polinv}). The total flux density varied on two occasions (2016 April 30 and May 1; Table~\ref{results_polinv}). PD and $\chi$ apparently remained rather stable; however, this could be  due to the relatively poor signal to noise ratio of these measurements. Considering the  long-term variability of 3C\,279, the total intensity changed by 50 percent, the} PD from 5 to 18 percent and $\chi$ by 100 deg, over the one year timespan covered in this study (Figure~\ref{longterm_polflux}).

\item{ {\bf PG\,1553$+$11:} This is a high frequency peaked BL Lac object \citep{Abdo10a} which remained `non-variable' during a single total flux density monitoring session (Table~\ref{results_finv}, Figure~\ref{fig_finv}). }

\item{ {\bf CTA\,102:} This blazar is classified as low-frequency peaked FSRQ \citep{Abdo10a}. The intra-night monitoring in flux density was carried out on two nights and polarization was measured on three nights. This blazar did not show variability in flux density (Tables~\ref{results_finv} and ~\ref{results_polinv}) or in spectral index (Table~\ref{results_cinv}). Simultaneous monitoring was conducted on 2015 October 13 and 15 with no changes detected in flux density, PD, and $\chi$ during either night (Figure~\ref{fig_simultaneous}). Nonetheless, the blazar flux density did show a $\approx$30 percent decrease over two days while the spectral index remained steady (Figure~\ref{longterm_spflux}). Our polarization data for CTA\,102 shows about a 25 percent change in total intensity with no changes in PD (around 5 percent) and $\chi$ (about 160 deg) over the long-term monitoring (Figure~\ref{longterm_polflux}). } 

\item{ {\bf 3C\,454.3:} This blazar also is classified as low-frequency peaked FSRQ \citep{Abdo10a}. It was monitored on a single occasion in total intensity where it showed `confirmed' variability in R--band and `probable' variability in V--band (Table~\ref{results_finv}); the colour remained steady between the two passbands (Table~\ref{results_cinv}). The polarization monitoring was carried out on two occasions and significant variability was noted in flux density, PD and $\chi$ on both nights (Table~\ref{results_polinv}). On 2015 October 16, the sampling was particularly good, so that the target was monitored continuously for the total duration of about six hours, during which a clear flux enhancement episode could be seen, including both the rising and the decaying phases. As a case study, we have analysed this particular dataset in Appendix~\ref{app:microflare} in more detail, attempting to separate the ``base emission component'' from the ``flaring emission component''. Unfortunately, none of the multiband total intensity and polarization monitoring sessions overlapped. On 2015 October 16 a hint of an increase in the PD as the flux density increases is noted in the PD--flux density plane, while $\chi$ remains stable at $\sim 210$ deg  (Figure~\ref{fig_poltrends}). During the polarization monitoring carried out on  2015 October 17, counter--clockwise loops in the PD--flux density and $\chi$--flux density planes, respectively, seem to be present, though no clear evolution in the PD--$\chi$ plane was seen (Figure~\ref{fig_poltrends}). Between those two consecutive nights the polarization data show a 10 percent intensity change, but also essentially constant PD ($\sim 7$ percent), along with a $\sim$30 deg change in $\chi$ (Figure~\ref{longterm_polflux}) }.

\end{enumerate}

\section{Discussion and Conclusions}
\label{sec:conclusions}

Microvariability in blazar sources has gathered special attention, partially because of the requirements of very efficient particle acceleration and very fast energy dissipation mechanisms necessary to produce it. Particularly challenging are the high-amplitude flares with  well-defined rise and decay profiles which allows for size estimation of the emission zone from flux  density doubling arguments. As such, these timescales occasionally violate the shortest `characteristic' variability timescale provided by the light-crossing timescale (t$_{lc}$) of the event horizon of the SMBH \citep[$\sim$15 min for 10$^8$M$_{\odot}$;][where t$_{lc}$= $r_g/$$c$ = $G$\,$M$/$c^3$ where $r_g$ is the gravitational radius, $G$ is the gravitational constant, $M$ is the mass of SMBH, and $c$ is the speed of light]{Begelman84}. 

Nonetheless, variations on intra-night timescales most often constitute monotonic increases or decreases in intensity, indicating sizes $>$\,10$^{15}$\,cm ($R \sim$\,$c$\,$\delta$\,$t_{var}$ for $\delta=$10 and taking $t_{var}$=\,5\,hr from the typical length of the intra-night monitoring session) at least in the simple, and frequently employed,  single emission zone scenario. Such sizes could then be related to the distance of the dissipation site ($r$) from the central SMBH by $R\, \simeq r\theta_j$ for a conical jet with an opening angle, $\theta_j$= $\theta_{obs} \sim 1/\Gamma$, where $\theta_{obs}$ is the angle between the jet axis and the observer's lines of sight and $\Gamma$ is the jet Lorentz factor \citep{Sikora94}. For the popular internal shock scenario of particle acceleration, shells of relativistic plasma, ejected intermittently from the center (with a larger dispersion in bulk velocity), collide with each other at a characteristic distance $r \sim r_g (\Gamma/10)^2$ \citep{Spada01,Moderski03}. The flux variability on hourly timescales noted here would then be compatible with particle acceleration by internal shocks. However, in such a scenario, a correlation between the flux and PD changes are expected, because of a compression of the jet plasma with initially tangled (by assumption) jet magnetic field at the shock front. Several observations for individual blazars on a night-to-night basis presented in the literature conforms to this model, possibly with some minor modifications or additions \citep[e.g.,][]{Hagen-Thorn08,Itoh13a, Covino15, Bhatta16}.

In our data set, we observe a variety of flux--PD variability patterns, including clear instances with a relatively straightforward anti-correlation (S5\,0716+714 on 2015 January 15 and 2017 March 23), correlation (3C\,454 on 2015 October 16), counter-clockwise looping (OJ\,287 on 2014 February 20), but also instances with no obvious correlation pattern (S5\,0716+714 on 2017 March 24, OJ\,287 on 2016 February 7, 3C\,454 on 2015 October 17). It is, however, worthwhile noting that whenever we see any order in the flux--PD changes, it is reflected at the same time in the flux--$\chi$ plane, or equivalently the PD--$\chi$ plane (see Figure\,\ref{fig_poltrends}). In particular, the anti-correlation between the flux and PD seen clearly in S5\,0716+714 on 2015 January 15 and 2017 March 23 are accompanied by an anti-correlation between the polarization angle $\chi$ and flux, while the counter-clockwise flux-PD looping behaviour noted for OJ\,287 on 2014 February 20 is accompanied by a clockwise looping in the flux-$\chi$ representation (a similar pattern of variability to what has been noticed for the blazar 3C\,454.3 in \citealt{GK19}, based on the extensive observations reported by \citealt{Gupta17}). This piece of evidence seems to suggest that in order to fully understand the physics behind the intra-night variability of blazar sources, a three-dimensional parameter space of a flux, polarization degree, and polarization angle has to be considered. This still could possibly be reconciled with the internal shock model, if one allows for example for a changes in the shock angle (i.e., the angle between the shock normal and the line of sight) during the shock evolution along the outflow \citep[see, e.g., the discussion in][]{Bhatta15}, and/or for a presence of a partly ordered magnetic field \citep[e.g., in a form of a helical distortion flowing through the shock; see][]{Perlman11}. 

It should be emphasized, however, that the results of polarization microvariability monitoring indicate variable PD and $\chi$ only on a few occasions (Table~\ref{results_polinv})). For about half of these instances, the quality of polarization measurements is poor (low signal to noise ratio, too few data points, or big gaps in the light curve). Therefore, we give the mean values of total flux density, PD and $\chi$ for the monitoring session considered (Columns 4, 5, 6 of Table~\ref{results_polinv}). The PD turns out to be between 2 and 23\% with a factor of few changes seen over the timescale of days/months for a given blazar source. The mean $\chi$ also showed significant variability over days/months timescales during the monitoring campaign (Table~\ref{results_polinv}, Figure~\ref{fig_polinv}).

Our study using quasi-simultaneous B, V, R, and I-band intra-night light curves allowed us to estimate the two-point spectral index for several intra-night monitoring sessions (see Table~\ref{results_cinv}). The mean $\alpha$ during our monitoring programme ranges from 0.68 up to 1.87. Moreover, out of 10 cases where statistically significant flux microvariability was detected (Table~\ref{results_finv}),  `confirmed' colour microvariability status is seen on four nights and a `probable' variable status is assigned on one night (Table~\ref{results_cinv}). This means that the `colour' is likely to be seen to vary on about half of the instances when significant flux density microvariability is detected in a sample, or, conversely, `achromatic' flux microvariability is expected on about half of the monitoring sessions, i.e., the colour microvariability DC is $\approx$50 percent. It is, however, important to note that `colour' or spectral index variability is a second order variation (as compared to flux density variations) and therefore is harder to detect due to bigger statistical errors on spectral index measurements arising solely from error propagation. Moreover, since we are severely limited by small number statistics in deriving a duty cycle (sample of 7 blazars monitored on 18 intra-night sessions, where flux density microvariability is noted on 10 occasions), this should only be considered as a crude estimate of a colour microvariability duty cycle for blazar sources. 

The reported spectral indices are typically $>0.5$, implying steep energy spectra of the emitting electrons (even taking into account spectral steepening due to radiative losses), consistently with what was established in the previous multiband optical (and optical-NIR) observations of the studied targets and other blazars of a similar type \citep[e.g.,][]{Agarwal15, Bachev15, Gupta16}. Such steep electron spectra with the corresponding power-law energy indices $>2$ are in principle what could be expected if the dominant electron acceleration mechanism is the Fermi type I process operating at mildly-relativistic/relativistic shock waves \citep[see][]{Sironi09,Sironi11}, in accord with the internal shock scenario. But again, within the framework of this scenario, various correlation patterns between the flux and the spectral index are expected depending on particular relations between the acceleration and cooling timescales; in particular, a clockwise looping in the flux-spectral index plane (``hard lagging'') is expected if both timescales are comparable, while a counter-clockwise looping (``soft-lagging'') is expected if the acceleration timescale is much shorter than the cooling timescale \citep{Kirk98}. Interestingly, for S5\,0716+714 observed on 2015 January 11, we clearly see a soft-lagging (the spectrum hardens during the flux rising phase, and softens when the flux decreases; see Figure\,\ref{fig_cmd}). A similar behaviour can be also noted for OJ\,287 on 2014 February 20, and this evolution, is notably accompanied by a counter-clockwise looping in the flux--PD plane, as well by a clock-wise looping in the flux-$\chi$ plane (see Figures\,\ref{fig_cmd} and \ref{fig_poltrends}). On the other hand, in the case of S5\,0716+714 on 2015 January 15, we see a a clear spectral flattening with the decreasing flux, which could be a part of a hard-looping cycle.      

When comparing data on these blazars taken over \emph{longer} timescales, a few percent to a factor of a few changes are noted in total flux densities (Figures~\ref{longterm_spflux}, ~\ref{longterm_polflux}). The two-point spectral index, however, shows a complex long-term behaviour with no preference of spectral state as a function of flux density (Figure~\ref{longterm_spflux}).  A detailed discussion of the linkage between the optical flux density and polarization variability of blazars is precluded at this stage due to the rather large variation in the data quality and temporal coverage achieved for the blazar sample covered in this study. Nonetheless, we would like to highlight that by far, the clearest example of chromatic variability found in the present study --- both on the short (intra-night) and also long timescales --- is for the intermediate-frequency peaked BL Lac object S5 0716+714: as noted in the previous paragraph, during its monitoring on 2015 January 15 its V-I spectral index showed a flattening from 1.6 to 1.45 as its V-flux  density decreased from 56.5 mJy to 55.5 mJy on hour-like time scale, and an analogous pattern was displayed by this source on month-like time scale, when its V-I spectral index flattened from 1.53 to 1.28 as its V-flux decreased from 32 to 17 mJy in the course of 25 days (Sect.~\ref{sec:results}, Figure~\ref{longterm_spflux}). The sense of this correlation is opposite to the bluer when brighter behaviour typically observed for BL Lac objects (Sect. 1).

Finally, we comment that for the observations analysed here, the computed flux microvariability DC, using 18 nights' monitoring of 7 blazars, turns out to be $\sim$45 percent (Table~\ref{results_finv}), similar to that previously obtained for the blazar source population, in general, of $\sim$40 percent \citep[][]{Goyal13b}, and much higher than that found for non-blazar AGN \citep{Goyal12}. In this respect, the targets selected in our monitoring are representative for the blazar population.

\clearpage
\section*{Acknowledgements}
We thank the referee for providing constructive comments on the manuscript. We thank the staff of ARIES and Kanata for carrying out some of the observations presented in this study. AG acknowledges support from the Polish National Science Centre (NCN) through the grant 2018/29/B/ST9/02298. {\L}S acknowledges the support of NCN via 2016/22/E/ST9/00061. SZ's work is supported by NCN via 2018/29/B/ST9/01793. GK would like to thank Indian National Science Academy for a Senior Scientist fellowship.The St. Petersburg University team acknowledges support from Russian Science Foundation grant 17-12-01029.


\begin{thebibliography}{}
\makeatletter
\relax
\def\mn@urlcharsother{\let\do\@makeother \do\$\do\&\do\#\do\^\do\_\do\%\do\~}
\def\mn@doi{\begingroup\mn@urlcharsother \@ifnextchar [ {\mn@doi@}
  {\mn@doi@[]}}
\def\mn@doi@[#1]#2{\def\@tempa{#1}\ifx\@tempa\@empty \href
  {http://dx.doi.org/#2} {doi:#2}\else \href {http://dx.doi.org/#2} {#1}\fi
  \endgroup}
\def\mn@eprint#1#2{\mn@eprint@#1:#2::\@nil}
\def\mn@eprint@arXiv#1{\href {http://arxiv.org/abs/#1} {{\tt arXiv:#1}}}
\def\mn@eprint@dblp#1{\href {http://dblp.uni-trier.de/rec/bibtex/#1.xml}
  {dblp:#1}}
\def\mn@eprint@#1:#2:#3:#4\@nil{\def\@tempa {#1}\def\@tempb {#2}\def\@tempc
  {#3}\ifx \@tempc \@empty \let \@tempc \@tempb \let \@tempb \@tempa \fi \ifx
  \@tempb \@empty \def\@tempb {arXiv}\fi \@ifundefined
  {mn@eprint@\@tempb}{\@tempb:\@tempc}{\expandafter \expandafter \csname
  mn@eprint@\@tempb\endcsname \expandafter{\@tempc}}}

\bibitem[\protect\citeauthoryear{{Abdo} et~al.,}{{Abdo}
  et~al.}{2010a}]{Abdo10a}
{Abdo} A.~A.,  et~al., 2010a, \mn@doi [\apj] {10.1088/0004-637X/715/1/429},
  \href {http://adsabs.harvard.edu/abs/2010ApJ...715..429A} {715, 429}

\bibitem[\protect\citeauthoryear{{Abdo} et~al.,}{{Abdo}
  et~al.}{2010b}]{Abdo10b}
{Abdo} A.~A.,  et~al., 2010b, \mn@doi [\apj] {10.1088/0004-637X/716/1/30},
  \href {http://adsabs.harvard.edu/abs/2010ApJ...716...30A} {716, 30}

\bibitem[\protect\citeauthoryear{{Abdo} et~al.,}{{Abdo}
  et~al.}{2010c}]{Abdo10c}
{Abdo} A.~A.,  et~al., 2010c, \mn@doi [\apj] {10.1088/0004-637X/722/1/520},
  \href {http://adsabs.harvard.edu/abs/2010ApJ...722..520A} {722, 520}

\bibitem[\protect\citeauthoryear{{Abdo} et~al.,}{{Abdo}
  et~al.}{2011a}]{Abdo11b}
{Abdo} A.~A.,  et~al., 2011a, \mn@doi [\apj] {10.1088/0004-637X/727/2/129},
  \href {http://adsabs.harvard.edu/abs/2011ApJ...727..129A} {727, 129}

\bibitem[\protect\citeauthoryear{{Abdo} et~al.,}{{Abdo}
  et~al.}{2011b}]{Abdo11a}
{Abdo} A.~A.,  et~al., 2011b, \mn@doi [\apj] {10.1088/0004-637X/736/2/131},
  \href {http://adsabs.harvard.edu/abs/2011ApJ...736..131A} {736, 131}

\bibitem[\protect\citeauthoryear{{Ackermann} et~al.,}{{Ackermann}
  et~al.}{2016}]{Ackermann16}
{Ackermann} M.,  et~al., 2016, \mn@doi [\apjl] {10.3847/2041-8205/824/2/L20},
  \href {http://adsabs.harvard.edu/abs/2016ApJ...824L..20A} {824, L20}

\bibitem[\protect\citeauthoryear{{Agarwal} et~al.,}{{Agarwal}
  et~al.}{2015}]{Agarwal15}
{Agarwal} A.,  et~al., 2015, \mn@doi [\mnras] {10.1093/mnras/stv1208}, \href
  {http://adsabs.harvard.edu/abs/2015MNRAS.451.3882A} {451, 3882}

\bibitem[\protect\citeauthoryear{{Agudo} et~al.,}{{Agudo}
  et~al.}{2011}]{Agudo11}
{Agudo} I.,  et~al., 2011, \mn@doi [\apjl] {10.1088/2041-8205/735/1/L10}, \href
  {http://adsabs.harvard.edu/abs/2011ApJ...735L..10A} {735, L10}

\bibitem[\protect\citeauthoryear{{Aharonian} et~al.,}{{Aharonian}
  et~al.}{2007}]{Aharonian07}
{Aharonian} F.,  et~al., 2007, \mn@doi [\apjl] {10.1086/520635}, \href
  {http://adsabs.harvard.edu/abs/2007ApJ...664L..71A} {664, L71}

\bibitem[\protect\citeauthoryear{{Angelakis} et~al.,}{{Angelakis}
  et~al.}{2016}]{Angelakis16}
{Angelakis} E.,  et~al., 2016, \mn@doi [\mnras] {10.1093/mnras/stw2217}, \href
  {https://ui.adsabs.harvard.edu/abs/2016MNRAS.463.3365A} {463, 3365}

\bibitem[\protect\citeauthoryear{{Bachev}}{{Bachev}}{2015}]{Bachev15}
{Bachev} R.,  2015, \mn@doi [\mnras] {10.1093/mnrasl/slv059}, \href
  {http://adsabs.harvard.edu/abs/2015MNRAS.451L..21B} {451, L21}

\bibitem[\protect\citeauthoryear{{Bachev}, {Strigachev}  \& {Semkov}}{{Bachev}
  et~al.}{2005}]{Bachev05}
{Bachev} R.,  {Strigachev} A.,   {Semkov} E.,  2005, \mn@doi [\mnras]
  {10.1111/j.1365-2966.2005.08708.x}, \href
  {http://cdsads.u-strasbg.fr/abs/2005MNRAS.358..774B} {358, 774}

\bibitem[\protect\citeauthoryear{{Begelman}, {Blandford}  \& {Rees}}{{Begelman}
  et~al.}{1984}]{Begelman84}
{Begelman} M.~C.,  {Blandford} R.~D.,   {Rees} M.~J.,  1984, \mn@doi [Reviews
  of Modern Physics] {10.1103/RevModPhys.56.255}, \href
  {http://adsabs.harvard.edu/abs/1984RvMP...56..255B} {56, 255}

\bibitem[\protect\citeauthoryear{{Begelman}, {Fabian}  \& {Rees}}{{Begelman}
  et~al.}{2008}]{Begelman08}
{Begelman} M.~C.,  {Fabian} A.~C.,   {Rees} M.~J.,  2008, \mn@doi [\mnras]
  {10.1111/j.1745-3933.2007.00413.x}, \href
  {http://adsabs.harvard.edu/abs/2008MNRAS.384L..19B} {384, L19}

\bibitem[\protect\citeauthoryear{{Bevington} \& {Robinson}}{{Bevington} \&
  {Robinson}}{2003}]{Bevington03}
{Bevington} P.~R.,  {Robinson} D.~K.,  2003, {Data reduction and error analysis
  for the physical sciences}.
3rd ed., by Philip R.~Bevington, and Keith D.~Robinson.~Boston, MA:
  McGraw-Hill, ISBN 0-07-247227-8, 2003

\bibitem[\protect\citeauthoryear{{Bhatta} et~al.,}{{Bhatta}
  et~al.}{2015}]{Bhatta15}
{Bhatta} G.,  et~al., 2015, \mn@doi [\apj] {10.1088/2041-8205/809/2/L27}, \href
  {https://ui.adsabs.harvard.edu/abs/2015ApJ...809L..27B} {809, L27}

\bibitem[\protect\citeauthoryear{{Bhatta} et~al.,}{{Bhatta}
  et~al.}{2016}]{Bhatta16}
{Bhatta} G.,  et~al., 2016, \mn@doi [\apj] {10.3847/0004-637X/832/1/47}, \href
  {http://adsabs.harvard.edu/abs/2016ApJ...832...47B} {832, 47}

\bibitem[\protect\citeauthoryear{{Blinov} et~al.,}{{Blinov}
  et~al.}{2016}]{Blinov16}
{Blinov} D.,  et~al., 2016, \mn@doi [\mnras] {10.1093/mnras/stw158}, \href
  {https://ui.adsabs.harvard.edu/abs/2016MNRAS.457.2252B} {457, 2252}

\bibitem[\protect\citeauthoryear{{Bonning} et~al.,}{{Bonning}
  et~al.}{2012}]{Bonning12}
{Bonning} E.,  et~al., 2012, \mn@doi [\apj] {10.1088/0004-637X/756/1/13}, \href
  {http://adsabs.harvard.edu/abs/2012ApJ...756...13B} {756, 13}

\bibitem[\protect\citeauthoryear{{B{\"o}ttcher}, {Reimer}, {Sweeney}  \&
  {Prakash}}{{B{\"o}ttcher} et~al.}{2013}]{Bottcher13}
{B{\"o}ttcher} M.,  {Reimer} A.,  {Sweeney} K.,   {Prakash} A.,  2013, \mn@doi
  [\apj] {10.1088/0004-637X/768/1/54}, \href
  {http://adsabs.harvard.edu/abs/2013ApJ...768...54B} {768, 54}

\bibitem[\protect\citeauthoryear{{Camenzind} \& {Krockenberger}}{{Camenzind} \&
  {Krockenberger}}{1992}]{Camenzind92}
{Camenzind} M.,  {Krockenberger} M.,  1992, \aap, \href
  {http://adsabs.harvard.edu/abs/1992A%26A...255...59C} {255, 59}

\bibitem[\protect\citeauthoryear{{Cellone}, {Romero}  \& {Araudo}}{{Cellone}
  et~al.}{2007}]{Cellone07}
{Cellone} S.~A.,  {Romero} G.~E.,   {Araudo} A.~T.,  2007, \mn@doi [\mnras]
  {10.1111/j.1365-2966.2006.11140.x}, \href
  {http://cdsads.u-strasbg.fr/abs/2007MNRAS.374..357C} {374, 357}

\bibitem[\protect\citeauthoryear{{Ciprini}, {Tosti}, {Raiteri}, {Villata},
  {Ibrahimov}, {Nucciarelli}  \& {Lanteri}}{{Ciprini} et~al.}{2003}]{Ciprini03}
{Ciprini} S.,  {Tosti} G.,  {Raiteri} C.~M.,  {Villata} M.,  {Ibrahimov} M.~A.,
   {Nucciarelli} G.,   {Lanteri} L.,  2003, \mn@doi [\aap]
  {10.1051/0004-6361:20030045}, \href
  {https://ui.adsabs.harvard.edu/abs/2003A&A...400..487C} {400, 487}

\bibitem[\protect\citeauthoryear{{Covino} et~al.,}{{Covino}
  et~al.}{2015}]{Covino15}
{Covino} S.,  et~al., 2015, \mn@doi [\aap] {10.1051/0004-6361/201525674}, \href
  {https://ui.adsabs.harvard.edu/abs/2015A&A...578A..68C} {578, A68}

\bibitem[\protect\citeauthoryear{{Dai}, {Wu}, {Zhu}, {Zhou}, {Ma}, {Yuan}  \&
  {Wang}}{{Dai} et~al.}{2013}]{Dai13}
{Dai} Y.,  {Wu} J.,  {Zhu} Z.-H.,  {Zhou} X.,  {Ma} J.,  {Yuan} Q.,   {Wang}
  L.,  2013, \mn@doi [\apjs] {10.1088/0067-0049/204/2/22}, \href
  {https://ui.adsabs.harvard.edu/abs/2013ApJS..204...22D} {204, 22}

\bibitem[\protect\citeauthoryear{{De Diego}}{{De Diego}}{2010}]{deDiego10}
{De Diego} J.~A.,  2010, \mn@doi [\aj] {10.1088/0004-6256/139/3/1269}, \href
  {http://cdsads.u-strasbg.fr/abs/2010AJ....139.1269D} {139, 1269}

\bibitem[\protect\citeauthoryear{{De Diego}}{{De Diego}}{2014}]{deDiego14}
{De Diego} J.~A.,  2014, \mn@doi [\aj] {10.1088/0004-6256/148/5/93}, \href
  {https://ui.adsabs.harvard.edu/abs/2014AJ....148...93D} {148, 93}

\bibitem[\protect\citeauthoryear{{Dom{\'\i}nguez} et~al.,}{{Dom{\'\i}nguez}
  et~al.}{2011}]{Dominguez11}
{Dom{\'\i}nguez} A.,  et~al., 2011, \mn@doi [\mnras]
  {10.1111/j.1365-2966.2010.17631.x}, \href
  {https://ui.adsabs.harvard.edu/abs/2011MNRAS.410.2556D} {410, 2556}

\bibitem[\protect\citeauthoryear{{Garcia}, {Sodr{\'e}}, {Jablonski}  \&
  {Terlevich}}{{Garcia} et~al.}{1999}]{Garcia99}
{Garcia} A.,  {Sodr{\'e}} L.,  {Jablonski} F.~J.,   {Terlevich} R.~J.,  1999,
  \mn@doi [Monthly Notices of the Royal Astronomical Society]
  {10.1046/j.1365-8711.1999.02884.x}, \href
  {https://ui.adsabs.harvard.edu/abs/1999MNRAS.309..803G} {309, 803}

\bibitem[\protect\citeauthoryear{{Gaur} et~al.,}{{Gaur} et~al.}{2012}]{Gaur12}
{Gaur} H.,  et~al., 2012, \mn@doi [\mnras] {10.1111/j.1365-2966.2012.21583.x},
  \href {https://ui.adsabs.harvard.edu/abs/2012MNRAS.425.3002G} {425, 3002}

\bibitem[\protect\citeauthoryear{{Gaur}, {Gupta}, {Bachev}, {Strigachev},
  {Semkov}, {Wiita}, {Gu}  \& {Ibryamov}}{{Gaur} et~al.}{2017}]{Gaur17}
{Gaur} H.,  {Gupta} A.,  {Bachev} R.,  {Strigachev} A.,  {Semkov} E.,  {Wiita}
  P.,  {Gu} M.,   {Ibryamov} S.,  2017, \mn@doi [Galaxies]
  {10.3390/galaxies5040094}, \href
  {https://ui.adsabs.harvard.edu/abs/2017Galax...5...94G} {5, 94}

\bibitem[\protect\citeauthoryear{{Gaur} et~al.,}{{Gaur} et~al.}{2019}]{Gaur19}
{Gaur} H.,  et~al., 2019, \mn@doi [\mnras] {10.1093/mnras/stz322}, \href
  {https://ui.adsabs.harvard.edu/abs/2019MNRAS.484.5633G} {484, 5633}

\bibitem[\protect\citeauthoryear{{Ghosh}, {Ramsey}, {Sadun}  \&
  {Soundararajaperumal}}{{Ghosh} et~al.}{2000}]{Ghosh10}
{Ghosh} K.~K.,  {Ramsey} B.~D.,  {Sadun} A.~C.,   {Soundararajaperumal} S.,
  2000, \mn@doi [\apjs] {10.1086/313313}, \href
  {http://adsabs.harvard.edu/abs/2000ApJS..127...11G} {127, 11}

\bibitem[\protect\citeauthoryear{{Glass}}{{Glass}}{1999}]{Glass99}
{Glass} I.~S.,  1999, {Handbook of Infrared Astronomy}.
Cambridge University Press, Aug 13, 1999, page 63

\bibitem[\protect\citeauthoryear{{Gonz{\'a}lez-P{\'e}rez}, {Kidger}  \&
  {Mart{\'\i}n-Luis}}{{Gonz{\'a}lez-P{\'e}rez} et~al.}{2001}]{Gonzalez-Perez01}
{Gonz{\'a}lez-P{\'e}rez} J.~N.,  {Kidger} M.~R.,   {Mart{\'\i}n-Luis} F.,
  2001, \mn@doi [\aj] {10.1086/322129}, \href
  {https://ui.adsabs.harvard.edu/abs/2001AJ....122.2055G} {122, 2055}

\bibitem[\protect\citeauthoryear{{Gopal-Krishna} \& {Wiita}}{{Gopal-Krishna} \&
  {Wiita}}{1992}]{Gopal-Krishna92}
{Gopal-Krishna} {Wiita} P.~J.,  1992, \aap, \href
  {http://adsabs.harvard.edu/abs/1992A%26A...259..109G} {259, 109}

\bibitem[\protect\citeauthoryear{{Gopal-Krishna}, {Sagar}  \&
  {Wiita}}{{Gopal-Krishna} et~al.}{1995}]{GK95}
{Gopal-Krishna} {Sagar} R.,   {Wiita} P.~J.,  1995, \mnras, \href
  {http://cdsads.u-strasbg.fr/abs/1995MNRAS.274..701G} {274, 701}

\bibitem[\protect\citeauthoryear{{Gopal-Krishna}, {Britzen}  \&
  {Wiita}}{{Gopal-Krishna} et~al.}{2019}]{GK19}
{Gopal-Krishna} {Britzen} S.,   {Wiita} P.,  2019, arXiv e-prints, \href
  {https://ui.adsabs.harvard.edu/abs/2019arXiv190611339G} {p. arXiv:1906.11339}

\bibitem[\protect\citeauthoryear{{Goyal}, {Gopal-Krishna}, {Wiita}, {Anupama},
  {Sahu}, {Sagar}  \& {Joshi}}{{Goyal} et~al.}{2012}]{Goyal12}
{Goyal} A.,  {Gopal-Krishna} {Wiita} P.~J.,  {Anupama} G.~C.,  {Sahu} D.~K.,
  {Sagar} R.,   {Joshi} S.,  2012, \mn@doi [\aap]
  {10.1051/0004-6361/201218888}, \href
  {http://cdsads.u-strasbg.fr/abs/2012A%26A...544A..37G} {544, A37}

\bibitem[\protect\citeauthoryear{{Goyal}, {Mhaskey}, {Gopal-Krishna}, {Wiita},
  {Stalin}  \& {Sagar}}{{Goyal} et~al.}{2013a}]{Goyal13a}
{Goyal} A.,  {Mhaskey} M.,  {Gopal-Krishna} {Wiita} P.~J.,  {Stalin} C.~S.,
  {Sagar} R.,  2013a, \mn@doi [Journal of Astrophysics and Astronomy]
  {10.1007/s12036-013-9183-7}, \href
  {http://adsabs.harvard.edu/abs/2013JApA...34..273G} {34, 273}

\bibitem[\protect\citeauthoryear{{Goyal}, {Gopal-Krishna}, {Stalin}  \&
  {Sagar}}{{Goyal} et~al.}{2013b}]{Goyal13b}
{Goyal} A.,  {Gopal-Krishna} {Wiita} P.~J.,  {Stalin} C.~S.,   {Sagar} R.,
  2013b, \mn@doi [\mnras] {10.1093/mnras/stt1373}, \href
  {http://adsabs.harvard.edu/abs/2013MNRAS.435.1300G} {435, 1300}

\bibitem[\protect\citeauthoryear{{Gu} \& {Ai}}{{Gu} \& {Ai}}{2011}]{Gu11}
{Gu} M.,  {Ai} Y.~L.,  2011, \mn@doi [Journal of Astrophysics and Astronomy]
  {10.1007/s12036-011-9051-2}, \href
  {https://ui.adsabs.harvard.edu/abs/2011JApA...32...87G} {32, 87}

\bibitem[\protect\citeauthoryear{{Gu}, {Lee}, {Pak}, {Yim}  \& {Fletcher}}{{Gu}
  et~al.}{2006}]{Gu06}
{Gu} M.~F.,  {Lee} C.~U.,  {Pak} S.,  {Yim} H.~S.,   {Fletcher} A.~B.,  2006,
  \mn@doi [\aap] {10.1051/0004-6361:20054271}, \href
  {https://ui.adsabs.harvard.edu/abs/2006A&A...450...39G} {450, 39}

\bibitem[\protect\citeauthoryear{{Gupta} et~al.,}{{Gupta}
  et~al.}{2016}]{Gupta16}
{Gupta} A.~C.,  et~al., 2016, \mn@doi [\mnras] {10.1093/mnras/stw377}, \href
  {http://adsabs.harvard.edu/abs/2016MNRAS.458.1127G} {458, 1127}

\bibitem[\protect\citeauthoryear{{Gupta} et~al.,}{{Gupta}
  et~al.}{2017}]{Gupta17}
{Gupta} A.~C.,  et~al., 2017, \mn@doi [\mnras] {10.1093/mnras/stx2072}, \href
  {https://ui.adsabs.harvard.edu/abs/2017MNRAS.472..788G} {472, 788}

\bibitem[\protect\citeauthoryear{{Gupta} et~al.,}{{Gupta}
  et~al.}{2019}]{Gupta19}
{Gupta} A.~C.,  et~al., 2019, \mn@doi [\aj] {10.3847/1538-3881/aafe7d}, \href
  {https://ui.adsabs.harvard.edu/abs/2019AJ....157...95G} {157, 95}

\bibitem[\protect\citeauthoryear{{Hagen-Thorn}, {Larionov}, {Jorstad},
  {Arkharov}, {Hagen-Thorn}, {Efimova}, {Larionova}  \&
  {Marscher}}{{Hagen-Thorn} et~al.}{2008}]{Hagen-Thorn08}
{Hagen-Thorn} V.~A.,  {Larionov} V.~M.,  {Jorstad} S.~G.,  {Arkharov} A.~A.,
  {Hagen-Thorn} E.~I.,  {Efimova} N.~V.,  {Larionova} L.~V.,   {Marscher}
  A.~P.,  2008, \mn@doi [\apj] {10.1086/523841}, \href
  {https://ui.adsabs.harvard.edu/abs/2008ApJ...672...40H} {672, 40}

\bibitem[\protect\citeauthoryear{{Hao}, {Wang}, {Jiang}  \& {Dai}}{{Hao}
  et~al.}{2010}]{Hao10}
{Hao} J.-M.,  {Wang} B.-J.,  {Jiang} Z.-J.,   {Dai} B.-Z.,  2010, \mn@doi
  [Research in Astronomy and Astrophysics] {10.1088/1674-4527/10/2/003}, \href
  {https://ui.adsabs.harvard.edu/abs/2010RAA....10..125H} {10, 125}

\bibitem[\protect\citeauthoryear{{Healey} et~al.,}{{Healey}
  et~al.}{2008}]{Healey08}
{Healey} S.~E.,  et~al., 2008, \mn@doi [\apjs] {10.1086/523302}, \href
  {https://ui.adsabs.harvard.edu/abs/2008ApJS..175...97H} {175, 97}

\bibitem[\protect\citeauthoryear{{Hughes}, {Aller}  \& {Aller}}{{Hughes}
  et~al.}{2011}]{Hughes11}
{Hughes} P.~A.,  {Aller} M.~F.,   {Aller} H.~D.,  2011, \mn@doi [\apj]
  {10.1088/0004-637X/735/2/81}, \href
  {http://adsabs.harvard.edu/abs/2011ApJ...735...81H} {735, 81}

\bibitem[\protect\citeauthoryear{{Ikejiri} et~al.,}{{Ikejiri}
  et~al.}{2011}]{Ikejiri11}
{Ikejiri} Y.,  et~al., 2011, \mn@doi [\pasj] {10.1093/pasj/63.3.327}, \href
  {http://adsabs.harvard.edu/abs/2011PASJ...63..639I} {63, 639}

\bibitem[\protect\citeauthoryear{{Isler}, {Urry}, {Coppi}, {Bailyn}, {Brady},
  {MacPherson}, {Buxton}  \& {Hasan}}{{Isler} et~al.}{2017}]{Isler17}
{Isler} J.~C.,  {Urry} C.~M.,  {Coppi} P.,  {Bailyn} C.,  {Brady} M.,
  {MacPherson} E.,  {Buxton} M.,   {Hasan} I.,  2017, \mn@doi [\apj]
  {10.3847/1538-4357/aa79fc}, \href
  {https://ui.adsabs.harvard.edu/abs/2017ApJ...844..107I} {844, 107}

\bibitem[\protect\citeauthoryear{{Itoh} et~al.,}{{Itoh} et~al.}{2013}]{Itoh13a}
{Itoh} R.,  et~al., 2013, \mn@doi [\apjl] {10.1088/2041-8205/768/2/L24}, \href
  {http://adsabs.harvard.edu/abs/2013ApJ...768L..24I} {768, L24}

\bibitem[\protect\citeauthoryear{{Jermak} et~al.,}{{Jermak}
  et~al.}{2016}]{Jermak16}
{Jermak} H.,  et~al., 2016, \mn@doi [\mnras] {10.1093/mnras/stw1770}, \href
  {https://ui.adsabs.harvard.edu/abs/2016MNRAS.462.4267J} {462, 4267}

\bibitem[\protect\citeauthoryear{{Jorstad} et~al.,}{{Jorstad}
  et~al.}{2007}]{Jorstad07}
{Jorstad} S.~G.,  et~al., 2007, \mn@doi [\aj] {10.1086/519996}, \href
  {http://adsabs.harvard.edu/abs/2007AJ....134..799J} {134, 799}

\bibitem[\protect\citeauthoryear{{Jorstad} et~al.,}{{Jorstad}
  et~al.}{2017}]{Jorstad17}
{Jorstad} S.~G.,  et~al., 2017, \mn@doi [\apj] {10.3847/1538-4357/aa8407},
  \href {https://ui.adsabs.harvard.edu/abs/2017ApJ...846...98J} {846, 98}

\bibitem[\protect\citeauthoryear{{Kawabata} et~al.,}{{Kawabata}
  et~al.}{2008}]{Kawabata08}
{Kawabata} K.~S.,  et~al., 2008, in Ground-based and Airborne Instrumentation
  for Astronomy II. p. 70144L, \mn@doi{10.1117/12.788569}

\bibitem[\protect\citeauthoryear{{Kirk}, {Rieger}  \& {Mastichiadis}}{{Kirk}
  et~al.}{1998}]{Kirk98}
{Kirk} J.~G.,  {Rieger} F.~M.,   {Mastichiadis} A.,  1998, \aap, \href
  {https://ui.adsabs.harvard.edu/abs/1998A&A...333..452K} {333, 452}

\bibitem[\protect\citeauthoryear{{Larionov} et~al.,}{{Larionov}
  et~al.}{2008}]{Larionov08}
{Larionov} V.~M.,  et~al., 2008, \mn@doi [\aap] {10.1051/0004-6361:200810937},
  \href {https://ui.adsabs.harvard.edu/abs/2008A&A...492..389L} {492, 389}

\bibitem[\protect\citeauthoryear{{Larionov} et~al.,}{{Larionov}
  et~al.}{2013}]{Larionov13}
{Larionov} V.~M.,  et~al., 2013, \mn@doi [\apj] {10.1088/0004-637X/768/1/40},
  \href {https://ui.adsabs.harvard.edu/abs/2013ApJ...768...40L} {768, 40}

\bibitem[\protect\citeauthoryear{{Larionov} et~al.,}{{Larionov}
  et~al.}{2016}]{Larionov16}
{Larionov} V.~M.,  et~al., 2016, \mn@doi [\mnras] {10.1093/mnras/stw1516},
  \href {https://ui.adsabs.harvard.edu/abs/2016MNRAS.461.3047L} {461, 3047}

\bibitem[\protect\citeauthoryear{{Li}, {Luo}, {Yang}, {Yang}, {Yang}  \&
  {Cai}}{{Li} et~al.}{2018}]{Li18}
{Li} X.-P.,  {Luo} Y.-H.,  {Yang} H.-T.,  {Yang} H.-Y.,  {Yang} C.,   {Cai} Y.,
   2018, \mn@doi [Research in Astronomy and Astrophysics]
  {10.1088/1674-4527/18/12/150}, \href
  {https://ui.adsabs.harvard.edu/abs/2018RAA....18..150L} {18, 150}

\bibitem[\protect\citeauthoryear{{Lynds}}{{Lynds}}{1967}]{Lynds67}
{Lynds} C.~R.,  1967, \mn@doi [\apj] {10.1086/149068}, \href
  {https://ui.adsabs.harvard.edu/abs/1967ApJ...147..837L} {147, 837}

\bibitem[\protect\citeauthoryear{{Madejski} \& {Sikora}}{{Madejski} \&
  {Sikora}}{2016}]{Madejski16}
{Madejski} G.~.,  {Sikora} M.,  2016, \mn@doi [\araa]
  {10.1146/annurev-astro-081913-040044}, \href
  {http://adsabs.harvard.edu/abs/2016ARA%26A..54..725M} {54, 725}

\bibitem[\protect\citeauthoryear{{Maraschi}, {Ghisellini}  \&
  {Celotti}}{{Maraschi} et~al.}{1992}]{Maraschi92}
{Maraschi} L.,  {Ghisellini} G.,   {Celotti} A.,  1992, \mn@doi [\apjl]
  {10.1086/186531}, \href {http://adsabs.harvard.edu/abs/1992ApJ...397L...5M}
  {397, L5}

\bibitem[\protect\citeauthoryear{{Marscher} \& {Gear}}{{Marscher} \&
  {Gear}}{1985}]{Marscher85}
{Marscher} A.~P.,  {Gear} W.~K.,  1985, \mn@doi [\apj] {10.1086/163592}, \href
  {http://adsabs.harvard.edu/abs/1985ApJ...298..114M} {298, 114}

\bibitem[\protect\citeauthoryear{{Marscher} et~al.,}{{Marscher}
  et~al.}{2008}]{Marscher08}
{Marscher} A.~P.,  et~al., 2008, \mn@doi [\nat] {10.1038/nature06895}, \href
  {http://adsabs.harvard.edu/abs/2008Natur.452..966M} {452, 966}

\bibitem[\protect\citeauthoryear{{Marziani}, {Sulentic}, {Dultzin-Hacyan},
  {Calvani}  \& {Moles}}{{Marziani} et~al.}{1996}]{Marziani96}
{Marziani} P.,  {Sulentic} J.~W.,  {Dultzin-Hacyan} D.,  {Calvani} M.,
  {Moles} M.,  1996, \mn@doi [\apjs] {10.1086/192291}, \href
  {https://ui.adsabs.harvard.edu/abs/1996ApJS..104...37M} {104, 37}

\bibitem[\protect\citeauthoryear{{Mead}, {Ballard}, {Brand}, {Hough}, {Brindle}
   \& {Bailey}}{{Mead} et~al.}{1990}]{Mead90}
{Mead} A.~R.~G.,  {Ballard} K.~R.,  {Brand} P.~W.~J.~L.,  {Hough} J.~H.,
  {Brindle} C.,   {Bailey} J.~A.,  1990, \aaps, \href
  {http://adsabs.harvard.edu/abs/1990A%26AS...83..183M} {83, 183}

\bibitem[\protect\citeauthoryear{{Meng}, {Wu}, {Webb}, {Zhang}  \&
  {Dai}}{{Meng} et~al.}{2017}]{Meng17}
{Meng} N.,  {Wu} J.,  {Webb} J.~R.,  {Zhang} X.,   {Dai} Y.,  2017, \mn@doi
  [\mnras] {10.1093/mnras/stx1055}, \href
  {https://ui.adsabs.harvard.edu/abs/2017MNRAS.469.3588M} {469, 3588}

\bibitem[\protect\citeauthoryear{{Meng}, {Zhang}, {Wu}, {Ma}  \& {Zhou}}{{Meng}
  et~al.}{2018}]{Meng18}
{Meng} N.,  {Zhang} X.,  {Wu} J.,  {Ma} J.,   {Zhou} X.,  2018, \mn@doi [\apjs]
  {10.3847/1538-4365/aacffe}, \href
  {https://ui.adsabs.harvard.edu/abs/2018ApJS..237...30M} {237, 30}

\bibitem[\protect\citeauthoryear{{Meyer}}{{Meyer}}{2018}]{Meyer18}
{Meyer} E.~T.,  2018, \mn@doi [Nature Astronomy] {10.1038/s41550-017-0349-0},
  \href {https://ui.adsabs.harvard.edu/abs/2018NatAs...2...32M} {2, 32}

\bibitem[\protect\citeauthoryear{{Moderski}, {Sikora}  \&
  {B{\l}a{\.z}ejowski}}{{Moderski} et~al.}{2003}]{Moderski03}
{Moderski} R.,  {Sikora} M.,   {B{\l}a{\.z}ejowski} M.,  2003, \mn@doi [\aap]
  {10.1051/0004-6361:20030794}, \href
  {https://ui.adsabs.harvard.edu/abs/2003A&A...406..855M} {406, 855}

\bibitem[\protect\citeauthoryear{{Moore} et~al.,}{{Moore}
  et~al.}{1982}]{Moore82}
{Moore} R.~L.,  et~al., 1982, \mn@doi [\apj] {10.1086/160266}, \href
  {https://ui.adsabs.harvard.edu/abs/1982ApJ...260..415M} {260, 415}

\bibitem[\protect\citeauthoryear{{Nilsson}, {Pursimo}, {Sillanp{\"a}{\"a}},
  {Takalo}  \& {Lindfors}}{{Nilsson} et~al.}{2008}]{Nilsson08}
{Nilsson} K.,  {Pursimo} T.,  {Sillanp{\"a}{\"a}} A.,  {Takalo} L.~O.,
  {Lindfors} E.,  2008, \mn@doi [\aap] {10.1051/0004-6361:200810310}, \href
  {https://ui.adsabs.harvard.edu/abs/2008A&A...487L..29N} {487, L29}

\bibitem[\protect\citeauthoryear{{Nilsson}, {Takalo}, {Lehto}  \&
  {Sillanp{\"a}{\"a}}}{{Nilsson} et~al.}{2010}]{Nilsson10}
{Nilsson} K.,  {Takalo} L.~O.,  {Lehto} H.~J.,   {Sillanp{\"a}{\"a}} A.,  2010,
  \mn@doi [\aap] {10.1051/0004-6361/201014198}, \href
  {http://adsabs.harvard.edu/abs/2010A%26A...516A..60N} {516, A60}

\bibitem[\protect\citeauthoryear{{Osterman Meyer}, {Miller}, {Marshall},
  {Ryle}, {Aller}, {Aller}  \& {Balonek}}{{Osterman Meyer}
  et~al.}{2009}]{Osterman-Meyer09}
{Osterman Meyer} A.,  {Miller} H.~R.,  {Marshall} K.,  {Ryle} W.~T.,  {Aller}
  H.,  {Aller} M.,   {Balonek} T.,  2009, \mn@doi [\aj]
  {10.1088/0004-6256/138/6/1902}, \href
  {https://ui.adsabs.harvard.edu/abs/2009AJ....138.1902O} {138, 1902}

\bibitem[\protect\citeauthoryear{{Pacholczyk}}{{Pacholczyk}}{1970}]{Pacholczyk70}
{Pacholczyk} A.~G.,  1970, {Radio astrophysics. Nonthermal processes in
  galactic and extragalactic sources}

\bibitem[\protect\citeauthoryear{{Padovani} et~al.,}{{Padovani}
  et~al.}{2017}]{Padovani17}
{Padovani} P.,  et~al., 2017, \mn@doi [\aapr] {10.1007/s00159-017-0102-9},
  \href {http://adsabs.harvard.edu/abs/2017A%26ARv..25....2P} {25, 2}

\bibitem[\protect\citeauthoryear{{Papadakis}, {Boumis}, {Samaritakis}  \&
  {Papamastorakis}}{{Papadakis} et~al.}{2003}]{Papadakis03}
{Papadakis} I.~E.,  {Boumis} P.,  {Samaritakis} V.,   {Papamastorakis} J.,
  2003, \mn@doi [\aap] {10.1051/0004-6361:20021581}, \href
  {http://adsabs.harvard.edu/abs/2003A%26A...397..565P} {397, 565}

\bibitem[\protect\citeauthoryear{{Perlman} et~al.,}{{Perlman}
  et~al.}{2011}]{Perlman11}
{Perlman} E.~S.,  et~al., 2011, \mn@doi [\apj] {10.1088/0004-637X/743/2/119},
  \href {http://adsabs.harvard.edu/abs/2011ApJ...743..119P} {743, 119}

\bibitem[\protect\citeauthoryear{{Raiteri}, {Villata}, {Lanteri}, {Cavallone}
  \& {Sobrito}}{{Raiteri} et~al.}{1998}]{Raiteri98}
{Raiteri} C.~M.,  {Villata} M.,  {Lanteri} L.,  {Cavallone} M.,   {Sobrito} G.,
   1998, \mn@doi [\aaps] {10.1051/aas:1998420}, \href
  {https://ui.adsabs.harvard.edu/abs/1998A&AS..130..495R} {130, 495}

\bibitem[\protect\citeauthoryear{{Rani} et~al.,}{{Rani} et~al.}{2010}]{Rani10}
{Rani} B.,  et~al., 2010, \mn@doi [\mnras] {10.1111/j.1365-2966.2010.16419.x},
  \href {https://ui.adsabs.harvard.edu/abs/2010MNRAS.404.1992R} {404, 1992}

\bibitem[\protect\citeauthoryear{{Richards} et~al.,}{{Richards}
  et~al.}{2011}]{Richards11}
{Richards} J.~L.,  et~al., 2011, \mn@doi [\apjs] {10.1088/0067-0049/194/2/29},
  \href {http://adsabs.harvard.edu/abs/2011ApJS..194...29R} {194, 29}

\bibitem[\protect\citeauthoryear{{Romero}, {Cellone}  \& {Combi}}{{Romero}
  et~al.}{1999}]{Romero99}
{Romero} G.~E.,  {Cellone} S.~A.,   {Combi} J.~A.,  1999, \mn@doi [\aaps]
  {10.1051/aas:1999184}, \href
  {http://cdsads.u-strasbg.fr/abs/1999A%26AS..135..477R} {135, 477}

\bibitem[\protect\citeauthoryear{{Rybicki} \& {Lightman}}{{Rybicki} \&
  {Lightman}}{1986}]{Rybicki86}
{Rybicki} G.~B.,  {Lightman} A.~P.,  1986, {Radiative Processes in
  Astrophysics}

\bibitem[\protect\citeauthoryear{{Sagar}}{{Sagar}}{1999}]{Sagar99}
{Sagar} R.,  1999, Current Science, \href
  {http://adsabs.harvard.edu/abs/1999CSci...77..643G} {77, 643}

\bibitem[\protect\citeauthoryear{{Sagar}, {Kumar}, {Omar}  \& {Pand
  ey}}{{Sagar} et~al.}{2010}]{Sagar10}
{Sagar} R.,  {Kumar} B.,  {Omar} A.,   {Pand ey} A.~K.,  2010, in Astronomical
  Society of India Conference Series. pp 203--210

\bibitem[\protect\citeauthoryear{{Saito}, {Stawarz}, {Tanaka}, {Takahashi},
  {Sikora}  \& {Moderski}}{{Saito} et~al.}{2015}]{Saito15}
{Saito} S.,  {Stawarz} {\L}.,  {Tanaka} Y.~T.,  {Takahashi} T.,  {Sikora} M.,
  {Moderski} R.,  2015, \mn@doi [\apj] {10.1088/0004-637X/809/2/171}, \href
  {http://adsabs.harvard.edu/abs/2015ApJ...809..171S} {809, 171}

\bibitem[\protect\citeauthoryear{{Sandrinelli}, {Covino}  \&
  {Treves}}{{Sandrinelli} et~al.}{2014}]{Sandrinelli14}
{Sandrinelli} A.,  {Covino} S.,   {Treves} A.,  2014, \mn@doi [\aap]
  {10.1051/0004-6361/201321558}, \href
  {https://ui.adsabs.harvard.edu/abs/2014A&A...562A..79S} {562, A79}

\bibitem[\protect\citeauthoryear{{Schmidt}}{{Schmidt}}{1965}]{Schmidt65}
{Schmidt} M.,  1965, \mn@doi [\apj] {10.1086/148217}, \href
  {https://ui.adsabs.harvard.edu/abs/1965ApJ...141.1295S} {141, 1295}

\bibitem[\protect\citeauthoryear{{Sikora}, {Begelman}  \& {Rees}}{{Sikora}
  et~al.}{1994}]{Sikora94}
{Sikora} M.,  {Begelman} M.~C.,   {Rees} M.~J.,  1994, \mn@doi [\apj]
  {10.1086/173633}, \href {http://adsabs.harvard.edu/abs/1994ApJ...421..153S}
  {421, 153}

\bibitem[\protect\citeauthoryear{{Sironi} \& {Spitkovsky}}{{Sironi} \&
  {Spitkovsky}}{2009}]{Sironi09}
{Sironi} L.,  {Spitkovsky} A.,  2009, \mn@doi [\apj]
  {10.1088/0004-637X/698/2/1523}, \href
  {https://ui.adsabs.harvard.edu/abs/2009ApJ...698.1523S} {698, 1523}

\bibitem[\protect\citeauthoryear{{Sironi} \& {Spitkovsky}}{{Sironi} \&
  {Spitkovsky}}{2011}]{Sironi11}
{Sironi} L.,  {Spitkovsky} A.,  2011, \mn@doi [\apj]
  {10.1088/0004-637X/726/2/75}, \href
  {https://ui.adsabs.harvard.edu/abs/2011ApJ...726...75S} {726, 75}

\bibitem[\protect\citeauthoryear{{Sironi}, {Petropoulou}  \&
  {Giannios}}{{Sironi} et~al.}{2015}]{Sironi15}
{Sironi} L.,  {Petropoulou} M.,   {Giannios} D.,  2015, \mn@doi [\mnras]
  {10.1093/mnras/stv641}, \href
  {http://adsabs.harvard.edu/abs/2015MNRAS.450..183S} {450, 183}

\bibitem[\protect\citeauthoryear{{Spada}, {Ghisellini}, {Lazzati}  \&
  {Celotti}}{{Spada} et~al.}{2001}]{Spada01}
{Spada} M.,  {Ghisellini} G.,  {Lazzati} D.,   {Celotti} A.,  2001, \mn@doi
  [\mnras] {10.1046/j.1365-8711.2001.04557.x}, \href
  {http://adsabs.harvard.edu/abs/2001MNRAS.325.1559S} {325, 1559}

\bibitem[\protect\citeauthoryear{{Stalin}, {Gopal Krishna}, {Sagar}  \&
  {Wiita}}{{Stalin} et~al.}{2004}]{Stalin04}
{Stalin} C.~S.,  {Gopal Krishna} {Sagar} R.,   {Wiita} P.~J.,  2004, \mn@doi
  [Journal of Astrophysics and Astronomy] {10.1007/BF02702287}, \href
  {http://cdsads.u-strasbg.fr/abs/2004JApA...25....1S} {25, 1}

\bibitem[\protect\citeauthoryear{{Stalin}, {Gopal-Krishna}, {Sagar}, {Wiita},
  {Mohan}  \& {Pandey}}{{Stalin} et~al.}{2006}]{Stalin06}
{Stalin} C.~S.,  {Gopal-Krishna} {Sagar} R.,  {Wiita} P.~J.,  {Mohan} V.,
  {Pandey} A.~K.,  2006, \mn@doi [\mnras] {10.1111/j.1365-2966.2005.09939.x},
  \href {https://ui.adsabs.harvard.edu/abs/2006MNRAS.366.1337S} {366, 1337}

\bibitem[\protect\citeauthoryear{{Urry} \& {Padovani}}{{Urry} \&
  {Padovani}}{1995}]{Urry95}
{Urry} C.~M.,  {Padovani} P.,  1995, \mn@doi [\pasp] {10.1086/133630}, \href
  {http://adsabs.harvard.edu/abs/1995PASP..107..803U} {107, 803}

\bibitem[\protect\citeauthoryear{{Villforth}, {Nilsson}, {{\O}stensen},
  {Heidt}, {Niemi}  \& {Pforr}}{{Villforth} et~al.}{2009}]{Villforth09}
{Villforth} C.,  {Nilsson} K.,  {{\O}stensen} R.,  {Heidt} J.,  {Niemi} S.-M.,
   {Pforr} J.,  2009, \mn@doi [\mnras] {10.1111/j.1365-2966.2009.14886.x},
  \href {http://adsabs.harvard.edu/abs/2009MNRAS.397.1893V} {397, 1893}

\bibitem[\protect\citeauthoryear{{Villforth} et~al.,}{{Villforth}
  et~al.}{2010}]{Villforth10}
{Villforth} C.,  et~al., 2010, \mn@doi [\mnras]
  {10.1111/j.1365-2966.2009.16133.x}, \href
  {http://adsabs.harvard.edu/abs/2010MNRAS.402.2087V} {402, 2087}

\bibitem[\protect\citeauthoryear{{Wagner} \& {Witzel}}{{Wagner} \&
  {Witzel}}{1995}]{Wagner95}
{Wagner} S.~J.,  {Witzel} A.,  1995, \mn@doi [\araa]
  {10.1146/annurev.aa.33.090195.001115}, \href
  {http://cdsads.u-strasbg.fr/abs/1995ARA%26A..33..163W} {33, 163}

\bibitem[\protect\citeauthoryear{{Wang}, {Tanaka}, {Wang}, {Kawabata},
  {Fukazawa}, {Itoh}  \& {Tziamtzis}}{{Wang} et~al.}{2015}]{Wang15}
{Wang} Z.,  {Tanaka} Y.~T.,  {Wang} C.,  {Kawabata} K.~S.,  {Fukazawa} Y.,
  {Itoh} R.,   {Tziamtzis} A.,  2015, \mn@doi [\apj]
  {10.1088/0004-637X/814/2/89}, \href
  {http://adsabs.harvard.edu/abs/2015ApJ...814...89W} {814, 89}

\bibitem[\protect\citeauthoryear{{Wierzcholska}, {Ostrowski}, {Stawarz},
  {Wagner}  \& {Hauser}}{{Wierzcholska} et~al.}{2015}]{Wierzcholska15}
{Wierzcholska} A.,  {Ostrowski} M.,  {Stawarz} {\L}.,  {Wagner} S.,   {Hauser}
  M.,  2015, \mn@doi [\aap] {10.1051/0004-6361/201423967}, \href
  {https://ui.adsabs.harvard.edu/abs/2015A&A...573A..69W} {573, A69}

\bibitem[\protect\citeauthoryear{{Wu}, {Zhou}, {Ma}  \& {Jiang}}{{Wu}
  et~al.}{2011}]{Wu11}
{Wu} J.,  {Zhou} X.,  {Ma} J.,   {Jiang} Z.,  2011, \mn@doi [\mnras]
  {10.1111/j.1365-2966.2011.19565.x}, \href
  {https://ui.adsabs.harvard.edu/abs/2011MNRAS.418.1640W} {418, 1640}

\bibitem[\protect\citeauthoryear{{Zhang}, {Wu}  \& {Meng}}{{Zhang}
  et~al.}{2018}]{Zhang18}
{Zhang} X.,  {Wu} J.,   {Meng} N.,  2018, \mn@doi [\mnras]
  {10.1093/mnras/sty1468}, \href
  {https://ui.adsabs.harvard.edu/abs/2018MNRAS.478.3513Z} {478, 3513}

\bibitem[\protect\citeauthoryear{{Zola} et~al.,}{{Zola} et~al.}{2012}]{Zola12}
{Zola} S.,  et~al., 2012, in Astronomical Society of India Conference Series.
  p.~239

\makeatother
\end{thebibliography}

\appendix
\section{Analysis of the `microflare' from  3C\,454.3 }
\label{app:microflare}

\begin{figure}
\centering
\hbox{
\includegraphics[width=0.45\textwidth]{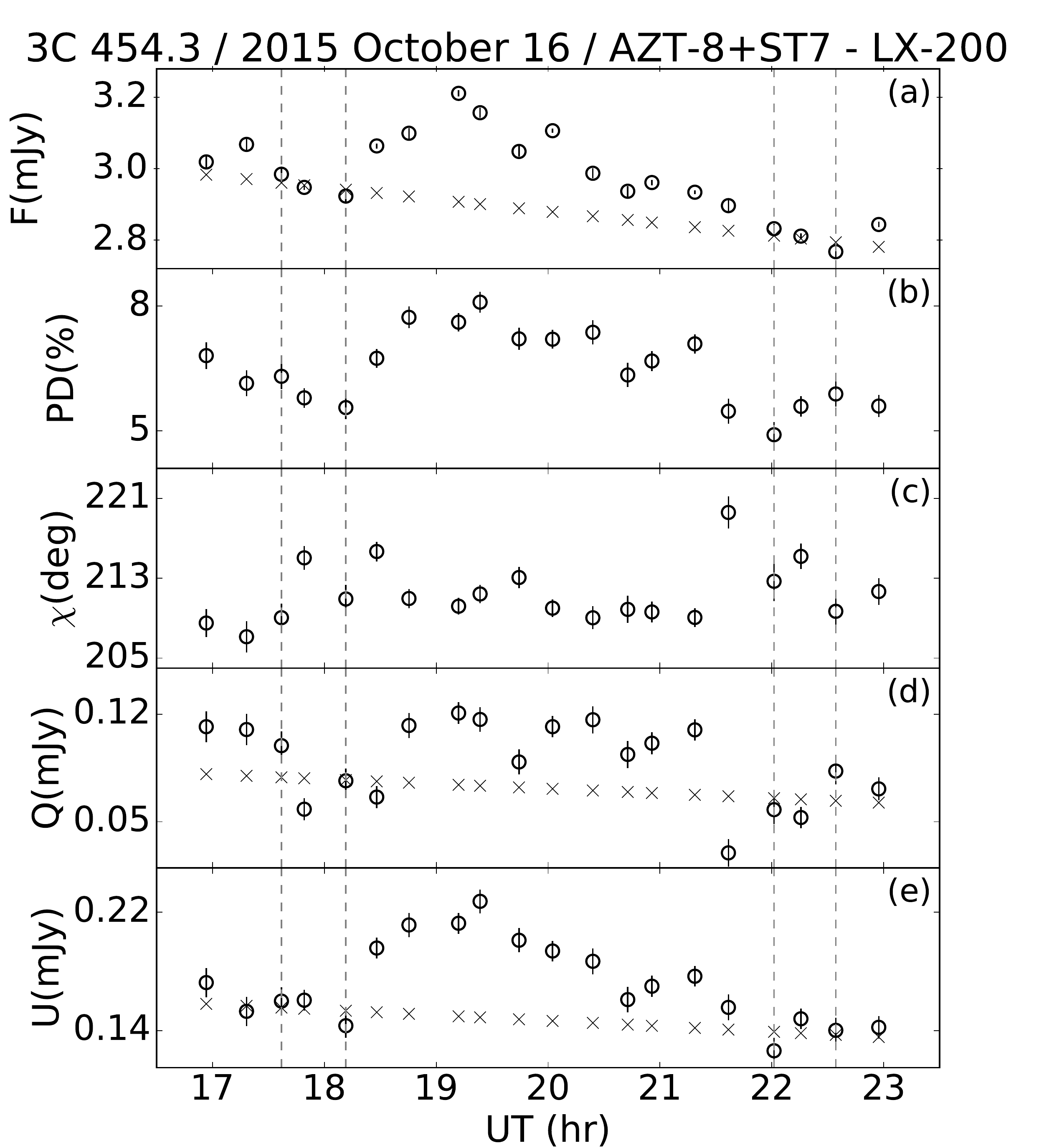}
}

\caption{Variations of total intensity, $PD$ and $\chi$ for the `microflare' for the blazar 3C\,454.3 observed on 2015 October 16: (a) total observed R-band flux density; (b) $PD$, (c) $\chi$, and (d and e) the Stokes parameters $Q$ and $U$, respectively. The two intervals between the outer pairs of dashed vertical lines are used to define the emission considered to arise from the base emission component for $F$ and $Q$ and $U$ intensities; crosses show the modeled emission of the base component. }
\label{fig_baseflare}%
\end{figure}

Figure~\ref{fig_baseflare} presents the polarization data set for the blazar 3C\,454.3 on 2015 October 16. On this occasion, a clear, approximately symmetric, microflare was observed with almost exponential rise and decay times of $\Delta t \approx$ 4.0\,hr. The total flux increased by $\sim$ 10 percent in $\approx$1.5\,hr. Here, we attempt to separate the microflare (shown within the inner set of dashed vertical lines in Figure~\ref{fig_baseflare}) emission from the slowly varying base emission from the blazar. We evaluate the parameters of this microflare emission, assumed to be a separate emission component superimposed on a slowly varying background component. The Stokes parameters are used to describe the polarized emission and are given as

\begin{eqnarray}
Q = PD \times F \cos(2 \chi), \\
U = PD \times F \sin(2 \chi) ,
\end{eqnarray}
where $F$ is the flux.

We assume that the observed emission is arising from two components in the blazar jet, the fast varying flare component (index 1) superposed on a slow varying background emission (index 0). To perform this decomposition we use the linearly additive properties of total flux and the 
Stokes Q and U intensities \citep{Pacholczyk70}, using the following relations

\begin{eqnarray}
\label{twocomp}
F & = & F_{0} + F_{1},  \\
Q & = & Q_{0} + Q_{1},  \\
U & = & U_{0} + U_{1}.  
\end{eqnarray}
The polarized flux (PF) is obtained as  $PF = PD \times F$

\begin{figure}
\centering
\hbox{
\includegraphics[width=0.45\textwidth]{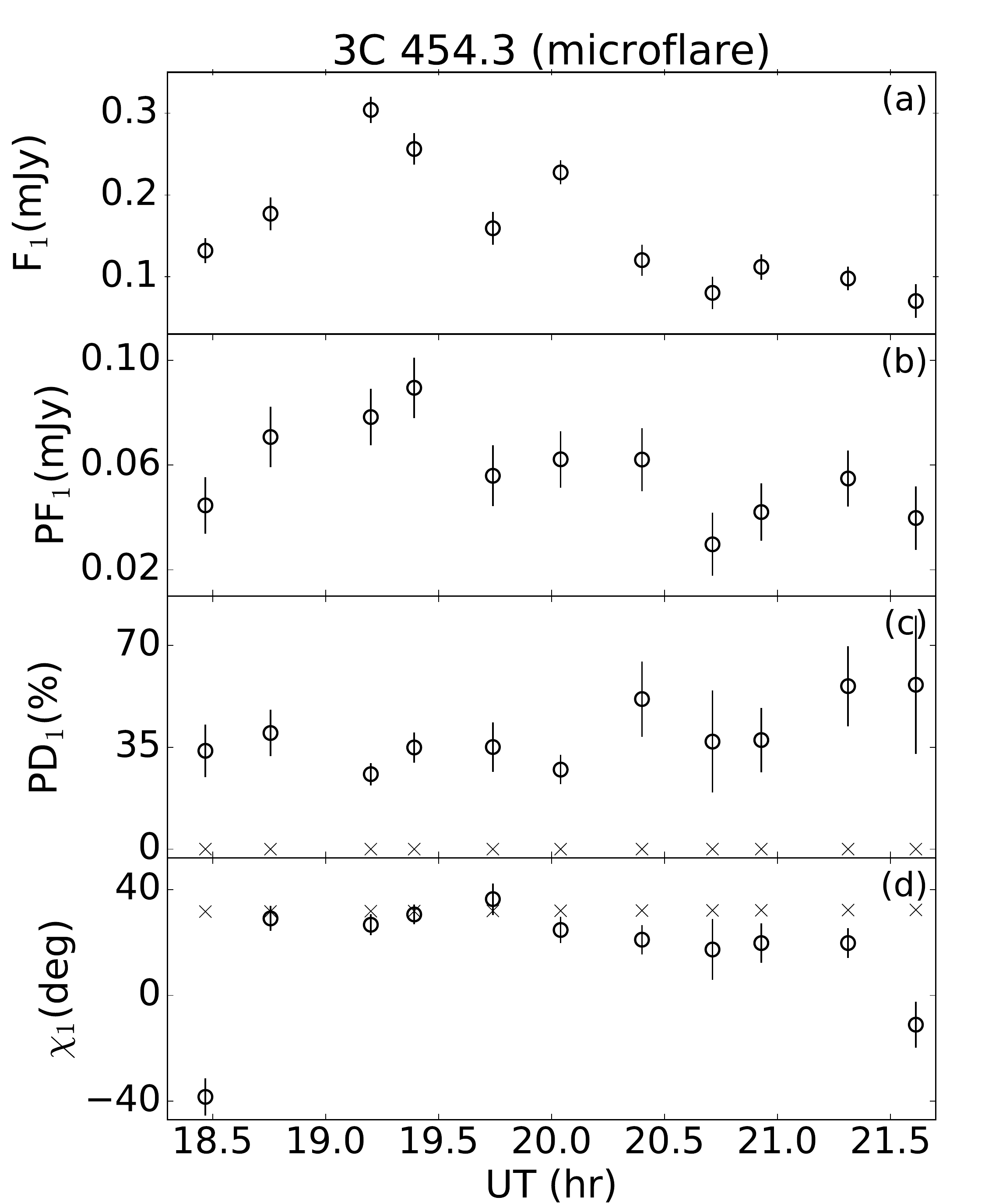}
}

\caption{The derived microflare (a) flux ($F_1$), (b) polarized flux ($PF_1$) 
(c) polarization degree ($PD_1$), (d) polarization angle ($\chi_1$); 
for comparison, $PD_0$ and $\chi_0$ of the slowly varying base component are presented with cross marks.  
}
\label{fig_shortflare}%
\end{figure}

\begin{figure}
\includegraphics[angle=0, width=0.5\textwidth]{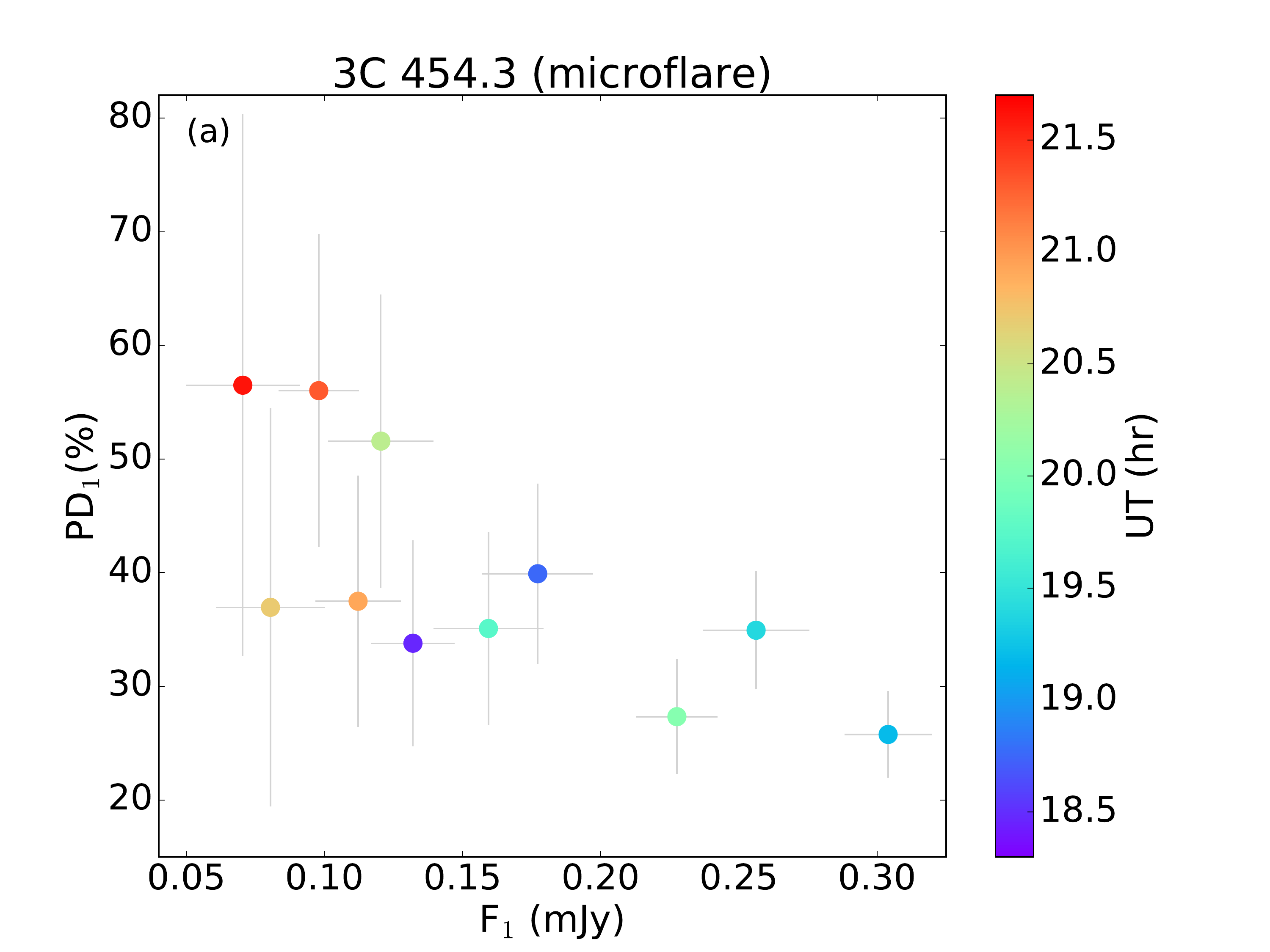}
\includegraphics[angle=0, width=0.5\textwidth]{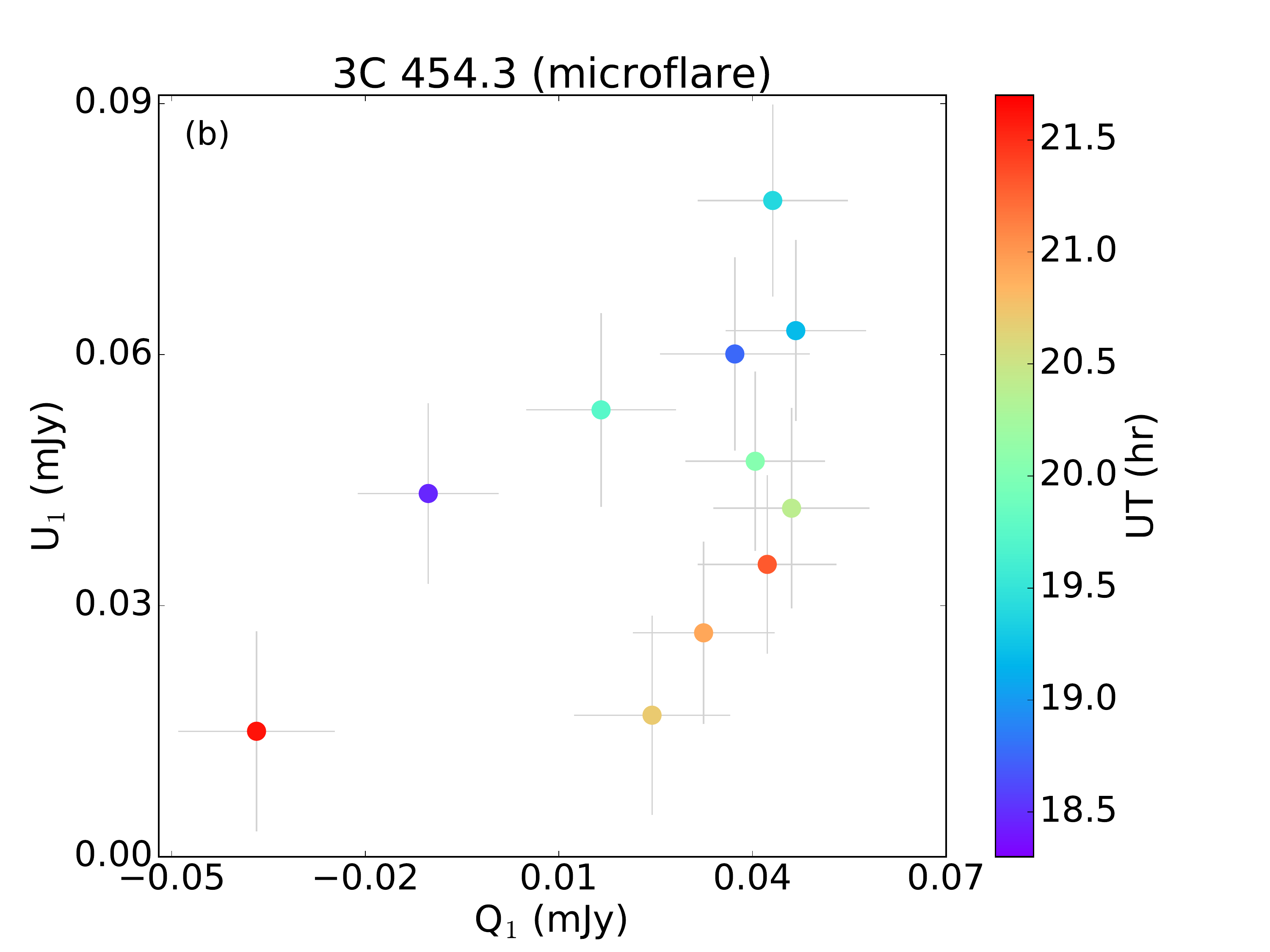}
 \caption{ Evolution of the microflare in the:  panel (a) ($PD, F$) plane; panel (b) ($Q, U$) plane.}
\label{fig_qu_var}
\end{figure}

In the modelling, all background emission components ($F, Q, U$) are fitted to the emissions observed before and after the microflare and it is assumed that the background is changing linearly within the microflare duration (Fig.~\ref{fig_baseflare}). Once the background intensities, $F_0$, $Q_0$ and $U_0$ are estimated, we subtract them from the total emission using the above expressions  to obtain $F_1$,  $Q_1$ and $U_1$ for the flaring component. With these parameters, we then evaluate the microflare parameters  $PF_1$, $PD_1$, and $\chi_1$  using the standard expressions and show them in Figure~\ref{fig_shortflare}. Additionally, in Fig.~\ref{fig_qu_var} we present the $PD_1$ vs.\ $F_1$ and $Q_1$ vs.\ $U_1$ plots for the microflare emission component. We note that derived microflare intensities hint of a systematic rotation of the polarized flux in the ($Q, U$) plane. The statistical uncertainties on the parameters of the microflare are derived using standard error propagation \citep{Bevington03}.

Our simple decomposition of the total intensity into slowly varying base and flaring components indicates that the the microflare component is highly polarized, with $PD_1 \approx 22\%-55\% $ ($\pm $ 8\%--22\%)  as compared to the much less polarized slowly varying base emission with $PD_0 \sim 5\% $($\pm $ 0.3\%). This implies the presence of a highly ordered magnetic field in the microflare emission region (Figure~\ref{fig_shortflare}). The polarization angle of the microflare component  is $\chi_1 \sim 18^{\circ}$ - $31^\circ$ ($\pm $ 3-11$^\circ$), slightly different from  the slowly varying base component with $\chi_0 \sim 31^\circ$ ($\pm $ 1$^\circ$). It is important to note that these $\chi_0$ and $\chi_1$ are very different from  the {\rm VLBI} jet position angle ($\approx$ $-$98$\pm$10$^\circ$; \citealt{Jorstad17}). From inspection of Fig.~\ref{fig_qu_var}, we note an indication of the microflare exhibiting loops  in the ($PD, F$) and ($Q, U$) planes. Therefore, the well-resolved microflare shown by 3C\,454.3  when polarization monitoring is available shows a highly polarized flare component, similar to that obtained for the blazar S5 0716+714 \citep[$PD\approx$50 percent;][]{Bhatta15}. The spatial scale of the microflare emitting region must be quite compact: smaller than $\Delta t \cdot c \approx {\rm few} \times  10^{14}$\,cm.


\bsp	
\label{lastpage}
\end{document}